\documentstyle[12pt,dina4p,epsfig]{article}
\title{\bf Some models of spin coherence and decoherence in storage rings
   \thanks{DESY 97-166}
}
\author{K. Heinemann\thanks{heineman@mail.desy.de} \\
 Deutsches Elektronen-Synchrotron  DESY ,\ Hamburg  
\\
             Notkestr.   85,\  22603 \ Hamburg  }
\date{}
\begin{document}
\newcounter{INDEX}
\setcounter{section}{0}
\setcounter{subsection}{0}
\setcounter{equation}{0}
\renewcommand
{\theequation}{\mbox{\thesection .\arabic{equation}\alph{INDEX}}}
\maketitle
%
\begin{abstract}
I present some simple exactly solvable models of spin diffusion caused by
synchrotron radiation noise in storage rings. I am able to use standard
stochastic differential equation and Fokker-Planck methods and I thereby
introduce, and exploit, the {\it polarization density}. This quantity obeys 
a linear evolution equation of the Bloch type, which is, like the 
Fokker-Planck equation, universal in the 
sense that it is independent of the state of the system.
I also briefly consider Bloch equations for other local 
polarization quantities derived from the polarization density. 
One of the models chosen is of relevance for some existing and proposed
low energy electron (positron) storage rings which need polarization. I present
numerical results for a ring with parameters typical of HERA and show
that, where applicable, the results of my approach are in satisfactory
agreement with calculations using SLIM. 
These calculations provide a numerical check of a basic tenet of the
conventional
method of calculating depolarization using the $\vec n$-axis. 
I also investigate the
equilibrium behaviour of the spin ensemble when there is no synchrotron 
radiation. Finally, I summarize other results which I have obtained using
the polarization density and which will be published separately.
\end{abstract}
\newpage
\tableofcontents
%
%
\setcounter{section}{0}
%
\section*{Introduction}
\addcontentsline{toc}{section}{Introduction}
\par This paper provides an introduction to the use of spin polarization 
transport equations of Fokker-Planck and Liouville type in electron (positron)
and proton storage rings. As vehicles for this study I use two exactly 
solvable but simple configurations called Machine I and Machine II.
\footnote{In appendices C and D I briefly consider machines III and IV
which are closely related to Machine I.}
\par Machine I is a smoothed planar ring with no vertical emittance and with
spins lying in the machine plane which diffuse as the result of the
stochastic nature of synchrotron radiation photon emission. 
It is therefore an extremely simple arrangement but it serves to introduce
the concepts without the latter being obscured by unnecessary
complication.
\par Machine II is similar to Machine I except that it contains a Siberian
Snake and is therefore relevant to some existing and proposed low energy
electron storage rings. By introducing the snake the equilibrium
polarization is constrained to the machine plane together with the spins 
and it therefore becomes possible to compare  the Fokker-Planck approach
with the conventional method of calculating the spin depolarization rate
and to comment on the validity of the latter.
\par A novel aspect of this work is the introduction of a phase space
dependent quantity called the polarization density. This
satisfies an evolution equation of the Bloch type 
\footnote{For Bloch equations in spin diffusion problems, see \cite{Abr61}.}
which provides a
causal azimuthal evolution of the polarization density. This linear equation
is universal, like the Fokker-Planck equation,
in the sense that it is independent of the state of the system.
\footnote{But it is model dependent.}
There is no equation which could provide a causal azimuthal evolution
for the polarization vector of a given ensemble, 
even in the simple cases of machines I and II.
Other local polarization quantities derived from the polarization density
also obey evolution equations of the Bloch type, but these equations
depend on the orbital state of the system.
\par The basic orbital 
formalism needed is introduced in
section 1. Then in section 2 the first model, Machine I, is studied. 
This configuration
was already used \cite{BBHMR94a,BBHMR94b} in a preliminary study of the rate
of decoherence of spins (= `horizontal spin diffusion') lying in the
machine plane of such a machine. As explained in those reports, the aim
was to estimate the difficulty of
\footnote{For an earlier  treatment, see also \cite{Kou91}.}
using a radial rf field to flip spins which had been previously
polarized into the vertical direction by the Sokolov-Ternov effect
\cite{ST64}. The approach adopted was to consider the effect of damped
stochastic synchrotron motion on the development of the distribution of
spins lying in the machine plane and some numerical results were given.
In this paper I will fill in some details omitted in
\cite{BBHMR94a,BBHMR94b} and develop some extra analytical tools.
\par The dynamics for Machine I can be studied for various initial
spin-orbit distributions. In this paper I
consider two scenarios, i.e. two stochastic processes.
The approach is pedagogical, step by step and exhaustive and uses elementary
methods for the solution of stochastic differential equations.
In scenario 1 I study the phase space
distribution for a starting distribution which is pointlike in both
spin and orbit space. In scenario 2 I begin with a pointlike spin
distribution but with an orbital distribution which is already in
equilibrium. The resulting phase space evolutions are referred to as
Process 1 and Process 2 respectively. By comparing the asymptotic spin
distributions I discover that they are not unique but that in
both cases there is no complete decoherence. I also study the effect
of switching off the synchrotron radiation. Process 1 has some
theoretical importance and Process 2 is more closely related to
physical situations. Numerical results for a machine similar to the
HERA electron ring are presented.
\par In section 3 I consider Machine II,
where in addition to the fields of Machine I the spin experiences the
influence of a pointlike Siberian Snake \cite{DK78}. The aim here is to
discover to what extent a snake can, by its tendency to cancel spin
perturbations, suppress decoherence. I consider three
scenarios, the stochastic processes 3,4,5. Process 3 has the same initial
spin-orbit distribution as in scenario 2 for Machine I and as one will see
the spin exhibits transients for the first few orbital damping times. 
Process 4 has no transients and for Process 5 I set the 
initial local polarization parallel to the $\vec n$-axis \cite{DK72,HH96}
at each point in phase space. By comparing these processes one finds that 
in the presence of radiation the asymptotic polarization direction is not
exactly parallel to the $\vec n$-axis. Furthermore in the presence of the snake
there is complete depolarization.
\par All of the noise processes studied in this report are processes of
Ornstein-Uhlenbeck type
\footnote{to be explained in sections 2.2 and 3.2}
and in particular they are Markovian diffusion processes. One can
determine their statistical properties in an explicit form.
\par Finally, in an Epilogue, I summarize further results involving the   
polarization density. The inclusion of a detailed account here
would make this paper too long. These aspects will appear in a separate
article.
\section{The orbital model}
\setcounter{equation}{0}
\setcounter{subsection}{0}
In this section I lay the foundations for the models to be discussed in 
sections 2 and 3.
\par The underlying mathematical model for  machines I and II 
\cite{BBHMR94a,BBHMR94b}
comprises a three-dimensional spin-orbit system for the two longitudinal
orbit variables and the angle describing the orientation of the spin in the
machine plane. In contrast to {\it discrete} stochastic processes used
occasionally in the literature my `time' parameter $s$ is {\it
continuous}, so that I can work with differential equations.
Only synchrotron motion is considered and the influence of the much
faster betatron oscillations is neglected as are the Stern-Gerlach 
forces (back reaction of the spin onto the orbit) and depolarizer fields
\cite{BHR94a,BHR94b,Hei96}. There are no vertical bends so that the design
orbit is planar.
Underlying machines I and II is the `smooth ring'
approximation. This takes advantage of the fact that the synchrotron tune
is usually very small so that the optical functions can be averaged around 
the ring. Then the combined orbital motion is described by the
following stochastic differential equation:
\footnote{The prime denotes the derivative w.r.t. $s$.}
\begin{eqnarray*}
   \frac{d}{ds}  \left( \begin{array}{c}
                \sigma(s)        \\
              \eta(s)
                \end{array}
         \right)
&\equiv&
  \left( \begin{array}{c}
                \sigma'(s)       \\
              \eta'(s)
                \end{array}
         \right)
               =
       \left( \begin{array}{cc}
      0 &   -\kappa                        \\
         \Omega_s^2 /\kappa    &  -2\cdot\alpha_s/L
                \end{array}
         \right)   \cdot
       \left( \begin{array}{c}
                \sigma(s)         \\
                 \eta(s)
                \end{array}
         \right)
    +                    \sqrt{\omega}\cdot
       \left( \begin{array}{c}
                 0             \\
                 \zeta(s)
                \end{array}
         \right)
                             \; ,
\end{eqnarray*}
where $s$ denotes the distance around the ring (the `azimuth'), $\sigma$ the
distance to the centre of the bunch, $\eta$ the fractional energy
deviation, and $\zeta$ simulates the noise due to the synchrotron
radiation. Also $\alpha_s$ is the one turn synchrotron damping
decrement and $\omega$ is the one turn averaged stochastic kick
strength where in terms of the curvature 
$K_{x}$ of the design orbit in the horizontal plane
\cite{Jow85,BHMR91}:
\footnote{Note that in contrast to the notation in
 \cite{BBHMR94a,BBHMR94b} I use the symbol $\eta$ instead of $p_{\sigma}$
and $\omega$ instead 
of $\tilde \omega$.}
\begin{eqnarray*}
\omega \equiv \biggl(
          |K_{x}|^{3}\cdot C_2 \biggr)_{average}
                                   \; , \qquad
        C_{2} \equiv
               \frac{55\cdot\sqrt{3}}{48}\cdot{C_{1}}\cdot
                       \Lambda\cdot
                       \gamma_{0}^{2}  \; , \qquad
C_{1} \equiv
                        \frac{2}{3}\cdot e^{2}\cdot
                       \frac{\gamma_{0}^{4}}{E_{0}} \; , \qquad
                      \Lambda \equiv \frac{\hbar}{m_{0}c_0} \; .
\end{eqnarray*}
Here $\gamma_0$ and   $E_0$ denote the design values of the Lorentz
factor and   the energy, $c_0$ the vacuum velocity of light,
and $e$ resp. $m_0$ denote the charge resp.
rest mass of the electron. 
\par The stochastic averages of the kicks $\zeta(s)$ are
\footnote{Note that $\delta$ denotes Dirac's delta function.}
\begin{eqnarray*}
&&                  <\zeta(s_1)\cdot\zeta(s_2)>
            =
                  \delta(s_1-s_2) \; , \qquad
                  <\zeta(s)>
           =
                  0 \; .
\end{eqnarray*}
Thus the stochastic part of the 
synchrotron radiation is
treated as a Gaussian white noise process. This is sufficient for my
purposes since the characteristic time for the emission of a photon is
very small compared with other time scales of the system.
Finally, $\kappa$ is the
compaction factor, $L$ is the length of the ring
and $\Omega_s =2\pi\cdot Q_s/L$ where $Q_s$ is the
synchrotron tune. The ring is perfectly aligned so that in the smooth
approximation the closed orbit and design orbit are identical. The
vertical emittance is taken to be zero. 
\setcounter{subsection}{0}
\setcounter{equation}{0}
\section{Machine I}
\subsection{  }
For `Machine I' the spin vectors are restricted
to the horizontal (machine) plane so that the spin vector $\vec{\xi}$
can be described by a single phase angle $\psi$.
\footnote{The extension to full three-dimensional spin motion, 
i.e. the inclusion
of vertical spin is briefly considered in appendices C and D.}
Although spin is a quantum mechanical phenomenon, in high energy
storage rings it can be treated at the semiclassical level using the
Thomas-BMT equation \cite{Tho27,BMT59}
\begin{eqnarray}
&& \vec{\xi}\ ' =
\vec{\Omega}_I\wedge \vec{\xi}   \; ,
\label{eq:1.2}
\end{eqnarray}
describing the precession of a classical spin $\vec{\xi}$ in electric
and magnetic fields. Alternatively I can take $\vec \xi$ to be the
spin expectation value of an electron in a pure state with spin along
$\vec \xi /||\vec \xi|| $.
                     The precession vector $\vec \Omega_I$ is a function
of the magnetic and electric fields and of the particle velocity and
energy. As is usual in this context I now write $\vec{\Omega}_I$
as a sum of a piece $\vec{\Omega}_{I,0}$ accounting for the fields on the
closed orbit
and a piece ${\vec \Omega}_{osc}$ accounting for 
synchrotron motion with respect to the closed orbit, i.e.
\begin{eqnarray*}
&&\vec{\Omega}_I = \vec{\Omega}_{I,0} + \vec{\Omega}_{osc} \; .
\end{eqnarray*}
Thus the Thomas-BMT equation on the closed orbit takes the form:
\begin{eqnarray}
&& \vec{\xi}\ ' =
\vec{\Omega}_{I,0}\wedge \vec{\xi}   \; .
\label{eq:1.3}
\end{eqnarray}
Since only motion in the
horizontal plane need be considered one can write
\begin{eqnarray*}
\vec{\Omega}_{I,0} &\equiv& ||\vec{\Omega}_{I,0}||\cdot\vec{e}_3 \; ,
\nonumber\\
\vec{\Omega}_{osc} &\equiv& \Omega_{osc}\cdot\vec{e}_3 \; ,
\end{eqnarray*}
where $\vec{e}_3$ points normal to the machine plane.
\footnote{Two additional unit-vectors $\vec e_1,\vec e_2$ are radial resp.
longitudinal w.r.t. the closed orbit.
Moreover $\vec{e}_1,\vec{e}_2,\vec{e}_3$ constitute
an orthonormal, right-handed dreibein on the closed orbit.
This defines the `machine frame'.}
By averaging (`smoothing') over one turn one obtains:
\begin{eqnarray*}
||\vec{\Omega}_{I,0}|| = (2 \pi \nu /L) \; ,
\end{eqnarray*}
where $\nu\equiv \gamma_0\cdot(g-2)/2$ is the number of spin
precessions per turn \cite{Cha81}.
Spin motion will be calculated conveniently with respect to a dreibein of
orthonormal axes 
              $\vec{m}_{0,I}(s),\vec{l}_{0,I}(s),\vec{n}_{0,I}(s)$,
which obey the Thomas-BMT equation on the closed orbit. 
By choice the vectors
$\vec{m}_{0,I},\vec{l}_{0,I}$
precess in the horizontal plane 
around the vertical dipole field according to 
(\ref{eq:1.3}). The vector $\vec{n}_{0,I}$ (=
$\vec{m}_{0,I}\wedge\vec{l}_{0,I}$) is vertical and therefore periodic
in $s$ with period $L$ (i.e. 1-turn periodic) in the machine frame.
\footnote{Thus $\vec{n}_{0,I}$ is the so called `$\vec n_0$-axis' of 
Machine I.}
The orthonormal axes can be chosen as: 
\begin{eqnarray*}
 \vec{m}_{0,I}(s)  &\equiv&
\sin(||\vec{\Omega}_{I,0}||\cdot s) \cdot \vec{e}_1-
\cos(||\vec{\Omega}_{I,0}||\cdot s) \cdot \vec{e}_2 \; , \nonumber\\
 \vec{l}_{0,I}(s)  &\equiv&
\cos(||\vec{\Omega}_{I,0}||\cdot s) \cdot \vec{e}_1+
\sin(||\vec{\Omega}_{I,0}||\cdot s) \cdot \vec{e}_2 \; , \nonumber\\
 \vec{n}_{0,I}(s) &\equiv& \vec e_3 \; .
\end{eqnarray*}
In dealing only with horizontal spin I introduce the spin
                                                      phase angle $\psi$ by
\begin{eqnarray*}
 \vec{\xi}  &\equiv& \frac{\hbar}{2}\cdot
\biggl( \vec{m}_{0,I}\cdot\cos(\psi)+
        \vec{l}_{0,I}\cdot\sin(\psi) \biggr)  \; .
\end{eqnarray*}
By only including synchrotron oscillations and 
averaging (smoothing) the instantaneous precession rate
over one turn the Thomas-BMT equation is equivalent to
\begin{eqnarray*}
&& \psi' = \Omega_{osc} =  (2 \pi \nu /L) \cdot \eta \; .
\end{eqnarray*}
Thus $\psi$ only
couples to and is only driven by $\eta$.
\par I also   introduce
the spin vector
\begin{eqnarray}
 \vec{S}  &\equiv& \frac{\hbar}{2}\cdot
  \left( \begin{array}{c}
                      \cos(\psi) \\
                      \sin(\psi) \\
                      0
                \end{array}
         \right)    \; ,
\label{eq:1.4}
\end{eqnarray}
describing the spin in the
$(\vec{m}_{0,I},\vec{l}_{0,I},\vec{n}_{0,I})$-frame,
so that the Thomas-BMT equation reads\\
as
\footnote{Note that $\vec\xi$, unlike $\vec S$, is the spin vector in an 
arbitrary frame. Thus (\ref{eq:1.2}), unlike (\ref{eq:1.5a}),
is valid in an arbitrary frame.}
\setcounter{INDEX}{1}
\begin{eqnarray}
&& \vec{S}'(s) =
\vec{W}_I\biggl(\eta(s)\biggr)\wedge \vec{S}(s)   \; ,
 \label{eq:1.5a}
\end{eqnarray}
\addtocounter{equation}{-1}
\addtocounter{INDEX}{1}%
with
\begin{eqnarray}
\vec{W}_I(\eta) &\equiv &
         \frac{2\pi\nu}{L}\cdot\eta\cdot
  \left( \begin{array}{c}
                      0          \\
                      0          \\
                      1
                \end{array}
         \right)    \; .
\label{eq:1.5b}
\end{eqnarray}
\setcounter{INDEX}{0}%
Thus for Machine I one has to deal with the following three-component
Langevin equation:
\begin{eqnarray}
&&       \left( \begin{array}{c}
                \sigma '(s)       \\
              \eta'(s)            \\
              \psi'(s)
                \end{array}
         \right)
               =
       \left( \begin{array}{ccc}
                      0 &   -\kappa   & 0         \\
  \Omega_s^2 /\kappa    &  -2\cdot\alpha_s/L &  0 \\
          0           &  2\pi\nu/L  &  0
                \end{array}
         \right)   \cdot
       \left( \begin{array}{c}
                \sigma(s)         \\
                 \eta(s)          \\
                 \psi(s)
                \end{array}
         \right)
    +                    \sqrt{\omega}\cdot
       \left( \begin{array}{c}
                 0             \\
                 \zeta(s)       \\
                 0
                \end{array}
         \right)
                             \; .
\label{eq:1.6}
\end{eqnarray}
One now sees that the noise  not only acts on the orbit motion
but  also indirectly on the spin via its coupling to $\eta$. It is this
coupling which will lead to the spin decoherence.
\subsection{The Langevin equation for Machine I}
\subsubsection*{2.2.1}
With the abbreviations:
\begin{eqnarray}
a \equiv -\kappa \; , \;\; b \equiv \Omega_s^2/\kappa 
=(4\pi^2 Q_s^2)/(\kappa L^2)
\; , \;\; c \equiv -2 \cdot \alpha_s/L\; , \;\;
d \equiv ||\vec{\Omega}_{I,0}||=2 \pi \nu /L  \; ,
 \label{eq:2.256}
\end{eqnarray}
the Langevin equation 
(\ref{eq:1.6}) can be rewritten as
\begin{eqnarray}
&& d\vec{x}(s) =  \underline{\cal A}_I\cdot\vec{x}(s)\cdot ds
               +\underline{\cal B}\cdot d\vec{\cal W}(s) \; ,
 \label{eq:2.1}
\end{eqnarray}
where
\begin{eqnarray}
\underline{\cal A}_I &\equiv&        \left( \begin{array}{ccc}
      0 & a & 0                            \\
      b & c & 0                            \\
      0 & d & 0
                \end{array}
         \right) \; , \qquad
    \underline{\cal B}  \equiv   \sqrt{\omega}\cdot
                              \left( \begin{array}{ccc}
   0 & 0  & 0                       \\
   0 & 1  & 0                        \\
   0 & 0  & 0
                \end{array}
         \right) \; ,  \qquad
 \vec{x}\equiv
       \left( \begin{array}{c}
                \sigma         \\
           \eta             \\
                \psi
                \end{array}
         \right)  \; ,
 \label{eq:2.264}
\end{eqnarray}
with
\begin{eqnarray*}
   d\vec{\cal W}(s) &\equiv &
       \left( \begin{array}{c}
             d{\cal W}_1(s)           \\
             d{\cal W}_2(s)             \\
             d{\cal W}_3(s)
                \end{array}
         \right)  \; .
\end{eqnarray*}
Here the ${\cal W}_k(s)$ are     Wiener processes
\cite{Gar85} related to the Gaussian white noise process $\zeta(s)$
formally by:
\begin{eqnarray}
&& d{\cal W}_k(s) = \zeta(s)\cdot ds \; .
 \label{eq:2.230}
\end{eqnarray}
For a practical storage ring:
\begin{eqnarray}
&& a<0 \;,\; b>0 \; , \; c<0\; , \; d>0\; ,\; \omega>0 \; .
 \label{eq:2.2}
\end{eqnarray}
Furthermore
\begin{eqnarray}
&& a\cdot b +c^2/4 < 0 \; ,
 \label{eq:2.212}
\end{eqnarray}
since $\alpha_s\ll Q_s$. The inequalities
(\ref{eq:2.2}),(\ref{eq:2.212}) are assumed throughout this paper. 
For the HERA electron ring running at about 27 GeV
the values are approximately:
$Q_s\approx 0.06,\\\alpha_s\approx 0.0032,\;
\kappa\approx 0.00069,\;\omega \approx 2\cdot 10^{-12}\, m^{-1},\;
L\approx 6300\, m, \;
d \approx 6.2\cdot 10^{-2} \; m^{-1}$, so that one has:
\begin{eqnarray}
&& a \approx -6.9\cdot 10^{-4} \; , \qquad b \approx 5.2\cdot 10^{-6}\; 
m^{-2} \; , \qquad
c \approx -1.0\cdot 10^{-6}\; m^{-1} \; , \nonumber\\
&&  d \approx 6.2\cdot 10^{-2} \; m^{-1} \;,
    \qquad \omega \approx 2.0\cdot 10^{-12}\; m^{-1} \;, \qquad
   L \approx 6.3\cdot 10^3\; m \; . \nonumber\\&&
 \label{eq:2.3}
\end{eqnarray}
The orbital damping `time' $\tau_{damp}$ of the system is given by
\begin{eqnarray*}
\tau_{damp} &\equiv& -\frac{2}{c} = \frac{L}{\alpha_s} \; ,
\end{eqnarray*}
so that $1/\tau_{damp}$ is the `orbital damping rate'.
\footnote{For more details on $\tau_{damp}$, see Appendix E.}
In particular I get
\begin{eqnarray*}
&& \tau_{damp}\approx 2.0\cdot 10^6\, m \; ,
\end{eqnarray*}
which corresponds to about 310 turns or about 6.6 milliseconds.
\par Note that by
(\ref{eq:1.5b}),
(\ref{eq:2.256}):
\begin{eqnarray}
\vec{W}_I(\eta) &\equiv &
         d\cdot\eta\cdot
  \left( \begin{array}{c}
                      0          \\
                      0          \\
                      1
                \end{array}
         \right)    \; .
\label{eq:2.257}
\end{eqnarray}
\subsubsection*{2.2.2}
Because $\underline{\cal A}_I$ and $\underline{\cal B}$ are matrices
which do not depend on $\vec x$
the Langevin equation 
(\ref{eq:2.1})
describes three-component 
processes of Ornstein-Uhlenbeck type \cite{Gar85}. 
Thus the stochastic integrations involved in the solution of
(\ref{eq:2.1})
can be either defined as Ito-integrations or
Stratonovich-integrations and lead by both methods to 
the Fokker-Planck equation 
(\ref{eq:2.8}). The analogous situation holds for Machine II.
\par If $s_0$ denotes the starting azimuth 
of a process
$\vec{x}(s)$ then $\vec{x}(s_0)$
is always assumed to be chosen so that $\vec{x}(s)$ is a 
Markovian diffusion process
\cite{Arn73,Gar85}.
\footnote{For the special processes 1 and 2 considered in detail I choose
$s_0=0$.}
\par The three-component differential equation
(\ref{eq:2.1})
has essentially only two nontrivial components. 
Writing 
(\ref{eq:2.1})
in more detail I get:
\setcounter{INDEX}{1}
\begin{eqnarray}
&& d\vec{z}(s) =  \underline{\cal A}_{orb}\cdot\vec{z}(s)\cdot ds
     +    \underline{\cal B}_{orb}\cdot d\vec{\cal W}_{orb}(s) \; ,
 \label{eq:2.233a} \\
\addtocounter{equation}{-1}
\addtocounter{INDEX}{1}
&& d\psi(s) =  d\cdot\eta(s)\cdot ds \; ,
 \label{eq:2.233b} 
\end{eqnarray}
\setcounter{INDEX}{0}%
where
\begin{eqnarray}
 \vec{z}\equiv
       \left( \begin{array}{c}
                \sigma         \\
           \eta
                \end{array}
         \right)  \; ,
 \label{eq:2.5}
\end{eqnarray}
and where:
\begin{eqnarray*}
\underline{\cal A}_{orb} &\equiv&        \left( \begin{array}{cc}
      0 & a                             \\
      b & c
                \end{array}
         \right) \; , \qquad
\underline{\cal B}_{orb}\equiv\sqrt{\omega}\cdot
                                                \left( \begin{array}{cc}
   0&  0                             \\
   0&  1
                \end{array}
         \right) \; , \qquad
    \vec{\cal W}_{orb}(s)  \equiv
       \left( \begin{array}{c}
              {\cal W}_1(s)           \\
              {\cal W}_2(s)
                \end{array}
         \right)  \; .
\end{eqnarray*}
The two-dimensionality is seen by transforming $\psi$ linearly to
a new variable:
\begin{eqnarray}
\tilde{\psi}\equiv\psi - \frac{d}{a}\cdot\sigma \; .
 \label{eq:2.213}
\end{eqnarray}
Then my Langevin equation 
(\ref{eq:2.1})
is equivalent to
\setcounter{INDEX}{1}
\begin{eqnarray}
&& d\vec{z}(s) =  \underline{\cal A}_{orb}\cdot\vec{z}(s)\cdot ds
     +    \underline{\cal B}_{orb}\cdot d\vec{\cal W}_{orb}(s) \; ,
 \label{eq:2.6a} \\
\addtocounter{equation}{-1}
\addtocounter{INDEX}{1}
&& d\tilde{\psi}(s) = 0 \; ,
\label{eq:2.6b}
\end{eqnarray}
\setcounter{INDEX}{0}%
so that $\tilde{\psi}(s)$ is $s$-independent.
\subsection{The Fokker-Planck equation for Machine I. Further properties of
Machine I}
\subsubsection*{2.3.1}
I abbreviate the stochastic average of a function
$f(\sigma,\eta,\psi)$ for a process $\vec{x}(s)$ by \\
$<f(\sigma(s),\eta(s),\psi(s))>$, so that:
\begin{eqnarray}
<f\biggl(\sigma(s),\eta(s),\psi(s)\biggr)>    &\equiv&
                  \int_{-\infty}^{+\infty} \; d\sigma
                             \int_{-\infty}^{+\infty} \; d\eta
                             \int_{-\infty}^{+\infty} \; d\psi \cdot
   w(\sigma,\eta,\psi;s)\cdot f(\sigma,\eta,\psi)\; ,
 \label{eq:2.10}
\end{eqnarray}
where the `probability density' $w$ characterizes the state of the system
at azimuth $s$. From (\ref{eq:2.10}) follows:
\begin{eqnarray}
&&   w(\sigma,\eta,\psi;s) = <\delta(\sigma-\sigma(s))\cdot
\delta(\eta-\eta(s))\cdot
\delta(\psi-\psi(s))> \; ,
 \label{eq:2.7}
\end{eqnarray}
so that the probability density is nonnegative and normalized by:
\begin{eqnarray}
   1 &=& \int_{-\infty}^{+\infty} \; d\sigma
         \int_{-\infty}^{+\infty} \; d\eta
         \int_{-\infty}^{+\infty} \; d\psi\cdot
 w(\sigma,\eta,\psi;s) \; .
 \label{eq:2.9}
\end{eqnarray}
One sees by
(\ref{eq:2.10}),(\ref{eq:2.9})
that the domains of the variables
$\sigma,\eta,\psi$ are chosen to be $(-\infty,+\infty)$, i.e. the real
numbers, and 
that the probability density obeys boundary conditions for each of the
variables $\sigma,\eta,\psi$ with
$w\rightarrow 0$ for $\sigma,\eta,\psi\rightarrow\pm\infty$. 
I call these `standard' boundary conditions.
Moreover I always assume that the stochastic
averages of the functions of interest are finite. 
For the motivation of these boundary
conditions see section 2.3.6. 
\par The key quantity  of interest when dealing with spin is
the `polarization vector' $\vec{P}_{tot}^w(s)$. This is 
the stochastic average of the normalized spin vector, i.e. it is given by
\begin{eqnarray}
\vec{P}_{tot}^w(s) &\equiv&
  \frac{2}{\hbar}\cdot <\vec{S}(s)> =
  \frac{2}{\hbar}\cdot \int_{-\infty}^{+\infty} \; d\sigma
                             \int_{-\infty}^{+\infty} \; d\eta
                             \int_{-\infty}^{+\infty} \; d\psi\cdot
   w(\sigma,\eta,\psi;s)
                                 \cdot \vec{S}
\nonumber\\
 &=& \int_{-\infty}^{+\infty} \; d\sigma
                             \int_{-\infty}^{+\infty} \; d\eta
                             \int_{-\infty}^{+\infty} \; d\psi\cdot
   w(\sigma,\eta,\psi;s)
\cdot  \left( \begin{array}{c}
                      \cos(\psi) \\
                      \sin(\psi) \\
                      0
                \end{array}
         \right)    \; ,
 \label{eq:2.11}
\end{eqnarray}
where:
\begin{eqnarray*}
 \vec{S}(s)  &\equiv& \frac{\hbar}{2}\cdot
  \left( \begin{array}{c}
        \cos(\psi(s)) \\
        \sin(\psi(s)) \\
                      0
                \end{array}
         \right)  \; .
\end{eqnarray*}
I define the `polarization' as the norm 
$||\vec{P}_{tot}^w||$ of the
polarization vector.
\subsubsection*{2.3.2}
The processes can be either described `directly' by
handling the stochastic averages $<f>$ or `indirectly' by
`ensembles' via the corresponding probabilities, e.g. the probability
density $w$. The latter obeys a Fokker-Planck
equation and for the Langevin equation 
(\ref{eq:2.1}) the Fokker-Planck equation has the form \cite{Gar85,Ris89}
\begin{eqnarray*}
 \frac{\partial w}{\partial s} &=&
-\sum_{j,k=1}^3 \frac{\partial }{\partial x_j}
     ({\cal A}_{I,jk}\cdot x_k\cdot w)
+\frac{1}{2}\cdot
 \sum_{j,k=1}^3 \frac{\partial^2}{\partial x_j\partial x_k}
 ( {\cal D}_{jk}\cdot w ) \; ,
\end{eqnarray*}
where
\begin{eqnarray*}
&& \underline{\cal D}\equiv
          \underline{\cal B}\cdot\underline{\cal B}^T
 =                             \left( \begin{array}{ccc}
   0      &    0 & 0 \\
   0      &   \omega & 0 \\
   0      &    0 &  0
                \end{array}
         \right)                           \:  .
\end{eqnarray*}
Therefore the Fokker-Planck equation can be written as
\begin{eqnarray}
   \frac{\partial w}{\partial s} &=&
 -\underbrace{
 a\cdot \eta \cdot
\frac{\partial w}{\partial \sigma}
- b\cdot\sigma \cdot \frac{\partial w}{\partial \eta}
                                                }_
               {{\rm synchrotron\; oscillation\;terms}} \;
 -\underbrace{
 d\cdot \eta
               \cdot \frac{\partial w}{\partial \psi}
                                                }_
               {{\rm spin \; precession \; term}}
 -\underbrace{
 c\cdot w
- c\cdot\eta \cdot \frac{\partial w}{\partial \eta}
                                                }_
               {{\rm damping \; terms}}
+ \underbrace{
          \frac{\omega}{2}\cdot
\frac{\partial^2 w}{\partial \eta^2}
                                                }_
               {{\rm diffusion \; term}}
\nonumber\\&&
\nonumber\\
&&  \equiv L_{FP,orb} \; w + L_{FP,I,spin} \; w
                    \; ,
 \label{eq:2.8}
\end{eqnarray}
where I used the abbreviations
\begin{eqnarray*}
  L_{FP,orb} &\equiv&
-a\cdot \eta \cdot
\frac{\partial}{\partial \sigma}
- b\cdot\sigma \cdot \frac{\partial}{\partial \eta}
 - c
- c\cdot\eta \cdot \frac{\partial}{\partial \eta}
+     \frac{\omega}{2}\cdot
\frac{\partial^2}{\partial \eta^2} \; ,
\nonumber\\
    L_{FP,I,spin} &\equiv &
-d\cdot \eta \cdot \frac{\partial }{\partial \psi} \; .
\end{eqnarray*}
\subsubsection*{2.3.3}
The `orbital part' $w_{orb}$ of a probability density $w$ is
defined by
\begin{eqnarray}
   w_{orb}(\sigma,\eta;s) &\equiv&
                             \int_{-\infty}^{+\infty} \; d\psi \cdot
   w(\sigma,\eta,\psi;s) \; .
 \label{eq:2.13}
\end{eqnarray}
Because $w$ denotes the probability density of a process $\vec x(s)$, one
observes that $w_{orb}$ is the probability density of the
corresponding orbital process $\vec z(s)$.
Given an orbital function $f(\sigma,\eta)$ one sees that its stochastic
average is determined by $w_{orb}$, i.e.
\footnote{Thus $w_{orb}$ describes the orbital distribution.}
\begin{eqnarray*}
 <f\biggl(\sigma(s),\eta(s)\biggr)>  &=&  \int_{-\infty}^{+\infty} \; d\sigma
                             \int_{-\infty}^{+\infty} \; d\eta \cdot
   w_{orb}(\sigma,\eta;s)\cdot f(\sigma,\eta)\; .
\end{eqnarray*}
Furthermore the orbital
part of the probability density is normalized by
\begin{eqnarray}
   1 &=& \int_{-\infty}^{+\infty} \; d\sigma
         \int_{-\infty}^{+\infty} \; d\eta\cdot
\; w_{orb}(\sigma,\eta;s) \; ,
 \label{eq:2.14}
\end{eqnarray}
which follows from 
(\ref{eq:2.9}). 
Because $w$ solves the Fokker-Planck equation
(\ref{eq:2.8}), 
$w_{orb}$ solves the `orbital Fokker-Planck equation'
\begin{eqnarray}
   \frac{\partial w_{orb}}{\partial s} &=&
                            L_{FP,orb} \; w_{orb} \; .
 \label{eq:2.15}
\end{eqnarray}
\subsubsection*{2.3.4}
The `spin part' $w_{spin}$ of a probability density $w$ is defined by
\begin{eqnarray}
   w_{spin}(\psi;s) &\equiv& \int_{-\infty}^{+\infty} \; d\sigma
                             \int_{-\infty}^{+\infty} \; d\eta \cdot
   w(\sigma,\eta,\psi;s) \; ,
 \label{eq:2.16}
\end{eqnarray}
and it is normalized by
\begin{eqnarray*}
   1 &=& \int_{-\infty}^{+\infty} \; d\psi\cdot
\; w_{spin}(\psi;s) \; ,
\end{eqnarray*}
which follows from 
(\ref{eq:2.9}). 
Given a function $f(\psi)$ depending only on
$\psi$ one sees that its stochastic average is
determined by $w_{spin}$\, , i.e.
\begin{eqnarray*}
 <f\biggl(\psi(s)\biggr)>  &=&  \int_{-\infty}^{+\infty} \; d\psi \cdot
   w_{spin}(\psi;s)\cdot f(\psi)\; .
\end{eqnarray*}
\subsubsection*{2.3.5}
With the standard boundary conditions one can
introduce via Fourier transformation 
a `characteristic function', namely $\Phi$
corresponding to $w$, defined by \cite{Gar85}:
\begin{eqnarray}
 \Phi(\vec{u};s) &=&
          \int_{-\infty}^{+\infty} \; d x_1
          \int_{-\infty}^{+\infty} \; d x_2
          \int_{-\infty}^{+\infty} \; d x_3\cdot
   \exp(i\cdot \vec{u}^T\cdot\vec{x})\cdot
   w(\vec{x};s) \; ,
 \label{eq:2.222}
\end{eqnarray}
so that
\begin{eqnarray}
 w(\vec{x};s) &=&  \frac{1}{8\pi^3}\cdot
          \int_{-\infty}^{+\infty} \; d u_1
          \int_{-\infty}^{+\infty} \; d u_2
          \int_{-\infty}^{+\infty} \; d u_3\cdot
   \exp(-i\cdot \vec{u}^T\cdot\vec{x})\cdot
   \Phi(\vec{u};s) \; .
 \label{eq:2.40}
\end{eqnarray}
Since   $w$ fulfills the
Fokker-Planck equation 
(\ref{eq:2.8}), one finds:
\begin{eqnarray}
 \frac{\partial \Phi}{\partial s} &=&
      \sum_{j,k=1}^3 \; {\cal A}_{I,kj}\cdot u_k\cdot
     \frac{\partial\Phi}{\partial u_j}
-\frac{1}{2}\cdot
 \sum_{j,k=1}^3 \;   {\cal D}_{jk}\cdot u_j\cdot u_k\cdot \Phi \; .
 \label{eq:2.41}
\end{eqnarray}
Analogously, for the orbital part one defines:
\begin{eqnarray}
 \Phi_{orb}(\vec{t};s) &\equiv&
          \int_{-\infty}^{+\infty} \; d z_1
          \int_{-\infty}^{+\infty} \; d z_2\cdot
   \exp(i\cdot \vec{t}^{\;T}\cdot\vec{z})\cdot
   w_{orb}(\vec{z};s) \; ,
 \label{eq:2.223}
\end{eqnarray}
from which follows
\begin{eqnarray}
 w_{orb}(\vec{z};s) &=&  \frac{1}{4\pi^2}\cdot
          \int_{-\infty}^{+\infty} \; d t_1
          \int_{-\infty}^{+\infty} \; d t_2\cdot
   \exp(-i\cdot \vec{t}^{\;T}\cdot\vec{z})\cdot
   \Phi_{orb}(\vec{t};s) \; .
\label{eq:2.24}
\end{eqnarray}
Because $w_{orb}$ fulfills the orbital
Fokker-Planck equation 
(\ref{eq:2.15}), I conclude:
\begin{eqnarray}
 \frac{\partial \Phi_{orb}}{\partial s} &=&
      \sum_{j,k=1}^2 \; {\cal A}_{I,kj}\cdot t_k\cdot
     \frac{\partial\Phi_{orb}}{\partial t_j}
-\frac{1}{2}\cdot
\sum_{j,k=1}^2\;{\cal D}_{jk}\cdot t_j\cdot t_k\cdot\Phi_{orb}\; .
\label{eq:2.25}
\end{eqnarray}
\subsubsection*{2.3.6}
My chosen boundary conditions (see 
(\ref{eq:2.9})
and the sentences
following) are very natural for $\sigma,\eta$. After all, the rms relative
energy spread for the values 
(\ref{eq:2.3})
is about $10^{-3}$ and the rms bunch length is
about 1 cm. On the contrary I will be dealing with spreads in $\psi$
of order $2\pi$ or more so that at first sight
it would seem unnatural to choose the domain $(-\infty,+\infty)$
for $\psi$. Indeed, if one writes the spin vector
$\vec S$ in spherical coordinates as:
\begin{eqnarray}
 \vec{S}  &=& \frac{\hbar}{2}\cdot
  \left( \begin{array}{c}
                      \cos(\psi)\cdot \sin(\theta) \\
                      \sin(\psi)\cdot \sin(\theta) \\
                      \cos(\theta)
                \end{array}
         \right)    \; ,
\label{eq:2.240}
\end{eqnarray}
then by (\ref{eq:1.4}) one can identify $\psi$ as the azimuthal angle, where
the polar angle $\theta$ equals $\pi/2$.
Since the values
$\psi=0$ resp. $\psi=2\pi$ are identified
it would then seem more appropriate to use a probability density $w_{per}$
which fulfills periodic boundary conditions in $\psi$:
\begin{eqnarray}
w_{per}(\sigma,\eta,\psi+2\pi;s) &=&
w_{per}(\sigma,\eta,\psi;s)  \; .
 \label{eq:2.205}
\end{eqnarray}
The normalization condition 
(\ref{eq:2.9})
would then be replaced by
\begin{eqnarray}
   1 &=& \int_{-\infty}^{+\infty} \; d\sigma
         \int_{-\infty}^{+\infty} \; d\eta
         \int_{0}^{2\pi} \; d\psi\cdot
\; w_{per}(\sigma,\eta,\psi;s) \; .
\label{eq:2.206}
\end{eqnarray}
However Process 2 considered in this section has, for $s>0$, a Gaussian 
probability density. Furthermore Process 1 has, for $s>0$,
a probability density 
which is a combination of a Gaussian function and a delta function. 
\footnote{Both processes have standard boundary conditions.}
Thus it is more convenient to adopt boundary conditions which allow one
to work with Gaussians as much as possible. 
Thus machines I and II are treated with the standard boundary conditions
and the periodic boundary conditions are only mentioned in passing.
\par The orbital part $w_{per,orb}$ of $w_{per}$ is defined by:
\begin{eqnarray}
   w_{per,orb}(\sigma,\eta;s) &\equiv& \int_{0}^{2\pi} \; d\psi\cdot
   w_{per}(\sigma,\eta,\psi;s) \; ,
\label{eq:2.208}
\end{eqnarray}
and the spin part $w_{per,spin}$ of $w_{per}$ is defined by:
\begin{eqnarray}
   w_{per,spin}(\psi;s) &\equiv& \int_{-\infty}^{+\infty} \; d\sigma
                             \int_{-\infty}^{+\infty} \; d\eta \cdot
   w_{per}(\sigma,\eta,\psi;s) \; ,
\label{eq:2.209}
\end{eqnarray}
which is normalized by
(\ref{eq:2.206}) as:
\begin{eqnarray}
   1 &=& \int_{0}^{2\pi} \; d\psi\cdot
\; w_{per,spin}(\psi;s) \; .
\label{eq:2.210}
\end{eqnarray}
Note that $w_{per,spin}$ is periodic in $\psi$:
\begin{eqnarray}
 w_{per,spin}(\psi+2\pi;s) &=&
 w_{spin}(\psi;s) \; .
\label{eq:2.211}
\end{eqnarray}
The polarization vector is defined by:
\begin{eqnarray}
\vec{P}_{tot}^w(s) &=&
  \int_{-\infty}^{+\infty} \; d\sigma
                             \int_{-\infty}^{+\infty} \; d\eta
                             \int_{0}^{2\pi} \; d\psi\cdot
   w_{per}(\sigma,\eta,\psi;s)
                                \cdot  \left( \begin{array}{c}
                      \cos(\psi) \\
                      \sin(\psi) \\
                      0
                \end{array}
         \right)    \; .
 \label{eq:2.238}
\end{eqnarray}
\par Given a process with standard boundary conditions with probability 
density $w$ and defining $w_{per}$ in one of the two following ways:
\begin{eqnarray}
&& w_{per}(\sigma,\eta,\psi;s) \equiv
\sum_{n=-\infty}^{\infty}\;
 w(\sigma,\eta,\psi+2\pi\cdot n;s) \; ,
\label{eq:2.241} \\
&& w_{per}(\sigma,\eta,\psi;s) \equiv
 \frac{1}{2\pi}\cdot\int_{-\infty}^{+\infty} \; d\psi_1\cdot
   w(\sigma,\eta,\psi_1;s) \nonumber\\&&\qquad
+ \frac{\sqrt{3}\cdot\cos(\psi)}{2\pi}\cdot
\int_{-\infty}^{+\infty} \; d\psi_1\cdot
   \cos(\psi_1)\cdot w(\sigma,\eta,\psi_1;s) \nonumber\\&&\qquad
+
\frac{\sqrt{3}\cdot\sin(\psi)}{2\pi}\cdot
\int_{-\infty}^{+\infty} \; d\psi_1\cdot
   \sin(\psi_1)\cdot w(\sigma,\eta,\psi_1;s) \; ,
\label{eq:2.242}
\end{eqnarray}
one observes that $w_{per}$ fulfills the above mentioned properties
and solves the Fokker-Planck equation
(\ref{eq:2.8}). Moreover one then finds that $w_{per,orb}=w_{orb}$.
\par The expression
(\ref{eq:2.242}) is of special interest if semiclassical considerations
come into play. In fact by adopting the spinning particle Wigner function 
formalism of 
\cite{Str57,GV88,GV89}
one originally deals with a Wigner function of the form
(\ref{eq:2.242}) and its evolution equation, and one can then in turn
try to construct the probability density $w$ and its underlying process,
i.e. design a model like Machine I from quantum mechanics.
\par Further remarks on the periodic boundary conditions are made in sections
2.7.3 and Appendix D.
For the effect of boundary conditions on Fokker-Planck equations, see also
\cite{Gar85} and for the effect on stochastic
differential equations, see \cite{GS71}.
\subsubsection*{2.3.7}
With the standard boundary conditions one can immediately
write down a differential equation for the covariance matrix of any
process running with Machine I. The covariance matrix of a process
is defined by
\begin{eqnarray}
    \underline{\sigma}(s)&\equiv &
                              \left( \begin{array}{ccc}
  \sigma_{11}(s)   &   \sigma_{12}(s)  &  \sigma_{13}(s)  \\
  \sigma_{21}(s)   &   \sigma_{22}(s)  &  \sigma_{23}(s)  \\
  \sigma_{31}(s)   &   \sigma_{32}(s)  &  \sigma_{33}(s)
                \end{array}
         \right)        \; ,
 \label{eq:2.17}
\end{eqnarray}
with
\begin{eqnarray*}
  \sigma_{11}(s)   &\equiv&
  <\biggl(\sigma(s)-<\sigma(s)>\biggr)^2>
 =<\biggl(\sigma(s)\biggr)^2>-\biggl(<\sigma(s)>\biggr)^2   \; ,
\nonumber\\
  \sigma_{12}(s)   &\equiv&
  <\biggl(\sigma(s)- <\sigma(s)>\biggr)\cdot
   \biggl(\eta(s)- <\eta(s)>\biggr)>
\nonumber\\
 &=&        <\sigma(s)\cdot \eta(s)>
          - <\sigma(s)>\cdot <\eta(s)> \; ,
\nonumber\\
  \sigma_{13}(s)   &\equiv&
  <\sigma(s)\cdot\psi(s)>-
            <\sigma(s)>\cdot <\psi(s)> \; ,
\nonumber\\
  \sigma_{21}(s)   &\equiv&  \sigma_{12}(s)  \; ,
\nonumber\\
  \sigma_{22}(s)   &\equiv&
  <\biggl(\eta(s)\biggr)^2>-\biggl(<\eta(s)>\biggr)^2   \; ,
\nonumber\\
  \sigma_{23}(s)   &\equiv&
  <\eta(s)\cdot\psi(s)>-
            <\eta(s)>\cdot <\psi(s)> \; ,
\nonumber\\
  \sigma_{31}(s)   &\equiv&  \sigma_{13}(s)  \; ,
\nonumber\\
  \sigma_{32}(s)   &\equiv&  \sigma_{23}(s)  \; ,
\nonumber\\
  \sigma_{33}(s)   &\equiv&
  <\biggl(\psi(s)\biggr)^2>-\biggl(<\psi(s)>\biggr)^2
                                          \; .
\end{eqnarray*}
Clearly, from their definition the diagonal elements are always
nonnegative. Furthermore the $\underline \sigma$ matrix is nonnegative 
definite
\footnote{This means that for every three-component vector $\vec v$
with real components one has the inequality:
\begin{eqnarray*}
&&\sum_{j,k=1}^3\; \sigma_{jk}\cdot v_j\cdot v_k \geq 0 \; .
\end{eqnarray*}
If $\underline{\sigma}$ is nonsingular, then it is positive
definite, i.e. the equal sign in the above inequality then only occurs
for $\vec v=0$.}
and symmetric.
\par Because $\vec{x}(s)$ is a process of Ornstein-Uhlenbeck type 
it may be shown
by using the standard boundary conditions that the covariance
matrix satisfies the following differential equation \cite{Van81}:
\begin{eqnarray}
\underline{\sigma}'&=&
\underline{\cal A}_I\cdot\underline{\sigma} +
\underline{\sigma}\cdot\underline{\cal A}_I^T+
                             \underline{{\cal D}} \; .
 \label{eq:2.18}
\end{eqnarray}
In component form 
(\ref{eq:2.18})
results in:
\begin{eqnarray*}
\sigma_{11}'&=& 2\cdot a\cdot\sigma_{12} \; ,
\nonumber\\
\sigma_{12}'&=& a\cdot\sigma_{22} +
                b\cdot\sigma_{11} +
                c\cdot\sigma_{12} \; ,
\nonumber\\
\sigma_{22}'&=& 2\cdot b\cdot\sigma_{12}
              + 2\cdot c\cdot\sigma_{22}
              + \omega \; ,
\nonumber\\
 \sigma_{13}'&=& a\cdot\sigma_{23}
                       +d\cdot\sigma_{12} \; ,
\nonumber\\
\sigma_{23}'&=& b\cdot\sigma_{13}
                       +c\cdot\sigma_{23}
                       +d\cdot\sigma_{22} \; ,
\nonumber\\
\sigma_{33}'&=& 2\cdot d\cdot\sigma_{23}\; .
\end{eqnarray*}
For the first moment vector $<\vec{x}(s)>$
one gets the following differential
equation:
\begin{eqnarray}
  <\vec{x}'(s)> &=& \underline{\cal A}_I\cdot
               <\vec{x}(s)> \; .
 \label{eq:2.19}
\end{eqnarray}
The differential equations 
(\ref{eq:2.18}),(\ref{eq:2.19})
can be easily derived from (\ref{eq:2.1}).
They are valid for all processes running with Machine I and are
particularly useful for processes whose  probability densities are
determined only by the covariance matrix and the first moment vector
such as processes 1 and 2. 
These equations also show
that for every process running with Machine I
the covariance matrix and the
first moment vector depend smoothly on $s$.
\par For the orbital part one gets:
\begin{eqnarray}
\underline{\sigma}_{orb}'&=&
\underline{\cal A}_{orb}\cdot\underline{\sigma}_{orb} +
\underline{\sigma}_{orb}\cdot\underline{\cal A}_{orb}^T+
                             \underline{{\cal D}}_{orb} \; ,
 \label{eq:2.224} \\
 <\vec{z}\ '(s)> &=& \underline{\cal A}_{orb}\cdot
             <\vec{z}(s)>\; ,
 \label{eq:2.225}
\end{eqnarray}
where:
\begin{eqnarray*}
    \underline{\sigma}_{orb}(s)&\equiv &
                              \left( \begin{array}{cc}
  \sigma_{11}(s)   &   \sigma_{12}(s)   \\
  \sigma_{21}(s)   &   \sigma_{22}(s)
                \end{array}
         \right)        \; , \qquad
\underline{{\cal D}}_{orb} \equiv \underline{\cal B}_{orb}\cdot
                            \underline{\cal B}_{orb}^T =
                               \left( \begin{array}{cc}
   0      &    0 \\
   0      &   \omega
                \end{array}
         \right) \; ,
\end{eqnarray*}
and where $\underline{\sigma}_{orb}$ denotes the `orbital covariance matrix'.
Note that one has by
(\ref{eq:2.224}):
\begin{eqnarray}
\biggl(\det(\underline{\sigma}_{orb})\biggr)' &=& 
2\cdot c\cdot
\det(\underline{\sigma}_{orb})
+\omega\cdot\sigma_{11} \; .
\label{eq:2.246}
\end{eqnarray}
Because one has for $j,k=1,2,3$:
\begin{eqnarray*}
<x_j(s)>    &=&
                  \int_{-\infty}^{+\infty} \; d\sigma
                             \int_{-\infty}^{+\infty} \; d\eta
                             \int_{-\infty}^{+\infty} \; d\psi \cdot
   w(\sigma,\eta,\psi;s)\cdot x_j = -i\cdot
\biggl(\frac{\partial\Phi}{\partial u_j}(\vec u;s)\biggr)_{\vec u=0}
\; , \nonumber\\
<x_j(s)\cdot x_k(s)>    &=&
                  \int_{-\infty}^{+\infty} \; d\sigma
                             \int_{-\infty}^{+\infty} \; d\eta
                             \int_{-\infty}^{+\infty} \; d\psi \cdot
   w(\sigma,\eta,\psi;s)\cdot x_j\cdot x_k = 
-\biggl(\frac{\partial^2\Phi}{\partial u_j\partial u_k}(\vec u;s
\biggr)_{\vec u=0}
\; , 
\end{eqnarray*}
the differential equations 
(\ref{eq:2.18}),(\ref{eq:2.19})
can be alternatively derived from
(\ref{eq:2.8})
or from 
(\ref{eq:2.41}).
\footnote{See also \cite{Van81}.
From the normalization (\ref{eq:2.9}) of $w$ it also follows
by (\ref{eq:2.40}):
\begin{eqnarray*}
 \Phi(\vec{u}=0;s) &=& 1 \; .
\end{eqnarray*}}
Note that the differential equations
(\ref{eq:2.18}),(\ref{eq:2.19})
in general do not hold for boundary conditions different from
the standard boundary conditions.
Since Machine II also gives rise to processes of Ornstein-Uhlenbeck type, 
relations analogous to equations 
(\ref{eq:2.18}) and (\ref{eq:2.19})
will apply.
\subsubsection*{2.3.8}
In Machine I
\footnote{The same is true for Machine II.}
the orbital motion is not influenced by the spin motion (see 
(\ref{eq:2.1})).
Thus once the orbital motion of a process has been determined, 
finding the spin  motion
reduces to solving the stochastic differential equation 
(\ref{eq:2.233b}) for $\psi$.
Equation (\ref{eq:2.233b}) for the process $\psi(s)$ 
is equivalent to the Thomas-BMT
equation (\ref{eq:1.5a})
for the process $\vec{S}(s)$.
The $s$-dependent vector $\vec{W}_I(\eta(s))$
is a stochastic process
whose properties are determined by the process
$\eta(s)$.
With 
(\ref{eq:1.5a})
one has moulded  the spin motion of a stochastic process into the
stochastic motion of the spin vector (`Brownian motion on the
2-sphere').
\footnote{See \cite{ACDO91} and the reference list therein.}
Instead of the spin variable $\psi$ this approach uses the variable
$\vec S$ which together with the orbit variables constitutes a
five-component spin-orbit vector 
\begin{eqnarray*}
&& \left( \begin{array}{c}
                \sigma       \\
                \eta       \\
                \vec{S}
               \end{array}
          \right)  \; ,
\end{eqnarray*}
whose fifth component vanishes 
in our case since the spin is horizontal.
For the models studied in the present article the three-component
vector $\vec{x}$ is more 
convenient.
\subsection{The probability density of Process 1}
\subsubsection*{2.4.1}
In this section I consider the outcome of scenario 1, which I call
`Process 1'. It is denoted by $\vec{x}^{(1)}(s)$ and I abbreviate:
\begin{eqnarray*}
&&\vec{x}^{(1)}(s) \equiv
       \left( \begin{array}{c}
                \sigma^{(1)}(s)       \\
                \eta^{(1)}(s)       \\
                \psi^{(1)}(s) 
               \end{array}
          \right)  \; .
\end{eqnarray*}
As explained in the
Introduction this process
corresponds to deterministic initial values which I
abbreviate as
\begin{eqnarray}
&&\vec{x}^{(1)}(0) =
       \left( \begin{array}{c}
                \sigma^{(1)}(0)       \\
                \eta^{(1)}(0)       \\
                \psi^{(1)}(0) 
               \end{array}
          \right) 
=       \left( \begin{array}{c}
                <\sigma^{(1)}(0)>       \\
                <\eta^{(1)}(0)>       \\
                <\psi^{(1)}(0)> 
               \end{array}
          \right)  \equiv
       \left( \begin{array}{c}
                \sigma_0       \\
           \eta_0            \\
                \psi_0
                \end{array}
         \right)    \; ,
 \label{eq:2.20}
\end{eqnarray}
where $\sigma_0,\eta_0,\psi_0$ denote arbitrary, but fixed,
real numbers. The process $\vec{x}^{(1)}(s)$ and the orbital process
\begin{eqnarray*}
 \vec{z}^{\,(1)}(s) &\equiv&
       \left( \begin{array}{c}
                \sigma^{(1)}(s)         \\
                \eta^{(1)}(s)
                \end{array}
         \right)
\end{eqnarray*}
are Markovian diffusion processes.
\par My main task in this section is to find the corresponding
probability density, $w_1$. It is easily shown that
\begin{eqnarray}
&&\exp(\underline{\cal A}_{orb}\cdot s)   =
 \frac{i}{2\cdot\lambda}
   \cdot                      \left( \begin{array}{cc}
    g_1(s)  &\qquad  -a\cdot g_2(s)  \\
-b\cdot  g_2(s)  &\qquad  - g_3(s)
                \end{array}
         \right)  \; ,
 \label{eq:2.231}
\end{eqnarray}
where
\begin{eqnarray}
g_1(s)&\equiv& \lambda_2\cdot \exp(\lambda_1\cdot s) - c.c.
 = i\cdot \exp(c\cdot s/2)\cdot\lbrack c\cdot\sin(\lambda\cdot s)
-2\cdot \lambda\cdot\cos(\lambda\cdot s)\rbrack \;,
\nonumber\\
   g_2(s)&\equiv& \exp(\lambda_1\cdot s) - c.c.
 = 2i\cdot \sin(\lambda\cdot s)\cdot\exp(c\cdot s/2) \; ,
\nonumber\\
 g_3(s)&\equiv& \lambda_1\cdot\exp(\lambda_1\cdot s) - c.c.
  = g_2'(s) = i\cdot \exp(c\cdot s/2)\cdot
    \lbrack c\cdot\sin(\lambda\cdot s)
    +2\cdot\lambda\cdot\cos(\lambda\cdot s)\rbrack \; , \nonumber\\&&
 \label{eq:2.232}
\end{eqnarray}
and where  $\lambda_1,\lambda_2$ are the eigenvalues of the
matrix $\underline{\cal A}_{orb}$ and are given by
\footnote{The symbol ${}^*$ denotes complex conjugation.}
\begin{eqnarray*}
  \lambda_1 &\equiv & i\cdot\sqrt{-a\cdot b - c^2/4} +\frac{c}{2}
             \equiv   i\cdot\lambda  +\frac{c}{2}
                                   \; ,\qquad
  \lambda_2  \equiv  \lambda_1^* \; .
\end{eqnarray*}
Note that the `orbital tune' is given by
\begin{eqnarray*}
 Q_{orb} &\equiv& \frac{\lambda\cdot L}{2\cdot\pi} \; ,
\end{eqnarray*}
reflecting the well known fact that the damping causes a small
shift in the orbital tune away from $Q_s$ via the term $c^2/4$.
Note also that $\lambda>0,\lambda_1\cdot\lambda_2>0$, which follows
from (\ref{eq:2.2}),
(\ref{eq:2.212}). If one specifies the constants according to 
(\ref{eq:2.3}), one gets
\begin{eqnarray*}
  \lambda \approx 6.0\cdot 10^{-5}\; m^{-1} \; .
\end{eqnarray*}
\par The first moment vector of Process 1 reads by
(\ref{eq:2.19}),(\ref{eq:2.20}) as:
\begin{eqnarray}
&&<\vec{x}^{(1)}(s)>  =
               \left( \begin{array}{c}
         <\sigma^{(1)}(s)>  \\
         <\eta^{(1)}(s)>    \\
         <\psi^{(1)}(s)>
                \end{array}
         \right)
  =                   \exp(\underline{\cal A}_{I}\cdot s)\cdot
                   <\vec{x}^{(1)}(0)>
\nonumber\\
 &=&
        \large{ \left( \begin{array}{c}
      \frac{i}{2\cdot\lambda}\cdot\biggl( \sigma_0\cdot g_1(s) -
  a\cdot\eta_0\cdot g_2(s) \biggr)    \\ \\
  \frac{i}{2\cdot\lambda}\cdot\biggl(
 -b\cdot\sigma_0\cdot g_2(s) - \eta_0\cdot g_3(s)\biggr)  \\ \\
             \psi_0 - \frac{d}{a}\cdot\sigma_0
 + \frac{i\cdot d}{2\cdot a\lambda}\cdot\biggl( \sigma_0\cdot g_1(s)
 - a \cdot \eta_0\cdot g_2(s)\biggr)
                \end{array}
         \right)}   \; .
 \label{eq:2.281}
\end{eqnarray}
One sees by 
(\ref{eq:2.281})
that $<\sigma^{(1)}(s)>$ and $<\eta^{(1)}(s)>$
damp away with the orbital damping rate $1/\tau_{damp}$.
\par Coming to the covariance matrix $\underline{\sigma}_1$ of Process 1,
one finds by the deterministic initial values
(\ref{eq:2.20}):
\begin{eqnarray}
 \underline{\sigma}_1(0) = 0 \; .
 \label{eq:2.282}
\end{eqnarray}
Therefore the differential equation
(\ref{eq:2.18})
for $\underline{\sigma}_{1}$ is solved by:
\begin{eqnarray}
\underline{\sigma}_{1}(s)&=&       \int_0^s\; ds_1\cdot
    \exp(\underline{\cal A}_{I}\cdot s_1)\cdot
    \underline{{\cal D}}\cdot
    \exp(\underline{\cal A}_{I}^T\cdot s_1)  
\nonumber\\
 &=&    -\frac{\omega}{8\cdot\lambda^2}\cdot
                               \left( \begin{array}{ccccc}
   2\cdot a^2\cdot g_4(s)            & \;\; &
 a\cdot g_2^2(s)&\;\;  & 2\cdot a\cdot d \cdot g_4(s) \\
      && \\
 a\cdot g_2^2(s)    & \; \; &
            2\cdot g_5(s) &\;\;      &     d\cdot g_2^2(s) \\
      && \\
 2\cdot a\cdot d \cdot g_4(s)  &\; \; &    d\cdot g_2^2(s)
                      &\;\;    &     2\cdot d^2\cdot g_4(s)
                \end{array}
         \right) \; ,
 \label{eq:2.283}
\end{eqnarray}
where
\begin{eqnarray}
   g_4(s)&\equiv & \int_0^s ds_1\cdot g_2^2(s_1)
\nonumber\\
 &=& -\frac{1}{abc}\cdot \exp(c\cdot s)
\cdot\lbrack c\cdot\lambda\cdot\sin(2\lambda\cdot s)
  -c^2\cdot\sin^2(\lambda\cdot s)-2\cdot\lambda^2\rbrack
  -\frac{2\lambda^2}{abc}  \; ,
 \label{eq:2.247} \\
 g_5(s) &\equiv &  \int_0^s ds_1\cdot g_3^2(s_1)
\nonumber\\
 &=&  \exp(c\cdot s)
\cdot\lbrack -2\cdot\lambda^2/c - \lambda\cdot
                         \sin(2\lambda\cdot s)
-c\cdot\sin^2(\lambda\cdot s)\rbrack
             +2\cdot\lambda^2/c \; .
 \label{eq:2.248} 
\end{eqnarray}
\begin{figure}[t]
\begin{center}
\epsfig{figure=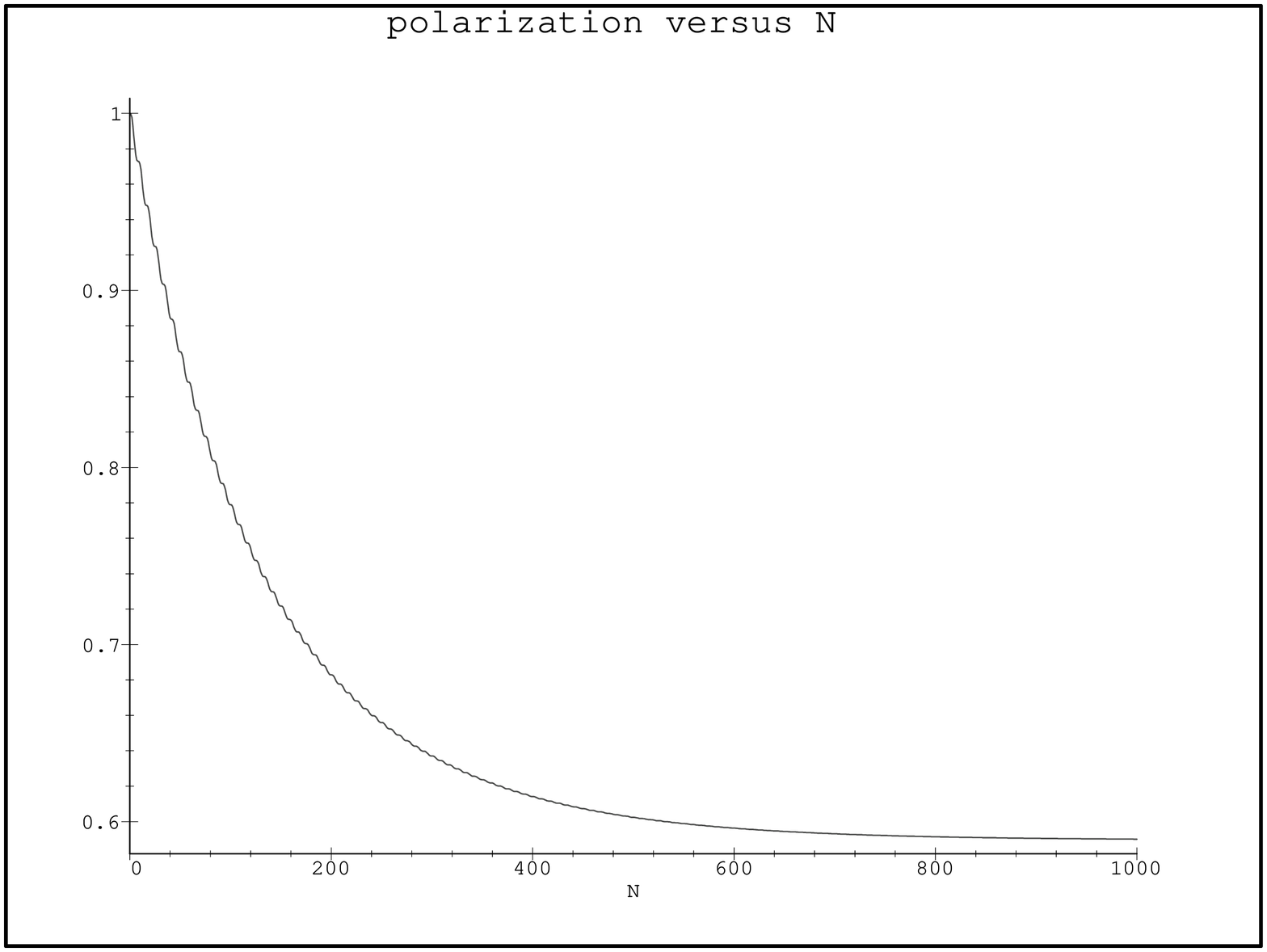,width=12cm,angle=-90}
\end{center}
\caption{Polarization $||\vec P_{tot}^{w_1}(NL)||$ 
of Process 1 for the first 1000 turns
assuming the HERA values 
(\ref{eq:2.3})}
\end{figure}
\subsubsection*{2.4.2}
Because Process 1 is initially deterministic, one finds by
(\ref{eq:2.7}),(\ref{eq:2.20}):
\begin{eqnarray}
 w_1(\sigma,\eta,\psi;0) &=&
\delta(\sigma-\sigma_0)\cdot\delta(\eta-\eta_0)
 \cdot\delta(\psi-\psi_0) \; .
 \label{eq:2.267}
\end{eqnarray}
From this follows by
(\ref{eq:2.13}):
\begin{eqnarray}
    w_{1,orb}(\sigma,\eta;0) &=& 
\delta(\sigma-\sigma_0)\cdot
 \delta(\eta-\eta_0) \; ,
 \label{eq:2.265}
\end{eqnarray}
where $w_{1,orb}$ denotes the orbital part of $w_1$.
Denoting the characteristic function of Process 1 by $\Phi_1$ I obtain via
(\ref{eq:2.222}),(\ref{eq:2.267}):
\begin{eqnarray}
 \Phi_1(\vec{u};0) &=&
   \exp\biggl(i\cdot\vec u^T\cdot<\vec x^{(1)}(0)>\biggr) \; .
 \label{eq:2.280}
\end{eqnarray}
Equations 
(\ref{eq:2.41}),
(\ref{eq:2.280})
pose an initial value problem and it
is easily checked by substitution and by using
(\ref{eq:2.18}),
(\ref{eq:2.19})
that its solution is given by:
\begin{eqnarray}
 \Phi_1(\vec{u};s) &=& \exp\biggl(-\frac{1}{2}\cdot
 \sum_{j,k=1}^3 \; \sigma_{1,jk}(s)\cdot u_j\cdot u_k
+i\cdot \vec{u}^T\cdot <\vec{x}^{(1)}(s)>\biggr) \; .
 \label{eq:2.269}
\end{eqnarray}
By (\ref{eq:2.283}) one observes:
\begin{eqnarray}
&& \sigma_{1,13}(s) = \frac{d}{a}\cdot\sigma_{1,11}(s) \; , \qquad
 \sigma_{1,23}(s) = \frac{d}{a}\cdot\sigma_{1,12}(s) \; , \qquad
 \sigma_{1,33}(s) = \frac{d^2}{a^2}\cdot\sigma_{1,11}(s) \; ,
 \label{eq:2.284}
\end{eqnarray}
so that from (\ref{eq:2.40}),(\ref{eq:2.269}) 
it follows that:
\begin{eqnarray}
 w_1(\sigma,\eta,\psi;s) &=&
 w_{1,orb}(\sigma,\eta;s) \cdot
\delta\biggl(\psi-\psi_0 -\frac{d}{a}\cdot(\sigma-\sigma_0)\biggr) \; .
 \label{eq:2.23}
\end{eqnarray}
By (\ref{eq:2.269}),
(\ref{eq:2.284})
the characteristic function $\Phi_{1,orb}$ corresponding
to $w_{1,orb}$ (see (\ref{eq:2.223})) reads as:
\begin{eqnarray}
 \Phi_{1,orb}(\vec{t};s) &=& \exp\biggl(-\frac{1}{2}\cdot
 \sum_{j,k=1}^2 \; \sigma_{1,orb,jk}(s)\cdot t_j\cdot t_k
+i\cdot \vec{t}^{\;T}\cdot <\vec{z}^{\,(1)}(s)>\biggr) \; ,
 \label{eq:2.226}
\end{eqnarray}
where $\underline{\sigma}_{1,orb}$ denotes the orbital covariance matrix 
of Process 1, 
which by (\ref{eq:2.283}) reads as:
\begin{eqnarray}
  &&  \underline{\sigma}_{1,orb}(s) =
                              \left( \begin{array}{cc}
  \sigma_{1,11}(s)   &   \sigma_{1,12}(s)   \\
  \sigma_{1,21}(s)   &   \sigma_{1,22}(s)
                \end{array}
         \right)  
 =
     -\frac{\omega}{8\cdot\lambda^2}\cdot
                               \left( \begin{array}{cc}
   2\cdot a^2\cdot g_4(s)            &
 a\cdot g_2^2(s)    \\
 a\cdot g_2^2(s)    &
            2\cdot g_5(s)
                \end{array}
         \right) \; .
 \label{eq:2.28}
\end{eqnarray}
By inserting the expression for   $\Phi_{1,orb}$
into (\ref{eq:2.24})
one sees  that $w_{1,orb}$   is Gaussian 
\footnote{I take the usual definition of `Gaussian', which
implies that the covariance matrix is nonsingular.}
in $\sigma,\eta$ of the form
\begin{eqnarray}
&&  w_{1,orb}(\sigma,\eta;s)
                        = \frac{1}{2\pi}\cdot
\det\biggl( \underline{\sigma}_{1,orb}(s)\biggr)^{-1/2}
\nonumber\\
 &&\qquad\cdot \exp\biggl(-\frac{1}{2}\cdot
       \left( \begin{array}{c}
            \sigma-<\sigma^{(1)}(s)> \\
            \eta-<\eta^{(1)}(s)>
                \end{array}
         \right)^T\cdot
    \underline{\sigma}_{1,orb}^{-1}(s)\cdot
       \left( \begin{array}{c}
            \sigma-<\sigma^{(1)}(s)> \\
            \eta-<\eta^{(1)}(s)>
                \end{array}
         \right) \biggr) \; ,
 \label{eq:2.27}
\end{eqnarray}
if $\underline{\sigma}_{1,orb}(s)$ is nonsingular.
\par I now show that
$\underline{\sigma}_{1,orb}(s)$ is nonsingular for $s>0$.
By (\ref{eq:2.282}) $\det(\underline{\sigma}_{1,orb}(0))$ vanishes, 
so that one obtains via
(\ref{eq:2.246}):
\begin{eqnarray}
 \det\biggl(\underline{\sigma}_{1,orb}(s)\biggr) &=&
\omega\cdot \int_{0}^s\; ds_1\cdot\exp\biggl(2\cdot c\cdot(s-s_1)\biggr)
\cdot\sigma_{11}(s_1) \; ,
\label{eq:2.249}
\end{eqnarray}
which by (\ref{eq:2.28}) simplifies to:
\begin{eqnarray}
 \det\biggl(\underline{\sigma}_{1,orb}(s)\biggr) &=&
-\frac{a^2\cdot\omega^2}{4\lambda^2}\cdot 
\int_{0}^s\; ds_1\cdot\exp\biggl(2\cdot c\cdot(s-s_1)\biggr)
\cdot g_4(s_1) \; .
\label{eq:2.250}
\end{eqnarray}
By (\ref{eq:2.232}) one sees that $g_2^2$ is nonpositive so that  
by (\ref{eq:2.247}) $g_4(s)$ is nonpositive and monotonically
decreasing for increasing $s$. Also one obtains by
(\ref{eq:2.247}):
\begin{eqnarray}
&& g_4(0)=g_4'(0)=g_4''(0) = 0 \; , \qquad g_4'''(0)= -8\cdot\lambda^2 < 0 \; .
\label{eq:2.251}
\end{eqnarray}
By the above mentioned properties of $g_4$ it is clear
that $g_4(s)<0$ for $s>0$ so that by
(\ref{eq:2.250}) $\det(\underline{\sigma}_{1,orb}(s))$
is positive for $s>0$. Hence the orbital covariance matrix 
(\ref{eq:2.28})
is nonsingular
for $s>0$ so that in fact $w_{1,orb}$ is Gaussian for $s>0$.
\par Finally from 
(\ref{eq:2.28})
and for $s>0$ it follows that:
\begin{eqnarray*}
\underline{\sigma}_{1,orb}^{-1}(s)&=&
    -\det\biggl(\underline{\sigma}_{1,orb}(s)\biggr)^{-1}
  \cdot\frac{\omega}{8\cdot \lambda^2}
       \cdot
                               \left( \begin{array}{cc}
            2\cdot g_5(s)            &
-a\cdot g_2^2(s)    \\
-a\cdot g_2^2(s)    &
   2\cdot a^2\cdot g_4(s)
                \end{array}
         \right)
\nonumber\\
&=&-\frac{8\cdot\lambda^2}{\omega\cdot a^2}
    \cdot\lbrack 4\cdot g_4(s)\cdot g_5(s) - g_2^4(s) \rbrack^{-1}
       \cdot
                               \left( \begin{array}{cc}
            2\cdot g_5(s)            &
-a\cdot g_2^2(s)    \\
-a\cdot g_2^2(s)    &
   2\cdot a^2\cdot g_4(s)
                \end{array}
         \right)  \; .
\end{eqnarray*}
Now I have made the probability density $w_1$ of  Process 1 explicit.
It is defined by (\ref{eq:2.23}), where
$w_{1,orb}$ is given for $s=0$ by (\ref{eq:2.265})
and for $s>0$ by (\ref{eq:2.27}).
The probability density $w_1$ fulfills the Fokker-Planck equation
(\ref{eq:2.8}) and the
normalization condition 
(\ref{eq:2.9}).
One sees that $w_1$ factors for $s>0$ into a Gaussian function and a delta
function. However 
$w_1$ is not Gaussian because, as follows from (\ref{eq:2.284}),
the covariance matrix of Process 1 is singular.
One thus observes the rather unusual feature that not every
process running with Machine I has a nonsingular covariance matrix. 
\par As mentioned in section 2.2.2
the three-component differential equation
(\ref{eq:2.1})
has only two nontrivial components. This is reflected by
Process 1 because the probability density 
(\ref{eq:2.23})
can be written as:
\begin{eqnarray*}
 w_1(\sigma,\eta,\psi;s) &=&
 w_{1,orb}(\sigma,\eta;s) \cdot
\delta\biggl(\tilde{\psi}-<\tilde{\psi}^{(1)}(0)> \biggr)\; ,
\end{eqnarray*}
where:
\begin{eqnarray*}
 && \tilde{\psi} \equiv \psi -\frac{d}{a}\cdot\sigma  \; , \qquad
 \tilde{\psi}^{(1)}(s) \equiv
   \psi^{(1)}(s) -\frac{d}{a}\cdot\sigma^{(1)}(s) \; .
\end{eqnarray*}
\subsection{Further properties of Process 1 and the transition
probability density for Machine I}
\subsubsection*{2.5.1}
For $s>0$
the spin part $w_{1,spin}$ of $w_1$ has by 
(\ref{eq:2.16}),
(\ref{eq:2.23}),
(\ref{eq:2.27})
the form:
\begin{eqnarray}
 w_{1,spin}(\psi;s) &=&
                             \int_{-\infty}^{+\infty} \; d\sigma
                             \int_{-\infty}^{+\infty} \; d\eta \cdot
   w_1(\sigma,\eta,\psi;s)
\nonumber\\
 &=& \biggl(2\pi\cdot
                       \sigma_{1,33}(s)\biggr)^{-1/2}
 \cdot\exp\biggl(
    -\frac{\biggl(\psi-<\psi^{(1)}(s)>\biggr)^2}{2\cdot\sigma_{1,33}(s)}
   \biggr) \; ,
 \label{eq:2.29}
\end{eqnarray}
and for $s=0$ one has by
(\ref{eq:2.16}),
(\ref{eq:2.267}):
\begin{eqnarray}
 w_{1,spin}(\psi;0) &=&
    \delta(\psi-\psi_0) \; .
 \label{eq:2.268}
\end{eqnarray}
With 
(\ref{eq:2.29}),
(\ref{eq:2.268})
one can easily calculate the polarization vector
for Process 1 
in the \\$(\vec{m}_{0,I},\vec{l}_{0,I},\vec{n}_{0,I})$-frame:
\begin{eqnarray}
&&  \vec{P}_{tot}^{w_1}(s) =
  \frac{2}{\hbar}\cdot <\vec{S}^{(1)}(s)>
 =                     \int_{-\infty}^{+\infty} \; d\sigma
                             \int_{-\infty}^{+\infty} \; d\eta
                             \int_{-\infty}^{+\infty} \; d\psi\cdot
   w_1(\sigma,\eta,\psi;s)\cdot
  \left( \begin{array}{c}
                      \cos(\psi) \\
                      \sin(\psi) \\
                      0
                \end{array}
         \right)
\nonumber\\
 &=&   \int_{-\infty}^{+\infty} \; d\psi\cdot
   w_{1,spin}(\sigma,\eta,\psi;s)\cdot
  \left( \begin{array}{c}
                      \cos(\psi) \\
                      \sin(\psi) \\
                      0
                \end{array}
         \right)
\nonumber\\
&=& \exp(-\sigma_{1,33}(s)/2)\cdot
  \left( \begin{array}{c}
                \cos\biggl(<\psi^{(1)}(s)>\biggr) \\
                \sin\biggl(<\psi^{(1)}(s)>\biggr) \\
                      0
                \end{array}
         \right)
                  \; ,
 \label{eq:2.30}
\end{eqnarray}
where:
\begin{eqnarray*}
 \vec{S}^{(1)}(s)  &\equiv& \frac{\hbar}{2}\cdot
  \left( \begin{array}{c}
        \cos(\psi^{(1)}(s)) \\
        \sin(\psi^{(1)}(s)) \\
                      0
                \end{array}
         \right)  \; .
\end{eqnarray*}
The polarization 
is thus
\begin{eqnarray}
  ||\vec P_{tot}^{w_1}(s)|| &=&
    \exp\biggl(-\sigma_{1,33}(s)/2\biggr) \;
 \label{eq:2.214}
\end{eqnarray}
and is consistent with the requirement that
\begin{eqnarray*}
  ||\vec P_{tot}^{w_1}(0)|| &=&   1 \; .
\end{eqnarray*}
\begin{figure}[t]
\begin{center}
\epsfig{figure=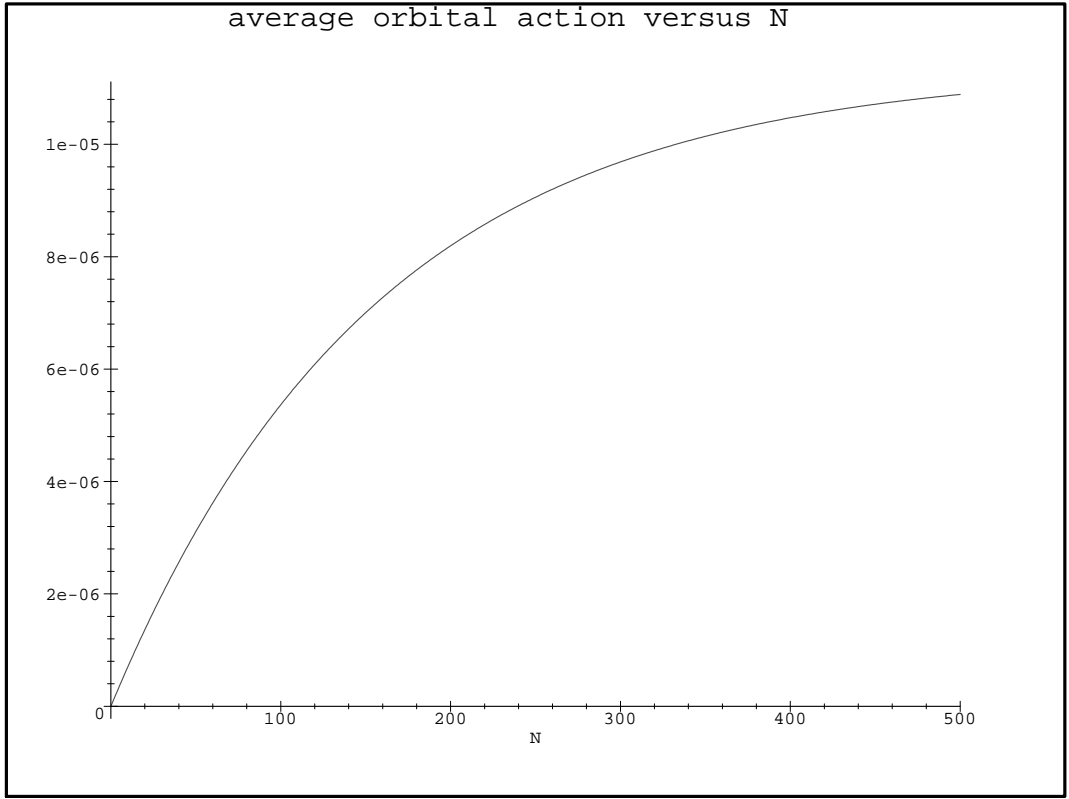,width=12cm,angle=-90}
\end{center}
\caption{The stochastic average $<J_{orb}^{(1)}(NL)>$ of the orbital action 
variable for the first 500 turns
of Process 1 assuming the HERA values 
(\ref{eq:2.3})
with $\sigma_0=\eta_0=0$}
\end{figure}
\subsubsection*{2.5.2}
For the far future, i.e. for $s=+\infty$, the covariance matrix
has by
(\ref{eq:2.283})
the form
\begin{eqnarray*}
\underline{\sigma}_1(+\infty)&=&
                     \large{          \left( \begin{array}{ccc}
  \sigma_{\sigma}^2  &   0   &
\frac{d}{a}\cdot\sigma_{\sigma}^2   \\
   0  &   \sigma_{\eta}^2    & 0 \\
\frac{d}{a}\cdot\sigma_{\sigma}^2
      &   0       &  \sigma_{\psi}^2
                \end{array}
         \right)} \; ,
\end{eqnarray*}
where
\footnote{If one specifies the constants according to 
(\ref{eq:2.3})
one gets
\begin{eqnarray*}
&&\sigma_{\eta}^2 \approx 1.0\cdot 10^{-6} \; , \qquad
\sigma_{\sigma}^2  \approx  0.00013 \, m^2 \; , \qquad
\sigma_{\psi}^2  \approx 1.06 
 \; .
\end{eqnarray*}
By this one also finds that $\tilde{\psi}$ is quite different from $\psi$,
because
\begin{eqnarray*}
&&\tilde{\psi} - \psi  =-\frac{d}{a}\cdot\sigma \approx
                       -\frac{d}{a}\cdot\sigma_{\sigma}
                       = \sigma_{\psi} \approx 1.03 
\; .
\end{eqnarray*}}
\begin{eqnarray*}
\sigma_{\sigma}^2 &\equiv& \frac{\omega\cdot a}{2bc} > 0
                                                             \; ,\qquad
\sigma_{\eta}^2 \equiv-\frac{\omega}{2c}
                   = \frac{\omega\cdot L}{4\alpha_s}
                  =   -\frac{b}{a}\cdot\sigma_{\sigma}^2  >  0 \; ,
\qquad
\sigma_{\psi}^2  \equiv
                     \frac{d^2}{a^2}\cdot\sigma_{\sigma}^2
      =  \frac{\nu^2\cdot\sigma_{\eta}^2}{Q_s^2}  >  0 \; .
\end{eqnarray*}
The equilibrium first moment vector reads as:
\begin{eqnarray*}
<\vec{x}^{(1)}(+\infty)>&=&(0,0,\psi_0-\frac{d}{a}\cdot\sigma_0)^T \;,
\end{eqnarray*}
which follows from 
(\ref{eq:2.281}).
Therefore, using
(\ref{eq:2.23}),
(\ref{eq:2.27}),
the probability density $w_1$ at $s=+\infty$ reads as
\begin{eqnarray}
&& w_1(\sigma,\eta,\psi;+\infty)  =
 w_{1,orb}(\sigma,\eta;+\infty)\cdot
\delta\biggl(\psi-\psi_0 -\frac{d}{a}\cdot(\sigma-\sigma_0)\biggr)\; ,
 \label{eq:2.31}
\end{eqnarray}
where
\begin{eqnarray}
 && w_{1,orb}(\sigma,\eta;+\infty)
                                 \equiv
   \frac{1}{2\pi\cdot\sigma_{\sigma}\cdot\sigma_{\eta}}\cdot
 \exp\biggl(-\sigma^2/2\sigma_{\sigma}^2-
            \eta^2/2\sigma_{\eta}^2 \biggr)
 \equiv w_{norm}(\sigma,\eta)
                                            \; .
\label{eq:2.215}
\end{eqnarray}
One thus sees that Process 1 reaches equilibrium, i.e. for $s\rightarrow
+\infty$ it approaches a stationary state determined by 
(\ref{eq:2.31}).
Because $\underline{\sigma}_1(+\infty)$ is singular this stationary
state is not Gaussian. In fact it is factored by 
(\ref{eq:2.31})
into a Gaussian
function and a delta function just as at finite $s$.
\par The polarization vector of Process 1 for $s=+\infty$ takes the form
\begin{eqnarray}
   \vec{P}_{tot}^{w_1}(+\infty)
&=& \exp\biggl(-\sigma_{1,33}(+\infty)/2\biggr)\cdot
  \left( \begin{array}{c}
               \cos\biggl(<\psi^{(1)}(+\infty)>\biggr) \\
               \sin\biggl(<\psi^{(1)}(+\infty)>\biggr) \\
                      0
                \end{array}
         \right)
\nonumber\\
&=& \exp(-\frac{d^2\cdot\sigma_{\sigma}^2}{2a^2})\cdot
  \left( \begin{array}{c}
        \cos(\psi_0-(d\cdot\sigma_0)/a) \\ \\
        \sin(\psi_0-(d\cdot\sigma_0)/a) \\ \\
                      0
                \end{array}
         \right)
                                               \; ,
 \label{eq:2.32}
\end{eqnarray}
where I used 
(\ref{eq:2.30}). Then
the polarization of Process 1 is given at $s=+\infty$ by
\begin{eqnarray}
 &&||\vec{P}_{tot}^{w_1}(+\infty)||  =
 \exp(-\frac{d^2\cdot\sigma_{\sigma}^2}{2a^2}) \; .
 \label{eq:2.33}
\end{eqnarray}
So for Process 1 the polarization does not decay completely, i.e.
there is no complete spin decoherence!
If one specifies the constants according to 
(\ref{eq:2.3})
one gets
\begin{eqnarray}
      ||\vec P_{tot}^{w_1}(+\infty)|| &\approx& 0.59 \; .
 \label{eq:2.34}
\end{eqnarray}
So one gets $59\%$ equilibrium polarization,
i.e. only a
moderate spin decoherence as already pointed out in \cite{BBHMR94a,BBHMR94b}.
The detailed $s$-dependence is shown in figure 1 where one sees that the
polarization reaches its asymptotic value after a few $\tau_{damp}$.
Careful inspection of the curve reveals a small ripple at twice the
synchrotron frequency. 
Furthermore one can show that $\sigma_{1,33}$ approaches
its equilibrium value on the scale of half the orbital
damping time $\tau_{damp}$.
\par Conventional wisdom has suggested that $\sigma_{\psi}$ should
increase asymptotically like $\sqrt{s}$ as for any simple diffusion
process. This is not the case as one has just seen. However, for the
simpler two-dimensional pure diffusion problem for $\eta$ and $\psi$
without synchrotron oscillations  the $\sqrt{s}$ growth does emerge and
for HERA it would result in a complete decoherence after a few orbital
damping times. So synchrotron motion is an essential ingredient.
\subsubsection*{2.5.3}
In the absence of synchrotron radiation 
($c=\omega=0$)
the orbital equations of
motion 
(\ref{eq:2.6a})
reduce to Hamiltonian equations of motion for the
Hamiltonian
\begin{eqnarray*}
H_{orb} &\equiv& - \frac{b}{2}\cdot\sigma^2
               + \frac{a}{2}\cdot\eta^2  \; .
\end{eqnarray*}
The Poisson bracket relation for $\sigma$ and $\eta$ is:
\begin{eqnarray*}
\lbrace \sigma,\eta \rbrace &=&1 \; .
\end{eqnarray*}
Introducing the abbreviations
\begin{eqnarray*}
 \lambda_0 &\equiv& \sqrt{-a\cdot b} \; ,
\end{eqnarray*}
one gets:
\begin{eqnarray*}
  Q_s &=& \frac{\lambda_0\cdot L}{2\cdot\pi} \; ,
\end{eqnarray*}
and the `orbital action' variable reads as
\begin{eqnarray*}
J_{orb} &\equiv&-\frac{L}{2\cdot\pi\cdot Q_s}\cdot H_{orb}
  =   \sqrt{-\frac{b}{4a}}\cdot \sigma^2
    + \sqrt{-\frac{a}{4b}}\cdot \eta^2
                                                \; .
\end{eqnarray*}
The corresponding orbital phase variable $\phi$ is defined by:
\begin{eqnarray*}
&&\sigma =
   (-\frac{a}{b})^{1/4}\cdot\sqrt{2J_{orb}}\cdot\cos(\phi)\;,\qquad
\eta  =  -(-\frac{b}{a})^{1/4}\cdot\sqrt{2J_{orb}}\cdot \sin(\phi)\; .
\end{eqnarray*}
Then
\begin{eqnarray*}
\lbrace \phi, J_{orb} \rbrace &=&1 \; ,
\end{eqnarray*}
so that $J_{orb},\phi$ are action-angle variables for the Hamiltonian
$H_{orb}$. In the presence of radiation the average action for Process 1 
takes the form
\begin{eqnarray*}
<J_{orb}^{(1)}(s)> &=& \sqrt{-\frac{b}{4a}}\cdot
                                 <\biggl(\sigma^{(1)}(s)\biggr)^2>
 + \sqrt{-\frac{a}{4b}}\cdot
                 <\biggl(\eta^{(1)}(s)\biggr)^2>
\nonumber\\
&=& \sqrt{-\frac{b}{4a}}\cdot\biggl\lbrack
   <\biggl(\sigma^{(1)}(s)\biggr)^2>
 - \frac{a}{b}\cdot <\biggl(\eta^{(1)}(s)\biggr)^2> \biggr\rbrack
\nonumber\\
&=& \sqrt{-\frac{b}{4a}}\cdot\biggl(
   \sigma_{1,11}(s)+<\sigma^{(1)}(s)>^2 -\frac{a}{b}\cdot
       \lbrack \sigma_{1,22}(s)+<\eta^{(1)}(s)>^2\rbrack\biggr)
\nonumber\\
&=& -\frac{1}{4\cdot\lambda^2}\cdot \sqrt{-\frac{b}{4a}} \cdot
                                     \biggl(
a^2\omega\cdot g_4(s) +\lbrack\sigma_0\cdot g_1(s)
 - a\cdot\eta_0\cdot g_2(s)\rbrack^2
-\frac{a\omega}{b}\cdot g_5(s)
\nonumber\\
&&\qquad -\frac{a}{b}\cdot\lbrack b\cdot\sigma_0\cdot g_2(s)
 +\eta_0\cdot g_3(s)\rbrack^2\biggr) \; .
\end{eqnarray*}
Note that the equilibrium value $<J_{orb}^{(1)}(+\infty)>$
is independent of $\sigma_0,\eta_0$:
\footnote{From section 2.9 it is clear that every process running with
Machine I has this equilibrium average value of $J_{orb}$.}
\begin{eqnarray*}
 <J_{orb}^{(1)}(+\infty)> &=&
\sqrt{-\frac{a}{b}} \cdot \sigma_{\eta}^2  \; .
\end{eqnarray*}
If one chooses $\sigma_0=\eta_0=0$ one gets:
\begin{eqnarray*}
<J_{orb}^{(1)}(s)> &=& -\frac{\omega}
                                 {4\cdot c\cdot\lambda^2}\cdot
\sqrt{-\frac{a}{b}} \cdot \biggl( -\exp(c\cdot s)\cdot\lbrack
2\cdot \lambda^2 + c^2\cdot\sin^2(\lambda\cdot s) \rbrack
+2\cdot \lambda^2 \biggr) \; .
\end{eqnarray*}
To illustrate the influence of the synchrotron radiation on the
orbital motion of Process 1, I display this
$<J_{orb}^{(1)}(s)>$ in figure 2 for the first 500 turns, where I
assume the HERA values 
(\ref{eq:2.3})
and $\sigma_0=\eta_0=0$. The stochastic average
$<J_{orb}^{(1)}(s)>$ reaches its asymptotic level after a few
$\tau_{damp}$ and with these parameters the $\sin^2(\lambda\cdot s)$
term gives a negligible contribution.
Note that with large $\sigma_0$ and $\eta_0$ the curve
could approach $<J_{orb}^{(1)}(+\infty)>$ from above.
\par In the radiationless case, i.e. in the limit, 
where $c,\omega\rightarrow 0$, the Fokker-Planck
equation (\ref{eq:2.8}) reduces to the Liouville equation:
\begin{eqnarray}
   \frac{\partial w}{\partial s} &=&
\lbrace H_{orb} \; , \; w \rbrace \; .
 \label{eq:2.234}
\end{eqnarray}
\subsubsection*{2.5.4}
Because Process 1 has deterministic initial values, its probability
density determines the transition probability density 
$w_{I,trans}$ \cite{Gar85} 
of all processes with standard boundary conditions, as shown below. 
In turn for every such process the probability density obeys
for $s_1\leq s$:
\begin{eqnarray}
&& w(\sigma,\eta,\psi;s) \nonumber\\&&\qquad
=  \int_{-\infty}^{+\infty} \; d\sigma_1
                             \int_{-\infty}^{+\infty} \; d\eta_1
                             \int_{-\infty}^{+\infty} \; d\psi_1\cdot
\; w_{I,trans}(\sigma,\eta,\psi;s|\sigma_1,\eta_1,\psi_1;s_1)
            \cdot w(\sigma_1,\eta_1,\psi_1;s_1) \; .
\nonumber\\
 \label{eq:2.216}
\end{eqnarray}
In particular the transition probability density
is nonnegative and normalized by:
\begin{eqnarray}
&& 1 =  \int_{-\infty}^{+\infty} \; d\sigma
                             \int_{-\infty}^{+\infty} \; d\eta
                             \int_{-\infty}^{+\infty} \; d\psi\cdot
 w_{I,trans}(\sigma,\eta,\psi;s|\sigma_1,\eta_1,\psi_1;s_1) \; .
\end{eqnarray}
In this case one has:
\begin{eqnarray*}
w_{I,trans}(\sigma,\eta,\psi;s|\sigma_0,\eta_0,\psi_0;0) &=&
w_1(\sigma,\eta,\psi;s) \; .
\end{eqnarray*}
Because the Langevin equation 
(\ref{eq:2.1})
is $s$-independent, the transition
probability density obeys: 
\begin{eqnarray}
w_{I,trans}(\sigma,\eta,\psi;s|\sigma_0,\eta_0,\psi_0;s_1) &=&
w_{I,trans}(\sigma,\eta,\psi;s-s_1|\sigma_0,\eta_0,\psi_0;0)\; ,
 \label{eq:2.35}
\end{eqnarray}
i.e.:
\begin{eqnarray*}
&&  w_{I,trans}(\sigma,\eta,\psi;s|\sigma_0,\eta_0,\psi_0;s_1)  =
w_1(\sigma,\eta,\psi;s-s_1) \; ,
\end{eqnarray*}
where $s_1\leq s$. From this it finally follows by
(\ref{eq:2.23}),
(\ref{eq:2.35}) that:
\begin{eqnarray}
&&  w_{I,trans}(\sigma,\eta,\psi;s|\sigma_0,\eta_0,\psi_0;s_1)  =
 w_{1,orb}(\sigma,\eta;s-s_1)
          \cdot
\delta\biggl(\psi-\psi_0-\frac{d}{a}\cdot(\sigma-\sigma_0)\biggr)\; .
 \label{eq:2.36}
\end{eqnarray}
Note that the transition probability density is only defined for $s_1\leq s$.
It also fulfills:
\begin{eqnarray}
   \frac{\partial w_{I,trans}}{\partial s} &=&
 - a\cdot \eta \cdot
\frac{\partial w_{I,trans}}{\partial \sigma}
- b\cdot\sigma \cdot \frac{\partial w_{I,trans}}{\partial \eta}
 - d\cdot \eta
               \cdot \frac{\partial w_{I,trans}}{\partial \psi}
\nonumber\\
&&
 -
 c\cdot w_{I,trans}
- c\cdot\eta \cdot \frac{\partial w_{I,trans}}{\partial \eta}
          + \frac{\omega}{2}\cdot
\frac{\partial^2 w_{I,trans}}{\partial \eta^2} \; ,
 \label{eq:2.217}
\end{eqnarray}
and the following initial condition:
\begin{eqnarray}
 w_{I,trans}(\sigma,\eta,\psi;s_1|\sigma_1,\eta_1,\psi_1;s_1) &=&
\delta(\sigma-\sigma_1)\cdot\delta(\eta-\eta_1)
 \cdot\delta(\psi-\psi_1) \; .
 \label{eq:2.218}
\end{eqnarray}
One sees by 
(\ref{eq:2.216})
that the probability density
has a causal azimuthal evolution, i.e. 
$w(\sigma,\eta,\psi;s_1)$ determines $w$ at a later
azimuth $s$. 
The transition probability density
$w_{I,trans}$ is independent of the process, 
and is hence a Green function for the Fokker-Planck equation 
(\ref{eq:2.8}) corresponding to the standard boundary conditions.
\par Given the probability density $w$ and the transition probability density
$w_{I,trans}$ the `joint probability density' $w_{joint}$ of the process 
is defined as:
\begin{eqnarray}
&& w_{joint}(\sigma,\eta,\psi;s;\sigma_1,\eta_1,\psi_1;s_1) \equiv
 w_{I,trans}(\sigma,\eta,\psi;s|\sigma_1,\eta_1,\psi_1;s_1)\cdot
 w(\sigma_1,\eta_1,\psi_1;s_1) \; . \qquad
 \label{eq:2.235}
\end{eqnarray}
Note that the joint probability density is only defined for $s_1\leq s$ and 
it is used in section 2.9.4.
\subsubsection*{2.5.5}
Statements analogous to those in the previous section can be made
about the orbital part of Machine I. Thus all of the statements
in section 2.5.4 are valid when $w$ is replaced by $w_{orb}$ and the
variable $\psi$ is omitted.  In particular the orbital transition
probability density $w_{orb,trans}$ for all processes with standard
boundary conditions reads as:
\begin{eqnarray}
 w_{orb,trans}(\sigma,\eta;s|\sigma_0,\eta_0;s_1) &=&
 w_{1,orb}(\sigma,\eta;s-s_1) \; ,
 \label{eq:2.37}
\end{eqnarray}
where $w_{1,orb}$ is given by 
(\ref{eq:2.265}),
(\ref{eq:2.27}) and where $s_1\leq s$. 
For the orbital part of a probability
density one obtains for $s_1\leq s$:
\begin{eqnarray}
&& w_{orb}(\sigma,\eta;s) =  \int_{-\infty}^{+\infty} \; d\sigma_1
                             \int_{-\infty}^{+\infty} \; d\eta_1
\; w_{orb,trans}(\sigma,\eta;s|\sigma_1,\eta_1;s_1)
            \cdot w_{orb}(\sigma_1,\eta_1;s_1) \; .
 \label{eq:2.219}
\end{eqnarray}
In particular the orbital transition probability density is nonnegative and 
normalized by:
\begin{eqnarray}
&& 1 =  \int_{-\infty}^{+\infty} \; d\sigma
                             \int_{-\infty}^{+\infty} \; d\eta\cdot
 w_{orb,trans}(\sigma,\eta;s|\sigma_1,\eta_1;s_1) \; .
\end{eqnarray}
Note that the orbital transition probability density
is only defined for $s_1\leq s$.
The orbital transition probability density fulfills:
\begin{eqnarray}
   \frac{\partial w_{orb,trans}}{\partial s} &=&
  L_{FP,orb} \; w_{orb,trans} \; ,
 \label{eq:2.220}
\end{eqnarray}
and the following initial conditions:
\begin{eqnarray}
 w_{orb,trans}(\sigma,\eta;s_1|\sigma_1,\eta_1;s_1) &=&
\delta(\sigma-\sigma_1)\cdot\delta(\eta-\eta_1) \; .
 \label{eq:2.221}
\end{eqnarray}
Note that:
\begin{eqnarray}
&& w_{orb,trans}(\sigma,\eta;s=+\infty|\sigma_0,\eta_0;s_1) =
 w_{1,orb}(\sigma,\eta;s=+\infty) =
 w_{norm}(\sigma,\eta)  \; .
 \label{eq:2.261}
\end{eqnarray}
One sees by 
(\ref{eq:2.219})
that the orbital probability density
has a causal azimuthal evolution, i.e. $w_{orb}(\sigma,\eta;s_1)$ 
determines $w_{orb}$ at a later
azimuth $s$. 
The orbital transition probability density
$w_{orb,trans}$ is independent of the process, 
and is hence a Green function for the orbital Fokker-Planck equation 
(\ref{eq:2.15}) corresponding to the standard boundary conditions.
\par Given $w_{orb}$ and the orbital transition probability density
the `orbital joint probability density' $w_{orb,joint}$ is defined as:
\begin{eqnarray}
&& w_{orb,joint}(\sigma,\eta;s;\sigma_1,\eta_1;s_1) \equiv
 w_{orb,trans}(\sigma,\eta;s|\sigma_1,\eta_1;s_1)\cdot
 w_{orb}(\sigma_1,\eta_1;s_1) \; .
 \label{eq:2.236}
\end{eqnarray}
Note that the orbital joint probability density is only defined for 
$s_1\leq s$ and it will be used in Appendix E.
\subsection{The probability density of Process 2}
\subsubsection*{2.6.1}
Although Process 1 has led to most of the methods needed for
problems of this kind it is too idealized; in an electron
 storage ring it is not possible to have an initial state with
deterministic orbital values, i.e. sharp orbital values at 
the initial azimuth $s=0$, and
complete polarization at the same azimuth 
since an injected beam or a beam at orbital equilibrium always occupies
a nonzero phase space volume.
\par Therefore in this section I consider another process, called
`Process 2', running with Machine I. It solves the Langevin equation
(\ref{eq:2.1})
and fulfills the standard boundary conditions. It is
denoted by $\vec{x}^{(2)}(s)$ and I abbreviate:
\begin{eqnarray*}
&&\vec{x}^{(2)}(s) \equiv
       \left( \begin{array}{c}
                \sigma^{(2)}(s)       \\
                \eta^{(2)}(s)       \\
                \psi^{(2)}(s) 
               \end{array}
          \right)  \; .
\end{eqnarray*}
However, in contrast to Process 1
the orbital variables are not deterministic at $s=0$ but have a Gaussian
distribution with the `equilibrium' probability density
$w_{norm}(\sigma,\eta)$, defined in 
(\ref{eq:2.215}). 
It describes an initial situation with
complete polarization and orbital equilibrium.
Denoting the probability density of Process 2 by $w_2$ one therefore has by
(\ref{eq:2.7}), 
(\ref{eq:2.13}):
\begin{eqnarray}
 w_2(\sigma,\eta,\psi;0) &\equiv&
 w_{norm}(\sigma,\eta)\cdot
\delta (\psi-\psi_0)\; .
 \label{eq:2.38}
\end{eqnarray}
This fulfills 
(\ref{eq:2.9})
and by applying the orbital transition probability density one
gets via 
(\ref{eq:2.219})
the  expected result that Process 2 is at `orbital equilibrium',
i.e. the orbital part $w_{2,orb}$ of $w_2$ has the form
\begin{eqnarray}
 w_{2,orb} &=&
 w_{norm} \; .
 \label{eq:2.39}
\end{eqnarray}
Thus for Process 2 the orbital variables remain  in equilibrium,
i.e. the orbital process
\begin{eqnarray*}
 \vec{z}^{\,(2)}(s) &\equiv&
       \left( \begin{array}{c}
                \sigma^{(2)}(s)         \\
                \eta^{(2)}(s)
                \end{array}
         \right)
\end{eqnarray*}
is stationary. 
Note also that $\vec{x}^{(2)}(s)$ and
$\vec{z}^{\,(2)}(s)$ are Markovian diffusion
processes.
\begin{figure}[t]
\begin{center}
\epsfig{figure=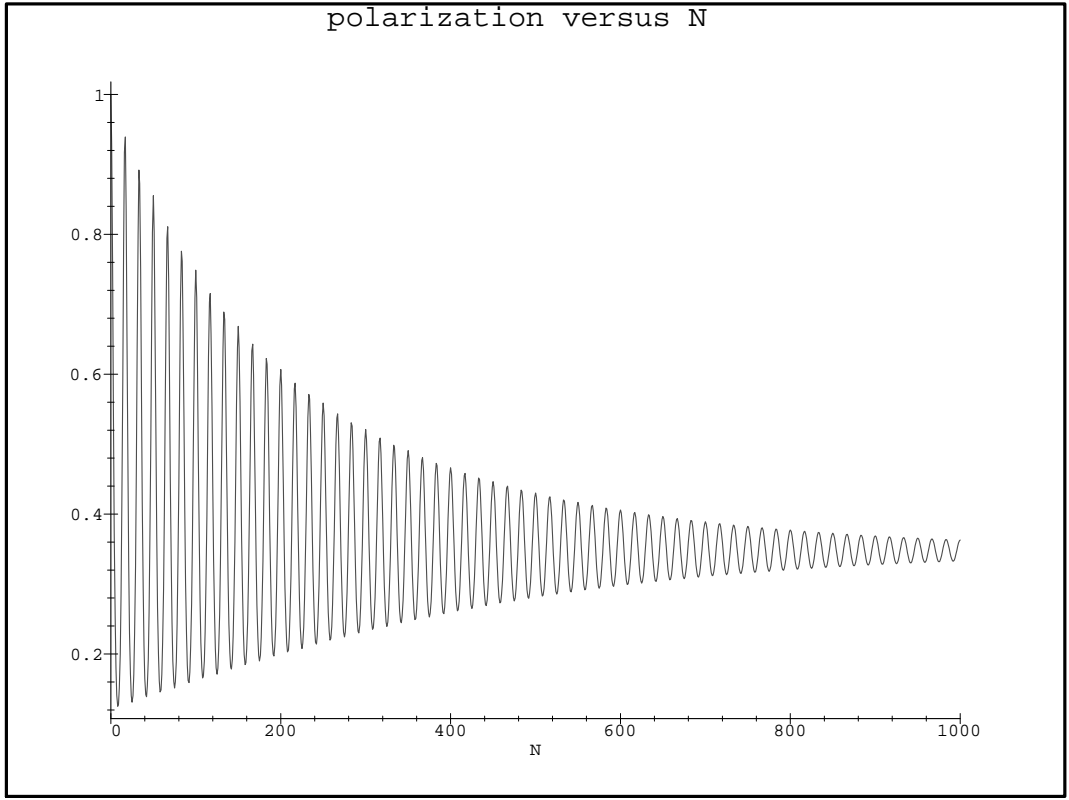,width=12cm,angle=-90}
\end{center}
\caption{Polarization $||\vec P_{tot}^{w_2}(NL)||$ 
of Process 2 for the first 1000 turns
assuming the HERA values
(\ref{eq:2.3})}
\end{figure}
\subsubsection*{2.6.2}
To obtain the probability density of Process 2 in explicit form
I again use the characteristic function.
Because of the initial conditions 
(\ref{eq:2.38})
one gets:
\begin{eqnarray}
 \Phi_2(\vec{u};0) &=&
   \exp\biggl(-\frac{1}{2}\cdot\sigma_{\sigma}^2\cdot u_1^2
    -\frac{1}{2}\cdot\sigma_{\eta}^2\cdot u_2^2
+i\cdot u_3\cdot \psi_0\biggr) \; .
 \label{eq:2.227}
\end{eqnarray}
Equations
(\ref{eq:2.41}),
(\ref{eq:2.227})
pose an initial value problem and it
is easily checked by 
substitution and by
(\ref{eq:2.18}),
(\ref{eq:2.19})
that its solution is given by:
\begin{eqnarray}
 \Phi_2(\vec{u};s) &=& \exp\biggl(-\frac{1}{2}\cdot
 \sum_{j,k=1}^3 \; \sigma_{2,jk}\cdot u_j\cdot u_k
+i\cdot \vec{u}^T\cdot <\vec{x}^{(2)}(s)>\biggr) \; ,
 \label{eq:2.228}
\end{eqnarray}
where $\underline{\sigma}_2$ denotes the covariance matrix of Process 2.
In addition 
(\ref{eq:2.18}),
(\ref{eq:2.19})
lead to
\begin{eqnarray}
 <\vec{x}^{(2)}(s)> &=& (0,0,\psi_0)^T \; ,
 \label{eq:2.42}
\end{eqnarray}
and:
\begin{eqnarray}
&&\underline{\sigma}_2(s) =
    \exp(\underline{\cal A}_I\cdot s)\cdot
\underline{\sigma}_2(0)\cdot
    \exp(\underline{\cal A}_I^T\cdot s)
+        \int_0^s\; ds_1\cdot
    \exp(\underline{\cal A}_I\cdot s_1)\cdot
    \underline{{\cal D}}\cdot
    \exp(\underline{\cal A}_I^T\cdot s_1)
\nonumber\\&&
\nonumber\\
  &=&
                             \large{  \left( \begin{array}{ccccc}
   \sigma_{\sigma}^2   &      &       0  & &
    \frac{d}{a}\cdot\sigma_{\sigma}^2 \cdot\lbrack  1
    -\frac{i}{2\lambda}\cdot g_1(s)\rbrack    \\
&& \\
   0  & & \sigma_{\eta}^2          & &
    \frac{i\cdot db}{2a\lambda}\cdot\sigma_{\sigma}^2\cdot g_2(s) \\
&& \\
    \frac{d}{a}\cdot\sigma_{\sigma}^2 \cdot\lbrack  1
    -\frac{i}{2\lambda}\cdot g_1(s)\rbrack  &\; &
 \frac{i\cdot db}{2a\lambda}\cdot\sigma_{\sigma}^2\cdot g_2(s) & \; &
     \frac{d^2}{a^2\lambda}\cdot\sigma_{\sigma}^2
    \cdot\lbrack 2\cdot\lambda - i\cdot g_1(s) \rbrack
                \end{array}
         \right)} \; .
\nonumber\\
 \label{eq:2.43}
\end{eqnarray}
Inserting the explicit form of $\Phi_2$ into 
(\ref{eq:2.40})
one finds that $w_2$ is given for $s>0$ by
\begin{eqnarray}
&& w_2(\sigma,\eta,\psi;s)  =
                      \sqrt{(2\pi)^{-3}\cdot
\det( \underline{\sigma}_{2}(s))^{-1}}\cdot
               \exp\biggl\lbrack -\frac{1}{2}\cdot
       \left( \begin{array}{c}
          \sigma \\
          \eta   \\
          \psi -\psi_{0}
                \end{array}
         \right)^T       \cdot
    \underline{\sigma}_{2}^{-1}(s)\cdot
       \left( \begin{array}{c}
          \sigma \\
          \eta   \\
          \psi -\psi_{0}
                \end{array}
         \right)   \biggr\rbrack                    \; , \nonumber\\&&
 \label{eq:2.44}
\end{eqnarray}
so that  $w_2$ is Gaussian for $s>0$ because the covariance matrix
$\underline{\sigma}_2$ is nonsingular for $s>0$.
\footnote{Its determinant has the form:
\begin{eqnarray*}
   \det\biggl( \underline{\sigma}_2(s)\biggr) &=&
\frac{a\cdot d^2\cdot\omega^3}{16\cdot b\cdot c^2\cdot\lambda^2}
\cdot g_4(s) \; ,
\end{eqnarray*}
which is positive for $s>0$ because, as shown in section 2.4.2,
$g_4(s)$ is negative for $s>0$.}
\par By 
(\ref{eq:2.38}),
(\ref{eq:2.44})
$w_2$ fulfills the
normalization condition 
(\ref{eq:2.9}).
\subsubsection*{2.6.3}
For $s=+\infty$ the first moment vector 
(\ref{eq:2.42})
reads as:
\begin{eqnarray*}
<\vec{x}^{(2)}(+\infty)>&=&(0,0,\psi_0)^T \;,
\end{eqnarray*}
and the covariance matrix has the form:
\begin{eqnarray*}
    \underline{\sigma}_2(+\infty)&=&
                               \left( \begin{array}{ccccc}
   \sigma_{\sigma}^2   &      &       0  & &
    (d\cdot\sigma_{\sigma}^2)/a \\
&& \\
   0  & & \sigma_{\eta}^2          & & 0 \\
&& \\
    (d\cdot\sigma_{\sigma}^2)/a
                                            &\; & 0
                                                               & \; &
     (2d^2\cdot\sigma_{\sigma}^2)/a^2
                \end{array}
         \right) \; .
\nonumber\\
\end{eqnarray*}
So Process 2 also reaches equilibrium. However, one sees that processes 1
and 2 approach different stationary states for $s\rightarrow
+\infty$, so that Machine I has no unique equilibrium state. In
particular the equilibrium state for Process 2 is Gaussian whereas for
Process 1 it is not.
\begin{figure}[t]
\begin{center}
\epsfig{figure=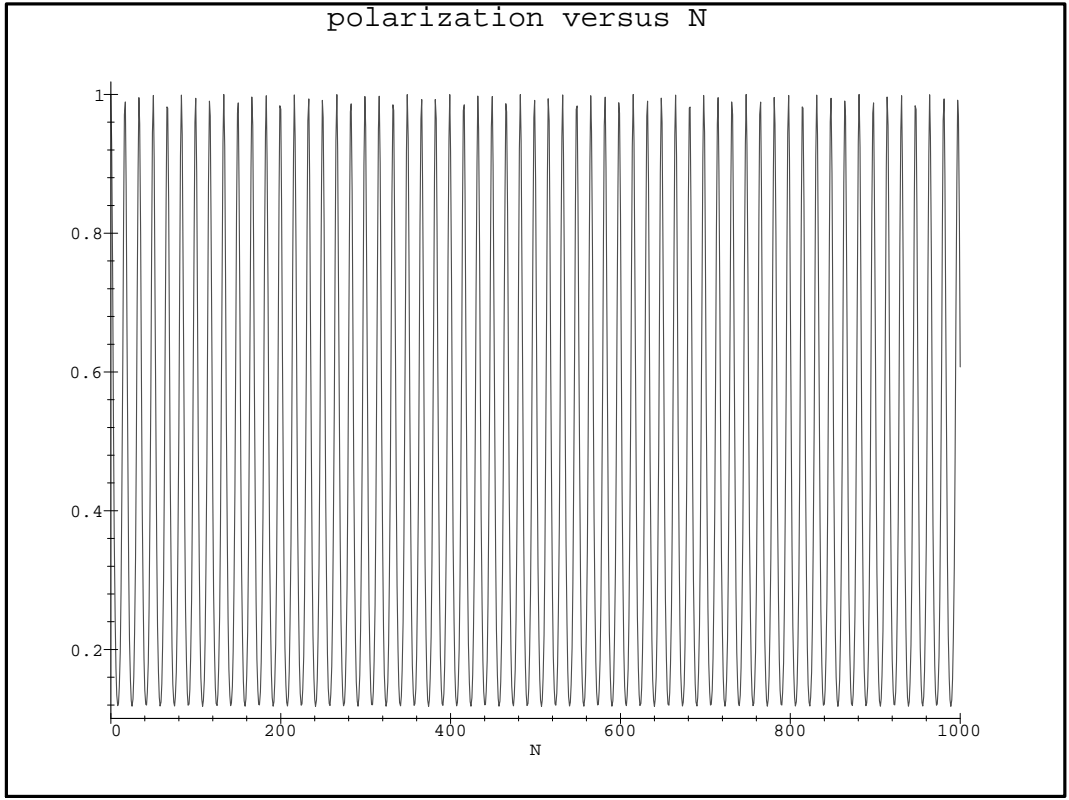,width=12cm,angle=-90}
\end{center}
\caption{Polarization $||\vec P_{tot}^{w_2}(NL)||$ 
of Process 2a for the first 1000 
turns assuming the HERA values 
(\ref{eq:2.3}), except that
$c,\omega\rightarrow 0$ with
$\omega/c=const=-2\cdot\sigma^2_{\eta}\approx -2.0\cdot 10^{-6}$.}
\end{figure}
\subsubsection*{2.6.4}
With 
(\ref{eq:2.38}),
(\ref{eq:2.44})
one can easily calculate the polarization vector for Process 2 in the\\
$(\vec{m}_{0,I},\vec{l}_{0,I},\vec{n}_{0,I})$-frame:
\begin{eqnarray}
&&\vec{P}_{tot}^{w_2}(s)  =
  \frac{2}{\hbar}\cdot <\vec{S}^{(2)}(s)>
 =                     \int_{-\infty}^{+\infty} \; d\sigma
                             \int_{-\infty}^{+\infty} \; d\eta
                             \int_{-\infty}^{+\infty} \; d\psi\cdot
   w_2(\sigma,\eta,\psi;s)\cdot
  \left( \begin{array}{c}
        \cos(\psi) \\
        \sin(\psi) \\
                      0
                \end{array}
         \right)
\nonumber\\
 &=&    \exp(-\sigma_{2,33}(s)/2)\cdot
  \left( \begin{array}{c}
        \cos(\psi_0) \\
        \sin(\psi_0) \\
                      0
                \end{array}
         \right)  \; ,
 \label{eq:2.57}
\end{eqnarray}
where:
\begin{eqnarray*}
 \vec{S}^{(2)}(s)  &\equiv& \frac{\hbar}{2}\cdot
  \left( \begin{array}{c}
        \cos(\psi^{(2)}(s)) \\
        \sin(\psi^{(2)}(s)) \\
                      0
                \end{array}
         \right)  \; .
\end{eqnarray*}
The polarization is then
\begin{eqnarray}
  ||\vec P_{tot}^{w_2}(s)|| &=&
    \exp\biggl(-\sigma_{2,33}(s)/2\biggr) \; ,
 \label{eq:2.204}
\end{eqnarray}
and of course
\begin{eqnarray*}
  ||\vec P_{tot}^{w_2}(0)|| &=&   1 \; .
\end{eqnarray*}
Comparing 
(\ref{eq:2.30}),
(\ref{eq:2.57})
one sees that the polarization vectors of
processes 1 and  2 are different.
\subsubsection*{2.6.5}
The polarization vector of Process 2 for
$s=+\infty$ takes the form
\begin{eqnarray*}
&& \vec{P}_{tot}^{w_2}(+\infty)
 =  \exp\biggl(-\sigma_{2,33}(+\infty)/2\biggr)\cdot
  \left( \begin{array}{c}
        \cos(\psi_0) \\
        \sin(\psi_0) \\
                      0
                \end{array}
         \right)
 =
      \exp(-\frac{d^2\cdot\sigma_{\sigma}^2}{a^2})\cdot
  \left( \begin{array}{c}
        \cos(\psi_0) \\
        \sin(\psi_0) \\
                      0
                \end{array}
         \right)   \; .
\end{eqnarray*}
The polarization of Process 2 at $s=+\infty$, i.e. the equilibrium
value of the polarization,
is therefore given by
\begin{eqnarray}
 &&||\vec{P}_{tot}^{w_2}(+\infty)|| =
 \exp(-\frac{d^2\cdot\sigma_{\sigma}^2}{a^2}) \; .
 \label{eq:2.58}
\end{eqnarray}
So also for Process 2 the polarization does not decay completely, i.e.
there is no complete spin decoherence.
If one specifies the constants according to 
(\ref{eq:2.3})
one gets
\begin{eqnarray*}
      ||\vec P_{tot}^{w_2}(+\infty)|| &\approx& 0.35 \; ,
\end{eqnarray*}
i.e. one gets $35\%$ equilibrium polarization, which is almost a factor
of two smaller than for Process 1.
\par One sees by 
(\ref{eq:2.43})
that
$\sigma_{2,33}(0)=0$ as required and that $\sigma_{2,33}(s)$
approaches its equilibrium value on the scale of the
orbital damping time $\tau_{damp}$, so that the equilibrium
polarization is approached more slowly than for Process 1.
\par The polarization of Process 2 is displayed for the HERA values
(\ref{eq:2.3})
in figure 3 for the first 1000 turns where one sees that
in contrast to Process 1 the spin equilibrium is
reached only after strong oscillations at the synchrotron frequency.
The reason for the difference is clear. In Process 2 the short time
behaviour is dominated by synchrotron motion and the beam has a prepared
energy spread. The damping and diffusion act on a longer time scale. But
in Process 1 there is no initial energy spread.
\par It is also interesting  to study how the polarization would
behave when starting with the equilibrium orbital distribution
but with no synchrotron radiation. I call this `Process 2a'. One could
use a solution based on the first three terms on the rhs of 
(\ref{eq:2.8})
but it is more
convenient to use the result 
(\ref{eq:2.57})
in the limit where
$c,\omega\rightarrow 0$ with
$\omega/c=const\approx -2\cdot\sigma^2_{\eta}\approx -2.0\cdot 10^{-6}$.
In this case the orbital phase space distribution remains unaltered
but the damping and diffusion forces have been turned off.
\footnote{Note that in this limit there is a very small shift
          in $\lambda$.}
The resulting polarization is displayed in figure 4 where one sees that
the polarization never reaches equilibrium and continues to oscillate
strongly at the synchrotron frequency. 
So although the orbital
distributions for processes 2 and 2a are identical the spins behave
very differently owing to the very different `hidden' orbital dynamics.
In Process 2 the spin motion is irreversible. In Process 2a the spins
tend to `remember' their initial distribution. 
\par This figure gives an impression of what could happen if one were
considering protons and
is reminiscent of
the long term polarization oscillations in figure 9 in \cite{HH96}.
In Appendix D I consider the nature of the
equilibrium distribution for $\psi$ in the radiationless case in
more detail.
\par This completes the detailed account of the analytical derivation
of the results for Machine I discussed in \cite{BBHMR94a,BBHMR94b}.
I now continue with further developments of the subject.
\subsection{The polarization density and its Bloch equation for Machine I}
%
%
\subsubsection*{2.7.1}
In this section I introduce the concept of `polarization density'.
\par Given a process with probability
density $w$, the polarization density $\vec{P}^w$ 
in the $(\vec{m}_{0,I},\vec{l}_{0,I},\vec{n}_{0,I})$-frame
is defined by
\begin{eqnarray}
 \vec{P}^w(\sigma,\eta;s) &\equiv&
                             \int_{-\infty}^{+\infty} \; d\psi \cdot
   w(\sigma,\eta,\psi;s)
            \cdot \frac{2}{\hbar}\cdot \vec{S}
\nonumber\\
 &=&                      \int_{-\infty}^{+\infty} \; d\psi \cdot
   w(\sigma,\eta,\psi;s) \cdot
  \left( \begin{array}{c}
        \cos(\psi) \\
        \sin(\psi) \\
                      0
                \end{array}
         \right)   \; .
 \label{eq:2.62}
\end{eqnarray}
One easily sees by 
(\ref{eq:2.11})
that the polarization vector $\vec{P}_{tot}^w$
satisfies
\begin{eqnarray}
 \vec{P}_{tot}^w(s) &=&
      \int_{-\infty}^{+\infty} \; d\sigma
                             \int_{-\infty}^{+\infty} \; d\eta\cdot
                                       \vec{P}^w(\sigma,\eta;s)
                                                 \; ,
 \label{eq:2.63}
\end{eqnarray}
hence the name `polarization density'.
\footnote{The terminology `polarization density' appears for example
in \cite{DK75}.}
$\vec{P}^w(\sigma,\eta;s)$ describes the contribution to the 
polarization vector from a point in the orbital phase space.
The standard boundary conditions of $w$ are
taken into account in 
(\ref{eq:2.62})
via the integration range of $\psi$.
Also one sees by (\ref{eq:2.63}) and the finiteness of the polarization
vector that the polarization density obeys
standard boundary conditions in the variables $\sigma,\eta$.
\par Note that by (\ref{eq:2.23}),
(\ref{eq:2.62}) the polarization density for Process 1 is given by:
\begin{eqnarray}
\vec{P}^{w_1}(\sigma,\eta;s) &=&
 w_{1,orb}(\sigma,\eta;s) \cdot
  \left( \begin{array}{c}
   \cos\biggl(\psi_0+(d\cdot(\sigma-\sigma_0))/a\biggr) \\
   \sin\biggl(\psi_0+(d\cdot(\sigma-\sigma_0))/a\biggr) \\
                      0
                \end{array}
         \right)          \; .
 \label{eq:2.68}
\end{eqnarray}
\par Using the Fokker-Planck equation 
(\ref{eq:2.8})
one finds that the polarization
density obeys the following equation of the Bloch type:
\begin{eqnarray*}
\frac{\partial\vec{P}^w} {\partial s}   &=&
       L_{FP,orb} \; \vec{P}^w +
\vec{W}_I\wedge \vec{P}^w \; ,
\end{eqnarray*}
which follows from 
(\ref{eq:2.8})
by partial integration. 
\footnote{The vector $\vec W_I$ is defined in 
(\ref{eq:2.257}).}
Explicitly one has
\begin{eqnarray}
&& \frac{\partial \vec{P}^w}{\partial s}  =
\underbrace{
-a\cdot \eta \cdot
\frac{\partial \vec{P}^w}{\partial \sigma}
- b\cdot\sigma \cdot \frac{\partial \vec{P}^w}{\partial \eta}
+ \vec{W}_I\wedge \vec{P}^w
                                                }_
               {{\rm radiationless \; part}}
-
  \underbrace{
 c\cdot \vec{P}^w
- c\cdot \eta \cdot \frac{\partial \vec{P}^w}{\partial \eta}
                                                }_
               {{\rm damping \; terms}}
         +
  \underbrace{
          \frac{\omega}{2}\cdot
\frac{\partial^2 \vec{P}^w}{\partial \eta^2}
                                                }_
               {{\rm diffusion \; term}}
                                           \; . \qquad\qquad
 \label{eq:2.64}
\end{eqnarray}
The radiationless Bloch equation reads as :
\begin{eqnarray}
   \frac{\partial \vec{P}^w}{\partial s}  =
-a\cdot \eta \cdot
\frac{\partial \vec{P}^w}{\partial \sigma}
- b\cdot\sigma \cdot \frac{\partial \vec{P}^w}{\partial \eta}
+ \vec{W}_I\wedge \vec{P}^w     \; .
 \label{eq:2.65}
\end{eqnarray}
\par There is an obvious connection between the radiationless Bloch
equation 
(\ref{eq:2.65})
and the Thomas-BMT equation
(\ref{eq:1.5a}). In fact the $s$-dependent vector:
\begin{eqnarray}
&&\vec{P}^w(\sigma(s),\eta(s);s)
 \label{eq:2.276}
\end{eqnarray}
fulfills (\ref{eq:1.5a}) because in the absence of radiation 
$\sigma(s),\eta(s)$ fulfill the following equations of motion:
\begin{eqnarray}
 && \sigma'(s) = a\cdot\eta(s) \; , \qquad
    \eta'(s) = b\cdot\sigma(s) \; .
 \label{eq:2.285}
\end{eqnarray}
One easily sees that this connection between the radiationless Bloch equation
and the Thomas-BMT equation also holds if
$\vec{P}^w$ in
(\ref{eq:2.276}) is replaced by any other quantity obeying 
(\ref{eq:2.65}).
An analogous connection holds for Machine II.
\subsubsection*{2.7.2}
Just as for the Fokker-Planck equation, the Bloch equation (\ref{eq:2.64})
for the polarization density
also has a causal azimuthal evolution, i.e. an initial polarization
density $\vec{P}^w(\sigma,\eta;s_0)$ determines $\vec{P}^w$ at a later
azimuth $s$. In particular there exists a function
$\underline{P}_I(\sigma,\eta;s|\sigma_1,\eta_1;s_1)$, which is a
$3\times 3$-matrix fulfilling for $s_1\leq s$:
\begin{eqnarray}
&& \vec{P}^w(\sigma,\eta;s) = \int_{-\infty}^{+\infty} \; d\sigma_1
                             \int_{-\infty}^{+\infty} \; d\eta_1\cdot
\; \underline{P}_I(\sigma,\eta;s|\sigma_1,\eta_1;s_1)
            \cdot\vec{P}^w(\sigma_1,\eta_1;s_1) \; ,
 \label{eq:2.66}
\end{eqnarray}
so that $\underline{P}_I$ `transports' a polarization
density from one azimuth to another. 
Note that  $\underline{P}_I$ is only defined for $s_1\leq s$.
The function
$\underline{P}_I$ is derived from the transition probability density
$w_{I,trans}$ and it may be shown that it can be written as:
\begin{eqnarray}
   \underline{P}_I(\sigma,\eta;s|\sigma_1,\eta_1;s_1)&\equiv&
   w_{orb,trans}(\sigma,\eta;s|\sigma_1,\eta_1;s_1)\cdot
\underline{R}_I(\sigma,\sigma_1) \; ,
 \label{eq:2.67}
\end{eqnarray}
where the $3\times 3$-matrix $\underline{R}_I$ has the form
\begin{eqnarray*}
\underline{R}_I(\sigma,\sigma_1) &\equiv& \left( \begin{array}{ccc}
  \cos\biggl((d\cdot(\sigma-\sigma_1))/a\biggr) &
- \sin\biggl((d\cdot(\sigma-\sigma_1))/a\biggr) & 0 \\
  \sin\biggl((d\cdot(\sigma-\sigma_1))/a\biggr) &
  \cos\biggl((d\cdot(\sigma-\sigma_1))/a\biggr) & 0 \\
      0 & 0 & 1
                \end{array}
         \right) \; .
\end{eqnarray*}
Using 
(\ref{eq:2.64}),
(\ref{eq:2.66})
one finds that $\underline{P}_I$ fulfills the
following equation:
\begin{eqnarray*}
\frac{\partial\underline{P}_I} {\partial s}   &=&
       L_{FP,orb} \; \underline{P}_I +
\underline{W}_I\cdot \underline{P}_I\; ,
\nonumber\\&&
\end{eqnarray*}
where
\begin{eqnarray*}
 \underline{W}_I&\equiv&
                 \left( \begin{array}{ccc}
 0                  &
 - d\cdot\eta &  0 \\
   d\cdot\eta &  0 & 0 \\
 0 & 0 & 0
              \end{array}
       \right)  \; .
\end{eqnarray*}
Also one finds that $\underline{P}_I$ fulfills the
following initial conditions:
\begin{eqnarray*}
   \underline{P}_I(\sigma,\eta;s_1|\sigma_1,\eta_1;s_1)&=&
\delta(\sigma-\sigma_1)\cdot\delta(\eta-\eta_1)\cdot
                 \left( \begin{array}{ccc}
 1  &  
 0  &
 0  \\
 0  &
 1  &
 0  \\
 0  &
 0  &
 1
              \end{array}
       \right)  \; .
\end{eqnarray*}
One sees that $\underline{P}_I$ is a Green function for the Bloch equation
(\ref{eq:2.64}) corresponding to the standard boundary conditions.
\par In the radiationless case the 
orbital transition probability density
$w_{orb,trans}$ modifies to \\
$w_{orb,trans,nrad}$, where:
\begin{eqnarray}
&& w_{orb,trans,nrad}(\sigma,\eta;s|\sigma_1,\eta_1;s_1) =
 w_{orb,trans,nrad}(\vec z;s|\vec z_1;s_1) \nonumber\\
&&\qquad=
\delta\biggl(\vec z-\exp((s-s_1)\cdot\underline{\cal A}_{orb,nrad})\cdot
\vec z_1\biggr) \; ,
 \label{eq:2.270}
\end{eqnarray}
with
\begin{eqnarray*}
&&\underline{\cal A}_{orb,nrad} \equiv        \left( \begin{array}{cc}
      0 & a                             \\
      b & 0
                \end{array}
         \right) \; , \qquad
  \vec z_1 \equiv        \left( \begin{array}{c}
      \sigma_1                             \\
      \eta_1
                \end{array}
         \right) \; .
\end{eqnarray*}
This follows from 
(\ref{eq:2.265}),(\ref{eq:2.27}),
(\ref{eq:2.37}). 
\footnote{Note that Process 1 is deterministic in this limit.}
Thus $\underline{P}_I$ modifies in this limit to
$\underline{P}_{I,nrad}$ with:
\begin{eqnarray*}
   \underline{P}_{I,nrad}(\sigma,\eta;s|\sigma_1,\eta_1;s_1)&\equiv&
   w_{orb,trans,nrad}(\sigma,\eta;s|\sigma_1,\eta_1;s_1)\cdot
\underline{R}_I(\sigma,\sigma_1) \; .
\end{eqnarray*}
\subsubsection*{2.7.3}
With the periodic boundary conditions discussed in section 2.3.6
one can express the polarization density in terms of $w_{per}$. By
(\ref{eq:2.62}) the polarization density reads 
for the two forms
(\ref{eq:2.241}),
(\ref{eq:2.242}),
of $w_{per}$ as: 
\begin{eqnarray}
 \vec{P}^w(\sigma,\eta;s) &=&
        \int_{0}^{2\pi} \; d\psi \cdot
   w_{per}(\sigma,\eta,\psi;s) \cdot
  \left( \begin{array}{c}
        \cos(\psi) \\
        \sin(\psi) \\
                      0
                \end{array}
         \right)   \; , \\
 \vec{P}^w(\sigma,\eta;s) &=&
   \frac{2}{\sqrt{3}}\cdot\int_{0}^{2\pi} \; d\psi \cdot
   w_{per}(\sigma,\eta,\psi;s) \cdot
  \left( \begin{array}{c}
        \cos(\psi) \\
        \sin(\psi) \\
                      0
                \end{array}
         \right)   \; .
 \label{eq:2.239}
\end{eqnarray}
\subsubsection*{2.7.4}
Having defined the polarization density I now introduce the `local
polarization vector' $\vec{P}_{loc}^w$ defined by
\begin{eqnarray}
 \vec{P}^w(\sigma,\eta;s) &\equiv&
   w_{orb}(\sigma,\eta,\psi;s) \cdot
 \vec{P}_{loc}^w(\sigma,\eta;s)  \; ,
 \label{eq:2.200}
\end{eqnarray}
and the `local polarization' defined by its norm
$||\vec{P}_{loc}^w||$.
Obviously
\begin{eqnarray*}
 \vec{P}_{tot}^w(s) &=&
      \int_{-\infty}^{+\infty} \; d\sigma
                             \int_{-\infty}^{+\infty} \; d\eta\cdot
         w_{orb}(\sigma,\eta;s)\cdot \vec{P}_{loc}^w(\sigma,\eta;s)
                                                 \; .
\end{eqnarray*}
$\vec P^w_{loc}$ is simply the spin polarization for an infinitesimal
packet of orbital phase space.
\par Clearly, I restrict myself to situations where the polarization
density  vanishes if $w_{orb}$ vanishes and where
 $0 \leq  ||\vec P^w_{loc}||  \leq 1 $. 
The direction $\vec{P}^w_{dir}$ of the local polarization is defined by
\begin{eqnarray}
 \vec{P}_{loc}^w &\equiv&
 ||\vec P_{loc}^w|| \cdot \vec{P}_{dir}^w \; .
\label{eq:2.203}
\end{eqnarray}
Hence by
(\ref{eq:2.68}) 
the local polarization vector and the local polarization direction
for Process 1 read as:
\begin{eqnarray*}
&&\vec{P}_{loc}^{w_1}(\sigma,\eta;s) =
\vec{P}_{dir}^{w_1}(\sigma,\eta;s) =
  \left( \begin{array}{c}
   \cos\biggl(\psi_0+(d\cdot(\sigma-\sigma_0))/a\biggr) \\
   \sin\biggl(\psi_0+(d\cdot(\sigma-\sigma_0))/a\biggr) \\
                      0
                \end{array}
         \right)          \; ,
\end{eqnarray*}
and the local polarization is:
\begin{eqnarray}
||\vec P_{loc}^{w_1}(\sigma,\eta;s)|| &=& 1 \; .
 \label{eq:2.69}
\end{eqnarray}
So for Process 1
each point in phase space is fully polarized and the $59\%$ is simply
due to the spread in $\vec P_{dir}^{w_1}$, not due to the value of
$||\vec P_{loc}^{w_1}||$.
\subsubsection*{2.7.5}
By 
(\ref{eq:2.15}),
(\ref{eq:2.64}),
(\ref{eq:2.200})
the local polarization vector obeys the following evolution equation
of Bloch type:
\begin{eqnarray}
&& \frac{\partial \vec{P}^w_{loc}}{\partial s}  =
\underbrace{
-a\cdot \eta \cdot
\frac{\partial \vec{P}^w_{loc}}{\partial \sigma}
- b\cdot\sigma \cdot \frac{\partial \vec{P}^w_{loc}}{\partial \eta}
+ \vec{W}_I\wedge \vec{P}^w_{loc}
                                                }_
               {{\rm radiationless \; part}}
-\underbrace{
 c\cdot \eta \cdot \frac{\partial \vec{P}^w_{loc}}{\partial \eta}
                                                }_
               {{\rm damping \; term}}
\nonumber\\&&\qquad
\underbrace{
+\omega\cdot\biggl(
\frac{1}{w_{orb}}\cdot\frac{\partial w_{orb}}{\partial\eta}
\cdot\frac{\partial \vec P^w_{loc}}{\partial\eta}
         +
          \frac{1}{2}\cdot
\frac{\partial^2 \vec{P}^w_{loc}}{\partial \eta^2} \biggr)
                                                }_
               {{\rm diffusion \; terms}}
                                           \; ,
 \label{eq:2.272}
\end{eqnarray}
which depends on $w_{orb}$ and is therefore not universal.
But for processes at orbital equilibrium, i.e. if $w_{orb}=w_{norm}$,
this simplifies by
(\ref{eq:2.215}),
to:
\begin{eqnarray}
&& \frac{\partial \vec{P}^w_{loc}}{\partial s}  =
\underbrace{
-a\cdot \eta \cdot
\frac{\partial \vec{P}^w_{loc}}{\partial \sigma}
- b\cdot\sigma \cdot \frac{\partial \vec{P}^w_{loc}}{\partial \eta}
+ \vec{W}_I\wedge \vec{P}^w_{loc}
                                                }_
               {{\rm radiationless \; part}}
+
  \underbrace{
 c\cdot \eta \cdot \frac{\partial \vec{P}^w_{loc}}{\partial \eta}
                                                }_
               {{\rm damping \; term}}
         +
  \underbrace{
          \frac{\omega}{2}\cdot
\frac{\partial^2 \vec{P}^w_{loc}}{\partial \eta^2}
                                                }_
               {{\rm diffusion \; term}}
                                           \; ,
 \label{eq:2.271}
\end{eqnarray}
which provides a causal azimuthal evolution, because 
(\ref{eq:2.64}) for the polarization 
density does.
Note that the damping terms in equations 
(\ref{eq:2.64}),
(\ref{eq:2.271})
have different structures.
However, as seen by
(\ref{eq:2.65}),
(\ref{eq:2.272}),
in the absence of radiation the local polarization vector obeys
the same radiationless Bloch equation as the polarization density.
Furthermore $||\vec P_{loc}^w||$ fulfills the Liouville
equation (\ref{eq:2.234}) in that case.
\subsubsection*{2.7.6}
By 
(\ref{eq:2.203}), 
(\ref{eq:2.272}) the local polarization direction obeys the following 
evolution equation
of Bloch type:
\begin{eqnarray}
&& \frac{\partial \vec{P}^w_{dir}}{\partial s}  =
\underbrace{
-a\cdot \eta \cdot
\frac{\partial \vec{P}^w_{dir}}{\partial \sigma}
- b\cdot\sigma \cdot \frac{\partial \vec{P}^w_{dir}}{\partial \eta}
+ \vec{W}_I\wedge \vec{P}^w_{dir}
                                                }_
               {{\rm radiationless \; part}}
-
\underbrace{
c\cdot \eta \cdot \frac{\partial \vec{P}^w_{dir}}{\partial \eta}
                                                }_
               {{\rm damping \; term}}
\nonumber\\&&\qquad
\underbrace{
+\omega\cdot\biggl(
\frac{1}{w_{orb}}\cdot\frac{\partial w_{orb}}{\partial\eta}
\cdot\frac{\partial \vec P^w_{dir}}{\partial\eta}
+\frac{1}{||\vec P^w_{loc}||}\cdot
\frac{\partial ||\vec P^w_{loc}||}{\partial\eta}
\cdot\frac{\partial \vec P^w_{dir}}{\partial\eta}
         -          \frac{1}{2}\cdot\lbrack
\vec{P}^w_{dir}\wedge(
\vec{P}^w_{dir}\wedge
\frac{\partial^2 \vec{P}^w_{dir}}{\partial \eta^2})\rbrack\biggr) 
                                                }_
               {{\rm diffusion \; terms}}
                                           \; , \nonumber\\&&
 \label{eq:2.273}
\end{eqnarray}
which depends on $w_{orb}$ and $||\vec P^w_{loc}||$ and
is therefore not universal. It is also nonlinear in $\vec P^w_{dir}$.
\par As seen by
(\ref{eq:2.65}),
(\ref{eq:2.273}),
in the absence of radiation the local polarization direction obeys
the same radiationless Bloch equation as the polarization density.
\subsubsection*{2.7.7}
The chief virtue of the polarization density stems
from the fact that it satisfies a {\it universal} and {\it linear}
differential equation (of the Bloch type). 
In the case of Machine I this equation is given by 
(\ref{eq:2.64}).
Furthermore this equation
provides a causal azimuthal evolution but this feature is not as important
as universality and linearity. 
\par One has seen that the local polarization vector and its direction
also obey Bloch 
equations but that these equations are not universal. 
Furthermore the equation for the local polarization direction
is in general nonlinear (see 
(\ref{eq:2.273})).
Clearly, in contrast to the full Fokker-Planck equation 
(\ref{eq:2.8}), all these Bloch equations 
enable one to study average spin behaviour without
having to look closely at the $\psi$ distribution $w_{spin}$.
The polarization
density, the local polarization vector and its direction 
only depend on orbital variables
and the effects of radiation are contained in damping and diffusion terms
of 
(\ref{eq:2.64}),
(\ref{eq:2.272}),
(\ref{eq:2.273}) which are associated with the orbital Fokker-Planck operator
$L_{FP,orb}$. Indeed, it is no accident that Bloch equations 
emerge for Machine I. See section 5. 
\par I make further comments about Bloch equations in section 2.8.4.
\subsection{The polarization properties of Machine I for G-processes}
\subsubsection*{2.8.1}
In this section I consider a special class of processes running
with Machine I and one aim is to consider the azimuthal evolution
of the polarization vector.
\par I consider  processes running with Machine I which have a general
Gaussian probability density in $\sigma,\eta,\psi$
for $s>s_0$. Thus for $s>s_0$:
\begin{eqnarray}
&& w(\sigma,\eta,\psi;s)  =
   w(\vec{x};s)  =  (2\pi)^{-3/2}\cdot
\det\biggl( \underline{\sigma}(s)\biggr)^{-1/2}
\nonumber\\
 &&\qquad\cdot \exp\biggl\lbrack -\frac{1}{2}\cdot
      \biggl(\vec{x}-<\vec{x}(s)>\biggr)^T \cdot
    \underline{\sigma}^{-1}(s)\cdot
\biggl(\vec{x}-<\vec{x}(s)>\biggr)\biggr\rbrack  \; ,
 \label{eq:2.70}
\end{eqnarray}
where $<\vec{x}(s)>$ and $\underline{\sigma}$ denote the first moment
vector and covariance matrix of the process. I call these
`G-processes'.
\footnote{I allow the covariance matrix of a G-process to be possibly
singular at the starting azimuth $s=s_0$.}
Hence Process 2 is a G-process.
By (\ref{eq:2.222}),(\ref{eq:2.70}) the characteristic function $\Phi$ of a 
G-process reads as:
\begin{eqnarray}
 \Phi(\vec{u};s) &=& \exp\biggl(-\frac{1}{2}\cdot
 \sum_{j,k=1}^3 \; \sigma_{jk}(s)\cdot u_j\cdot u_k
+i\cdot \vec{u}^T\cdot <\vec{x}(s)>\biggr) \; .
\label{eq:2.229}
\end{eqnarray}
Because the first moment vector and the covariance matrix depend
continuously on $s$, (\ref{eq:2.229}) holds even at $s=s_0$, so that the
characteristic function is especially convenient for those G-processes whose
covariance matrix is singular at $s=s_0$.
\par Due to 
(\ref{eq:2.16})
the spin part $w_{spin}$ of $w$ is also Gaussian for $s>s_0$, i.e.:
\begin{eqnarray}
w_{spin}(\psi;s) &\equiv& \biggl( 2\pi\cdot
          \sigma_{33}(s)\biggr)^{-1/2}
 \cdot \exp \biggl( -\frac{(\psi-<\psi(s)>)^2}{2\cdot
        \sigma_{33}(s)}
   \biggr) \; .
 \label{eq:2.71}
\end{eqnarray}
The corresponding polarization vector has  the simple form:
\begin{eqnarray}
&&\vec{P}_{tot}^w(s)  =
  \frac{2}{\hbar}\cdot <\vec{S}(s)>
 =                     \int_{-\infty}^{+\infty} \; d\sigma
                             \int_{-\infty}^{+\infty} \; d\eta
                             \int_{-\infty}^{+\infty} \; d\psi\cdot
   w(\sigma,\eta,\psi;s)\cdot
  \left( \begin{array}{c}
        \cos(\psi) \\
        \sin(\psi) \\
                      0
                \end{array}
         \right)
\nonumber\\
&&=   \int_{-\infty}^{+\infty} \; d\psi\cdot
   w_{spin}(\sigma,\eta,\psi;s) \cdot
  \left( \begin{array}{c}
        \cos(\psi) \\
        \sin(\psi) \\
                      0
                \end{array}
         \right)
 =  \exp(-\sigma_{33}(s)/2)\cdot
  \left( \begin{array}{c}
        \cos(<\psi(s)>) \\
        \sin(<\psi(s)>) \\
                      0
                \end{array}
         \right)  \; ,
\nonumber\\
 \label{eq:2.72}
\end{eqnarray}
and the polarization:
\begin{eqnarray}
      ||\vec P_{tot}^w(s)|| &=&
    \exp\biggl(-\sigma_{33}(s)/2\biggr)  \; .
 \label{eq:2.73}
\end{eqnarray}
Note that
(\ref{eq:2.72}),(\ref{eq:2.73}) hold even at $s=s_0$ because
$\sigma_{33}(s)$ and $<\psi(s)>$ depend continuously on $s$.
\subsubsection*{2.8.2}
By (\ref{eq:2.225}) the stochastic averages 
$<\sigma(s)>,<\eta(s)>$ of the orbital 
variables have a causal azimuthal evolution, i.e. they are determined
by the initial values $<\sigma(s_0)>,<\eta(s_0)>$:
\footnote{Note that (\ref{eq:2.252}) not only holds for G-processes.}
\begin{eqnarray}
&&\left( \begin{array}{c}
               <\sigma(s)>   \\
               <\eta(s)>
                \end{array}
         \right)
   =   \exp\biggl(\underline{\cal A}_{orb}\cdot(s-s_0)\biggr)\cdot
\left( \begin{array}{c}
               <\sigma(s_0)>   \\
               <\eta(s_0)>
                \end{array}
         \right) \; .
 \label{eq:2.252}
\end{eqnarray}
However, for the spin vector such a causal behaviour does not prevail and
this already shows up for G-processes.
\par By (\ref{eq:2.72})
one sees that two G-processes which have the same values for
$<\psi(s_0)>$ and $\sigma_{33}(s_0)$ have the same initial polarization
vector. However it does not follow from this that both processes have the
same polarization vector 
for $s>s_0$, because by using the differential equations
(\ref{eq:2.18}),
(\ref{eq:2.19}) and by using the freedom of choice of
$<\sigma(s_0)>,<\eta(s_0)>,
\sigma_{11}(s_0),\\
\sigma_{12}(s_0),
\sigma_{13}(s_0),
\sigma_{22}(s_0),
\sigma_{23}(s_0)$
one easily finds that the two processes in general
have different polarization vectors for $s>s_0$.
This holds even if both processes are in the same orbital state, i.e.
have the same $w_{orb}$. As an example I compare the initial conditions:
\begin{eqnarray}
&&<\sigma(s_0)> = <\eta(s_0)> = <\psi(s_0)> = 0 \; , \qquad
\sigma_{11}(s_0) = \sigma_{\sigma}^2 \; , \qquad
   \sigma_{22}(s_0) = \sigma_{\eta}^2 \; , \nonumber\\
&&   \sigma_{33}(s_0) = 2\cdot(d^2/a^2)\cdot\sigma_{\sigma}^2 \; , \qquad
    \sigma_{12}(s_0) = 
    \sigma_{13}(s_0) = 
    \sigma_{23}(s_0) = 0 \; , 
 \label{eq:2.253}
\end{eqnarray}
with the initial conditions:
\begin{eqnarray}
&&<\sigma(s_0)> = <\eta(s_0)> = <\psi(s_0)> = 0 \; , \qquad
\sigma_{11}(s_0) = \sigma_{\sigma}^2 \; , \qquad
   \sigma_{22}(s_0) = \sigma_{\eta}^2 \; , \nonumber\\
&&   \sigma_{33}(s_0) = 2\cdot(d^2/a^2)\cdot\sigma_{\sigma}^2 \; , \qquad
     \sigma_{13}(s_0) = (d/a)\cdot\sigma_{\sigma}^2 \; , \qquad
    \sigma_{12}(s_0) = 
    \sigma_{23}(s_0) = 0 \; , \qquad
 \label{eq:2.254}
\end{eqnarray}
where each set of initial conditions defines a specific G-process.
Both processes are in orbital equilibrium, so they are in the same orbital
state with $w_{orb}=w_{norm}$. In particular they have the same orbital
stochastic
averages $<\sigma(s)>=<\eta(s)>=0$. Also both processes have same initial 
polarization vector:
\begin{eqnarray}
&&\vec{P}_{tot}^w(s_0)  =
 \exp(-\frac{d^2\cdot\sigma_{\sigma}^2}{a^2})
\cdot\left( \begin{array}{c}
        1 \\
         0 \\
                      0
                \end{array}
         \right)  \; .
 \label{eq:2.255}
\end{eqnarray}
However using
(\ref{eq:2.18})
one quickly finds that the polarization vectors for the two 
processes evolve in
different ways. In particular the equilibrium polarization 
$||\vec P_{tot}^w(+\infty)||$ is $\exp(-(2\cdot 
d^2\cdot\sigma_{\sigma}^2)/a^2)$
for the process (\ref{eq:2.253}) and 
it is $\exp(-(d^2\cdot\sigma_{\sigma}^2)/(a^2))$
for the process (\ref{eq:2.254}).
\footnote{Note also that process (\ref{eq:2.254}), unlike process
(\ref{eq:2.253}), is stationary.}
\par Hence the initial values
$<\sigma(s_0)>,<\eta(s_0)>,\vec P_{tot}^w(s_0)$
do not determine the future behaviour of the polarization vector.
In particular there exists no differential equation 
for the azimuthal evolution of the polarization vector, which could
provide such a causal azimuthal evolution.
Indeed by
differentiating 
(\ref{eq:2.72})
and by using
(\ref{eq:2.18}),
(\ref{eq:2.19}) one obtains for a G-process the differential equation:
\begin{eqnarray}
   \biggl(\vec{P}_{tot}^{w}(s)\biggr)' 
&=& d\cdot<\eta(s)>\cdot\biggl(
  \left( \begin{array}{c}
                      0          \\
                      0          \\
                      1
                \end{array}
         \right)   \wedge  \vec{P}_{tot}^w(s)\biggr)
 - d\cdot\sigma_{23}(s) \cdot
   \vec{P}_{tot}^w(s)    \; ,
 \label{eq:2.74}
\end{eqnarray}
which at first sight appears to be an appropriate evolution equation for the
polarization vector. However it is not a universal 
equation because $\sigma_{23}(s)$ depends on the process, confirming 
the above conclusions.
Note also that by 
(\ref{eq:2.18}),
(\ref{eq:2.72}),
one can write 
(\ref{eq:2.74}) as:
\begin{eqnarray}
   \biggl(\vec{P}_{tot}^{w}(s)\biggr)' 
&=& d\cdot<\eta(s)>\cdot\biggl(
  \left( \begin{array}{c}
                      0          \\
                      0          \\
                      1
                \end{array}
         \right)   \wedge  \vec{P}_{tot}^w(s)\biggr)
  + \frac{\vec{P}_{tot}^w(s)}{2\cdot||\vec{P}_{tot}^w(s)||^2}\cdot
    \biggl(||\vec{P}_{tot}^w(s)||^2\biggr)' \; .
\end{eqnarray}
\par Concluding this section I have seen that, at least without further
approximation, 
no (Bloch) equation for the polarization
vector exists in Machine I
which would provide a causal evolution for the stochastic average:
\begin{eqnarray*}
  \left( \begin{array}{c}
      <\sigma(s)> \\
      <\eta(s)>   \\
      (\hbar/2)\cdot\vec{P}_{tot}^{w}(s)
\end{array}
         \right) 
\end{eqnarray*}
of the five-component spin-orbit vector.
However there is a universal Bloch equation (\ref{eq:2.64})
giving a causal azimuthal evolution of the polarization density.
\subsubsection*{2.8.3}
To calculate the local polarization quantities of a G-process
one first observes for $s>s_0$ that:
\begin{eqnarray*}
&& \int_{-\infty}^{+\infty} \; d\psi\cdot
   w(\sigma,\eta,\psi;s)\cdot\exp(i\cdot\psi)=
   w_{orb}(\sigma,\eta;s)
\nonumber\\
   && \cdot
               \exp\biggl(
-\frac{2\cdot i\cdot\sigma_{inv,13}(s)\cdot\biggl(\sigma-<\sigma(s)>\biggr)
+2\cdot i\cdot\sigma_{inv,23}(s)\cdot\biggl(\eta-<\eta(s)>\biggr)
+1}{2\cdot\sigma_{inv,33}(s)}
                                        \biggr)
\nonumber\\
   && \cdot
                             \exp\biggl(i\cdot<\psi(s)>\biggr) \; ,
\end{eqnarray*}
where $\underline{\sigma}_{inv}$ denotes the inverse of the
covariance matrix.
From this follows for $s>s_0$:
\begin{eqnarray}
&& \vec{P}^w(\sigma,\eta;s) =
   w_{orb}(\sigma,\eta;s)\cdot
               \exp\biggl(-\frac{1}{2\cdot\sigma_{inv,33}(s)}
    \biggr)
\nonumber\\ &&
   \cdot \large{
\left( \begin{array}{c}
     \cos\biggl(
       - \frac{\sigma_{inv,13}(s)}{\sigma_{inv,33}(s)}\cdot
                                 (\sigma-<\sigma(s)>)
       - \frac{\sigma_{inv,23}(s)}{\sigma_{inv,33}(s)}\cdot
                                 (\eta-<\eta(s)>) +<\psi(s)>
                                           \biggr) \\
     \sin\biggl(
       - \frac{\sigma_{inv,13}(s)}{\sigma_{inv,33}(s)}\cdot
                                 (\sigma-<\sigma(s)>)
       - \frac{\sigma_{inv,23}(s)}{\sigma_{inv,33}(s)}\cdot
                                 (\eta-<\eta(s)>) +<\psi(s)>
                                           \biggr) \\
                      0
                \end{array}
         \right)}       \; .
\nonumber\\
 \label{eq:2.76}
\end{eqnarray}
Therefore the local polarization vector reads for $s>s_0$ as:
\begin{eqnarray}
&& \vec{P}_{loc}^{w}(\sigma,\eta;s) =
          \exp\biggl( -\frac{1}{2\cdot\sigma_{inv,33}(s)}\biggr)
\nonumber\\ &&
   \cdot \large{\left( \begin{array}{c}
     \cos\biggl(
       - \frac{\sigma_{inv,13}(s)}{\sigma_{inv,33}(s)}\cdot
                                 (\sigma-<\sigma(s)>)
       - \frac{\sigma_{inv,23}(s)}{\sigma_{inv,33}(s)}\cdot
                                 (\eta-<\eta(s)>) +<\psi(s)>
                                           \biggr) \\
     \sin\biggl(
       - \frac{\sigma_{inv,13}(s)}{\sigma_{inv,33}(s)}\cdot
                                 (\sigma-<\sigma(s)>)
       - \frac{\sigma_{inv,23}(s)}{\sigma_{inv,33}(s)}\cdot
                                 (\eta-<\eta(s)>) +<\psi(s)>
                                           \biggr) \\
                      0
                \end{array}
         \right)}        \; ,
\nonumber\\
 \label{eq:2.202}
\end{eqnarray}
and the local polarization for $s>s_0$ is:
\begin{eqnarray}
&& ||\vec P_{loc}^{w}(\sigma,\eta;s)|| =
          \exp\biggl( -\frac{1}{2\cdot\sigma_{inv,33}(s)}\biggr)
                        \; .
 \label{eq:2.77}
\end{eqnarray}
One sees that for $s>s_0$ the local polarization of every G-process is 
uniform across phase space and
that $0\leq ||\vec P_{loc}^w||\leq 1$.
\footnote{In fact $\underline{\sigma}_{inv}(s)$ is positive definite
for $s>s_0$ because $\underline{\sigma}(s)$ is. From this it follows that:
$\sigma_{inv,33}(s)>0$ for $s>s_0$, which proves the assertion.}
Of course, the polarization density
(\ref{eq:2.76}) obeys the Bloch equation
(\ref{eq:2.64}).
\par As an example one gets for Process 2:
\begin{eqnarray}
&& \vec{P}^{w_2}(\sigma,\eta;s) =
   w_{norm}(\sigma,\eta)\cdot
      \exp\biggl(\frac{d^2\cdot\omega}{8\lambda^2}\cdot g_4(s)
    \biggr)
\nonumber\\ &&
   \cdot\large{ \left( \begin{array}{c}
     \cos\biggl(
  \frac{d}{2\cdot a\cdot\lambda}\cdot \lbrack
    2\cdot\lambda -i\cdot g_1(s) \rbrack \cdot \sigma
 - \frac{i\cdot d}{2\cdot \lambda}\cdot g_2(s) \cdot\eta
           +\psi_0
                                           \biggr) \\
     \sin\biggl(
  \frac{d}{2\cdot a\cdot\lambda}\cdot \lbrack
    2\cdot\lambda -i\cdot g_1(s) \rbrack \cdot \sigma
 - \frac{i\cdot d}{2\cdot \lambda}\cdot g_2(s) \cdot\eta
           +\psi_0
                                           \biggr) \\
                      0
                \end{array}
         \right)}        \; ,
 \label{eq:2.78}
\end{eqnarray}
from which follows:
\begin{eqnarray}
&& ||\vec P_{loc}^{w_2}(\sigma,\eta;s)|| =
      \exp\biggl(\frac{d^2\cdot\omega}{8\lambda^2}\cdot g_4(s)
                                         \biggr)
                        \; .
 \label{eq:2.79}
\end{eqnarray}
Thus the local polarization of Process 2 
starts from the value 1 at $s=0$ and decreases monotonically with
increasing azimuth. It approaches the following equilibrium value:
\begin{eqnarray*}
   ||\vec P_{loc}^{w_2}(\sigma,\eta;+\infty)|| &=&
      \exp\biggl(-\frac{d^2\cdot\sigma_{\sigma}^2}{2a^2}\biggr)
                        \; .
\end{eqnarray*}
With the HERA values 
(\ref{eq:2.3})
the local polarization
value for the equilibrium of Process 2 is 0.59. Contrast this with 
(\ref{eq:2.69}).
\subsubsection*{2.8.4}
For G-processes at orbital equilibrium, i.e. for $w_{orb}=w_{norm}$,
the Bloch equation 
(\ref{eq:2.273})
simplifies by
(\ref{eq:2.215}),
(\ref{eq:2.77}) to:
\begin{eqnarray}
&& \frac{\partial \vec{P}^w_{dir}}{\partial s}  =
\underbrace{
-a\cdot \eta \cdot
\frac{\partial \vec{P}^w_{dir}}{\partial \sigma}
- b\cdot\sigma \cdot \frac{\partial \vec{P}^w_{dir}}{\partial \eta}
+ \vec{W}_I\wedge \vec{P}_{dir}^w
                                                }_
               {{\rm radiationless \; part}}
+  \underbrace{
 c\cdot \eta \cdot \frac{\partial \vec{P}^w_{dir}}{\partial \eta}    
                                                }_
               {{\rm damping \; term}} \; ,
\label{eq:2.275}
\end{eqnarray}
which provides a causal azimuthal evolution, because 
(\ref{eq:2.271}) ensures this for the local polarization vector  
and because the local polarization is uniform.
Note that the damping terms in equations 
(\ref{eq:2.64}),
(\ref{eq:2.275})
have different structures and that
there is no diffusion term in (\ref{eq:2.275}).
\subsection{Miscellaneous equilibrium properties of Machine I}
\subsubsection*{2.9.1}
With the examples of processes 1,2 one has already seen that Machine I has
no unique equilibrium state. Therefore in this section I study the
asymptotic  behaviour of arbitrary processes running with
Machine I.
\par To come to that I conclude first of all from 
(\ref{eq:2.216}), 
(\ref{eq:2.36}), 
(\ref{eq:2.37}), 
(\ref{eq:2.261}): 
\begin{eqnarray}
   w(\sigma,\eta,\psi;+\infty) &=&
                             \int_{-\infty}^{+\infty} \; d\sigma_0
                             \int_{-\infty}^{+\infty} \; d\eta_0
                             \int_{-\infty}^{+\infty} \; d\psi_0
\nonumber\\
&&        \cdot
   w_{I,trans}(\sigma,\eta,\psi;+\infty|\sigma_0,\eta_0,\psi_0;s_0)
           \cdot w(\sigma_0,\eta_0,\psi_0;s_0)
\nonumber\\
   &=&                       \int_{-\infty}^{+\infty} \; d\sigma_0
                             \int_{-\infty}^{+\infty} \; d\eta_0
                             \int_{-\infty}^{+\infty} \; d\psi_0\cdot
   w_{orb,trans}(\sigma,\eta;+\infty|\sigma_0,\eta_0;s_0)
\nonumber\\
&&\qquad   \cdot w(\sigma_0,\eta_0,\psi_0;s_0)\cdot
\delta\biggl(\psi-\psi_0 -\frac{d}{a}\cdot(\sigma-\sigma_0)\biggr)
\nonumber\\
&=&  w_{norm}(\sigma,\eta)\cdot
                          \int_{-\infty}^{+\infty} \; d\sigma_0
                             \int_{-\infty}^{+\infty} \; d\eta_0
                             \int_{-\infty}^{+\infty} \; d\psi_0\cdot
\; w(\sigma_0,\eta_0,\psi_0;s_0)
\nonumber\\
\qquad && \cdot
\delta\biggl(\psi-\psi_0 -\frac{d}{a}\cdot(\sigma-\sigma_0)\biggr)
\nonumber\\
&& =    w_{norm}(\sigma,\eta)\cdot
                          \int_{-\infty}^{+\infty} \; d\sigma_0
                             \int_{-\infty}^{+\infty} \; d\eta_0\cdot
\;
w\biggl(\sigma_0,\eta_0,\psi
      -\frac{d}{a}\cdot(\sigma-\sigma_0);s_0 \biggr) 
\nonumber\\
&& =          w_{norm}(\sigma,\eta) \cdot
      w_{aver}(\psi-\frac{d}{a}\cdot\sigma;s_0) \; ,
 \label{eq:2.83}
\end{eqnarray}
where:
\begin{eqnarray}
  w_{aver}(\psi;s) &\equiv &
    \int_{-\infty}^{+\infty} \; d\sigma
                             \int_{-\infty}^{+\infty} \; d\eta\cdot
           w(\sigma,\eta,\psi+\frac{d}{a}\cdot\sigma;s) \; .
 \label{eq:2.244}
\end{eqnarray}
The equilibrium probability density (\ref{eq:2.83})
is not the same for every process,
reflecting  the fact that Machine I has no unique equilibrium state.
From 
(\ref{eq:2.83}),
(\ref{eq:2.244}) it is also clear
that only average
\footnote{For example the $\eta$-dependence of the initial state
is completely integrated out.}
information about the initial state is needed in order to determine the
corresponding equilibrium state, i.e. for a given equilibrium state
there are many different initial states which all approach the same
equilibrium.
\par From 
(\ref{eq:2.244}) it follows that $w_{aver}$ is normalized:
\begin{eqnarray}
 \int_{-\infty}^{+\infty} \; d\psi \cdot  w_{aver}(\psi;s)
&=&
                             \int_{-\infty}^{+\infty} \; d\sigma
                             \int_{-\infty}^{+\infty} \; d\eta
                             \int_{-\infty}^{+\infty} \; d\psi\cdot
           w(\sigma,\eta,\psi+\frac{d}{a}\cdot\sigma;s)
\nonumber\\
&=&
                             \int_{-\infty}^{+\infty} \; d\sigma
                             \int_{-\infty}^{+\infty} \; d\eta
                             \int_{-\infty}^{+\infty} \; d\psi\cdot
           w(\sigma,\eta,\psi;s)
 =  1 \; .
 \label{eq:2.245}
\end{eqnarray}
From 
(\ref{eq:2.13}),
(\ref{eq:2.83}),
(\ref{eq:2.245}) follows:
\begin{eqnarray}
 w_{orb}(\sigma,\eta;+\infty)&=&
   w_{norm}(\sigma,\eta)\cdot
                             \int_{-\infty}^{+\infty} \; d\psi
    \cdot w_{aver}(\psi-\frac{d}{a}\cdot\sigma;s_0) 
\nonumber\\
&=&
   w_{norm}(\sigma,\eta)\cdot
                             \int_{-\infty}^{+\infty} \; d\psi
    \cdot w_{aver}(\psi;s_0) 
=   w_{norm}(\sigma,\eta) \; .
 \label{eq:2.82}
\end{eqnarray}
This confirms again that every process running with Machine I leads to
the same orbital equilibrium characterized by $w_{norm}$.
\par The equilibrium probability density 
(\ref{eq:2.83})
not only fulfills the Fokker-Planck
equation 
(\ref{eq:2.8})
but also the radiationless Fokker-Planck equation
\begin{eqnarray*}
   \frac{\partial w      }{\partial s} &=&
-a\cdot \eta \cdot
\frac{\partial w      }{\partial \sigma}
- b\cdot\sigma \cdot \frac{\partial w      }{\partial \eta}
-d\cdot \eta
               \cdot \frac{\partial w}{\partial \psi}
                      \; .
\end{eqnarray*}
So at equilibrium the damping and diffusion balance each other and
the Fokker-Planck equation effectively reduces to a Liouville
equation, i.e. at equilibrium one effectively has a Hamiltonian flow of the
probability density. Furthermore, since Machine I is smooth, the
asymptotic $w$  is independent of $s$ so that
  $\partial w/\partial s=0$.
\par For the orbital part one gets analogously:
\begin{eqnarray*}
   \frac{\partial w_{orb}}{\partial s} &=&
-a\cdot \eta \cdot
\frac{\partial w_{orb}}{\partial \sigma}
- b\cdot\sigma \cdot \frac{\partial w_{orb}}{\partial \eta}
= \lbrace H_{orb}, w_{orb} \rbrace  = 0 \; .
\end{eqnarray*}
Using the fact that the orbital equilibrium is unique with
$w_{orb}=w_{norm}$, one thus has:
\begin{eqnarray*}
  \lbrace H_{orb}, w_{norm} \rbrace &=&0 \; .
\end{eqnarray*}
This relation is obviously fulfilled because:
\begin{eqnarray*}
w_{norm}(\sigma,\eta) &=&
 \frac{1}{2\pi\cdot\sigma_{\sigma}\cdot\sigma_{\eta}}\cdot
    \exp(-\frac{H_{orb}}{a\cdot\sigma_{\eta}^2}) \; .
\end{eqnarray*}
For more details on the Hamiltonian description, see Appendix D.
\subsubsection*{2.9.2}
Having obtained a tractable formula for the equilibrium states of
Machine I one can now consider the equilibrium polarization properties.
The spin part of the equilibrium probability density has the form:
\begin{eqnarray}
 w_{spin}(\psi;+\infty) &=& \int_{-\infty}^{+\infty} \; d\sigma
     \int_{-\infty}^{+\infty} \; d\eta \cdot
   w_{norm}(\sigma,\eta)\cdot
      w_{aver}(\psi-\frac{d}{a}\cdot\sigma;s_0)
\nonumber\\
 &=&-\frac{a}{d}\cdot \int_{-\infty}^{+\infty} \; d\eta
     \int_{-\infty}^{+\infty} \; d\psi_1 \cdot
   w_{norm}\biggl(\frac{a}{d}\cdot(\psi-\psi_1),\eta\biggr)\cdot
      w_{aver}(\psi_1;s_0) \; .
\nonumber\\
 &\equiv&
    -\frac{a}{d}\cdot
     \int_{-\infty}^{+\infty} \; d\psi_1 \cdot
   w_{norm,red}\biggl(\frac{a}{d}\cdot(\psi-\psi_1)\biggr)\cdot
      w_{aver}(\psi_1;s_0) \; ,
 \label{eq:2.84}
\end{eqnarray}
where I introduced the abbreviation
\begin{eqnarray*}
  w_{norm,red}(\sigma) &\equiv &
                             \int_{-\infty}^{+\infty} \; d\eta\cdot
                  w_{norm}(\sigma,\eta)
 =  (2\pi)^{-1/2}\cdot\sigma_{\sigma}^{-1}\cdot
 \exp(-\sigma^2/2\sigma_{\sigma}^2) \; .
\end{eqnarray*}
To determine the equilibrium polarization vector I introduce the
auxiliary constant
\begin{eqnarray}
 h^w &\equiv&  
   |h^w|\cdot\exp(i\cdot\chi^w) \equiv
\int_{-\infty}^{+\infty} \; d\psi \cdot
 w_{spin}(\psi;+\infty)\cdot \exp(i\cdot\psi)
\nonumber\\
&=&
   -\frac{a}{d}\cdot
     \int_{-\infty}^{+\infty} \; d\psi_1
     \int_{-\infty}^{+\infty} \; d\psi \cdot
   w_{norm,red}\biggl(\frac{a}{d}\cdot(\psi-\psi_1)\biggr)\cdot
      w_{aver}(\psi_1;s_0) \cdot \exp(i\cdot\psi)
\nonumber\\
&=&
   -\frac{a}{d}\cdot
    (2\pi)^{-1/2}\cdot\sigma_{\sigma}^{-1}\cdot
     \int_{-\infty}^{+\infty} \; d\psi_1
     \int_{-\infty}^{+\infty} \; d\psi \cdot
\exp\biggl(
    -\frac{a^2}{2d^2\sigma_{\sigma}^2}\cdot(\psi-\psi_1)^2\biggr)
\nonumber\\
&& \qquad
\cdot w_{aver}(\psi_1;s_0) \cdot \exp(i\cdot\psi)
\nonumber\\
&=&
 \exp(-\frac{d^2\cdot\sigma_{\sigma}^2}{2a^2}) \cdot
     \int_{-\infty}^{+\infty} \; d\psi_1 \cdot
      w_{aver}(\psi_1;s_0) \cdot \exp(i\cdot\psi_1)
                          \; .
 \label{eq:2.85}
\end{eqnarray}
By 
(\ref{eq:2.11}),
(\ref{eq:2.85})
the equilibrium polarization is given by
\begin{eqnarray}
 && ||\vec{P}^w_{tot}(+\infty)||^2  =
   | \int_{-\infty}^{+\infty} \; d\psi \cdot
 w_{spin}(\psi;+\infty)\cdot \cos(\psi)|^2
\nonumber\\
&&\qquad   +| \int_{-\infty}^{+\infty} \; d\psi \cdot
 w_{spin}(\psi;+\infty)\cdot \sin(\psi)|^2
\nonumber\\
 &&=  | \int_{-\infty}^{+\infty} \; d\psi \cdot
 w_{spin}(\psi;+\infty)\cdot \exp(i\cdot\psi)|^2
 = |h^w |^2 \nonumber\\
&& =
 \exp(-\frac{d^2\cdot\sigma_{\sigma}^2}{a^2}) \cdot
   | \int_{-\infty}^{+\infty} \; d\psi \cdot
      w_{aver}(\psi;s_0) \cdot \exp(i\cdot\psi)|^2 \; .
 \label{eq:2.86}
\end{eqnarray}
An interesting application of 
(\ref{eq:2.86})
is that it allows the
determination of the maximum equilibrium polarization
possible for Machine I. First of all one observes by
(\ref{eq:2.244}),
(\ref{eq:2.245})
that $w_{aver}$ is
nonnegative and normalized.
From this follows the inequality
\begin{eqnarray*}
   | \int_{-\infty}^{+\infty} \; d\psi \cdot
      w_{aver}(\psi;s_0) \cdot \exp(i\cdot\psi)| &\leq & 1 \; .
\end{eqnarray*}
Inserting this into 
(\ref{eq:2.86})
leads to
\begin{eqnarray*}
  ||\vec P^w_{tot}(+\infty)||   &\leq&
 \exp(-\frac{d^2\cdot\sigma_{\sigma}^2}{2a^2})
\equiv P_{max}                                  \; ,
\end{eqnarray*}
so that the upper limit for the equilibrium polarization is
$\exp(-(d^2\cdot\sigma_{\sigma}^2)/(2a^2))$. In fact  Process 1
reaches this value (see 
(\ref{eq:2.33})). So if one specifies the constants
according to 
(\ref{eq:2.3}), then no process running with Machine I has an
equilibrium polarization greater than the 0.59 of Process 1. 
This is not surprising since Process 1 is fully ordered at the start.
However Process 1 is not
the only possible process having this equilibrium value.
Another example is given by the stationary process
with the probability density:
\begin{eqnarray}
 && w(\sigma,\eta,\psi;s) = 
   w_{norm}(\sigma,\eta)\cdot\delta(\psi-\frac{d}{a}\cdot\sigma) \; .
\label{eq:2.274}
\end{eqnarray}
\par Coming to the local polarization quantities at equilibrium, I conclude
from (\ref{eq:2.83}),
(\ref{eq:2.85}):
\begin{eqnarray}
&& \vec{P}^w(\sigma,\eta;+\infty) =
          w_{norm}(\sigma,\eta) \cdot
                             \int_{-\infty}^{+\infty} \; d\psi \cdot
      w_{aver}(\psi-\frac{d}{a}\cdot\sigma;s_0) \cdot
  \left( \begin{array}{c}
        \cos(\psi) \\
        \sin(\psi) \\
                      0
                \end{array}
         \right)
\nonumber\\
&=&\Re e\biggl\lbrace   w_{norm}(\sigma,\eta) \cdot
                             \int_{-\infty}^{+\infty} \; d\psi \cdot
      w_{aver}(\psi-\frac{d}{a}\cdot\sigma;s_0) \cdot
  \left( \begin{array}{c}
          1        \\
          -i       \\
                      0
                \end{array}
         \right)
  \cdot\exp(i\cdot\psi) \biggr\rbrace
\nonumber\\
&=&\Re e\biggl\lbrace   w_{norm}(\sigma,\eta) \cdot
                             \int_{-\infty}^{+\infty} \; d\psi \cdot
      w_{aver}(\psi;s_0) \cdot
  \left( \begin{array}{c}
          1        \\
          -i       \\
                      0
                \end{array}
         \right)
  \cdot\exp\biggl(i\cdot(\psi+\frac{d}{a}\cdot\sigma)\biggr)
\biggr\rbrace
\nonumber\\
&=&\Re e\biggl\lbrace   w_{norm}(\sigma,\eta) \cdot
  \left( \begin{array}{c}
          1        \\
          -i       \\
                      0
                \end{array}
         \right)
  \cdot\exp(i\cdot\frac{d}{a}\cdot\sigma) \cdot
                             \int_{-\infty}^{+\infty} \; d\psi \cdot
      w_{aver}(\psi;s_0) \cdot \exp(i\cdot\psi)
\biggr\rbrace
\nonumber\\
&=&\Re e\biggl\lbrace   w_{norm}(\sigma,\eta) \cdot
  \left( \begin{array}{c}
          1        \\
          -i       \\
                      0
                \end{array}
         \right)
  \cdot\exp(i\cdot\frac{d}{a}\cdot\sigma) \cdot
   h^w \cdot P_{max}^{-1}
\biggr\rbrace
\nonumber\\
&=& |h^w|\cdot P_{max}^{-1}\cdot w_{norm}(\sigma,\eta) \cdot
 \Re e\biggl\lbrace
  \left( \begin{array}{c}
          1        \\
          -i       \\
                      0
                \end{array}
         \right)
  \cdot\exp(i\cdot\frac{d}{a}\cdot\sigma) \cdot
       \exp(i\cdot\chi^w)
\biggr\rbrace
\nonumber\\
&=& |h^w|\cdot P_{max}^{-1}\cdot w_{norm}(\sigma,\eta) \cdot
  \left( \begin{array}{c}
     \cos((d\cdot\sigma)/a+\chi^w) \\  \\
     \sin((d\cdot\sigma)/a+\chi^w) \\  \\
                      0
                \end{array}
         \right)  \; .
 \label{eq:2.87}
\end{eqnarray}
Having obtained the simple form 
(\ref{eq:2.87})
of the equilibrium polarization
density one can now also write down the other equilibrium polarization 
quantities:
\begin{eqnarray}
&& \vec{P}^w_{loc}(\sigma,\eta;+\infty) =
 |h^w|\cdot P_{max}^{-1}\cdot
  \left( \begin{array}{c}
     \cos((d\cdot\sigma)/a+\chi^w) \\  \\
     \sin((d\cdot\sigma)/a+\chi^w) \\  \\
                      0
                \end{array}
         \right)  \; , 
\label{eq:2.277} \\
&& \vec{P}^w_{dir}(\sigma,\eta;+\infty) =
  \left( \begin{array}{c}
     \cos((d\cdot\sigma)/a+\chi^w) \\  \\
     \sin((d\cdot\sigma)/a+\chi^w) \\  \\
                      0
                \end{array}
         \right)  \; , 
\label{eq:2.278} \\
&& \vec{P}^w_{tot}(+\infty) =
    |h^w|\cdot
  \left( \begin{array}{c}
     \cos(\chi^w) \\
     \sin(\chi^w) \\
                      0
                \end{array}
         \right)  \; .
\label{eq:2.279}
\end{eqnarray}
Thus one has found that the polarization quantities at
equilibrium are characterized by the complex constant
$h^w$, which is easily determined (see 
(\ref{eq:2.244}),
(\ref{eq:2.85}))
if one knows the
initial state:
\begin{eqnarray*}
   h^w &=&   \exp(-\frac{d^2\cdot\sigma_{\sigma}^2}{2a^2}) \cdot
     \int_{-\infty}^{+\infty} \; d\sigma
     \int_{-\infty}^{+\infty} \; d\eta
     \int_{-\infty}^{+\infty} \; d\psi \cdot
           w(\sigma,\eta,\psi+\frac{d}{a}\cdot\sigma;s_0) \cdot
                                 \exp(i\cdot\psi) \; .
\end{eqnarray*}
Note that by 
(\ref{eq:2.87}),
(\ref{eq:2.277}),
(\ref{eq:2.278}) one finds that at equilibrium
the polarization density,
the local polarization vector and its
direction, besides fulfilling their Bloch equations 
(\ref{eq:2.64}),
(\ref{eq:2.271}),
(\ref{eq:2.273}), 
also fulfill the
radiationless Bloch equation 
(\ref{eq:2.65}). 
Since Machine I is smooth the asymptotic polarization quantities
are independent of $s$. So $\partial \vec{P}^w/\partial s 
=\partial \vec{P}^w_{loc}/\partial s 
=\partial \vec{P}^w_{dir}/\partial s = 0$.
\subsubsection*{2.9.3}
Now I apply the differential equations 
(\ref{eq:2.18}),
(\ref{eq:2.19})
to find the first
and second moments for the equilibrium of an arbitrary
process running with Machine I. First of all I get
\begin{eqnarray*}
&& \sigma_{11}(+\infty) = \sigma_{\sigma}^2 \; ,
\qquad
   \sigma_{22}(+\infty) = \sigma_{\eta}^2 \; ,
\qquad
   \sigma_{12}(+\infty) =
   \sigma_{21}(+\infty) = 0 \; ,
\end{eqnarray*}
which follows from 
(\ref{eq:2.82}).
\par Applying 
(\ref{eq:2.18})
one then gets:
\begin{eqnarray*}
&& 0 = \sigma_{13}'(+\infty)
     = a\cdot \sigma_{23}(+\infty) +d\cdot \sigma_{12}(+\infty)
     = a\cdot \sigma_{23}(+\infty)  \; ,
\nonumber\\
&& 0 = \sigma_{23}'(+\infty)
     =   b\cdot \sigma_{13}(+\infty) +  c\cdot \sigma_{23}(+\infty)
+d\cdot \sigma_{22}(+\infty)
\nonumber\\
&& = b\cdot \sigma_{13}(+\infty) +d\cdot \sigma_{22}(+\infty)
   = b\cdot \sigma_{13}(+\infty) +d\cdot \sigma_{\eta}^2 \; ,
\end{eqnarray*}
where I also used:
\begin{eqnarray*}
0 &=& \underline{\sigma}'(+\infty)\;,
\end{eqnarray*}
which follows from the fact that every process running with Machine I
approaches equilibrium. Also from 
(\ref{eq:2.18}) it
follows that:
\begin{eqnarray*}
   0 &=& \sigma_{33}'  +\frac{d^2}{a^2}\cdot\sigma_{11}'
          -\frac{2d}{a}\cdot \sigma_{13}'  \; ,
\end{eqnarray*}
i.e.:
\begin{eqnarray*}
 && \sigma_{33}(s)+\frac{d^2}{a^2}\cdot\sigma_{11}(s)
  -\frac{2d}{a}\cdot\sigma_{13}(s) =
\sigma_{33}(s_0)+\frac{d^2}{a^2}\cdot\sigma_{11}(s_0)
  -\frac{2d}{a}\cdot\sigma_{13}(s_0) \; .
\end{eqnarray*}
Hence the equilibrium covariance matrix has the form
\begin{eqnarray}
&&    \underline{\sigma}(+\infty) =
                               \left( \begin{array}{ccc}
   \sigma_{\sigma}^2          &       0  &
    (d\cdot\sigma_{\sigma}^2)/a \\
&& \\
   0  &   \sigma_{\eta}^2  & 0   \\
&& \\
    (d\cdot\sigma_{\sigma}^2)/a  &
      0 &
   (d^2/a^2)\cdot\sigma_{\sigma}^2 +
\sigma_{33}(s_0)+(d^2/a^2)\cdot\sigma_{11}(s_0)
  -(2d/a)\cdot\sigma_{13}(s_0)                 \end{array}
         \right) \; . \nonumber\\&&
 \label{eq:2.88}
\end{eqnarray}
This is the equilibrium covariance matrix for an arbitrary process
running with Machine I. One sees that the $(33)$-element is simply
determined by the initial covariance matrix. Stating it differently:
two processes running with Machine I have equilibrium covariance
matrices which can only differ by the $(33)$-element. Of course, the
equilibrium covariance matrices of processes 1 and 2 have the form 
(\ref{eq:2.88}). Also one finds that the equilibrium covariance matrices
of the processes defined by
(\ref{eq:2.253}),
(\ref{eq:2.254}) are different, confirming the results of section 2.8.2.
\par By (\ref{eq:2.213}) one has:
\begin{eqnarray*}
 && \sigma_{33}(s)+\frac{d^2}{a^2}\cdot\sigma_{11}(s)
  -\frac{2d}{a}\cdot\sigma_{13}(s)  =
<\biggl(\tilde\psi(s)-<\tilde\psi(s)>\biggr)^2> \; ,
\end{eqnarray*}
so that by the nonnegativity of this expression
and by (\ref{eq:2.88})
the minimum value possible for
$\sigma_{33}(+\infty)$ is given by
$(d^2/a^2)\cdot\sigma_{\sigma}^2$.
Note that the determinant of 
(\ref{eq:2.88})
only vanishes in this case, i.e. the
equilibrium covariance matrix is singular if and only if
\begin{eqnarray*}
 0 &=& \sigma_{33}(s_0)+\frac{d^2}{a^2}\cdot\sigma_{11}(s_0)
  -\frac{2d}{a}\cdot\sigma_{13}(s_0) \; .
\end{eqnarray*}
An example is Process 1. Another example is given by the process
with the probability density (\ref{eq:2.274}).
\par Using 
(\ref{eq:2.19})
one easily finds the equilibrium first moment vector
of an arbitrary process $\vec{x}(s)$ running with Machine I:
\begin{eqnarray}
 <\vec{x}(+\infty)> &=&
   \biggl(0,0,<\psi(s_0)>-\frac{d}{a}\cdot<\sigma(s_0)>\biggr)^T \; .
\label{eq:2.262}
\end{eqnarray}
The equilibrium first moment vector obeys:
\begin{eqnarray}
&& \underline{\cal A}_I\cdot
 <\vec{x}(+\infty)> = 0 \; ,
 \label{eq:2.263}
\end{eqnarray}
which of course can be concluded directly from
(\ref{eq:2.19}) even without knowing the explicit form 
(\ref{eq:2.262}).
Note that $<\vec{x}(+\infty)>$ is not uniquely determined by
(\ref{eq:2.263}), because
the determinant of $\underline{\cal A}_I$ vanishes (see (\ref{eq:2.264})).
Thus the nonuniqueness of the equilibrium
state of Machine I follows from the singular nature of the matrix
$\underline{\cal A}_I$.
\par For G-processes (see section 2.8) the equilibrium states are just
determined by the two numbers $<\psi(+\infty)>$ resp.
$\sigma_{33}(+\infty)$ so that the family
of those equilibrium states is two-parametric.
\footnote{In particular the family of stationary G-processes is 
two-parametric.}
\subsubsection*{2.9.4}
A stationary process 
fulfills the condition of `detailed balance' \cite{Gar85}, if one has:
\begin{eqnarray}
 w_{joint}(\vec{x};s;\vec{x}_1;s_1)
              &=&
 w_{joint}(\underline{\varepsilon}\cdot\vec{x}_1;s;
\underline{\varepsilon}\cdot\vec{x};s_1) \; ,
 \label{eq:2.260}
\end{eqnarray}
where $w_{joint}$ denotes the joint probability density and where the matrix
\begin{eqnarray*}
 \underline{\varepsilon} &\equiv &
                               \left( \begin{array}{ccc}
   \varepsilon_1 & 0 & 0 \\
    0 & \varepsilon_2 & 0  \\
    0 & 0 & \varepsilon_3
                \end{array}
         \right)
\end{eqnarray*}
defines the time reversal operation with:
\begin{eqnarray*}
&& \varepsilon_1^2 =  \varepsilon_2^2 =  \varepsilon_3^2 = 1 \; .
\end{eqnarray*}
Using the probability density and the transition probability density this can 
be expressed via (\ref{eq:2.235}) by:
\begin{eqnarray*}
 w_{I,trans}(\vec{x};s|\vec{x}_1;s_1)\cdot
 w(\vec{x}_1;s_1) &=&
 w_{I,trans}(\underline{\varepsilon}\cdot\vec{x}_1;s|
\underline{\varepsilon}\cdot\vec{x};s_1)\cdot
 w(\underline{\varepsilon}\cdot\vec{x};s_1) \; .
\end{eqnarray*}
The condition of detailed balance roughly means that for the stationary
process described by $w$ each possible transition
\begin{eqnarray*}
&& (\vec{x}_1;s_1) \rightarrow  (\vec{x};s)
\end{eqnarray*}
is balanced by the `reverse' transition:
\begin{eqnarray*}
&& (\underline{\varepsilon}\cdot \vec{x};s_1) \rightarrow
(\underline{\varepsilon}\cdot \vec{x}_1;s) \; .
\end{eqnarray*}
Choosing the matrix $\underline{\varepsilon}$ so that
\footnote{Thus I choose $\eta$ as a `velocity' variable.}:
\begin{eqnarray*}
&& 1 = \varepsilon_1  = \varepsilon_3   = -\varepsilon_2 \; ,
\end{eqnarray*}
I will show for Machine I that every stationary process
fulfills the condition of detailed balance.
\par By (\ref{eq:2.83}) the probability density of a
stationary process can be written in the form
\begin{eqnarray}
   w(\sigma,\eta,\psi;s) &=&
          w_{norm}(\sigma,\eta) \cdot
      w_{aver}(\psi-\frac{d}{a}\cdot\sigma;s_0) \; ,
 \label{eq:2.243}
\end{eqnarray}
where $w_{aver}$ is given by 
(\ref{eq:2.244}).
Thus by using 
(\ref{eq:2.36}),
(\ref{eq:2.235}),
(\ref{eq:2.37}),
(\ref{eq:2.236}),
(\ref{eq:2.243})
the joint probability density of a stationary process can be written as:
\begin{eqnarray}
   w_{joint}(\sigma,\eta,\psi;s;\sigma_1,\eta_1,\psi_1;s_1) &=&
   w_{orb,joint}(\sigma,\eta;s;\sigma_1,\eta_1;s_1)\cdot
 \delta\biggl(\psi-\psi_1-\frac{d}{a}\cdot(\sigma-\sigma_1)\biggr)
\nonumber\\&&\qquad
\cdot
w_{aver}(\psi_1-\frac{d}{a}\cdot\sigma_1;s_0) \; ,
 \label{eq:2.89}
\end{eqnarray}
where $w_{orb,joint}$ denotes the orbital joint probability density, 
which via
(\ref{eq:2.236}),
(\ref{eq:2.243}) is given by:
\begin{eqnarray}
&& w_{orb,joint}(\sigma,\eta;s;\sigma_1,\eta_1;s_1) =
 w_{orb,trans}(\sigma,\eta;s|\sigma_1,\eta_1;s_1)\cdot
 w_{norm}(\sigma_1,\eta_1) \; .
 \label{eq:2.258}
\end{eqnarray}
From
(\ref{eq:2.215}),
(\ref{eq:2.37}),
(\ref{eq:2.258})
follows:
\begin{eqnarray}
 w_{orb,joint}(\sigma,\eta;s;\sigma_1,\eta_1;s_1)
              &=&
 w_{orb,joint}(\sigma_1,-\eta_1;s;\sigma,-\eta;s_1) \; .
 \label{eq:2.259}
\end{eqnarray}
Combining 
(\ref{eq:2.89}),
(\ref{eq:2.259}) one observes that 
(\ref{eq:2.260}) holds
so that I have proven that for Machine I every stationary process
fulfills the condition of detailed balance.
\par For Machine I, as for many stochastic systems whose stationary states obey
detailed balance, the `Onsager relations' \cite{Gar85} hold. 
By these relations the covariance matrix
$\underline{\sigma}$ for         a stationary process fulfills:
\begin{eqnarray}
&& \underline{\varepsilon}\cdot \underline{\cal A}_I\cdot
 \underline{\sigma} =
  \biggl(\underline{\varepsilon}\cdot \underline{\cal A}_I\cdot
 \underline{\sigma}\biggr)^T \; .
 \label{eq:2.90}
\end{eqnarray}
In fact, the covariance matrix of a stationary process is the equilibrium
covariance matrix and the latter, given by (\ref{eq:2.88}), fulfills
the Onsager relations 
(\ref{eq:2.90}). This proves that the Onsager relations 
hold for every stationary process of Machine I.
\subsection{Resonant spin flip}
The study of Machine I was
originally motivated by a wish to know whether 
it is possible to flip vertical polarization from up to down by perturbing
the spins with an oscillating radial magnetic field running at a frequency
close to $\nu$ and hence almost in resonance with the precession of the
spin basis $\vec{m}_{0,I}(s),\vec{l}_{0,I}(s),\vec{n}_{0,I}(s)$. My 
calculation suggests that for the smoothed machine the horizontal spin 
components would partially 
decohere within a few orbital damping times. Perhaps in reality there would
be complete decoherence 
\cite{BBHMR94a,BBHMR94b}. In any case it looks as if the spin flip
procedure
should be completed within a fraction of the orbital damping time. However it 
must be bourne in mind that I have neglected the oscillating field in my 
calculations. I am pursuing this topic further.
\subsection{Recapitulation of Machine I}
Although the stochastic spin-orbit system of Machine I is very simple
it has served to illustrate the application of standard stochastic
differential equation theory to such systems. Moreover, this study
is a useful introduction to the treatment of a more complicated system,
namely Machine II.
To orient the reader I recapitulate the main
results here:
\begin{itemize}
\item For Machine I all processes reach equilibrium.
\item However the equilibrium is not unique.
\item There is no equation which provides a causal
azimuthal evolution for the stochastic average of the
five-component
spin-orbit vector, i.e. there is
no appropriate Bloch equation for the polarization vector in Machine I.
\item But there is a universal Bloch equation which provides a causal
azimuthal evolution for the polarization density.
\item The Bloch equations for the local polarization vector and its 
direction provide causal
azimuthal evolution only under certain circumstances.
\item The value of the local polarization is uniform across phase space for 
Process 1 and G-processes (e.g. Process 2).
\item There is an upper limit to the equilibrium polarization, namely the
equilibrium value of Process 1.
\end{itemize}
\setcounter{subsection}{0}
\setcounter{equation}{0}
\section{Machine II}
\subsection{  }
In section 2 I studied the spin distribution w.r.t. a pair of (usually) 
nonperiodic vectors (${\vec m}_0,{\vec l}_0$) lying in the horizontal
plane and found that this distribution always reaches equilibrium. 
However, this is not an equilibrium w.r.t. the directions ${\vec
e}_1,{\vec e}_2$. On the contrary, the equilibrium spin direction 
on the closed orbit is the $\vec n_0$-axis, i.e. that
solution ${\vec n}_0$ to the Thomas-BMT equation on the closed
orbit which is 1-turn periodic in the machine frame and
in Machine I this is vertical. Thus it would be interesting to
use my formalism to study spin diffusion w.r.t. an equilibrium spin
direction which is also constrained to lie in the horizontal plane. This
can be arranged by including a Siberian Snake \cite{DK78} in the smoothed
optic of Machine I to create `Machine II'. Siberian snakes are devices that
rotate a spin on the closed orbit by an angle of $\pi$ around a fixed axis
which usually lies in the horizontal plane. With such a snake the
${\vec n}_0$-axis is horizontal. 
\par This layout is of great practical interest for some existing or proposed
electron storage rings (e.g the MIT-Bates, AmPs and  BTCF rings) 
\footnote{See \cite{Bar96} and the reference list therein.}
where
horizontal spin polarization is required at the interaction points but
whose energy is too low for a useful Sokolov-Ternov \cite{ST64}
polarization rate to be achieved. These rings use Siberian Snakes to 
ensure that the ${\vec n}_0$-axis lies in the horizontal plane. A polarized 
electron beam is injected with its polarization vector parallel to the
${\vec n}_0$-axis at the injection point and as well as determining the
${\vec n}_0$-axis
the snakes are supposed to suppress the spin diffusion that one
naively expects when spins lie in the horizontal plane so that useful 
polarization lifetimes can be achieved. 
\footnote{But recall from
Machine I that it is not so clear that there will be complete depolarization.}
\par In this section I use my model to calculate the spin decoherence in the
presence of a pointlike snake whose rotation axis is radial 
\footnote{In the language of \cite{Mon84} this is
a `type 2' snake. See section 6.3 thereof.
The depolarization time determined below (see eq. (\ref{eq:3.267}))
would be the same if the rotation axis were longitudinal.}
As we will see, this will
allow a comparison with calculations using the SLIM formalism \cite{Cha81}
which is based on a linearized description of spin motion and will allow
a basic assumption underlying conventional treatments 
\cite{DK72,DK73,Man87,BHMR91} 
to be checked from scratch - at least for this model.
\par The Thomas-BMT equation for Machine II reads as
\begin{eqnarray}
 \vec{\xi}\ ' &=&
\vec{\Omega}_{II}\wedge \vec{\xi}   \; ,
\label{eq:3.1}
\end{eqnarray}
with
\begin{eqnarray}
\vec{\Omega}_{II} &\equiv&
\vec{\Omega}_{II,0} + \vec{\Omega}_{osc} \; ,
\nonumber\\
\vec{\Omega}_{II,0} &\equiv&
\Omega_{II,0,dipole}\cdot\vec{e}_3
+\Omega_{II,0,snake}\cdot\delta_{L,per}\cdot\vec{e}_1 \; ,
\nonumber\\
\Omega_{II,0,dipole} &\equiv&
 ||\vec{\Omega}_{I,0}|| = d \; , \nonumber\\&&
 \label{eq:3.2}
\end{eqnarray}
where the machine frame dreibein
${\vec e}_1,{\vec e}_2,\vec e_3$ has the same meaning as for Machine I.
Here $\Omega_{II,0,dipole},\Omega_{II,0,snake}$ are constant,
$\delta_{L,per}(s)$ denotes the periodic delta function
with period $L$
\footnote{Note that:
\begin{eqnarray*}
\delta_{L,per}(s) &\equiv& 
\sum_{n=-\infty}^{\infty}\;
 \delta(s- n\cdot L) \; .
\end{eqnarray*}
For more details on step functions and delta functions see \cite{Lig59}.}
and $\vec\Omega_{osc}$ is defined in section 2.1.
The snake is located at $s=0$ and:
\begin{eqnarray*}
 \Omega_{II,0,snake}     &=& \pi \; .
\end{eqnarray*}
The $\vec n_0$-axis for Machine II
is given in Appendix A and reads as:
\begin{eqnarray}
 \vec{n}_{0,II}(s)  &\equiv&
\cos\biggl(g_6(s)\biggr) \cdot \vec{e}_1+
\sin\biggl(g_6(s)\biggr) \cdot \vec{e}_2 \; ,
 \label{eq:3.3}
\end{eqnarray}
with
\begin{eqnarray*}
 g_6(s) &\equiv& d\cdot
    \biggl(s-L/2-L\cdot {\cal G}(s/L)\biggr)  \; ,
\end{eqnarray*}
where the step function $\cal G$ is defined by:
\begin{eqnarray*}
 {\cal G}(s) &\equiv&
   N  \qquad\qquad {\rm if} \; N < s < N+1 \; ,
\end{eqnarray*}
and where $N$, as always in this section, denotes an integer.
Note that $\vec{n}_{0,II}(s)$ is 1-turn periodic in the machine frame.
I also define the vectors
$\vec{m}_{0,II}(s),\vec{l}_{0,II}(s)$:
\begin{eqnarray}
&& \vec{l}_{0,II}(s)  \equiv \theta_{2L,per}(s)\cdot\vec{e}_3 \; , \qquad
 \vec{m}_{0,II}(s) \equiv \vec{l}_{0,II}(s)\wedge \vec{n}_{0,II}(s) \; ,
 \label{eq:3.244}
\end{eqnarray}
where $\theta_{2L,per}$ denotes a 2-turn periodic step function 
and where the step is located at the snake. Explicitly one has:
\begin{eqnarray*}
   \theta_{2L,per}(s)  &\equiv&
                  \left\{ \begin{array}{l}
 1  \qquad {\rm if} \; 2NL < s <    (2N+1)L\\
-1  \qquad {\rm if} \; (2N+1)L < s  <   2NL+ 2L  \; .
                          \end{array}
                  \right.
\end{eqnarray*}
Note that:
\begin{eqnarray*}
&&   \theta_{2L,per}(s) = (-1)^{{\cal G}(s/L)} \; .
\end{eqnarray*}
The vectors $\vec{m}_{0,II}(s),\vec{l}_{0,II}(s)$
            are 2-turn periodic in $s$ in the machine frame. 
The
vectors $\vec{n}_{0,II},\vec{m}_{0,II}$
                              precess in the horizontal plane around the
vertical dipole field.
\footnote{Because $\vec{m}_{0,II}(s),\vec{l}_{0,II}(s)$ are 2-turn periodic in
$s$ in the machine frame, the fractional part of the closed-orbit
spin tune equals $1/2$. This is the trademark of a snake
\cite{Mon84}.}
As before 
I deal only with horizontal spin 
and therefore define the phase angle $\psi$ by
\begin{eqnarray*}
 \vec{\xi}  &\equiv& \frac{\hbar}{2}\cdot
\biggl( \vec{n}_{0,II}\cdot\cos(\psi)+
        \vec{m}_{0,II}\cdot\sin(\psi) \biggr)  \; .
\end{eqnarray*}
Hence the Thomas-BMT equation (\ref{eq:3.1}), reexpressed in the
$(\vec{n}_{0,II},\vec{m}_{0,II},\vec{l}_{0,II})$-frame, is
equivalent to
\begin{eqnarray*}
 \psi ' &=& (2 \pi \nu /L) \cdot \eta \cdot \theta_{2L,per}
         = d \cdot \eta \cdot \theta_{2L,per} = \eta\cdot
\hat{d} \; ,
\end{eqnarray*}
where:
\begin{eqnarray}
  \hat{d}(s) &\equiv& d\cdot \theta_{2L,per}(s) \; .
\end{eqnarray}
I also   introduce
the spin vector
\begin{eqnarray*}
 \vec{S}  &\equiv& \frac{\hbar}{2}\cdot
  \left( \begin{array}{c}
     \cos(\psi) \\
     \sin(\psi) \\
                      0
                \end{array}
         \right)  \; ,
\end{eqnarray*}
describing the spin in the
$(\vec{n}_{0,II},\vec{m}_{0,II},\vec{l}_{0,II})$-frame. Then, as shown in 
Appendix A, the Thomas-BMT equation reads as
\begin{eqnarray}
 \vec{S}'(s) &=&
\vec{W}_{II}\biggl(\eta(s);s\biggr)\wedge \vec{S}(s)   \; ,
 \label{eq:3.4}
\end{eqnarray}
with
\begin{eqnarray}
\vec{W}_{II}(\eta;s) &=&  \hat{d}(s)\cdot\eta\cdot
  \left( \begin{array}{c}
           0       \\
           0       \\
           1
                \end{array}
         \right)  \; .
 \label{eq:3.243}
\end{eqnarray}
\subsection{The Langevin equation and the Fokker-Planck equation
            for        Machine II}
\subsubsection*{3.2.1}
The Langevin equation for Machine II is given by
\begin{eqnarray}
 d\vec{x}(s) &=&  \underline{\cal A}_{II}\cdot\vec{x}(s)\cdot ds
               +\underline{\cal B}\cdot d\vec{\cal W}(s) \; ,
 \label{eq:3.5}
\end{eqnarray}
where
\begin{eqnarray*}
\underline{\cal A}_{II} &\equiv&        \left( \begin{array}{ccc}
      0 & a & 0                            \\
      b & c & 0                            \\
      0 & \hat{d} & 0
                \end{array}
         \right) \; .
\end{eqnarray*}
Because $\underline{\cal A}_{II}$ and $\underline{\cal B}$ are 
matrices independent of $\vec x$,
the Langevin equation 
(\ref{eq:3.5})
describes three-component processes of Ornstein-Uhlenbeck type \cite{Gar85}. 
If $s_0$ denotes the starting azimuth 
of a process $\vec{x}(s)$ then as for Machine I $\vec{x}(s_0)$
is always assumed to be chosen so that $\vec{x}(s)$ is a 
Markovian diffusion process.
\footnote{For the special processes 3,4 and 5 considered in detail I choose
the starting azimuth as $s=0$.}
\subsubsection*{3.2.2}
The Fokker-Planck equation corresponding to the Langevin equation
(\ref{eq:3.5})
has the form \cite{Gar85,Ris89}
\begin{eqnarray*}
 \frac{\partial w}{\partial s} &=&
-\sum_{j,k=1}^3\frac{\partial}{\partial x_j}( {\cal A}_{II,jk}\cdot 
 x_k\cdot w )
+\frac{1}{2}\cdot
 \sum_{j,k=1}^3 \frac{\partial^2}{\partial x_j\partial x_k}
 ( {\cal D}_{jk}\cdot w ) \; .
\end{eqnarray*}
Therefore the Fokker-Planck equation can be written
\begin{eqnarray}
   \frac{\partial w}{\partial s} &=&
 -\underbrace{
 a\cdot \eta \cdot
\frac{\partial w}{\partial \sigma}
- b\cdot\sigma \cdot \frac{\partial w}{\partial \eta}
                                                }_
               {{\rm synchrotron\; oscillation\;terms}} \;
 -\underbrace{
 \hat{d}\cdot \eta
               \cdot \frac{\partial w}{\partial \psi}
                                                }_
               {{\rm spin \; precession \; term}}
 -\underbrace{
      c\cdot w
- c\cdot\eta \cdot \frac{\partial w}{\partial \eta}
                                                }_
               {{\rm damping \; terms}}
+ \underbrace{
          \frac{\omega}{2}\cdot
\frac{\partial^2 w}{\partial \eta^2}
                                                }_
               {{\rm diffusion \; term}}
\nonumber\\&&
\nonumber\\
 && \equiv
          L_{FP,orb} \; w + L_{FP,II,spin} \; w
                    \; ,
 \label{eq:3.7}
\end{eqnarray}
where I used the abbreviation
\begin{eqnarray*}
    L_{FP,II,spin} &\equiv &
-\hat{d}\cdot \eta \cdot \frac{\partial }{\partial \psi}  \; .
\end{eqnarray*}
As for Machine I use standard boundary conditions in all three variables
$\sigma,\eta,\psi$ so that the probability densities of the 
processes considered are
normalized by (\ref{eq:2.9}).
\par By the
Fokker-Planck equation 
(\ref{eq:3.7}), the characteristic function $\Phi$ corresponding to a
probability density $w$ 
(see (\ref{eq:2.222}))
obeys:
\begin{eqnarray}
 \frac{\partial \Phi}{\partial s} &=&
      \sum_{j,k=1}^3 \; {\cal A}_{II,kj}\cdot u_k\cdot
     \frac{\partial\Phi}{\partial u_j}
-\frac{1}{2}\cdot
 \sum_{j,k=1}^3 \;   {\cal D}_{jk}\cdot u_j\cdot u_k\cdot \Phi \; .
 \label{eq:3.251}
\end{eqnarray}
\subsection{The polarization density and its Bloch equation for Machine II}
The polarization density is defined in the same way as for Machine I
(see (\ref{eq:2.62})). 
From the Fokker-Planck equation 
(\ref{eq:3.7})
one obtains:
\begin{eqnarray*}
\frac{\partial\vec{P}^w} {\partial s}   &=&
       L_{FP,orb} \; \vec{P}^w +
\vec{W}_{II}\wedge \vec{P}^w \; ,
\end{eqnarray*}
which is the Bloch equation for Machine II w.r.t. the
$(\vec{n}_{0,II},\vec{m}_{0,II},\vec{l}_{0,II})$-frame.
Writing it out explicitly one gets
\begin{eqnarray}
&&   \frac{\partial \vec{P}^w}{\partial s}  =
\underbrace{
-a\cdot \eta \cdot
\frac{\partial \vec{P}^w}{\partial \sigma}
- b\cdot\sigma \cdot \frac{\partial \vec{P}^w}{\partial \eta}
+ \vec{W}_{II}\wedge \vec{P}^w
                                                }_
               {{\rm radiationless \; part}}
-
  \underbrace{
     c\cdot \vec{P}^w
- c\cdot \eta \cdot \frac{\partial \vec{P}^w}{\partial \eta}
                                                }_
               {{\rm damping \; terms}}
         +
  \underbrace{
          \frac{\omega}{2}\cdot
\frac{\partial^2 \vec{P}^w}{\partial \eta^2}
                                                }_
               {{\rm diffusion \; term}}
                                           \; .
\nonumber\\&&
 \label{eq:3.10}
\end{eqnarray}
The radiationless Bloch equation underlying
Machine II reads as:
\begin{eqnarray}
   \frac{\partial \vec{P}^w}{\partial s} &=&
-a\cdot \eta \cdot
\frac{\partial \vec{P}^w}{\partial \sigma}
- b\cdot\sigma \cdot \frac{\partial \vec{P}^w}{\partial \eta}
+ \vec{W}_{II}\wedge \vec{P}^w    \; .
 \label{eq:3.11}
\end{eqnarray}
\subsection{Further properties of Machine II}
\subsubsection*{3.4.1}
\par With the standard boundary conditions one can immediately
write down a differential equation for the covariance matrix 
$\underline{\sigma}$ of any
process running with Machine II:
\begin{eqnarray}
\underline{\sigma}'&=&
\underline{\cal A}_{II}\cdot\underline{\sigma} +
\underline{\sigma}\cdot\underline{\cal A}_{II}^T+
\underline{{\cal D}} \; .
\label{eq:3.8}
\end{eqnarray}
Writing out the components results in:
\begin{eqnarray*}
\sigma_{11}'&=& 2\cdot a\cdot\sigma_{12} \; ,
\nonumber\\
\sigma_{12}'&=& a\cdot\sigma_{22} +
                b\cdot\sigma_{11} +
                c\cdot\sigma_{12} \; ,
\nonumber\\
\sigma_{22}'&=& 2\cdot b\cdot\sigma_{12}
              + 2\cdot c\cdot\sigma_{22}
              + \omega \; ,
\nonumber\\
 \sigma_{13}'&=& a\cdot\sigma_{23}
                       +\hat{d}\cdot\sigma_{12} \; ,
\nonumber\\
\sigma_{23}'&=& b\cdot\sigma_{13}
                       +c\cdot\sigma_{23}
                       +\hat{d}\cdot\sigma_{22} \; ,
\nonumber\\
\sigma_{33}'&=& 2\cdot \hat{d}\cdot\sigma_{23}\; .
\end{eqnarray*}
For the first moment vector one gets the following differential
equation:
\begin{eqnarray}
 <\vec{x}'(s)>&=&\underline{\cal A}_{II}(s)\cdot
               <\vec{x}(s)> \; .
\label{eq:3.9}
\end{eqnarray}
The differential equations 
(\ref{eq:3.8}),(\ref{eq:3.9})
can be easily derived from any of the equations
(\ref{eq:3.5}),(\ref{eq:3.7}) or
(\ref{eq:3.251}) in analogy with (\ref{eq:2.18}),
(\ref{eq:2.19}).
\subsubsection*{3.4.2}
In this section I consider further properties arising for
processes which are at orbital equilibrium, i.e. for which $w_{orb}=w_{norm}$
\footnote{The statements of section 3.4.2 are only valid for processes
at orbital equilibrium.}.
Firstly:
\begin{eqnarray}
&& <\sigma(s)> = <\eta(s)> = 0 \; .
 \label{eq:3.12}
\end{eqnarray}
Then from (\ref{eq:3.9}),(\ref{eq:3.12}) one gets:
\begin{eqnarray}
&& <\psi'(s)> = 0 \; .
\label{eq:3.13}
\end{eqnarray}
From (\ref{eq:3.12}),(\ref{eq:3.13}) follows:
\begin{eqnarray}
<\vec{x}(s)> &=&   (0,0,<\psi(s_0)>)^T \; .
\label{eq:3.219}
\end{eqnarray}
The orbital matrix elements of the covariance matrix read as:
\begin{eqnarray}
&&  \sigma_{11}   = \sigma_{\sigma}^2 \; ,
\qquad
  \sigma_{12}    =   \sigma_{21} = 0 \; ,
\qquad
  \sigma_{22}    =   \sigma_{\eta}^2 \; ,
\label{eq:3.14}
\end{eqnarray}
so that the determinant of the covariance matrix is:
\begin{eqnarray}
 \det(\underline{\sigma}) &=&
    \sigma^2_{\eta}\cdot\sigma^2_{\sigma}\cdot\sigma_{33}
- \sigma^2_{\sigma}\cdot\sigma_{23}^2
- \sigma^2_{\eta}\cdot\sigma_{13}^2 \; .
\label{eq:3.15}
\end{eqnarray}
From this follows by using the differential equation
(\ref{eq:3.8}):
\begin{eqnarray}
\biggl(\det(\underline{\sigma})\biggr)' &=&
- 2\cdot c\cdot\sigma^2_{\sigma}\cdot\sigma_{23}^2 \; ,
\label{eq:3.200}
\end{eqnarray}
so that:
\begin{eqnarray}
 \det\biggl(\underline{\sigma}(s)\biggr) &=&
\det\biggl(\underline{\sigma}(s_0)\biggr)
- 2\cdot c\cdot\sigma^2_{\sigma}\cdot \int_{s_0}^s\; ds_1
                       \cdot\sigma_{23}^2(s_1) \; .
\label{eq:3.201}
\end{eqnarray}
Also one obtains by 
(\ref{eq:3.8}),(\ref{eq:3.14}):
\begin{eqnarray}
&&                \left( \begin{array}{c}
 \sigma_{13}' \\
 \sigma_{23}'
                \end{array}
         \right)     =
\underline{\cal A}_{orb}\cdot
               \left( \begin{array}{c}
 \sigma_{13}  \\
 \sigma_{23}
                \end{array}
         \right)
+  \sigma^2_{\eta}\cdot      \left( \begin{array}{c}
     0          \\
 \hat{d}
                \end{array}
         \right) \; , 
\label{eq:3.202} \\
&& \sigma_{33}' = 2\cdot \hat{d}\cdot\sigma_{23}\; .
\label{eq:3.203}
\end{eqnarray}
Note that 
(\ref{eq:3.202}),(\ref{eq:3.203}) are formally solved by:
\begin{eqnarray}
&&                  \left( \begin{array}{c}
 \sigma_{13}(s)  \\
 \sigma_{23}(s)
                \end{array}
         \right)    =
    \exp(\underline{\cal A}_{orb}\cdot (s-s_0))\cdot
                  \left( \begin{array}{c}
 \sigma_{13}(s_0)  \\
 \sigma_{23}(s_0)
                \end{array}
         \right)
\nonumber\\
   && \qquad      +
  \sigma^2_{\eta}\cdot
  \int_{s_0}^s\; ds_1\cdot
    \exp\biggl(\underline{\cal A}_{orb}\cdot(s-s_1)\biggr)\cdot
                              \left( \begin{array}{c}
     0          \\
 \hat{d}(s_1)
                \end{array}
         \right) \; , \nonumber\\
&& \sigma_{33}(s)  =  \sigma_{33}(s_0)  
+   2\cdot \int_{s_0}^s\; ds_1
                  \cdot\sigma_{23}(s_1)\cdot\hat{d}(s_1) \; . \nonumber\\&&
\label{eq:3.225}
\end{eqnarray}
By (\ref{eq:3.219}),(\ref{eq:3.14}),
(\ref{eq:3.225}) one sees that 
the first moment vector and the covariance matrix depend continuously on $s$.
However the dependence is not smooth because
the discontinuity of
$\hat{d}(s)$ at $s=NL$ causes 
(see (\ref{eq:3.202}))
$\sigma_{23}'(s)$ to be discontinuous at $s=NL$ ($N$=integer).
\par In addition to the 
Bloch equation for the polarization density, at orbital equilibrium
a Bloch equation
for the local polarization vector holds. In fact from 
(\ref{eq:2.215}),
(\ref{eq:2.200}),
(\ref{eq:3.10})
follows:
\begin{eqnarray}
&& \frac{\partial \vec{P}^w_{loc}}{\partial s}  =
-a\cdot \eta \cdot
\frac{\partial \vec{P}^w_{loc}}{\partial \sigma}
- b\cdot\sigma \cdot \frac{\partial \vec{P}^w_{loc}}{\partial \eta}
+ \vec{W}_{II}\wedge \vec{P}^w_{loc}
+ c\cdot \eta \cdot \frac{\partial \vec{P}^w_{loc}}{\partial \eta}
         +
          \frac{\omega}{2}\cdot
\frac{\partial^2 \vec{P}^w_{loc}}{\partial \eta^2}
                                           \; .
 \label{eq:3.249}
\end{eqnarray}
\subsubsection*{3.4.3}
In the remainder of section 3 I study three different G-processes at
orbital equilibrium which I call processes 3,4,5.
\footnote{The statements of section 3.4.3 are only valid for G-processes
at orbital equilibrium.}
For Machine II G-processes are defined as for Machine I; see section
2.8. 
Moreover by (\ref{eq:3.14}) 
the inverse 
$\underline{\sigma}_{inv}$ of the covariance matrix 
fulfills:
\begin{eqnarray}
    \sigma_{inv,13} &=&
 -\frac{\sigma_{\eta}^2\cdot \sigma_{13}}
    {\det(\underline{\sigma})} \; ,
\qquad
    \sigma_{inv,23}  =
 -\frac{\sigma_{\sigma}^2\cdot \sigma_{23}}
    {\det(\underline{\sigma})} \; ,
\qquad
    \sigma_{inv,33}  =
  \frac{\sigma_{\sigma}^2\cdot \sigma_{\eta}^2}
    {\det(\underline{\sigma})} \; .
 \label{eq:3.213}
\end{eqnarray}
Using sections 2.8 and 3.4.2 one finds:
\begin{eqnarray}
&&\vec{P}_{tot}^w(s)  =
   \exp(-\sigma_{33}(s)/2)\cdot
  \left( \begin{array}{c}
        \cos(<\psi(s_0)>) \\
        \sin(<\psi(s_0)>) \\
                      0
                \end{array}
         \right)  \; ,
 \label{eq:3.214}
\\
&&      ||\vec P_{tot}^w(s)|| =
    \exp(-\sigma_{33}(s)/2)  \; , 
 \label{eq:3.215}
\\
&& ||\vec P_{loc}^{w}(\sigma,\eta;s)|| =
               \exp\biggl(
-\frac{\det(\underline{\sigma})}{2\cdot\sigma_{\sigma}^2\cdot
 \sigma_{\eta}^2} \biggr) \; , \label{eq:3.217}
\\
&& \vec{P}^{w}_{dir}(\sigma,\eta;s) =
   \large{      \left( \begin{array}{c}
     \cos\biggl(
        \frac{\sigma_{13}(s)}{\sigma_{\sigma}^2}\cdot
                                 \sigma
      + \frac{\sigma_{23}(s)}{\sigma_{\eta}^2}\cdot
                                 \eta +<\psi(s_0)>
                                           \biggr) \\
     \sin\biggl(
        \frac{\sigma_{13}(s)}{\sigma_{\sigma}^2}\cdot
                                 \sigma
      + \frac{\sigma_{23}(s)}{\sigma_{\eta}^2}\cdot
                                 \eta +<\psi(s_0)>
                                           \biggr) \\
                      0
                \end{array}
         \right)}        \; ,
 \label{eq:3.218} \\
&& \vec{P}^{w}(\sigma,\eta;s) =
   w_{norm}(\sigma,\eta)\cdot ||\vec P_{loc}^{w}(\sigma,\eta;s)||
\cdot\vec{P}^{w}_{dir}(\sigma,\eta;s) \; ,
 \label{eq:3.216} \\
&&\Phi(\vec{u};s) = \exp\biggl(-\frac{1}{2}\cdot
 \sum_{j,k=1}^3 \; \sigma_{jk}(s)\cdot u_j\cdot u_k
+i\cdot<\psi(s_0)>\cdot u_3
 \biggr) \; ,
\label{eq:3.252}
\end{eqnarray}
where I also used
(\ref{eq:2.229}),(\ref{eq:3.213}). Because
the first moment vector and the covariance matrix 
depend continuously on $s$, one observes that 
the quantities in
(\ref{eq:3.214}),
(\ref{eq:3.215}),
(\ref{eq:3.217}),
(\ref{eq:3.218}),
(\ref{eq:3.216}),
(\ref{eq:3.252}) 
depend continuously on $s$ so that these equations
even hold at $s=s_0$.
Of course $0\leq ||\vec P_{loc}^w||\leq 1$ and as with all G-processes
$||\vec P_{loc}^w||$ is uniform across phase space.
\par In addition to the 
Bloch equation (\ref{eq:3.10})
for the polarization density and the
Bloch equation 
(\ref{eq:3.249})
for the local polarization vector
a Bloch equation
for the local polarization direction holds. In fact from 
(\ref{eq:2.203}),
(\ref{eq:3.249}),
(\ref{eq:3.217})
follows:
\begin{eqnarray}
&& \frac{\partial \vec{P}^w_{dir}}{\partial s}  =
-a\cdot \eta \cdot
\frac{\partial \vec{P}^w_{dir}}{\partial \sigma}
- b\cdot\sigma \cdot \frac{\partial \vec{P}^w_{dir}}{\partial \eta}
+ \vec{W}_{II}\wedge \vec{P}^w_{dir}
+ c\cdot \eta \cdot \frac{\partial \vec{P}^w_{dir}}{\partial \eta}
                                           \; .
 \label{eq:3.250}
\end{eqnarray}
Naturally one could describe processes 3,4 and 5 with the aid of
the Bloch equations 
(\ref{eq:3.10}),
(\ref{eq:3.249}),
(\ref{eq:3.250})
which one would solve by standard methods
(e.g. using Green functions or method of characteristics) but for such
simple G-processes the first and second moments are easily obtained, 
so that on this occasion the Bloch equations are not needed.
\subsection{The probability density of Process 3}
\subsubsection*{3.5.1}
Now I consider the process $\vec{x}^{(3)}(s)$,
called `Process 3' and I abbreviate:
\begin{eqnarray*}
&&\vec{x}^{(3)}(s) \equiv
       \left( \begin{array}{c}
                \sigma^{(3)}(s)       \\
                \eta^{(3)}(s)       \\
                \psi^{(3)}(s) 
               \end{array}
          \right)  \; .
\end{eqnarray*}
This process is characterized by the following
three conditions:
\begin{itemize}
\item It is a G-process at orbital equilibrium and its starting
azimuth is $s_0=0$.
\item The initial local polarization direction is parallel to the
$\vec{n}_0$-axis of Machine II.
\item Its initial local polarization is 1.
\end{itemize}
By the second and third conditions one has:
\begin{eqnarray}
&& ||\vec P_{loc}^{w_3}(\sigma,\eta;0)|| = 1 \; , \qquad
 \vec{P}^{w_3}_{dir}(\sigma,\eta;0) =
         \left( \begin{array}{c}
                           1 \\
                           0  \\
                           0
                \end{array}
         \right)        \; , 
 \label{eq:3.230}
\end{eqnarray}
where the probability density of Process 3 is denoted by $w_3$.
Thus one has by
(\ref{eq:3.217}),
(\ref{eq:3.218}):
\begin{eqnarray}
&&  <\psi^{(3)}(0)> = \psi_{0,m} \; , \label{eq:3.231} \\
&&  \sigma_{3,13}(0)   = \sigma_{3,23}(0)   = 0 \; , \label{eq:3.232} \\
&& \det\biggl(\underline{\sigma}_3(0)\biggr) = 0 \; ,
\label{eq:3.233}
\end{eqnarray}
where the covariance matrix of Process 3 is denoted by $\underline{\sigma}_3$
and where:
\begin{eqnarray}
&&  \psi_{0,m} = 2\pi\cdot m \; ,
\label{eq:3.237}
\end{eqnarray}
$m$ being an arbitrary integer.
\footnote{The physical properties of Process 3 are independent of 
the value of $m$, as will become clear below. Thus without loss of generality
one could set $m=0$. The same holds for processes 4 and 5.}
From the first condition on Process 3 and from 
(\ref{eq:3.15}),
(\ref{eq:3.232}),
(\ref{eq:3.233})
it follows that:
\begin{eqnarray}
&&  \sigma_{3,33}(0)  = 0 \; .
\label{eq:3.234}
\end{eqnarray}
Also from 
(\ref{eq:3.219}),
(\ref{eq:3.231})
follows:
\begin{eqnarray}
&&<\vec{x}^{(3)}(s)> =  
<\vec{x}^{(3)}(0)> =  
 (0,0,\psi_{0,m})^T \; .
\label{eq:3.19}
\end{eqnarray}
By 
(\ref{eq:3.252}),
(\ref{eq:3.231}) one obtains:
\begin{eqnarray}
 \Phi_3(\vec{u};s) &=& \exp\biggl(-\frac{1}{2}\cdot
 \sum_{j,k=1}^3 \; \sigma_{3,jk}(s)\cdot u_j\cdot u_k
+i\cdot\psi_{0,m}\cdot u_3
 \biggr) \; ,
\label{eq:3.236}
\end{eqnarray}
where $\Phi_3$ denotes the characteristic function corresponding to $w_3$
and which for $s=0$ simplifies by 
(\ref{eq:3.14}),
(\ref{eq:3.232}),
(\ref{eq:3.234})
to:
\begin{eqnarray}
 \Phi_3(\vec{u};0) &=& \exp\biggl(
-\frac{1}{2}\cdot\sigma_{\sigma}^2\cdot u_1^2
-\frac{1}{2}\cdot\sigma_{\eta}^2\cdot u_2^2 
+i\cdot\psi_{0,m}\cdot u_3\biggr) \; .
\label{eq:3.235}
\end{eqnarray}
From (\ref{eq:2.40}),
(\ref{eq:3.235}) it follows that the initial probability density
of Process 3 takes the expected form:
\begin{eqnarray}
 w_3(\sigma,\eta,\psi;0) &=&
 w_{norm}(\sigma,\eta)\cdot
\delta (\psi-\psi_{0,m})\; .
 \label{eq:3.17}
\end{eqnarray}
By 
(\ref{eq:2.40}),
(\ref{eq:3.236}) one sees that $w_3$ is Gaussian if
$\underline{\sigma}_3$ is nonsingular.
Because the integral:
\begin{eqnarray*}
\int_0^s\; ds_1\cdot\sigma_{3,23}^2(s_1)
\end{eqnarray*}
is positive for $s>0$,
\footnote{This can be concluded from:
\begin{eqnarray*}
  \sigma_{3,23}'(0)  &=& d\cdot \sigma^2_{\sigma} > 0 \;,
\end{eqnarray*}
which follows from 
(\ref{eq:3.202}),(\ref{eq:3.232}).}
the determinant of the covariance matrix of Process 3 is positive for
$s>0$. Thus for $s>0$ the probability density is Gaussian, confirming
that Process 3 is a G-process, and one obtains for $s>0$:
\begin{eqnarray}
&& w_3(\sigma,\eta,\psi;s)  =
                      \sqrt{(2\pi)^{-3}\cdot
\det( \underline{\sigma}_{3}(s))^{-1}}\cdot
               \exp\biggl\lbrack -\frac{1}{2}\cdot
       \left( \begin{array}{c}
          \sigma \\
          \eta   \\
          \psi -\psi_{0,m}
                \end{array}
         \right)^T       \cdot
    \underline{\sigma}_{3}^{-1}(s)\cdot
       \left( \begin{array}{c}
          \sigma \\
          \eta   \\
          \psi -\psi_{0,m}
                \end{array}
         \right)   \biggr\rbrack \; . 
\nonumber\\&&
\label{eq:3.18}
\end{eqnarray}
By (\ref{eq:3.17}),
(\ref{eq:3.18})
$w_3$ fulfills the
normalization condition 
(\ref{eq:2.9}). Also it follows for the orbital part of $w_3$ that:
\begin{eqnarray*}
 w_{3,orb} &=&
 w_{norm}  \; ,
\end{eqnarray*}
confirming that Process 3 is at orbital equilibrium.
Note that $\vec{x}^{(3)}(s)$ is a Markovian diffusion
process.
\subsubsection*{3.5.2}
Now I continue the calculation of the covariance matrix
$\underline{\sigma}_3$ of Process 3 and by 
(\ref{eq:3.8}) this is basically an
integration problem. In this section I determine the matrix elements
$\sigma_{3,13},\sigma_{3,23}$ which fulfill the
differential equation 
(\ref{eq:3.202}).
To do this I first obtain a 2-turn periodic special solution (denoted by
$(g_{7},g_{8})$) of (\ref{eq:3.202}), i.e.
\begin{eqnarray}
&&                \left( \begin{array}{c}
  g_{7}' \\
  g_{8}'
                \end{array}
         \right)     =
\underline{\cal A}_{orb}\cdot
               \left( \begin{array}{c}
  g_{7}  \\
  g_{8}
                \end{array}
         \right)
+  \sigma^2_{\eta}\cdot      \left( \begin{array}{c}
     0          \\
 \hat{d}
                \end{array}
         \right) \; .
\label{eq:3.24}
\end{eqnarray}
Because the difference of two such
solutions solves the homogeneous equation corresponding
to 
(\ref{eq:3.24})
one observes that $(g_{7},g_{8})$ is unique since the 2-turn periodic
solution to the homogeneous equation vanishes.
\footnote{This follows from the form of the matrix $\underline{\cal A}_{orb}$.}
I can then write:
\begin{eqnarray}
&&
                  \left( \begin{array}{c}
 \sigma_{3,13}(s)  \\
 \sigma_{3,23}(s)
                \end{array}
         \right)    \equiv
                  \left( \begin{array}{c}
 g_{7}(s)  \\
 g_{8}(s)
                \end{array}
         \right)
 +                \left( \begin{array}{c}
 g_{9}(s)  \\
 g_{10}(s)
                \end{array}
         \right)  \; ,
\label{eq:3.26}
\end{eqnarray}
where $(g_{9},g_{10})$ is a solution of the corresponding
homogeneous equation, i.e.
\begin{eqnarray}
&&                \left( \begin{array}{c}
 g_{9}' \\
 g_{10}'
                \end{array}
         \right)     =
\underline{\cal A}_{orb}\cdot
               \left( \begin{array}{c}
 g_{9}  \\
 g_{10}
                \end{array}
         \right)  \; .
\label{eq:3.210}
\end{eqnarray}
With 
(\ref{eq:3.210})
one finds that after a few orbital damping times
$g_9,g_{10}$ fade away so that by 
(\ref{eq:3.26})
$\sigma_{3,13}(s),\sigma_{3,23}(s)$
become 2-turn periodic in $s$.
Hence after a few orbital damping times one gets:
\begin{eqnarray}
&& \left( \begin{array}{c}
 \sigma_{3,13}(s)  \\
 \sigma_{3,23}(s)
                \end{array}
         \right)    \approx
                  \left( \begin{array}{c}
 g_{7}(s)  \\
 g_{8}(s)
                \end{array}
         \right)   \; .
\label{eq:3.28}
\end{eqnarray}
One finds that $(g_{7},g_{8})$ is given explicitly by:
\begin{eqnarray}
&&
  g_{7}(s) =
 \frac{\hat{d}(s)\cdot\sigma^2_{\eta}}{2\cdot\lambda\cdot b}
   \cdot \biggl(i\cdot g_1(s-L\cdot{\cal G}(s/L))
               \cdot g_{11}\cdot\exp(-c\cdot L/2)
\nonumber\\&& \qquad
            + i\cdot g_1(s-L\cdot{\cal G}(s/L)-L)
               \cdot g_{11}\cdot\exp(c\cdot L/2)-2\cdot\lambda\biggr)
                               \; ,
\nonumber\\
&&
  g_{8}(s) =
 - \frac{\hat{d}(s)\cdot\sigma^2_{\eta}}{2\cdot\lambda}
   \cdot \biggl(i\cdot g_2(s-L\cdot{\cal G}(s/L))
               \cdot g_{11}\cdot\exp(-c\cdot L/2)
\nonumber\\&&\qquad
            + i\cdot g_2(s-L\cdot{\cal G}(s/L)-L)
               \cdot g_{11}\cdot\exp(c\cdot L/2) \biggr)
                               \; , \nonumber\\&&
\label{eq:3.36}
\end{eqnarray}
where
\begin{eqnarray*}
 && g_{11} \equiv
         \frac{1}{\cosh(c\cdot L/2)+\cos(\lambda\cdot L)} \; .
\end{eqnarray*}
This can be checked by showing that the expressions in
(\ref{eq:3.36})
solve
(\ref{eq:3.24}) and are 2-turn periodic in $s$.
Using
(\ref{eq:3.232}),
(\ref{eq:3.26}) and
(\ref{eq:3.36}) to fix
$(g_{9},g_{10})$ at $s=0$ one then obtains:
\begin{eqnarray}
&&                \left( \begin{array}{c}
  g_{9}(s)  \\
  g_{10}(s)
                \end{array}
         \right)   
\nonumber\\
&=& 
 \frac{i\cdot d\cdot g_{11}\cdot\sigma^2_{\eta}}{2\cdot\lambda\cdot b}
\cdot
              \left( \begin{array}{c}
  g_1(s)\cdot
     \lbrack \sinh(c\cdot L/2)+\cos(\lambda\cdot L)\rbrack
            -  g_1(s-L)\cdot \exp(c\cdot L/2) 
                                              \\ \\
 -b\cdot g_2(s)\cdot
     \lbrack \sinh(c\cdot L/2)+\cos(\lambda\cdot L)\rbrack
            +  b\cdot g_2(s-L)\cdot \exp(c\cdot L/2)   \biggr)
                \end{array}
         \right)
                               \; ,
\nonumber\\
\label{eq:3.37}
\end{eqnarray}
and it is simple to confirm that the expression in
(\ref{eq:3.37})
solves
(\ref{eq:3.210}).
Combining 
(\ref{eq:3.26}),
(\ref{eq:3.36}),
(\ref{eq:3.37})
one finally has the explicit forms:
\begin{eqnarray}
&& \sigma_{3,13}(s)  =
 \frac{\sigma^2_{\eta}}{2\cdot\lambda\cdot b}\cdot\biggl(
   i\cdot\hat{d}(s)\cdot g_1(s-L\cdot{\cal G}(s/L))
               \cdot g_{11}\cdot\exp(-c\cdot L/2)
\nonumber\\&&
            + i\cdot\hat{d}(s)\cdot g_1(s-L\cdot{\cal G}(s/L)-L)
               \cdot g_{11}\cdot\exp(c\cdot L/2)
            - 2\cdot\lambda\cdot\hat{d}(s)
\nonumber\\&&
              - i\cdot g_1(s) \cdot g_{11}\cdot\exp(-c\cdot L/2)
            - i\cdot g_1(s-L)\cdot g_{11}\cdot\exp(c\cdot L/2)
            + i\cdot g_1(s) \biggr)
                                                             \; ,
\nonumber\\&&
\nonumber\\
&& \sigma_{3,23}(s)  =
 \frac{\sigma^2_{\eta}}{2\cdot\lambda}\cdot\biggl(
  -i\cdot\hat{d}(s)\cdot g_2(s-L\cdot{\cal G}(s/L))
               \cdot g_{11}\cdot\exp(-c\cdot L/2)
\nonumber\\&&
            - i\cdot\hat{d}(s)\cdot g_2(s-L\cdot{\cal G}(s/L)-L)
               \cdot g_{11}\cdot\exp(c\cdot L/2)
\nonumber\\&&
              + i\cdot g_2(s) \cdot g_{11}\cdot\exp(-c\cdot L/2)
            + i\cdot g_2(s-L)\cdot g_{11}\cdot\exp(c\cdot L/2)
            - i\cdot g_2(s) \biggr)
                                                             \; .
\nonumber\\&&
\label{eq:3.38}
\end{eqnarray}
This can be checked by showing that the expressions
(\ref{eq:3.38}) solve
(\ref{eq:3.202}) and obey:
\begin{eqnarray*}
&& \lim_{0<s\rightarrow 0}\lbrack \sigma_{3,13}(s)\rbrack  =
   \lim_{0<s\rightarrow 0}\lbrack \sigma_{3,23}(s)\rbrack  = 0 \; .
\end{eqnarray*}
Note also that for $0\leq s\leq L$ one has:
\begin{eqnarray*}
&& \sigma_{3,13}(s)  =   \sigma_{2,13}(s)  \; , \qquad
   \sigma_{3,23}(s)  =   \sigma_{2,23}(s)  \; ,
\end{eqnarray*}
which also follows from the fact that processes 2 and 3 are identical
for $0\leq s\leq L$.
\subsubsection*{3.5.3}
Coming finally to $\sigma_{3,33}$ I first of all get from
(\ref{eq:3.202}),
(\ref{eq:3.203}),
(\ref{eq:3.234}):
\begin{eqnarray}
&& \sigma_{3,33}(s)  =  \sigma_{3,33}(0)  + \int_0^s\; ds_1\cdot
\sigma_{3,33}'(s_1)
= \int_0^s\; ds_1\cdot\sigma_{3,33}'(s_1) =     2\cdot \int_0^s\; ds_1
                  \cdot\sigma_{3,23}(s_1)\cdot\hat{d}(s_1)  \; , \nonumber\\&&
\label{eq:3.39}
\end{eqnarray}
i.e.:
\begin{eqnarray}
&& \sigma_{3,33}(s)  =
           g_{12}(s) + g_{13}(s)  \; ,
\label{eq:3.56}
\end{eqnarray}
where
\begin{eqnarray}
&& g_{12}(s)  = 2\cdot \int_0^s\; ds_1
\cdot g_{8}(s_1)\cdot\hat{d}(s_1)  \; , \qquad
g_{13}(s)  = 2\cdot \int_0^s\; ds_1
\cdot g_{10}(s_1)\cdot\hat{d}(s_1)  \; .
\label{eq:3.228}
\end{eqnarray}
By straightforward integrations one then obtains:
\begin{eqnarray}
&& g_{12}(s)  =  g_{14}(s) + s \cdot  g_{15} \; , 
\label{eq:3.229} \\
&&g_{13}(s)= \frac{i\cdot d\cdot \sigma^2_{\eta}}{\lambda}
   \cdot \frac{1-\exp(\lambda_1\cdot L)}
              {1+\exp(\lambda_1\cdot L)}
\cdot g_{16}(s) + c.c. \; ,
\label{eq:3.211} 
\end{eqnarray}
where:
\begin{eqnarray}
&& g_{14}(s) =
    \frac{i\cdot d^2\cdot g_{11}\cdot
                         \sigma^2_{\eta}}{\lambda\cdot a\cdot b}
   \cdot \biggl(g_1(s- {\cal G}(s/L))\cdot\exp(-c\cdot L/2)
            + g_1(s-{\cal G}(s/L)-L)\cdot\exp(c\cdot L/2)
\nonumber\\&&
               -g_1(0)\cdot \exp(-c\cdot L/2)
            - g_1(-L)\cdot \exp(c\cdot L/2)
                                                              \biggr)
       - \lbrack s-{\cal G}(s/L)  \rbrack   \cdot  g_{15} \; , 
\label{eq:3.212} \\
&& g_{15} \equiv
\frac{2\cdot d^2\cdot g_{11}\cdot\sigma_{\eta}^2}
   {a\cdot b\cdot L\cdot\lambda}
   \cdot \biggl(
                  2\cdot\lambda\cdot\sinh(c\cdot L/2)
   -         c\cdot\sin(\lambda\cdot L)  \biggr) \; , 
\label{eq:3.221} \\
&& g_{16}(s)              =
    \frac{1}{\lambda_1}\cdot\hat{d}(s)\cdot\exp(\lambda_1\cdot s)
-\frac{d}{\lambda_1}
\nonumber\\&&
 + \frac{2\cdot d}{\lambda_1} \cdot
          \exp(\lambda_1\cdot L) \cdot
   \frac{1-\exp\biggl(2\cdot L\cdot\lambda_1\cdot{\cal G}(s/2L)\biggr)}
        {1+\exp(\lambda_1\cdot L)}
\nonumber\\&&
          + \frac{2\cdot d}{\lambda_1} \cdot
                                \lbrack
            {\cal G}(s/2L+1/2) -  {\cal G}(s/2L) \rbrack
 \cdot \exp\biggl(\lambda_1\cdot(2L\cdot {\cal G}(s/2L)+L)\biggr) \; .
\label{eq:3.42}  
\end{eqnarray}
One sees by 
(\ref{eq:3.42}) that after a few orbital damping times $g_{16}$
becomes constant so that by 
(\ref{eq:3.211}) 
$g_{13}$ becomes constant, too.
Also one observes that $g_{14}(s)$ is 1-turn periodic in $s$ because the
function ${\cal G}(s/L)-s$ is 1-turn periodic in $s$.
Hence after a few orbital damping times one gets:
\begin{eqnarray}
&& \sigma_{3,33}(s)  \approx
           g_{13}(+\infty) + g_{14}(s) + s \cdot  g_{15} \; ,
\label{eq:3.57}
\end{eqnarray}
so that $\sigma_{3,33}(s)$ quickly splits up
additively into a term increasing linearly with $s$ plus a term
1-turn periodic in $s$.
The expression (\ref{eq:3.56}) can be checked by showing that it
solves (\ref{eq:3.203}) and obeys:
\begin{eqnarray*}
&& \lim_{0<s\rightarrow 0} \lbrack\sigma_{3,33}(s)\rbrack  = 0 \; .
\end{eqnarray*}
\par This completes the calculation of the covariance matrix, whose
matrix elements are given by 
(\ref{eq:3.14}),(\ref{eq:3.38}),(\ref{eq:3.56}).
Note that due to ${\cal G}(s)$ the
functions $\sigma_{3,13}(s),\sigma_{3,23}(s),
\sigma_{3,33}(s)$ at first sight are undetermined at the points where 
$s/L=$integer. Nevertheless due to their continuity in $s$ 
(see section 3.4.2) they are well defined at these points.
\par With 
(\ref{eq:3.14}),(\ref{eq:3.19}),(\ref{eq:3.38}),(\ref{eq:3.56}) I have 
determined the
first moment vector and the covariance matrix of Process 3 so that the
probability density is fixed.
\subsubsection*{3.5.4}
From
(\ref{eq:3.214}),(\ref{eq:3.231}) follows:
\begin{eqnarray}
&&\vec{P}_{tot}^{w_3}(s) 
                        =  \exp(-\sigma_{3,33}(s)/2)\cdot
  \left( \begin{array}{c}
        1 \\
        0 \\
                      0
                \end{array}
         \right)  \; . 
\label{eq:3.20}
\end{eqnarray}
By 
(\ref{eq:3.56}),(\ref{eq:3.20})
the polarization of Process 3 is given by
\begin{eqnarray}
&&  ||\vec P_{tot}^{w_3}(s)|| =
    \exp\biggl(-\frac{\sigma_{3,33}(s)}{2}\biggr)=
    \exp\biggl(-\frac{g_{13}(s)}{2}\biggr)
\cdot\exp\biggl(-\frac{g_{14}(s)}{2}\biggr)
\cdot\exp\biggl(- \frac{s\cdot g_{15}}{2}\biggr) \; . \qquad
\label{eq:3.58}
\end{eqnarray}
By (\ref{eq:3.58})
one sees that after a few orbital damping times the polarization
reads as:
\begin{eqnarray*}
  ||\vec P_{tot}^{w_3}(s)|| &\approx&
    \exp\biggl(-g_{13}(+\infty)/2\biggr)
\cdot\exp\biggl(-g_{14}(s)/2\biggr)
\cdot\exp\biggl(- s\cdot g_{15}/2\biggr) \; .
\end{eqnarray*}
At long times this is the product of a factor which has the
period of the ring and an exponentially decaying factor. 
Because the factor $\exp(- g_{13}(s)/2)$ is not constant
from the beginning, $||\vec P_{tot}^{w_3}||$ 
contains a `transient' contribution 
which later damps away.
\par Due to the factor $\exp(- s\cdot g_{15}/2)$ one
observes:
\begin{eqnarray*}
\lim_{s\rightarrow +\infty} \lbrack\sigma_{3,33}(s)\rbrack &=&  +\infty \; ,
\end{eqnarray*}
i.e. one has complete spin decoherence of Process 3:
\begin{eqnarray*}
      ||\vec P_{tot}^{w_3}(+\infty)|| &=& 0 \; .
\end{eqnarray*}
I define the depolarization rate as:
\begin{eqnarray}
   \frac{1}{\tau_{spin}} &\equiv& \frac{g_{15}}{2} > 0 \; .
 \label{eq:3.59}
\end{eqnarray}
If one specifies the constants according to 
(\ref{eq:2.3})
one gets
\begin{eqnarray}
  \tau_{spin} &\approx& 7.6\cdot 10^7\, m \; ,
 \label{eq:3.267}
\end{eqnarray}
which corresponds to 12000 turns, i.e. about 260 milliseconds.
If on the other hand $Q_s$ (and therefore $\lambda$) were close to half of 
an integer $g_{15}$ would, because of its factor $g_{11}$, become very large
and $\tau_{spin}$ would be very small. This is exactly what one expects
when sitting close to a spin-orbit resonance \cite{BHMR91}.
\par Process 3 is a rough model for the behaviour of the polarization
after injection into the rings mentioned in section 3.1 and it is therefore
interesting to study the transient behaviour in (\ref{eq:3.58}).
To come to that I study the
complicated azimuthal dependence of the polarization `turn by turn', i.e.
I consider its behaviour with increasing number of turns and thus
investigate the sequence $\sigma_{3,33}(2NL)$, where $N$ is a nonnegative
integer.
My `observation point'
is  at the azimuth $s=2NL$, i.e. at the snake after every second turn.
\footnote{I consider the sequence
$\sigma_{3,33}(2NL)$ instead of $\sigma_{3,33}(NL)$ because of
mathematical convenience.}
\begin{figure}[t]
\begin{center}
\epsfig{figure=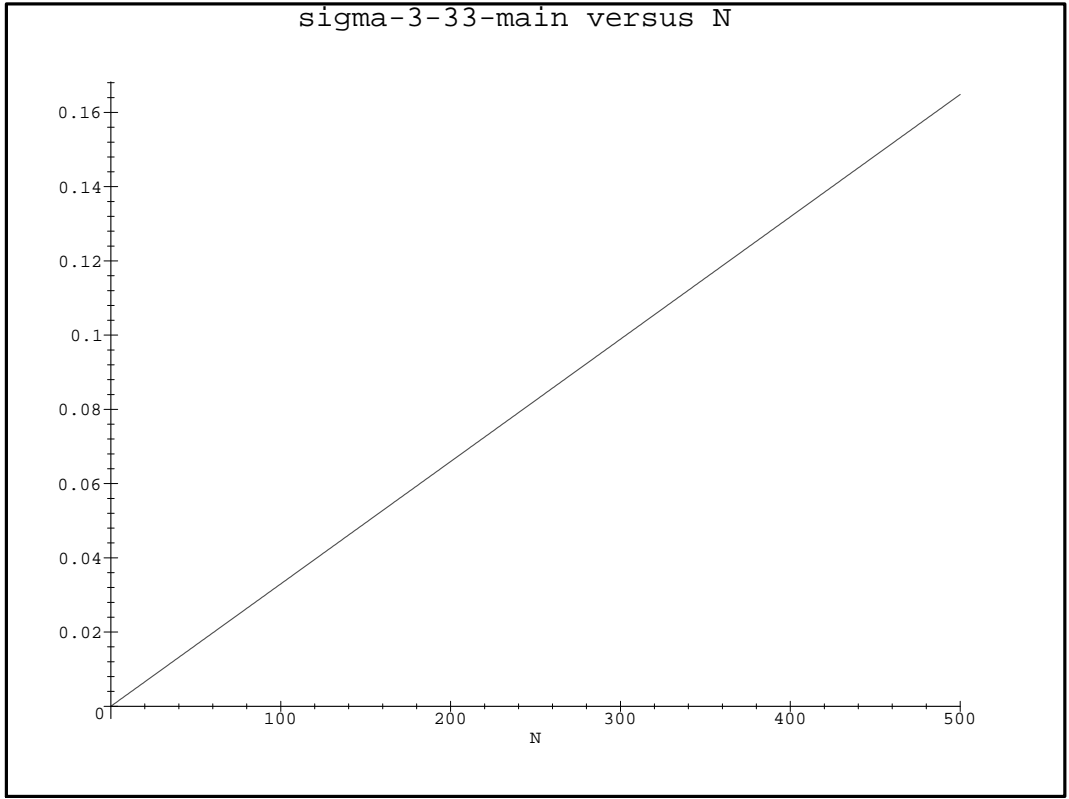,width=12cm,angle=-90}
\end{center}
\caption{The main term $\sigma_{3,33,main}(2NL)$ of
$\sigma_{3,33}(2NL)$ for the first 1000 turns of Process 3
assuming the HERA values 
(\ref{eq:2.3})}
\end{figure}
\par First of all one gets
\begin{eqnarray}
&& \sigma_{3,33}(2NL)  =
           g_{13}(2NL) + g_{14}(2NL) + 2NL \cdot  g_{15} \; .
\label{eq:3.60}
\end{eqnarray}
This can be simplified because $g_{14}(s)$ is 1-turn periodic in $s$ so that:
\begin{eqnarray*}
&& g_{14}(2NL) = g_{14}(0) \; .
\end{eqnarray*}
Also one has by 
(\ref{eq:3.212}):
\begin{eqnarray*}
&& g_{14}(0) = 0 \; ,
\end{eqnarray*}
so that (\ref{eq:3.60})
simplifies to:
\begin{eqnarray}
&& \sigma_{3,33}(2NL)  =
   \sigma_{3,33,inter}(2NL)  +
         \sigma_{3,33,main}(2NL)  \; ,
\label{eq:3.61}
\end{eqnarray}
where
\footnote{inter$\equiv$ `intermediate' which expresses that it affects the
polarization only at the beginning of Process 3.}
\begin{eqnarray}
&& \sigma_{3,33,main}(s) \equiv  s\cdot  g_{15} \; , \qquad
 \sigma_{3,33,inter}(s) \equiv g_{13}(s) \; .
\label{eq:3.62}
\end{eqnarray}
\begin{figure}[t]
\begin{center}
\epsfig{figure=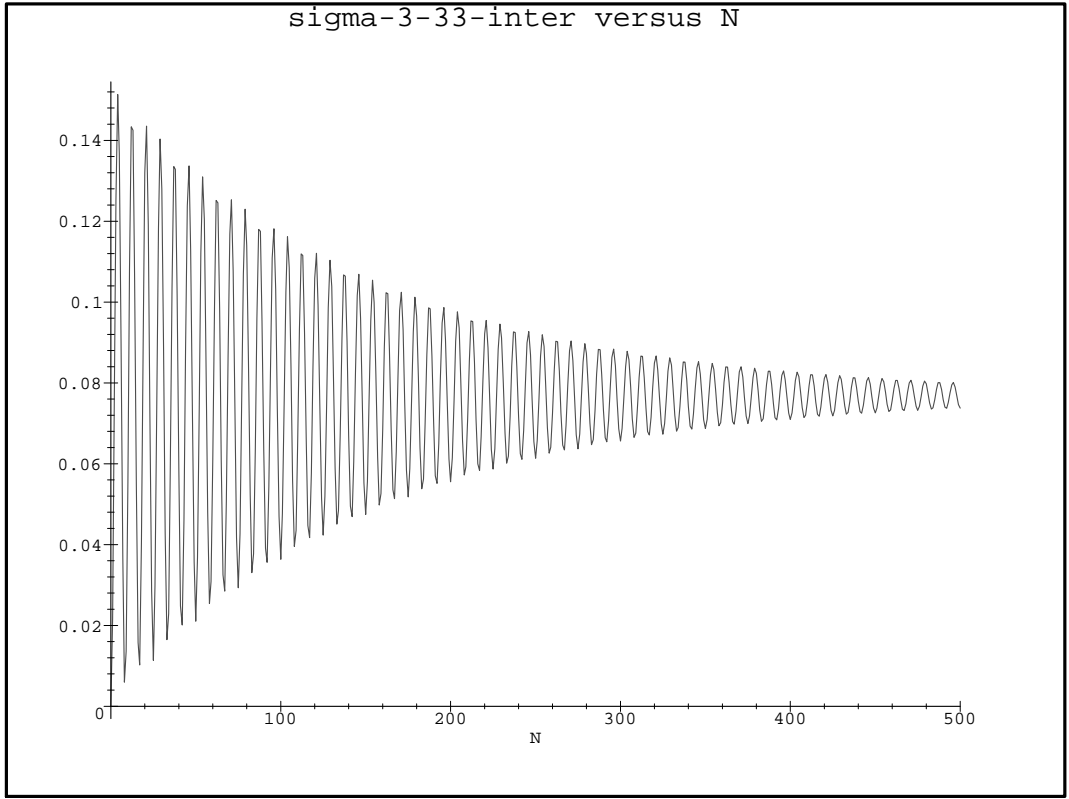,width=12cm,angle=-90}
\end{center}
\caption{The intermediate term $\sigma_{3,33,inter}(2NL)$ of
$\sigma_{3,33}(2LN)$ for the first 1000 turns of Process 3
assuming the HERA values 
(\ref{eq:2.3})}
\end{figure}
Inserting 
(\ref{eq:3.61}) into 
(\ref{eq:3.58})
yields:
\begin{eqnarray}
  ||\vec P_{tot}^{w_3}(2NL)|| &=&
    \exp\biggl(-\sigma_{3,33,inter}(2NL)/2\biggr)\cdot
    \exp\biggl(-\sigma_{3,33,main}(2NL)/2\biggr)
  \; .
\label{eq:3.63}
\end{eqnarray}
All transient behaviour is contained in the first factor on the rhs
of (\ref{eq:3.63}) and as one has seen this term converges to a finite
value after a few orbital damping times. 
Unfortunately, polarimeters are usually not fast enough to measure the
transients.
To look at $\sigma_{3,33,inter}(2NL)$ in more detail I use 
(\ref{eq:3.42}) to calculate:
\begin{eqnarray}
&& g_{16}(2NL)              =
    \frac{d}{\lambda_1}\cdot\exp(\lambda_1\cdot 2NL)
-\frac{d}{\lambda_1}
 + \frac{2\cdot d}{\lambda_1} \cdot
          \exp(\lambda_1\cdot L) \cdot
   \frac{1-\exp(\lambda_1\cdot 2NL)}
        {1+\exp(\lambda_1\cdot L)}
\nonumber\\
&=& \frac{d}{\lambda_1}\cdot\biggl(
   \exp(\lambda_1\cdot 2NL)\cdot \lbrack 1-
   \frac{2\cdot\exp(\lambda_1\cdot L)}
        {1+\exp(\lambda_1\cdot L)} \rbrack
   -1 + \frac{2\cdot\exp(\lambda_1\cdot L)}
        {1+\exp(\lambda_1\cdot L)} \biggr)
\nonumber\\
&=& \frac{d}{\lambda_1}\cdot
   \frac{1-\exp(\lambda_1\cdot L)}
        {1+\exp(\lambda_1\cdot L)}\cdot
   \lbrack  \exp(\lambda_1\cdot 2NL) -1  \rbrack  \; ,
\label{eq:3.64}
\end{eqnarray}
from which it follows by 
(\ref{eq:3.211}) that:
\begin{eqnarray}
&& \sigma_{3,33,inter}(2NL) =   g_{13}(2NL)
=   \frac{i\cdot d\cdot\sigma_{\eta}^2}{\lambda}
   \cdot\frac{1-\exp(\lambda_1\cdot L)}
        {1+\exp(\lambda_1\cdot L)}\cdot  g_{16}(2NL)  + c.c.
\nonumber\\
&=& \frac{i\cdot d^2\cdot\sigma_{\eta}^2}{\lambda\cdot\lambda_1}
   \cdot\biggl(\frac{1-\exp(\lambda_1\cdot L)}
        {1+\exp(\lambda_1\cdot L)}\biggr)^2\cdot
   \lbrack  \exp(\lambda_1\cdot 2NL) -1  \rbrack   + c.c.
\nonumber\\
&=& \frac{d^2\cdot\sigma_{\eta}^2}{\lambda}
\cdot\biggl( i \cdot g_{17}\cdot
\lbrack 1 - \exp(c\cdot NL)\cdot \cos(\lambda\cdot 2NL)  \rbrack
\nonumber\\&&
- g_{18} \cdot
                \exp(c\cdot NL)\cdot \sin(\lambda\cdot 2NL)\rbrack
\biggr)   \; ,
 \label{eq:3.65}
\end{eqnarray}
where
\begin{eqnarray*}
&&g_{17} \equiv
   -\frac{i}{a\cdot b}\cdot
                           \lbrack
   \cosh(c\cdot L/2)
+ \cos(\lambda\cdot L)\rbrack^{-2}
\nonumber\\
&&
   \cdot  \biggl(   \lambda\cdot \cosh(c\cdot L)
 +\lambda\cdot\cos(\lambda\cdot 2L)
 -2\cdot\lambda
   -2\cdot c\cdot \sinh(c\cdot L/2)\cdot\sin(\lambda\cdot L)
                               \biggr)  \; ,
\nonumber\\
&&g_{18} \equiv
   -\frac{1}{2\cdot a\cdot b}\cdot
                           \lbrack
   \cosh(c\cdot L/2)
+ \cos(\lambda\cdot L)\rbrack^{-2}
\nonumber\\
&&
   \cdot
\biggl(
             c \cdot
                                \cosh(c\cdot L)
 +c\cdot\cos(\lambda\cdot 2L)
 -2\cdot c
   +8\cdot \lambda\cdot \sinh(c\cdot L/2)\cdot\sin(\lambda\cdot L)
\biggr) \; .
\end{eqnarray*}
\begin{figure}[t]
\begin{center}
\epsfig{figure=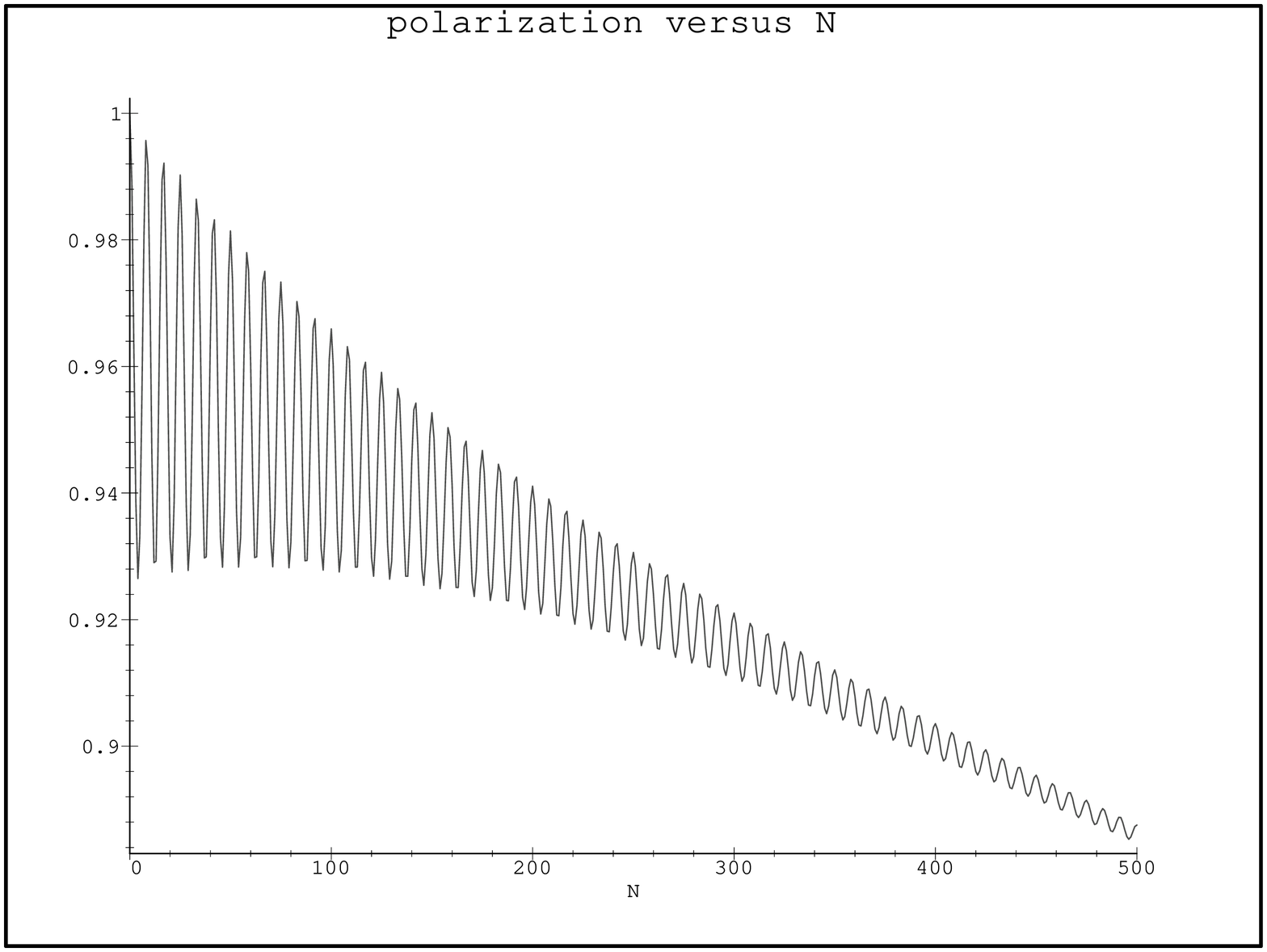,width=12cm,angle=-90}
\end{center}
\caption{Polarization $||\vec P_{tot}^{w_3}(2NL)||$ of Process 3 for the first 
1000 turns assuming the HERA values 
(\ref{eq:2.3})}
\end{figure}
\par For the HERA values 
(\ref{eq:2.3}) 
the main term $\sigma_{3,33,main}(2NL)$
and the intermediate term $\sigma_{3,33,inter}(2NL)$ are displayed
in figure 5 and   figure 6 for the first 1000 turns. One sees by
comparing figure 5 with figure 6 that for the first few hundreds of
turns the intermediate term dominates the polarization. However the
intermediate term is so small that it never seriously degrades the
polarization. The strongest effect is a degradation of the
polarization value to $0.93$ after 8 turns. This can be seen in figure
7 where the polarization is displayed for the first 1000 turns.
The polarization is also displayed for the first 20000 turns in
figure 8. Finally in figure 9 the polarization is displayed for the
limiting case, where $c,\omega\rightarrow 0$ with
$\omega/c=const=-2\cdot\sigma^2_{\eta}\approx -2.0\cdot 10^{-6}$.
I call this `Process 3a'.
\begin{figure}[t]
\begin{center}
\epsfig{figure=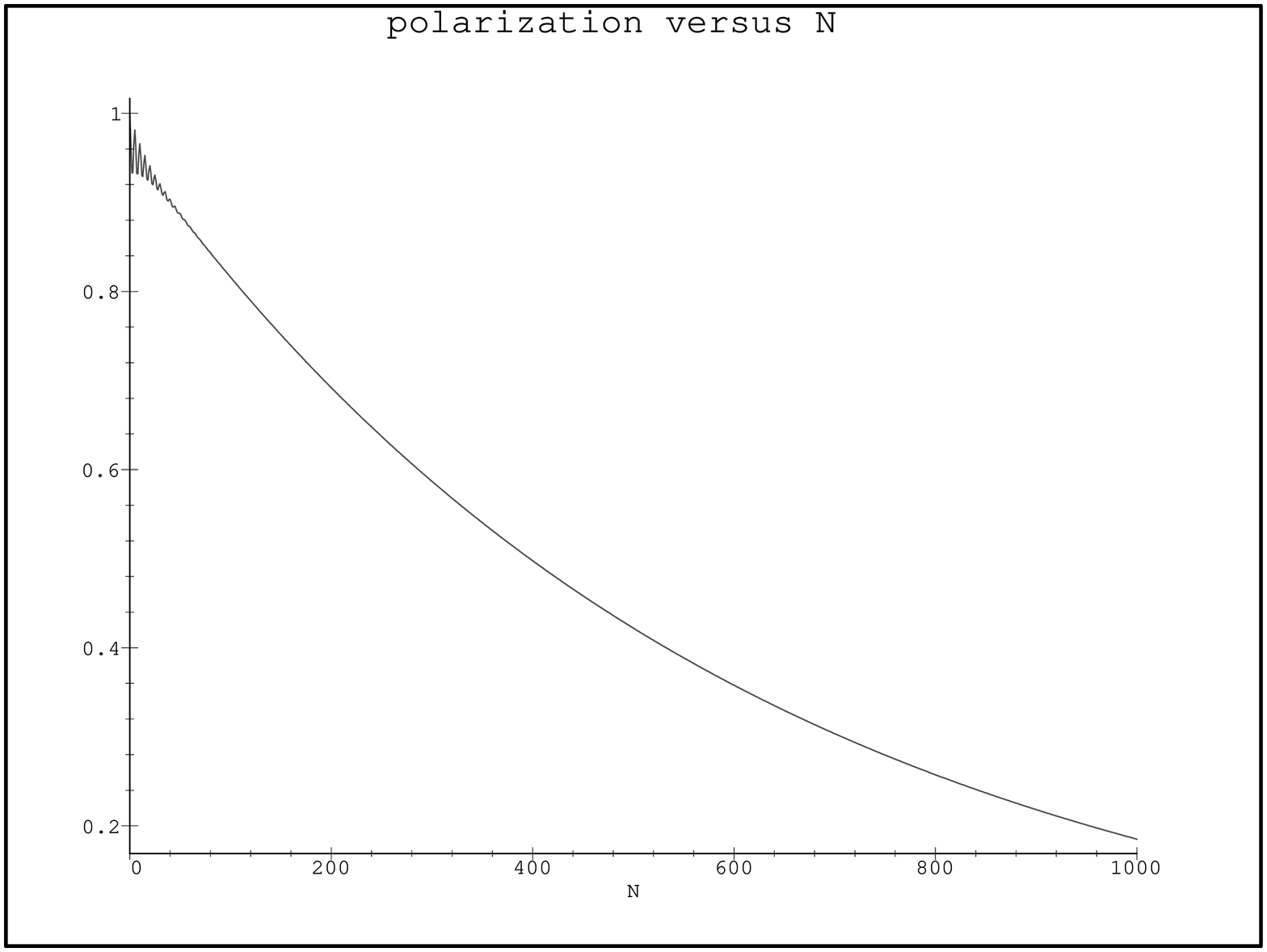,width=12cm,angle=-90}
\end{center}
\caption{Polarization $||\vec P_{tot}^{w_3}(20NL)||$ 
of Process 3 for the first 
20000 turns assuming the HERA values 
(\ref{eq:2.3})}
\end{figure}
Figure 9 shows that in the absence of radiation and when only synchrotron 
motion is considered
the snake holds the polarization within narrow limits.
This is a kind of spin echo effect \cite{Abr61}.
\par Comparing figure 7 and figure 8 with figure 3 one also sees that for
the first few thousands of turns the polarization of Process 3
is much larger than for Process 2 showing that as expected the snake
{\it can} strongly suppress oscillations in the spin distribution.
However, and this is at first unexpected when recalling the equilibrium
reached by Process 2,
in the end there is complete
decoherence for Process 3. But, on the other hand, one should not
be surprised when one recalls that the calculations with SLIM \cite{Cha81}
for a perfectly aligned flat ring with a pointlike radial snake in which
only spin diffusion due to synchrotron motion generated in the arcs is
included, also predict complete depolarization \cite{Bar97}.
Similar calculations also show that spin diffusion due to horizontal
betatron motion in the arcs
is very much less than that due to synchrotron motion.
\begin{figure}[t]
\begin{center}
\epsfig{figure=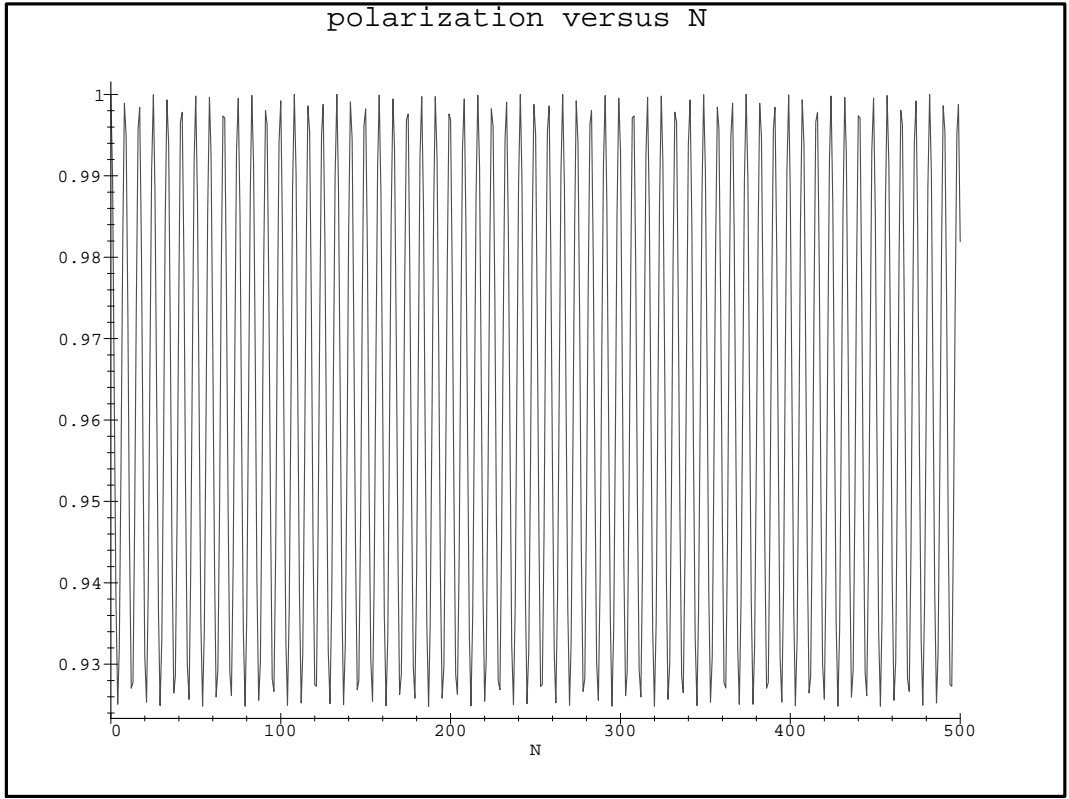,width=12cm,angle=-90}
\end{center}
\caption{Polarization $||\vec P_{tot}^{w_{3a}}(2NL)||$
of Process 3a for the first 1000 turns
assuming the HERA values 
(\ref{eq:2.3}), except that
$c,\omega\rightarrow 0$ with
$\omega/c=const=-2\cdot\sigma^2_{\eta}\approx -2.0\cdot 10^{-6}$.}
\end{figure}
\subsubsection*{3.5.5}
Another illustration of the transient behaviour of Process 3 is
provided by calculating the polarization after one turn. During the
first turn Process 3 is identical to Process 2 so that by 
(\ref{eq:2.204}) one gets:
\begin{eqnarray*}
&&  ||\vec P_{tot}^{w_3}(L)|| =
  ||\vec P_{tot}^{w_2}(L)||  =
    \exp\biggl(-\sigma_{2,33}(L)/2\biggr) =
    \exp\biggl(
    -\frac{d^2}{2a^2\lambda}\cdot\sigma_{\sigma}^2
    \cdot\lbrack 2\cdot\lambda - i\cdot g_1(L) \rbrack \biggr)\; .
\end{eqnarray*}
Thus due to its transient behaviour the polarization of Process 3
behaves during the first turn as if it decays exponentially
with the naive depolarization time given by
\begin{eqnarray*}
     \frac{2a^2\lambda L}{d^2\cdot\sigma_{\sigma}^2}
    \cdot\biggl( 2\cdot\lambda - i\cdot g_1(L) \biggr)^{-1} \; ,
\end{eqnarray*}
which is quite different from $\tau_{spin}$. In fact assuming the
HERA values 
(\ref{eq:2.3}) 
it results in
\begin{eqnarray*}
     \frac{2 a^2\lambda L}{d^2\cdot\sigma_{\sigma}^2}
  \cdot\biggl( 2\cdot\lambda - i\cdot g_1(L) \biggr)^{-1}
                           \approx   85000\, m  \; .
\end{eqnarray*}
\subsubsection*{3.5.6}
By 
(\ref{eq:3.15}),
(\ref{eq:3.217}),(\ref{eq:3.218}),(\ref{eq:3.216}),
(\ref{eq:3.19})
the polarization density, the local polarization, and the direction of the
local polarization of Process 3 read as:
\begin{eqnarray}
&& ||\vec P_{loc}^{w_3}(\sigma,\eta;s)|| =
               \exp\biggl(
-\frac{\det(\underline{\sigma}_3)}{2\cdot\sigma_{\sigma}^2\cdot
 \sigma_{\eta}^2} \biggr)
=              \exp\biggl(
-\frac{\sigma_{3,33}(s)}{2}  \biggr) \cdot
               \exp\biggl(
   \frac{\sigma_{3,13}^2(s)}{2\cdot\sigma^2_{\sigma}}
+\frac{\sigma_{3,23}^2(s)}{2\cdot\sigma^2_{\eta}} \biggr)
               \; , \qquad
\label{eq:3.69} \\
&& \vec{P}^{w_3}_{dir}(\sigma,\eta;s) =
   \large{      \left( \begin{array}{c}
     \cos\biggl(
        \frac{\sigma_{3,13}(s)}{\sigma_{\sigma}^2}\cdot
                                 \sigma
      + \frac{\sigma_{3,23}(s)}{\sigma_{\eta}^2}\cdot
                                 \eta  \biggr) \\
     \sin\biggl(
        \frac{\sigma_{3,13}(s)}{\sigma_{\sigma}^2}\cdot
                                 \sigma
      + \frac{\sigma_{3,23}(s)}{\sigma_{\eta}^2}\cdot
                                 \eta      \biggr) \\
                      0
                \end{array}
         \right)}        \; ,
\label{eq:3.70} \\
&& \vec{P}^{w_3}(\sigma,\eta;s) =
   w_{norm}(\sigma,\eta)\cdot ||\vec P_{loc}^{w_3}(\sigma,\eta;s)|| 
               \cdot\vec{P}^{w_3}_{dir}(\sigma,\eta;s) \; .
\label{eq:3.68}
\end{eqnarray}
The local polarization starts from the value 1 at
$s=0$ and decreases towards its vanishing equilibrium value.
\par With the local polarization quantities at hand one can reconsider the
transient behaviour of Process 3 in more detail.
After a few orbital damping times the transient behaviour
disappears so that the local polarization quantities acquire certain
periodicity properties w.r.t. $s$.
In fact from section 3.5.2 it is clear that 
on this time scale
$\sigma_{3,13}(s),\sigma_{3,23}(s)$ become
2-turn periodic in $s$ and 
change sign from turn to turn:
\begin{eqnarray*}
&& g_7(s) = -g_7(s+L) \; , \qquad 
 g_8(s) = -g_8(s+L) \; .
\end{eqnarray*}
Hence by (\ref{eq:3.70}) the local polarization direction
becomes 2-turn periodic
in $s$ in the\\
$(\vec{n}_{0,II},\vec{m}_{0,II},\vec{l}_{0,II})$-frame.
Since in the machine frame 
$\vec{m}_{0,II}$ also changes sign from turn to turn, the local polarization
direction becomes 1-turn periodic
in the machine frame after a few orbital damping times.
Note that at the phase space point where
$\sigma= \sigma_{\sigma}$ and $\eta= \sigma_{\eta}$ 
and for the HERA values
(\ref{eq:2.3})
the $\vec n_0$-axis
deviates at the snake by about 200
milliradians from the asymptotic local polarization direction.
\par Moreover in section 3.5.3 I observed that
$\sigma_{3,33}(s)$ quickly splits up
additively into a term increasing linearly with $s$ plus a term
1-turn periodic in $s$. Hence by 
(\ref{eq:3.69}) the local polarization factors
into an exponentially decaying part and a part
1-turn periodic in $s$.
\footnote{In fact one can use these properties of the local polarization
and its direction as the definition of a
transient free process.}
\par I now round off section 3 by considering two processes with
contrasting transient behaviour, one of which
shows no transients and one which will turn out to illustrate very nicely
the validity of a tenet at the basis of the standard
method of calculating the rate of depolarization.
Both processes are G-processes at orbital equilibrium.
\subsection{The probability density of Process 4}
\subsubsection*{3.6.1}
For Process 3 I found
that during the first few orbital damping times transient
behaviour prevents an exponential decay of the
polarization. Knowing this it is now simple to define a
modification of Process 3 which shows no transient behaviour of the 
polarization properties at any $s$ and which
from the beginning has those periodicity properties which Process 3
acquires only after a few orbital damping times. For 
this process $\vec{x}^{(4)}(s)$,
called `Process 4', the
exponential decay of the polarization shows up right from the beginning
because no transient terms destroy the exponential structure.
I abbreviate:
\begin{eqnarray*}
&&\vec{x}^{(4)}(s) \equiv
       \left( \begin{array}{c}
                \sigma^{(4)}(s)       \\
                \eta^{(4)}(s)       \\
                \psi^{(4)}(s) 
               \end{array}
          \right)  \; .
\end{eqnarray*}
Process 4 is characterized by the following four conditions:
\begin{itemize}
\item It is a G-process at orbital equilibrium and its starting
azimuth is $s_0=0$.
\item The direction of the local polarization is 2-turn periodic
in $s$ in the
$(\vec{n}_{0,II},\vec{m}_{0,II},\vec{l}_{0,II})$-frame and 1-turn periodic
in the machine frame.
\item The direction of the local polarization on the closed orbit is parallel
to the $\vec{n}_0$-axis of Machine II.
\item Its initial local polarization is 1.
\end{itemize}
The second condition ensures that the process is free of transients.
\par By the third condition one has:
\begin{eqnarray}
&& \vec{P}^{w_4}_{dir}(0,0;s) =
         \left( \begin{array}{c}
                           1 \\
                           0  \\
                           0
                \end{array}
         \right)        \; , 
 \label{eq:3.238}
\end{eqnarray}
where the probability density of Process 4 is denoted by $w_4$.
Thus one has by
(\ref{eq:3.218}),
(\ref{eq:3.238}):
\begin{eqnarray}
&&  <\psi^{(4)}(0)> = \psi_{0,m} \; ,
\label{eq:3.239}
\end{eqnarray}
where $\psi_{0,m}$ is given by
(\ref{eq:3.237}).
From 
(\ref{eq:3.219}),
(\ref{eq:3.239})
follows:
\begin{eqnarray}
&&<\vec{x}^{(4)}(s)> =
<\vec{x}^{(4)}(0)> =
 (0,0, \psi_{0,m} )^T \; .
\label{eq:3.204}
\end{eqnarray}
Also, due to the second condition on Process 4 one observes by
(\ref{eq:3.218}) that
$\sigma_{4,13}(s),\sigma_{4,23}(s)$ are 2-turn periodic in $s$, 
where $\underline{\sigma}_{4}$ denotes the covariance matrix of
Process 4. Because $\sigma_{4,13}(s),\sigma_{4,23}(s)$ obey
(\ref{eq:3.202}) and are 2-turn periodic in $s$, one concludes by
section 3.5.2 that:
\begin{eqnarray}
&&
                  \left( \begin{array}{c}
 \sigma_{4,13}(s)  \\
 \sigma_{4,23}(s)
                \end{array}
         \right)   =
                  \left( \begin{array}{c}
 g_{7}(s)  \\
 g_{8}(s)
                \end{array}
         \right) \; ,
\label{eq:3.71}
\end{eqnarray}
where $g_7,g_8$ are given by
(\ref{eq:3.36}).
\par Coming to $\sigma_{4,33}$ one first of all gets
by using 
(\ref{eq:3.203}),
(\ref{eq:3.228}),
(\ref{eq:3.229}),
(\ref{eq:3.71}):
\begin{eqnarray}
&& \sigma_{4,33}(s)  =  \sigma_{4,33}(0)  + \int_0^s\; ds_1\cdot
\sigma_{4,33}'(s_1) =
   \sigma_{4,33}(0)  + 2\cdot  \int_0^s \; ds_1\cdot
   \sigma_{4,23}(s_1)\cdot\hat{d}(s_1) \nonumber\\
&&=  \sigma_{4,33}(0)  + 2\cdot  \int_0^s \; ds_1\cdot
    g_{8}(s_1)\cdot\hat{d}(s_1) \nonumber\\
&&=
  \sigma_{4,33}(0)  +   g_{14}(s)
                      + s\cdot g_{15} =
 \sigma_{4,33}(0)  +   g_{14}(s)
+ \frac{2\cdot s}{\tau_{spin}} \; ,
\label{eq:3.222}
\end{eqnarray}
so that $\sigma_{4,33}(s)$ separates additively into a part 1-turn periodic in
$s$ and a part linear in $s$. Now I have exploited the
first three conditions on Process 4 
and to fix $\sigma_{4,33}$ I now impose the fourth condition
which
by (\ref{eq:3.217}) reads as:
\begin{eqnarray}
 \det\biggl(\underline{\sigma}_4(0)\biggr) &=&  0 \; .
\label{eq:3.76}
\end{eqnarray}
\begin{figure}[t]
\begin{center}
\epsfig{figure=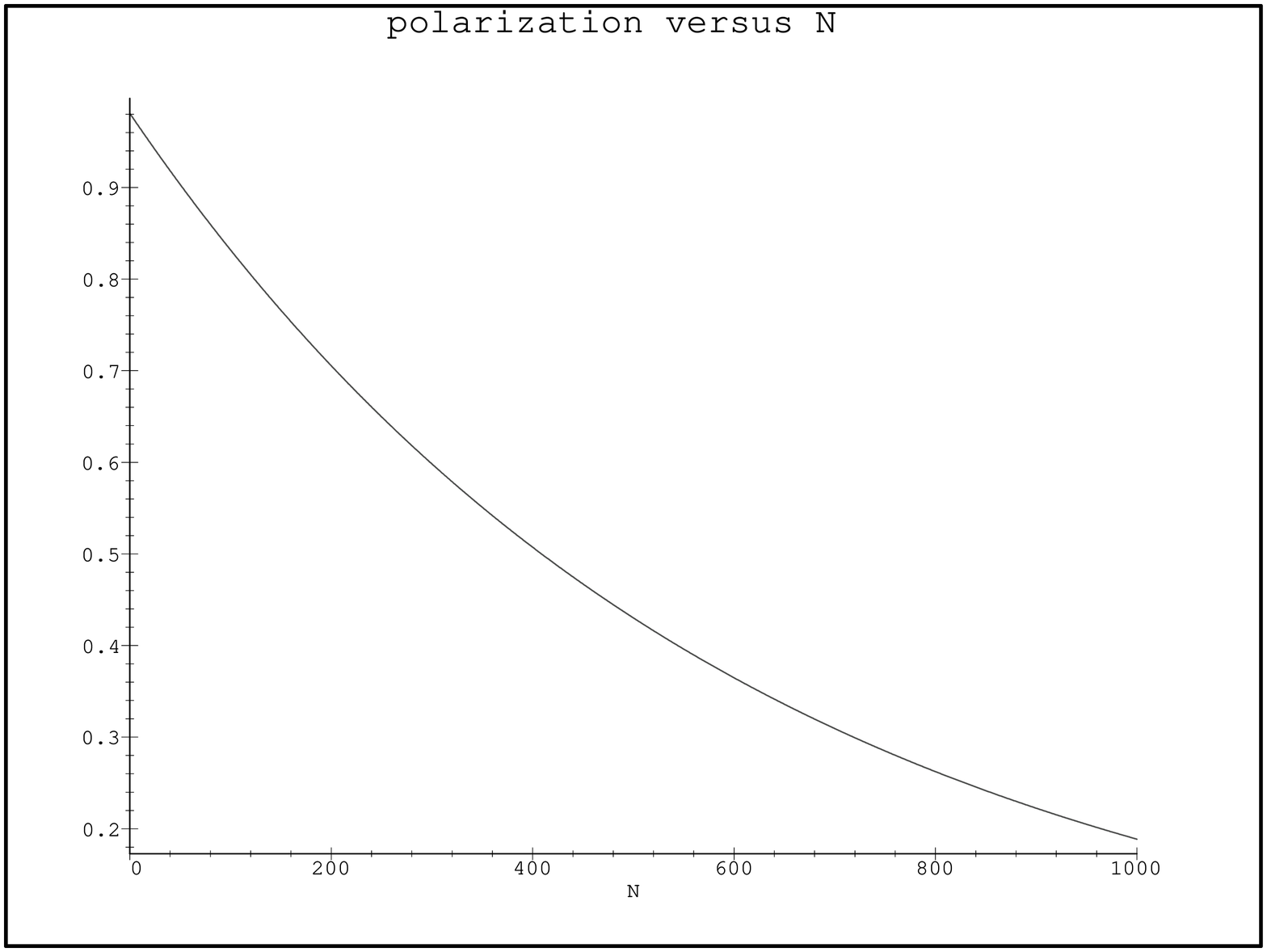,width=12cm,angle=-90}
\end{center}
\caption{Polarization $||\vec P_{tot}^{w_4}(20NL)||$ 
of Process 4 for the first 
20000 turns assuming the HERA values 
(\ref{eq:2.3})}
\end{figure}
From 
(\ref{eq:3.15}), 
(\ref{eq:3.71}), 
(\ref{eq:3.76}) follows:
\begin{eqnarray}
&&  \sigma_{4,33}(0)= \frac{\sigma_{4,13}^2(0)}{\sigma^2_{\sigma}}
                 + \frac{\sigma_{4,23}^2(0)}{\sigma^2_{\eta}}
          = \frac{g_{7}^2(0)}{\sigma^2_{\sigma}}
                 + \frac{g_{8}^2(0)}{\sigma^2_{\eta}}\; ,
\label{eq:3.75}
\end{eqnarray}
which fixes $\sigma_{4,33}(0)$. Inserting this into 
(\ref{eq:3.222}) yields:
\begin{eqnarray}
&& \sigma_{4,33}(s)  
          = \frac{g_{7}^2(0)}{\sigma^2_{\sigma}}
                 + \frac{g_{8}^2(0)}{\sigma^2_{\eta}} +   g_{14}(s)
                      + s\cdot g_{15} 
 = \frac{g_{7}^2(0)}{\sigma^2_{\sigma}}
                 + \frac{g_{8}^2(0)}{\sigma^2_{\eta}} +   g_{14}(s)
+ \frac{2\cdot s}{\tau_{spin}} \; .
\label{eq:3.74}
\end{eqnarray}
With 
(\ref{eq:3.14}), 
(\ref{eq:3.204}), 
(\ref{eq:3.71}),
(\ref{eq:3.74})
I have determined the first
moment vector and the covariance matrix of Process 4. 
By (\ref{eq:3.252}),
(\ref{eq:3.239})
the characteristic function $\Phi_4$
corresponding to $w_4$ reads as: 
\begin{eqnarray}
 \Phi_4(\vec{u};s) &=& \exp\biggl(-\frac{1}{2}\cdot
 \sum_{j,k=1}^3 \; \sigma_{4,jk}(s)\cdot u_j\cdot u_k
+i\cdot \psi_{0,m}\cdot u_3 \biggr) \; .
\label{eq:3.240}
\end{eqnarray}
Therefore to prove that the above construction constitutes a G-process I 
just have to show that the covariance
matrix is nonsingular for $s>0$.
This can be done in
analogy to Process 3. In fact from 
(\ref{eq:3.201}), 
(\ref{eq:3.76}) it follows that:
\begin{eqnarray}
 \det\biggl(\underline{\sigma}_4(s)\biggr) &=&
- 2\cdot c\cdot\sigma^2_{\sigma}\cdot \int_0^s\; ds_1
                       \cdot\sigma_{4,23}^2(s_1)  \; ,
\label{eq:3.77}
\end{eqnarray}
and by 
(\ref{eq:3.71}) one has
\begin{eqnarray*}
&& \sigma_{4,23}(0) \neq 0    \; .
\end{eqnarray*}
Hence by 
(\ref{eq:3.77}) $\underline{\sigma}_{4}$ is nonsingular for $s>0$,
confirming that Process 4 is a G-process.
Thus the probability density reads for $s>0$ as:
\begin{eqnarray}
&& w_4(\sigma,\eta,\psi;s)  =
                      \sqrt{(2\pi)^{-3}\cdot
\det( \underline{\sigma}_{4}(s))^{-1}}\cdot
               \exp\biggl\lbrack -\frac{1}{2}\cdot
       \left( \begin{array}{c}
          \sigma \\
          \eta   \\
          \psi -\psi_{0,m}
                \end{array}
         \right)^T       \cdot
    \underline{\sigma}_{4}^{-1}(s)\cdot
       \left( \begin{array}{c}
          \sigma \\
          \eta   \\
          \psi -\psi_{0,m}
                \end{array}
         \right)   \biggr\rbrack \; , \nonumber\\&&
\label{eq:3.73}
\end{eqnarray}
and by using 
(\ref{eq:2.40}),(\ref{eq:3.240}) 
and the expressions for the first moment vector and the covariance
matrix it reads at $s=0$ as:
\begin{eqnarray}
&& w_4(\sigma,\eta,\psi;0)  =
   w_{norm}(\sigma,\eta)\cdot\delta\biggl(\psi
-\frac{\sigma_{4,13}(0)}{\sigma_{\sigma}^2}\cdot\sigma
-\frac{\sigma_{4,23}(0)}{\sigma_{\eta}^2}\cdot\eta - \psi_{0,m} \biggr) \; .
\label{eq:3.245}
\end{eqnarray}
By (\ref{eq:3.73}),
(\ref{eq:3.245})
$w_4$ fulfills the
normalization condition 
(\ref{eq:2.9}) and the orbital part of $w_4$ obeys:
\begin{eqnarray*}
 w_{4,orb} &=&
 w_{norm}  \; ,
\end{eqnarray*}
confirming that Process 4 is at orbital equilibrium.
Note that $\vec{x}^{(4)}(s)$ is a Markovian diffusion
process.
\begin{figure}[t]
\begin{center}
\epsfig{figure=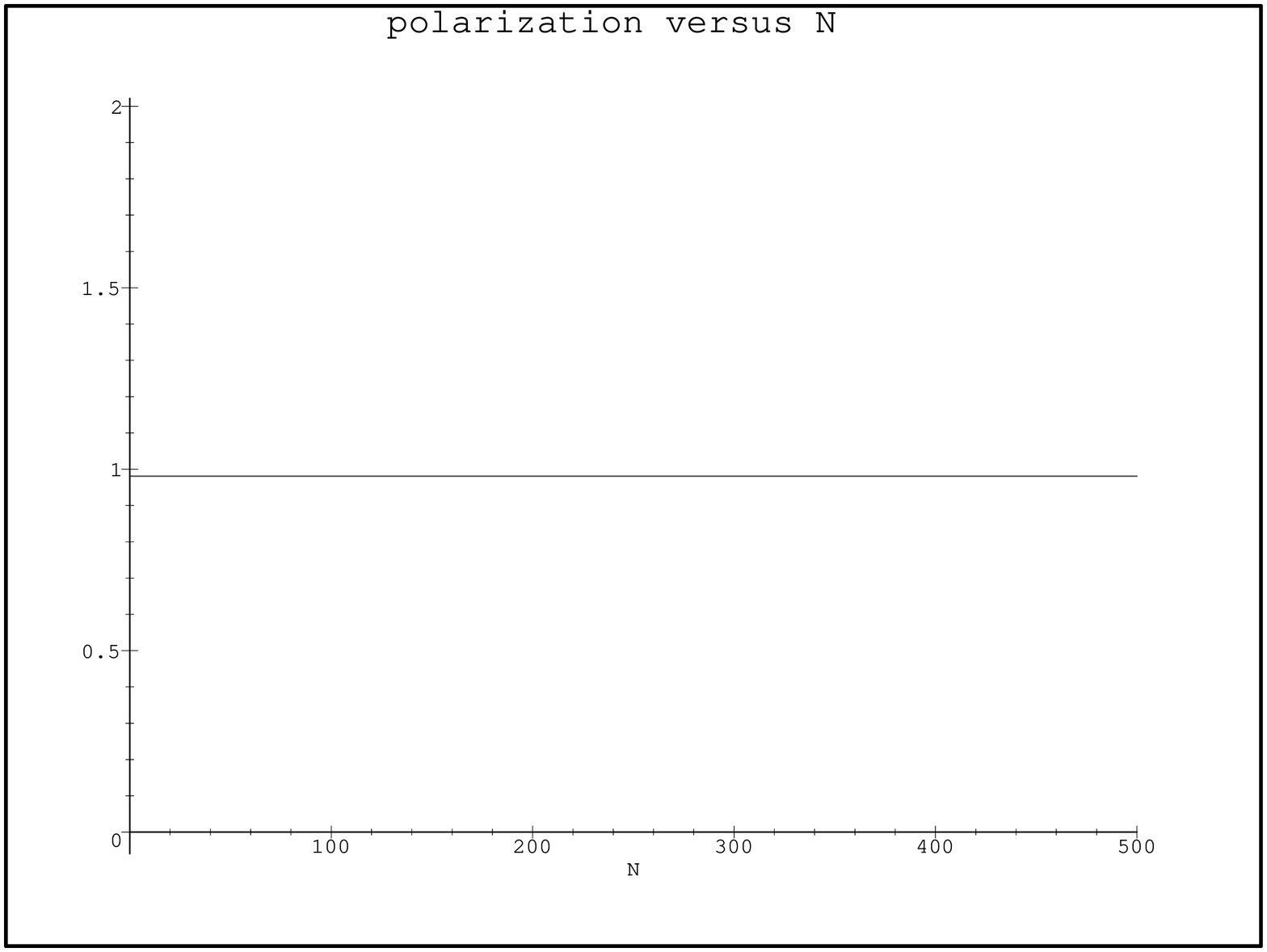,width=12cm,angle=-90}
\end{center}
\caption{Polarization $||\vec P_{tot}^{w_{4a}}(2NL)||$
of Process 4a for the first 1000 turns
assuming the HERA values 
(\ref{eq:2.3}), except that
$c,\omega\rightarrow 0$ with
$\omega/c=const=-2\cdot\sigma^2_{\eta}\approx -2.0\cdot 10^{-6}$.}
\end{figure}
\subsubsection*{3.6.2}
By (\ref{eq:3.214}),(\ref{eq:3.204})
the polarization vector reads as:
\begin{eqnarray}
&&\vec{P}_{tot}^{w_4}(s)  =
  \exp(-\sigma_{4,33}(s)/2)\cdot
  \left( \begin{array}{c}
        1 \\
        0 \\
                      0
                \end{array}
         \right)  \; ,
\label{eq:3.79}
\end{eqnarray}
so that by 
(\ref{eq:3.222})
the polarization is:
\begin{eqnarray}
  ||\vec P_{tot}^{w_4}(s)|| &=&
    \exp\biggl(-\sigma_{4,33}(s)/2\biggr)=
    \exp\biggl(-\sigma_{4,33}(0)/2-g_{14}(s)/2\biggr)
\cdot\exp\biggl(-s/\tau_{spin}\biggr) \; ,
\label{eq:3.80}
\end{eqnarray}
where $\sigma_{4,33}(0)$ is given by
(\ref{eq:3.75}).
Thus Process 4 exhibits complete spin docoherence:
\begin{eqnarray}
      ||\vec P_{tot}^{w_4}(+\infty)|| &=& 0 \; .
 \label{eq:3.81}
\end{eqnarray}
Note that by (\ref{eq:3.80}) the initial polarization of Process 4 is
not complete, so that the initial direction of the local polarization
is not uniform. This will be confirmed below.
\par By
(\ref{eq:3.15}), 
(\ref{eq:3.217}), 
(\ref{eq:3.71}),
(\ref{eq:3.222})
the local polarization reads as:
\begin{eqnarray}
&& ||\vec P_{loc}^{w_4}(\sigma,\eta;s)|| =
               \exp\biggl(
-\frac{\det\biggl(\underline{\sigma}_4(s)\biggr)}
      {2\cdot\sigma_{\sigma}^2\cdot\sigma_{\eta}^2} \biggr)
=              \exp\biggl(
-\frac{\sigma_{4,33}(s)}{2}  \biggr) \cdot
               \exp\biggl(
   \frac{\sigma_{4,13}^2(s)}{2\cdot\sigma^2_{\sigma}}
+\frac{\sigma_{4,23}^2(s)}{2\cdot\sigma^2_{\eta}} \biggr)
\nonumber\\
   &=&         \exp\biggl(
-\frac{s}{\tau_{spin}}\biggr)
   \cdot       \exp\biggl(
   \frac{\sigma_{4,13}^2(s)}{2\cdot\sigma^2_{\sigma}}
+\frac{\sigma_{4,23}^2(s)}{2\cdot\sigma^2_{\eta}}
-\frac{\sigma_{4,33}(0)}{2}-\frac{g_{14}(s)}{2}\biggr)
\nonumber\\
   &=&         \exp\biggl(
-\frac{s}{\tau_{spin}}\biggr)
   \cdot       \exp\biggl(
   \frac{g_{7}^2(s)}{2\cdot\sigma^2_{\sigma}}
+\frac{g_{8}^2(s)}{2\cdot\sigma^2_{\eta}}
-\frac{\sigma_{4,33}(0)}{2}-\frac{g_{14}(s)}{2}\biggr)
                        \; .
\label{eq:3.83}
\end{eqnarray}
As for every G-process the local polarization of Process 4 is uniform. 
It starts from the value 1 at
$s=0$ and decreases towards its vanishing equilibrium value.
\par By 
(\ref{eq:3.218}),
(\ref{eq:3.239}),
(\ref{eq:3.71})
the direction of the local polarization reads as:
\begin{eqnarray}
&& \vec{P}^{w_4}_{dir}(\sigma,\eta;s) =
   \large{      \left( \begin{array}{c}
     \cos\biggl(
        \frac{g_7(s)}{\sigma_{\sigma}^2}\cdot
                                 \sigma
      + \frac{g_8(s)}{\sigma_{\eta}^2}\cdot
                                 \eta \biggr) \\
     \sin\biggl(
        \frac{g_7(s)}{\sigma_{\sigma}^2}\cdot
                                 \sigma
      + \frac{g_8(s)}{\sigma_{\eta}^2}\cdot
                                 \eta      \biggr) \\
                      0
                \end{array}
         \right)}             \; ,
\label{eq:3.84}
\end{eqnarray}
which is 2-turn periodic in $s$.
Also it is 1-turn periodic in the machine frame.
By 
(\ref{eq:3.216}),
(\ref{eq:3.83}),
(\ref{eq:3.84})
the polarization density is given by:
\begin{eqnarray}
&& \vec{P}^{w_4}(\sigma,\eta;s) =
   w_{norm}(\sigma,\eta)\cdot
   \exp\biggl(
-\frac{s}{\tau_{spin}}\biggr)
   \cdot       \exp\biggl(
   \frac{g_{7}^2(s)}{2\cdot\sigma^2_{\sigma}}
+\frac{g_{8}^2(s)}{2\cdot\sigma^2_{\eta}}
-\frac{\sigma_{4,33}(0)}{2}-\frac{g_{14}(s)}{2}\biggr) \nonumber\\&&\qquad
\cdot
   \large{      \left( \begin{array}{c}
     \cos\biggl(
        \frac{g_7(s)}{\sigma_{\sigma}^2}\cdot
                                 \sigma
      + \frac{g_8(s)}{\sigma_{\eta}^2}\cdot
                                 \eta \biggr) \\
     \sin\biggl(
        \frac{g_7(s)}{\sigma_{\sigma}^2}\cdot
                                 \sigma
      + \frac{g_8(s)}{\sigma_{\eta}^2}\cdot
                                 \eta      \biggr) \\
                      0
                \end{array}
         \right)}             \; .
 \label{eq:3.247}
\end{eqnarray}
\par The observed periodicity properties of the local polarization
quantities of Process 4 show the lack of any transient behaviour
of Process 4.
\subsubsection*{3.6.3}
To compare processes 3 and 4 in more detail one observes by 
(\ref{eq:3.28}),
(\ref{eq:3.71})
that after a few orbital damping times one gets:
\begin{eqnarray}
&& \sigma_{4,13}(s) \approx   \sigma_{3,13}(s) , \qquad
   \sigma_{4,23}(s) \approx   \sigma_{3,23}(s)  \; .
\label{eq:3.85}
\end{eqnarray}
Applying this to 
(\ref{eq:3.218}) one sees that on this time scale
the direction of the local
polarization of Process 4 becomes the same as that of Process 3.
\par Also by 
(\ref{eq:3.56}),
(\ref{eq:3.74}) one observes that:
\begin{eqnarray}
&&\sigma_{4,33}(s) - \sigma_{3,33}(s) \approx
     \sigma_{4,33}(0) - g_{13}(+\infty)
   = \sigma_{4,33}(0)
  -\frac{i\cdot d^2\cdot\sigma^2_{\eta}}{\lambda}\cdot g_{17} \; .
\label{eq:3.86}
\end{eqnarray}
Combining 
(\ref{eq:3.15}),
(\ref{eq:3.85}),
(\ref{eq:3.86}) one gets:
\begin{eqnarray}
&& \det\biggl(\underline{\sigma}_{4}(s)\biggr)
-\det\biggl(\underline{\sigma}_{3}(s)\biggr) \approx
     \sigma^2_{\eta}\cdot \sigma^2_{\sigma}\cdot
                               \lbrack
   \sigma_{4,33}(s)  -  \sigma_{3,33}(s) \rbrack
\nonumber\\
&&\qquad\approx \sigma^2_{\eta}\cdot \sigma^2_{\sigma}\cdot  \sigma_{4,33}(0)
    - \frac{i\cdot d^2\cdot\sigma^4_{\eta}\cdot
                           \sigma^2_{\sigma}}{\lambda}\cdot g_{17} \; ,
\label{eq:3.87}
\end{eqnarray}
so that by 
(\ref{eq:3.217}),
(\ref{eq:3.87}) one gets after some orbital damping times:
\begin{eqnarray}
&& ||\vec P_{loc}^{w_4}(\sigma,\eta;s)|| \approx
\exp\biggl(
   -\frac{\sigma_{4,33}(0)}{2}
    +       \frac{i\cdot d^2\cdot\sigma^2_{\eta}}{2\cdot\lambda}\cdot
    g_{17}\biggr)  \cdot
         ||\vec P_{loc}^{w_3}(\sigma,\eta;s)|| \; .
\label{eq:3.88}
\end{eqnarray}
Hence after a few orbital damping times one finds that the local
polarization of Process 3 is proportional to the local
polarization of Process 4. 
One thus sees that the local polarization, as well as the polarization,
is the product of a 1-turn periodic factor and an exponentially decaying factor
with the same depolarization rate $1/\tau_{spin}$ as was
observed for the long term behaviour of Process 3.
Furthermore the polarization densities are
proportional:
\begin{eqnarray}
&& \vec{P}^{w_4}(\sigma,\eta;s) \approx
\exp\biggl(
   -\frac{\sigma_{4,33}(0)}{2}
    +       \frac{i\cdot d^2\cdot\sigma^2_{\eta}}{2\cdot\lambda}\cdot
    g_{17}\biggr) \cdot
   \vec{P}^{w_3}(\sigma,\eta;s) \; .
\label{eq:3.89}
\end{eqnarray}
\par For the HERA values 
(\ref{eq:2.3})
the polarization of Process 4 is
displayed in figure 10 for the first 20000 turns.
In figure 11 the polarization is displayed for the
limiting case, where $c,\omega\rightarrow 0$ with
$\omega/c=const=-2\cdot\sigma^2_{\eta}\approx -2.0\cdot 10^{-6}$.
I call this `Process 4a'.
Note that the polarization is constant from turn to turn, i.e. 1-turn periodic.
I return to this point at the end of section 3. 
\subsubsection*{3.6.4}
One easily observes that all processes which fulfill the first three conditions
for Process 4 are devoid of transient behaviour. In fact one observes for
those processes:
\begin{eqnarray*}
&& ||\vec P_{loc}^{w}(\sigma,\eta;s)|| =
          \exp\biggl(
-\frac{s}{\tau_{spin}}\biggr)
   \cdot       \exp\biggl(
   \frac{g_{7}^2(s)}{2\cdot\sigma^2_{\sigma}}
+\frac{g_{8}^2(s)}{2\cdot\sigma^2_{\eta}}
-\frac{\sigma_{33}(0)}{2}-\frac{g_{14}(s)}{2}\biggr) \; , \\
&& \vec{P}^{w}_{dir}(\sigma,\eta;s) =
   \large{      \left( \begin{array}{c}
     \cos\biggl(
        \frac{g_7(s)}{\sigma_{\sigma}^2}\cdot
                                 \sigma
      + \frac{g_8(s)}{\sigma_{\eta}^2}\cdot
                                 \eta \biggr) \\
     \sin\biggl(
        \frac{g_7(s)}{\sigma_{\sigma}^2}\cdot
                                 \sigma
      + \frac{g_8(s)}{\sigma_{\eta}^2}\cdot
                                 \eta      \biggr) \\
                      0
                \end{array}
         \right)}                \; .
\end{eqnarray*}
In this class of processes Process 4 is the one with the largest polarization,
i.e. the polarization $||\vec P_{tot}^w||$ of those processes obeys:
\begin{eqnarray*}
&& ||\vec P_{tot}^{w}(s)|| \leq ||\vec P_{tot}^{w_4}(s)|| \; .
\end{eqnarray*}
\subsection{The probability density of Process 5}
\subsubsection*{3.7.1}
In this section I consider the process $\vec{x}^{(5)}(s)$, called
`Process 5', whose initial state was
used in several studies of the past \cite{DK72,BHMR91}.
I abbreviate:
\begin{eqnarray*}
&&\vec{x}^{(5)}(s) \equiv
       \left( \begin{array}{c}
                \sigma^{(5)}(s)       \\
                \eta^{(5)}(s)       \\
                \psi^{(5)}(s) 
               \end{array}
          \right)  \; .
\end{eqnarray*}
Process 5 is characterized by the following three conditions:
\begin{itemize}
\item It is a G-process at orbital equilibrium and its starting
azimuth is $s_0=0$.
\item The direction of the initial local polarization is given by the
$\vec{n}$-axis of Machine II \cite{DK72,HH96}.
\item Its initial local polarization is 1.
\end{itemize}
For my two-dimensional orbital phase space the $\vec n$-axis, denoted
by $\vec n_{II}$ (see Appendix B),
is a unit-vector periodic solution of the radiationless Bloch equation 
(\ref{eq:B.1}) satisfying the condition:
\begin{eqnarray*}
&& \vec n_{II}(\sigma,\eta;s) = \vec n_{II}(\sigma,\eta;s+L) \; ,
\end{eqnarray*}
in the machine frame
and it is a key component of the standard method for calculating the 
rate of depolarization by perturbation theory in real rings. 
\footnote{In the $(\vec{n}_{0,II},\vec{m}_{0,II},\vec{l}_{0,II})$-frame the
$\vec n$-axis is denoted
by $\hat{\vec n}_{II}$ 
(see Appendix B).
It obeys the
radiationless Bloch equation 
(\ref{eq:3.11})
and satisfies the condition
\begin{eqnarray*}
&& \hat{\vec n}_{II}(\sigma,\eta;s) = \hat{\vec n}_{II}(\sigma,\eta;s+2L) \; .
\end{eqnarray*}}
In the 
absence of radiation an ensemble, for which the local polarization direction
is parallel
to the $\vec n$-axis and for which the local polarization equals $1$, 
remains in this state.
See the discussion in \cite{HH96}. 
\footnote{Note that in this treatment the $\vec n$-axis 
enters the picture as the
periodic solution of the radiationless Bloch equation. 
This is in contrast to other treatments, where the $\vec n$-axis 
appears in the 
diagonalization of the combined spin-orbit Hamiltonian (see Appendix D)
\cite{DK73,Yok86,BHR94b}. For the relation of the $\vec n$-axis 
to the Thomas-BMT 
equation, see section 2.7.1.}
In the presence of radiation it is 
usually assumed \cite{DK73,Man87,BHMR91} that the local polarization direction 
remains 
parallel to the $\vec n$-axis at each point in phase space and it is this 
assumption that provides a way to calculate the depolarization rate for 
real rings using perturbation theory. In this section I will use my 
exact analytical methods to check these assumptions for Machine II and to
show analytically that in the absence of radiation an ensemble initially 
polarized along the $\vec n$-axis is in spin equilibrium.
\par Process 5 differs from Process 4 by the second condition.
By the first condition one has via (\ref{eq:3.219}):
\begin{eqnarray}
&&<\vec{x}^{(5)}(s)> =
<\vec{x}^{(5)}(0)> =
 (0,0, <\psi^{(5)}(0)>)^T \; .
\label{eq:3.90}
\end{eqnarray}
A constraint on $<\psi^{(5)}(0)>$ 
will be derived below using the second condition.
\par Because of the first condition on Process 5 I can apply
(\ref{eq:3.217}),(\ref{eq:3.218}) by which one gets:
\begin{eqnarray}
&& ||\vec P_{loc}^{w_5}(\sigma,\eta;s)|| =
               \exp\biggl(
-\frac{\det(\underline{\sigma}_5)}{2\cdot\sigma_{\sigma}^2\cdot
 \sigma_{\eta}^2} \biggr) \; , \label{eq:3.223}
\\
&& \vec{P}^{w_5}_{dir}(\sigma,\eta;s) =
   \large{      \left( \begin{array}{c}
     \cos\biggl(
        \frac{\sigma_{5,13}(s)}{\sigma_{\sigma}^2}\cdot
                                 \sigma
      + \frac{\sigma_{5,23}(s)}{\sigma_{\eta}^2}\cdot
                                 \eta +<\psi^{(5)}(0)>
                                           \biggr) \\
     \sin\biggl(
        \frac{\sigma_{5,13}(s)}{\sigma_{\sigma}^2}\cdot
                                 \sigma
      + \frac{\sigma_{5,23}(s)}{\sigma_{\eta}^2}\cdot
                                 \eta +<\psi^{(5)}(0)>
                                           \biggr) \\
                      0
                \end{array}
         \right)}        \; ,
 \label{eq:3.224}
\end{eqnarray}
where $\underline{\sigma}_{5}$ denotes the covariance matrix of
Process 5.
Applying the second condition on Process 5 one gets:
\begin{eqnarray}
&& \vec{P}^{w_5}_{dir}(\sigma,\eta;0) =
\lim_{0<s\rightarrow 0} \lbrack \hat{\vec{n}}_{II}(\sigma,\eta;s) \rbrack \; ,
\label{eq:3.93}
\end{eqnarray}
where $\hat{\vec{n}}_{II}$ denotes the $\vec{n}$-axis in the 
$(\vec{n}_{0,II},\vec{m}_{0,II},\vec{l}_{0,II})$-frame which is given
by (\ref{eq:B.6}).
From 
(\ref{eq:3.224}),
(\ref{eq:3.93}),
(\ref{eq:B.6}) follows:
\begin{eqnarray}
&&\sigma_{5,13}(0)= \sigma_{\sigma}^2\cdot 
\lim_{0<s\rightarrow 0} \lbrack 
g_{19}(s) \rbrack = 0 \; ,
\nonumber\\
&&  \sigma_{5,23}(0) = \sigma_{\eta}^2\cdot 
\lim_{0<s\rightarrow 0} \lbrack 
g_{20}(s)\rbrack
   =-\frac{d\cdot\sigma_{\eta}^2}{\lambda_0}\cdot
     \frac{\sin(\lambda_0\cdot L)}{1+\cos(\lambda_0\cdot L)} \; ,
\nonumber\\&&
\label{eq:3.97}
\end{eqnarray}
and:
\begin{eqnarray}
&&  <\psi^{(5)}(0)> = \psi_{0,m}   \; ,
\label{eq:3.98}
\end{eqnarray}
where $\psi_{0,m}$ is given by
(\ref{eq:3.237}).
With
(\ref{eq:3.98}) one can simplify
(\ref{eq:3.224}) to:
\begin{eqnarray}
&& \vec{P}^{w_5}_{dir}(\sigma,\eta;s) =
   \large{      \left( \begin{array}{c}
     \cos\biggl(
        \frac{\sigma_{5,13}(s)}{\sigma_{\sigma}^2}\cdot
                                 \sigma
      + \frac{\sigma_{5,23}(s)}{\sigma_{\eta}^2}\cdot
                                 \eta \biggr) \\
     \sin\biggl(
        \frac{\sigma_{5,13}(s)}{\sigma_{\sigma}^2}\cdot
                                 \sigma
      + \frac{\sigma_{5,23}(s)}{\sigma_{\eta}^2}\cdot
                                 \eta      \biggr) \\
                      0
                \end{array}
         \right)}        \; ,
 \label{eq:3.241}
\end{eqnarray}
so that by 
(\ref{eq:3.216}),
the polarization density reads as:
\begin{eqnarray}
&& \vec{P}^{w_5}(\sigma,\eta;s) =
   w_{norm}(\sigma,\eta)\cdot ||\vec P_{loc}^{w_5}(\sigma,\eta;s)|| 
               \cdot
   \large{      \left( \begin{array}{c}
     \cos\biggl(
        \frac{\sigma_{5,13}(s)}{\sigma_{\sigma}^2}\cdot
                                 \sigma
      + \frac{\sigma_{5,23}(s)}{\sigma_{\eta}^2}\cdot
                                 \eta \biggr) \\
     \sin\biggl(
        \frac{\sigma_{5,13}(s)}{\sigma_{\sigma}^2}\cdot
                                 \sigma
      + \frac{\sigma_{5,23}(s)}{\sigma_{\eta}^2}\cdot
                                 \eta      \biggr) \\
                      0
                \end{array}
         \right)}        \; . \qquad
 \label{eq:3.248}
\end{eqnarray}
From (\ref{eq:3.90}),(\ref{eq:3.98}) follows:
\begin{eqnarray}
<\vec{x}^{(5)}(s)> &=& (0,0,\psi_{0,m})^T \; .
\label{eq:3.96}
\end{eqnarray}
Now I have exploited the
first two conditions on Process 5 and one 
sees that they do not fix $\sigma_{5,33}$. Therefore I impose
the third condition, namely:
\begin{eqnarray*}
&& ||\vec P^{w_5}_{loc}(\sigma,\eta;0)|| = 1 \; ,
\end{eqnarray*}
so that by (\ref{eq:3.223}) one has:
\begin{eqnarray}
 \det\biggl(\underline{\sigma}_5(0)\biggr) &=&  0 \; .
\label{eq:3.95}
\end{eqnarray}
By 
(\ref{eq:3.15}) this leads to:
\begin{eqnarray}
&&  \sigma_{5,33}(0) =
   \frac{\sigma_{5,13}^2(0)}{\sigma^2_{\sigma}}
 + \frac{\sigma_{5,23}^2(0)}{\sigma^2_{\eta}} \; ,
\label{eq:3.99}
\end{eqnarray}
and from
(\ref{eq:3.97}),
(\ref{eq:3.99}) follows:
\begin{eqnarray}
&&  \sigma_{5,33}(0) 
=   \frac{d^2\cdot\sigma_{\eta}^2}{\lambda_0^2}\cdot\biggl(
 \frac{\sin(\lambda_0\cdot L)}{1+\cos(\lambda_0\cdot L)} \biggr)^2 \; .
\label{eq:3.94}
\end{eqnarray}
With 
(\ref{eq:3.202}),
(\ref{eq:3.203}),
(\ref{eq:3.97}),
(\ref{eq:3.94})
one has an initial value problem which determines
$\sigma_{5,13},\sigma_{5,23},\\\sigma_{5,33}$.
\begin{figure}[t]
\begin{center}
\epsfig{figure=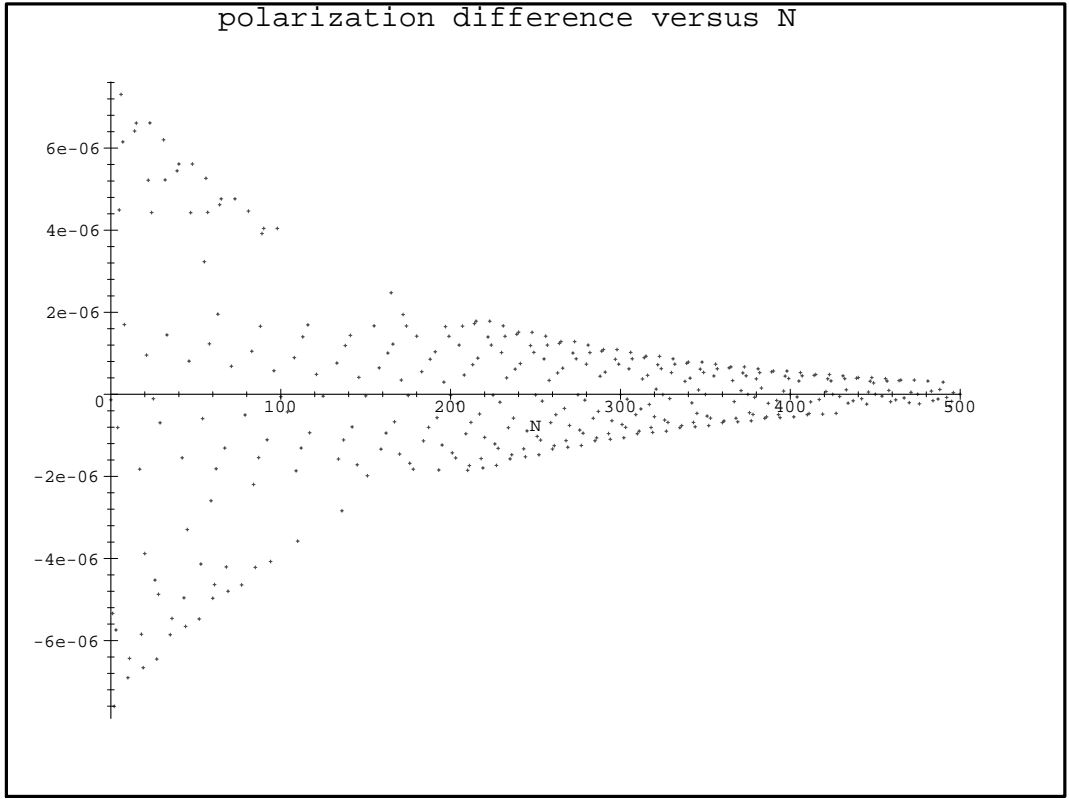,width=12cm,angle=-90}
\end{center}
\caption{Difference $||\vec P_{tot}^{w_5}(2NL)||-||\vec P_{tot}^{w_4}(2NL)||$ 
of the polarization of processes 4,5 for the first 1000 turns
assuming the HERA values
(\ref{eq:2.3})}
\end{figure}
\subsubsection*{3.7.2}
Coming to the calculation of $\sigma_{5,13},\sigma_{5,23}$ for $s>0$
one has by 
(\ref{eq:3.202}),
(\ref{eq:3.97}):
\begin{eqnarray}
&&                \left( \begin{array}{c}
 \sigma_{5,13}(s)  \\
 \sigma_{5,23}(s)
                \end{array}
         \right)     =
               \left( \begin{array}{c}
 \sigma_{3,13}(s)  \\
 \sigma_{3,23}(s)
                \end{array}
         \right)     +
 \left( \begin{array}{c}
     g_{21}(s) \\
     g_{22}(s)
                \end{array}
         \right)  =
 \left( \begin{array}{c}
     g_{7}(s) \\
     g_{8}(s)
                \end{array}
         \right)  +
 \left( \begin{array}{c}
     g_{9}(s) \\
     g_{10}(s)
                \end{array}
         \right) +  
 \left( \begin{array}{c}
     g_{21}(s) \\
     g_{22}(s)
                \end{array}
         \right)   \; , \nonumber\\&&
\label{eq:3.102}
\end{eqnarray}
where:
\begin{eqnarray}
&& g_{21}(s) =
\frac{i\cdot a\cdot d\cdot\sigma_{\eta}^2}{2\cdot\lambda_0\cdot\lambda}\cdot
     \frac{\sin(\lambda_0\cdot L)}{1+\cos(\lambda_0\cdot L)} \cdot g_2(s) \; ,
\nonumber\\
&& g_{22}(s) =
\frac{i\cdot d\cdot\sigma_{\eta}^2}{2\cdot\lambda_0\cdot\lambda}\cdot
     \frac{\sin(\lambda_0\cdot L)}{1+\cos(\lambda_0\cdot L)} \cdot g_3(s) \; .
\nonumber\\&&
\label{eq:3.227}
\end{eqnarray}
This can be checked by showing that the expression in
(\ref{eq:3.102}) obeys 
(\ref{eq:3.202}), 
(\ref{eq:3.97}). 
With (\ref{eq:3.102}) one finds that after a few orbital damping times
$\sigma_{5,13}(s),\sigma_{5,23}(s)$ become 2-turn periodic in $s$ with:
\begin{eqnarray}
&& \left( \begin{array}{c}
 \sigma_{5,13}(s)  \\
 \sigma_{5,23}(s)
                \end{array}
         \right)    \approx
                  \left( \begin{array}{c}
 g_{7}(s)  \\
 g_{8}(s)
                \end{array}
         \right)   \; .
\label{eq:3.226}
\end{eqnarray}
By (\ref{eq:3.28}),
(\ref{eq:3.85}),
(\ref{eq:3.226}) one sees that the asymptotic local polarization
direction of processes 3,4,5 are the same.
Coming finally to $\sigma_{5,33}$ I first of all get from
(\ref{eq:3.203}),
(\ref{eq:3.39}),
(\ref{eq:3.102}):
\begin{eqnarray}
&& \sigma_{5,33}(s)  =
  \sigma_{5,33}(0)  + \int_0^s\; ds_1\cdot
\sigma_{5,33}'(s_1) =
   \sigma_{5,33}(0)  +  2\cdot \int_0^s\; ds_1
                  \cdot\sigma_{5,23}(s_1)\cdot\hat{d}(s_1)
\nonumber\\
&&=  \sigma_{5,33}(0)  +  2\cdot \int_0^s\; ds_1
                  \cdot\sigma_{3,23}(s_1)\cdot\hat{d}(s_1)
+\frac{i\cdot d\cdot\sigma_{\eta}^2}{\lambda_0\cdot\lambda}\cdot
     \frac{\sin(\lambda_0\cdot L)}{1+\cos(\lambda_0\cdot L)} \cdot
     \int_0^s\; ds_1\cdot g_3(s_1)\cdot\hat{d}(s_1)
\nonumber\\
&\equiv&  \sigma_{5,33}(0)  +  \sigma_{3,33}(s)
                       +  g_{23}(s)  \; ,
\label{eq:3.103}
\end{eqnarray}
where
\begin{eqnarray}
&& g_{23}(s)  =
 \frac{i\cdot d\cdot\sigma_{\eta}^2}{\lambda_0\cdot\lambda}\cdot
     \frac{\sin(\lambda_0\cdot L)}{1+\cos(\lambda_0\cdot L)} \cdot
   \lambda_1\cdot g_{16}(s) + c.c.
\label{eq:3.104}
\end{eqnarray}
This can be checked by showing that the expression in
(\ref{eq:3.103}) obeys 
(\ref{eq:3.203}),
(\ref{eq:3.94}). 
With 
(\ref{eq:3.14}), 
(\ref{eq:3.96}), 
(\ref{eq:3.102}),
(\ref{eq:3.103})
I have determined the first
moment vector and the covariance matrix of Process 5. 
By (\ref{eq:3.252}),
(\ref{eq:3.98})
the characteristic function $\Phi_5$
corresponding to $w_5$ reads as: 
\begin{eqnarray}
 \Phi_5(\vec{u};s) &=& \exp\biggl(-\frac{1}{2}\cdot
 \sum_{j,k=1}^3 \; \sigma_{5,jk}(s)\cdot u_j\cdot u_k
+i\cdot \psi_{0,m}\cdot u_3 \biggr) \; .
\label{eq:3.242}
\end{eqnarray}
Therefore to prove that the above construction constitutes a G-process I 
just have to show that the covariance
matrix is nonsingular for $s>0$.
This can be done
analogously to processes 3 and 4. In fact it follows from 
(\ref{eq:3.201}), 
(\ref{eq:3.95}) that:
\begin{eqnarray}
 \det\biggl(\underline{\sigma}_5(s)\biggr) &=&
- 2\cdot c\cdot\sigma^2_{\sigma}\cdot \int_0^s\; ds_1
                       \cdot\sigma_{5,23}^2(s_1)  \; ,
\label{eq:3.100}
\end{eqnarray}
and by 
(\ref{eq:3.102}) one has
\begin{eqnarray*}
&& \sigma_{5,23}(0) \neq 0    \; .
\end{eqnarray*}
Hence by 
(\ref{eq:3.100}) $\underline{\sigma}_{5}$ is nonsingular for $s>0$,
confirming that Process 5 is a G-process.
Thus for $s>0$ the probability density reads as:
\begin{eqnarray}
&& w_5(\sigma,\eta,\psi;s)  =
                      \sqrt{(2\pi)^{-3}\cdot
\det( \underline{\sigma}_{5}(s))^{-1}}\cdot
               \exp\biggl\lbrack -\frac{1}{2}\cdot
       \left( \begin{array}{c}
          \sigma \\
          \eta   \\
          \psi -\psi_{0,m}
                \end{array}
         \right)^T       \cdot
    \underline{\sigma}_{5}^{-1}(s)\cdot
       \left( \begin{array}{c}
          \sigma \\
          \eta   \\
          \psi -\psi_{0,m}
                \end{array}
         \right)   \biggr\rbrack \; , \nonumber\\&&
\label{eq:3.101}
\end{eqnarray}
and by using 
(\ref{eq:2.40}),(\ref{eq:3.242}) 
and the expressions for the first moment vector and the covariance
matrix it reads at $s=0$ as:
\begin{eqnarray}
&& w_5(\sigma,\eta,\psi;0)  =
   w_{norm}(\sigma,\eta)\cdot\delta\biggl(\psi
-\frac{\sigma_{5,23}(0)}{\sigma_{\eta}^2}\cdot\eta - \psi_{0,m} \biggr) \; .
\label{eq:3.246}
\end{eqnarray}
By (\ref{eq:3.101}),
(\ref{eq:3.246})
$w_5$ fulfills the
normalization condition 
(\ref{eq:2.9}) and the orbital part of $w_5$ obeys:
\begin{eqnarray*}
 w_{5,orb} &=&
 w_{norm}  \; ,
\end{eqnarray*}
confirming that Process 5 is at orbital equilibrium.
Note that $\vec{x}^{(5)}(s)$ is a Markovian diffusion
process.
\begin{figure}[t]
\begin{center}
\epsfig{figure=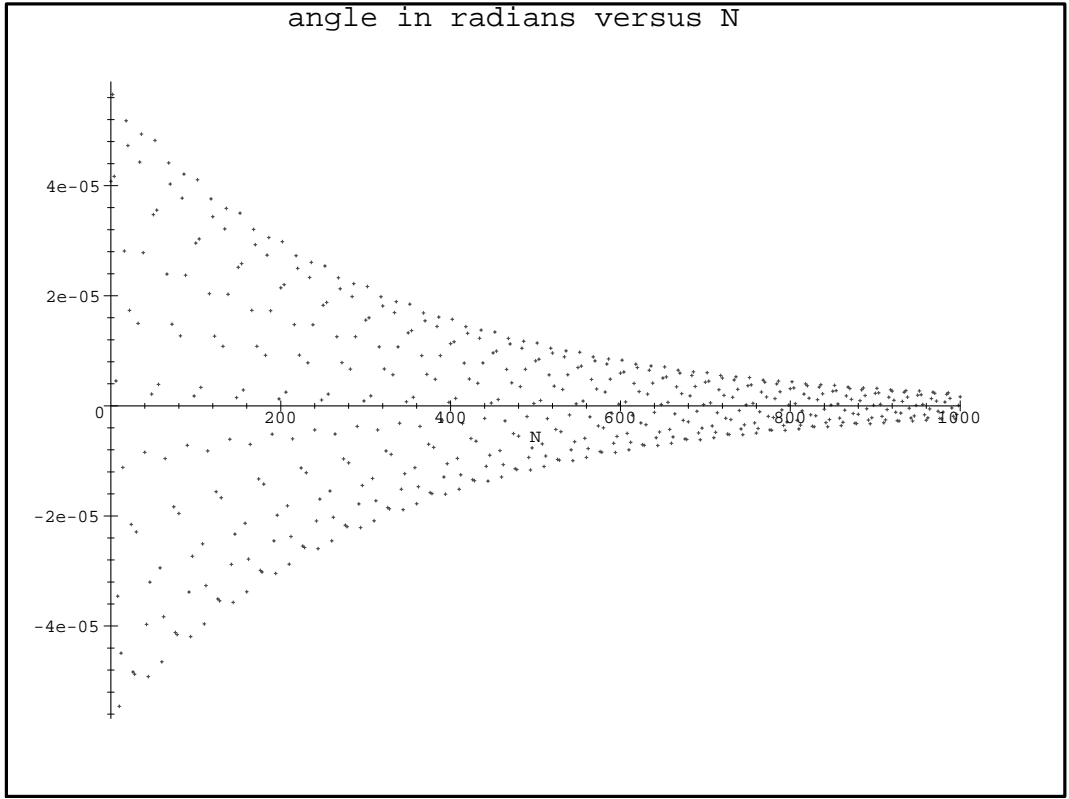,width=12cm,angle=-90}
\end{center}
\caption{Angle in radians 
between the local polarization directions of processes 4,5
at the snake for the
phase space point where
$\sigma= \sigma_{\sigma}$ and $\eta= \sigma_{\eta}$
versus the number $N$ of turns
assuming the HERA values 
(\ref{eq:2.3})}
\end{figure}
\subsubsection*{3.7.3}
By (\ref{eq:3.214}), 
(\ref{eq:3.98}) the
polarization vector reads as:
\begin{eqnarray}
&&\vec{P}_{tot}^{w_5}(s)  =
   \exp(-\sigma_{5,33}(s)/2)\cdot
  \left( \begin{array}{c}
              1      \\
              0      \\
                      0
                \end{array}
         \right)   ,   \;  \;  \;
\label{eq:3.107}
\end{eqnarray}
so that by 
(\ref{eq:3.103}) the polarization is:
\begin{eqnarray}
  ||\vec P_{tot}^{w_5}(s)|| &=&
    \exp\biggl(-\sigma_{5,33}(s)/2\biggr)=
    \exp\biggl(-\sigma_{5,33}(0)/2-
                \sigma_{3,33}(s)/2-g_{23}(s)/2\biggr)
                                      \; .
\label{eq:3.110}
\end{eqnarray}
Thus one has complete spin decoherence of Process 5:
\begin{eqnarray}
      ||\vec P_{tot}^{w_5}(+\infty)|| &=& 0 \; .
 \label{eq:3.108}
\end{eqnarray}
Since $g_{16}(s)$
becomes constant after a few orbital damping times, 
$g_{23}(s)$ becomes constant too.
Hence I get:
\begin{eqnarray*}
&& \sigma_{5,33}(s)  \approx
          \sigma_{5,33}(0)  +  \sigma_{3,33}(s)
                       +  g_{23}(+\infty)
\nonumber\\ &&
  \approx
 \sigma_{5,33}(0)  +  g_{13}(+\infty) + g_{14}(s) + s \cdot  g_{15}
                       +  g_{23}(+\infty)
           \; ,
\end{eqnarray*}
i.e. after a few orbital damping times $\sigma_{5,33}(s)$ splits up
additively into a term increasing linearly with $s$ plus a term
2-turn periodic in $s$.
\par To study the
azimuthal dependence of the polarization `turn by turn', 
I consider the sequence $\sigma_{5,33}(2NL)$.
\par First of all one gets by 
(\ref{eq:3.103}):
\begin{eqnarray}
&& \sigma_{5,33}(2NL)  =
   \sigma_{5,33}(0) + \sigma_{3,33}(2NL)
                       + g_{23}(2NL)  \; .
\label{eq:3.115}
\end{eqnarray}
This can be simplified by calculating
\begin{eqnarray}
&&  g_{23}(2NL)  =
\frac{i\cdot d^2\cdot g_{11}\cdot\sigma_{\eta}^2}
        {2\cdot\lambda_0\cdot\lambda}
\cdot\frac{\sin(\lambda_0\cdot L)}{1+\cos(\lambda_0\cdot L)} \cdot
\lbrack g_2(2NL)
\nonumber\\&&\qquad
-4\cdot i\cdot\exp\biggl(c\cdot L\cdot(N+1/2)\biggr)
\cdot  \sin(\lambda\cdot L)\cdot
  \cos(\lambda\cdot 2NL)  - \exp(c\cdot L)\cdot g_2(2NL)
\nonumber\\&&\qquad
+4\cdot i\cdot\exp(c\cdot L/2)\cdot  \sin(\lambda\cdot L)
                                                   \rbrack
                                                        \; ,
\label{eq:3.111}
\end{eqnarray}
where I used 
(\ref{eq:3.64}). This converges after a few orbital damping times
to a constant whereas $\sigma_{3,33}(2NL)$ grows exponentially.
It is now clear that processes 3,4 and 5 all have the same
depolarization time $\tau_{spin}$ and the same asymptotic local polarization
direction. 
\footnote{One can easily show that this not only holds for processes 3,4
and 5 but for all G-processes at orbital equilibrium.}
\par A plot of the azimuthal dependence of the polarization
for Process 5 for the HERA values
(\ref{eq:2.3})
is visually
indistinguishable from the plot in figure 10 for Process 4 and is
therefore not presented here.  
\footnote{Note that the initial polarization of processes 4 and 5 is
not complete because the initial direction of the local polarization
is not uniform - see 
(\ref{eq:3.84}),(\ref{eq:3.241}).}
However the polarizations of the
two processes are {\it not} identical as can be seen in figure 12 where I
plot the difference. This is tiny and shows oscillating transient
behaviour which must be due to transient behaviour in Process 5 since
Process 4 is transient free. In figure 13 I plot the azimuthal dependence of
the angle between the local polarization direction
of processes 4 and 5 at the phase space point where
$\sigma= \sigma_{\sigma}$ and $\eta= \sigma_{\eta}$. One again sees
oscillating transient behaviour with an angle difference of typically
$0.05$ milliradians whereas the angle between the $\vec n_0$-axis and the
$\vec n$-axis at the same point is about
200 milliradians.
\par The interpretation of these findings is straightforward; although, as
discussed above, the $\vec n$-axis should give the local polarization 
direction in the
absence of radiation, this is no longer exactly true in the presence of
radiation. We have already seen for Process 3 that when the initial
polarization direction deviates typically by 200
milliradians from the asymptotic local polarization direction, 
polarization
fluctuations of several percent occur which eventually damp away while at
the same time the local polarization direction
approaches its asymptotic distribution. A similar
thing happens with Process 5 except that the difference between the 
$\vec n$-axis and the asymptotic local polarization direction is very small 
so that the polarization
fluctuations are correspondingly small. It can be shown using 
(\ref{eq:3.84}),
(\ref{eq:B.2})
that the angle between the $\vec n$-axis and the asymptotic
local polarization direction
at the phase space point where
$\sigma= \sigma_{\sigma}$ and $\eta= \sigma_{\eta}$
is given at the snake by:
\begin{eqnarray*}
&& 
-\frac{1}{\sigma_{\eta}}\cdot g_{8}(0)
-\frac{1}{\sigma_{\sigma}}\cdot g_{7}(0)
+\sigma_{\eta}\cdot
\lim_{0<s\rightarrow 0} \lbrack g_{20}(s) \rbrack \; ,
\end{eqnarray*}
which vanishes in the limit where
$c$ and $\omega$ go to zero with 
$\omega/c=const=-2\cdot\sigma^2_{\eta}\approx -2.0\cdot 10^{-6}$
and which for small $c$ is given approximately by:
\begin{eqnarray*}
&& \frac{d\cdot c\cdot\sigma_{\eta}}{2\cdot\lambda_0^2}\cdot
\frac{\lambda_0\cdot L-\sin(\lambda_0\cdot L)}
{1+\cos(\lambda_0\cdot L)} \; .
\end{eqnarray*}
Thus I have shown that the $\vec n$-axis does not describe the direction of 
the local polarization in Machine II in the presence of radiation. However, the
relative difference in the directions is extremely small and can, for
practical purposes be ignored for Machine II.
\par The depolarization rate obtained using SLIM \cite{Cha81} for a real
perfectly aligned flat HERA lattice with a pointlike radial snake 
and when only spin diffusion due to synchrotron motion generated in the 
arcs is included, is in satisfactory agreement with $\tau_{spin}^{-1}$
\cite{Bar97}. 
\par As one can see from 
(\ref{eq:B.7}), the SLIM approximation to 
the $\vec n$-axis is quite good near phase space points where
$\sigma= \sigma_{\sigma}$ and $\eta= \sigma_{\eta}$. However, it
becomes progressively worse towards the edges of phase space.
\par In the absence of radiation the local 
polarization direction should be
parallel to the $\vec n$-axis and the 
polarization of an ensemble in orbital equilibrium set
up in this state should be constant from turn to turn.  
This can be confirmed 
analytically by putting $c$ and $\omega$ to zero with 
$\omega/c=const=-2\cdot\sigma^2_{\eta}=-2\cdot 10^{-6}$,
in which limit Process 5 modifies to
`Process 5a'.
Using 
(\ref{eq:3.103}),
(\ref{eq:3.107})
the polarization at the snake is then given by:
\begin{eqnarray*}
&& ||\vec P_{tot}^{w_{5a}}(NL)|| =
\exp\biggl(-\frac{d^2\cdot\sigma_{\eta}^2\cdot\sin^2(\lambda_0\cdot L)}
{2\lambda_0^2\cdot(1+\cos(\lambda_0\cdot L))^2}\biggr) \; ,
\end{eqnarray*}
which is indeed constant from turn to turn. 
Also the local polarization is constant from turn to turn.
For the HERA values (\ref{eq:2.3}) this expression gives about $0.98$.
It also coincides with the constant 
polarization seen in Process 4a. 
One can now see why the polarization of Process 4a is constant
from turn to turn; if, as in
Process 4, I require that the local polarization direction 
is already asymptotic at $s=0$ and switch 
off the radiation, this local polarization direction 
must coincide with the $\vec n$-axis since it now fulfills the characteristic
properties of the $\vec n$-axis.
Then, just as 
in Process 5a, the polarization must be constant from turn to turn and
Process 4a is identical with Process 5a.
\subsection{The polarization density and spin matching for Machine II}
Although the synchrotron radiation parameters $c$ and $\omega$ are 
$s$-independent the present 
formalism can be used to analyze more complex rings.
For example, the Green function for the polarization density, in particular
the radiationless Green function, can be used to analyze the effect of
`lumped' radiators such as asymmetric wigglers \cite{Mon84}. 
\footnote{Asymmetric wigglers can be designed so that the overall spin phase
advance is zero and so that no dispersion is generated overall.}
If, for example, a wiggler is placed diametrically opposite the snake,
then one expects on the basis of standard `spin matching' concepts
that the spin diffusion due to the extra excitation of $\eta$ would be almost
cancelled \cite{BKRRS85}.
This can be made quantitative by considering the `$\underline{G}$ matrix', the 
$2\times 6$ spin-orbit coupling matrix of the SLIM formalism. Writing this
in the form
\begin{eqnarray*}
 \underline{G} = (\underline{g}_{\,x},\underline{g}_{\,z},\underline{g}_{\,s})
 \; ,
\end{eqnarray*}
where the $\underline{g}$'s are $2\times 2$ matrices it is simple to show  
\cite{BHR94b} that $\underline{g}_{\,s}$ for 
the interval from $s=L/2$ to $s=3L/2$ in Machine II is:
\begin{eqnarray}
&& \underline{g}_s(3L/2;L/2) 
= 2\cdot d\cdot\biggl(\cos(\lambda_0\cdot L/2)-1 \biggr)\cdot
 \large{\left( \begin{array}{cc}
  \frac{1}{a}\cdot\cos(\lambda_0\cdot L/2)  &  
  \frac{1}{\lambda_0}\cdot\sin(\lambda_0\cdot L/2) \\
 0 &  0
         \end{array}
         \right)}   \; . \nonumber\\&&
\label{eq:3.263} 
\end{eqnarray}
The nonzero matrix elements of $\underline{g}_{\,s}$ vanish as $\lambda_0$ goes
to zero and then in linear approximation a spin travelling from
$s=L/2$ to $s=3L/2$ on any synchrotron orbit is unperturbed overall.
This interval is then said to be `spin transparent' and the depolarization
due to the wiggler is cancelled. For the HERA values
(\ref{eq:2.3}) $\underline{g}_{\,s}$ does not exactly vanish but since 
$Q_s$ is small $\underline{g}_{\,s}$ is still
small enough to ensure
that the depolarization due to the wiggler should be largely suppressed.
\par However, in realistic machines like HERA, there is still
excitation in the remainder of the ring. So even if the radiation power
from the wiggler were dominant, the radiation in the remainder of the ring
would still cause depolarization on the `time' scale $\tau_{spin}$.
These conjectures are confirmed by numerical calculations with SLIM 
\cite{Bar97}. 
\par Spin transparency can be discussed using the polarization density.
Since in this section I am only considering the radiation from the wiggler,
this condition can be investigated using the polarization density
for the radiationless case. 
The causality properties of the azimuthal evolution for
Machine II are the same as for Machine I, so that in particular one has
a Green function for the polarization density.
For the radiationless case the
Green function $\underline{P}_{II,nrad}$ for the radiationless Bloch equation 
(\ref{eq:3.11}) is given by:
\begin{eqnarray}
   \underline{P}_{II,nrad}(\sigma,\eta;s|\sigma_1,\eta_1;s_1)&\equiv&
   w_{orb,trans,nrad}(\sigma,\eta;s|\sigma_1,\eta_1;s_1)\cdot
\underline{R}_{II,nrad}(\sigma_1,\eta_1;s;s_1) \; , \qquad
\label{eq:3.256} 
\end{eqnarray}
where $w_{orb,trans,nrad}$ is given by 
(\ref{eq:2.270}) and where: 
\begin{eqnarray}
\underline{R}_{II,nrad} &\equiv& \left( \begin{array}{ccc}
  \cos(\rho) &
- \sin(\rho) & 0 \\
  \sin(\rho) &
  \cos(\rho) & 0 \\
      0 & 0 & 1
                \end{array}
         \right) \; ,
\label{eq:3.257} 
\end{eqnarray}
with:
\begin{eqnarray}
&& \rho(\sigma_1,\eta_1;s;s_1) \equiv 
\frac{\sigma_1}{a}\cdot\rho_1(s;s_1)
+\frac{\eta_1}{\lambda_0}\cdot\rho_2(s;s_1) \; , 
\label{eq:3.258} \\
&& \rho_1(s;s_1) \equiv
 \hat{d}(s)\cdot\cos\biggl(\lambda_0\cdot(s-s_1)\biggr)
 -\frac{d}{1+\cos(\lambda_0\cdot L)}\cdot\biggl\lbrace
\cos\biggl(\lambda_0\cdot\lbrack
2\cdot L\cdot{\cal G}(s/2L)-s_1\rbrack\biggr)\nonumber\\&&\qquad
+\cos\biggl(\lambda_0\cdot\lbrack
2\cdot L\cdot{\cal G}(s/2L)+L-s_1\rbrack\biggr)\biggr\rbrace \nonumber\\&&
\qquad
+\biggl(d-\hat{d}(s)\biggr)
\cdot\cos\biggl(\lambda_0\cdot\lbrack
2\cdot L\cdot{\cal G}(s/2L)+L-s_1\rbrack\biggr)
-\hat{d}(s_1) \nonumber\\&&
 +\frac{d}{1+\cos(\lambda_0\cdot L)}\cdot\biggl\lbrace
\cos\biggl(\lambda_0\cdot\lbrack
2\cdot L\cdot{\cal G}(s_1/2L)-s_1\rbrack\biggr)
+\cos\biggl(\lambda_0\cdot\lbrack
2\cdot L\cdot{\cal G}(s_1/2L)+L-s_1\rbrack\biggr)\biggr\rbrace \nonumber\\&&
\qquad
-\biggl(d-\hat{d}(s_1)\biggr)
\cdot\cos\biggl(\lambda_0\cdot\lbrack
2\cdot L\cdot{\cal G}(s_1/2L)+L-s_1\rbrack\biggr) \; ,
\label{eq:3.253}  \\
&& \rho_2(s;s_1) \equiv
 \hat{d}(s)\cdot\sin\biggl(\lambda_0\cdot(s-s_1)\biggr)
 -\frac{d}{1+\cos(\lambda_0\cdot L)}\cdot\biggl\lbrace
\sin\biggl(\lambda_0\cdot\lbrack
2\cdot L\cdot{\cal G}(s/2L)-s_1\rbrack\biggr)\nonumber\\&&\qquad
+\sin\biggl(\lambda_0\cdot\lbrack
2\cdot L\cdot{\cal G}(s/2L)+L-s_1\rbrack\biggr)\biggr\rbrace \nonumber\\&&
\qquad
+\biggl(d-\hat{d}(s)\biggr)
\cdot\sin\biggl(\lambda_0\cdot\lbrack
2\cdot L\cdot{\cal G}(s/2L)+L-s_1\rbrack\biggr)
 \nonumber\\&&
 +\frac{d}{1+\cos(\lambda_0\cdot L)}\cdot\biggl\lbrace
\sin\biggl(\lambda_0\cdot\lbrack
2\cdot L\cdot{\cal G}(s_1/2L)-s_1\rbrack\biggr)
+\sin\biggl(\lambda_0\cdot\lbrack
2\cdot L\cdot{\cal G}(s_1/2L)+L-s_1\rbrack\biggr)\biggr\rbrace \nonumber\\&&
\qquad
-\biggl(d-\hat{d}(s_1)\biggr)
\cdot\sin\biggl(\lambda_0\cdot\lbrack
2\cdot L\cdot{\cal G}(s_1/2L)+L-s_1\rbrack\biggr) \; .
\label{eq:3.254} 
\end{eqnarray}
Moreover:
\begin{eqnarray}
&& \vec{P}^w(\sigma,\eta;s) = \int_{-\infty}^{+\infty} \; d\sigma_1
                             \int_{-\infty}^{+\infty} \; d\eta_1\cdot
\; \underline{P}_{II,nrad}(\sigma,\eta;s|\sigma_1,\eta_1;s_1)
            \cdot\vec{P}^w(\sigma_1,\eta_1;s_1) \; ,
\label{eq:3.255} 
\end{eqnarray}
where $\vec{P}^w$ is an arbitrary solution of
(\ref{eq:3.11}). Therefore, and
by using the connection (see section 2.7.1) between (\ref{eq:3.11})
and the Thomas-BMT equation 
(\ref{eq:3.4}), one finds for a Thomas-BMT solution $\vec{S}(s)$ on
a given orbit $\sigma(s),\eta(s)$:
\begin{eqnarray}
 \vec{S}(s) &=& \underline{R}_{II,nrad}\biggl(\sigma(s_1),\eta(s_1);s;s_1
\biggr)\cdot\vec{S}(s_1) \nonumber\\
&=& \left( \begin{array}{ccc}
  \cos\biggl(\rho(\sigma(s_1),\eta(s_1);s;s_1)\biggr) &
  -\sin\biggl(\rho(\sigma(s_1),\eta(s_1);s;s_1)\biggr) & 0 \\
   \sin\biggl(\rho(\sigma(s_1),\eta(s_1);s;s_1)\biggr) &
   \cos\biggl(\rho(\sigma(s_1),\eta(s_1);s;s_1)\biggr) & 0 \\
    0 & 0 & 1
                \end{array}
         \right)\cdot\vec{S}(s_1)
 \; , \qquad
\label{eq:3.259} 
\end{eqnarray}
where by (\ref{eq:3.258}):
\begin{eqnarray}
&&  \rho(\sigma(s_1),\eta(s_1);s;s_1) =
\frac{\sigma(s_1)}{a}\cdot\rho_1(s;s_1)
+\frac{\eta(s_1)}{\lambda_0}\cdot\rho_2(s;s_1) \; .
\label{eq:3.265} 
\end{eqnarray}
Note that, due to the absence of radiation, $\sigma(s),\eta(s)$ fulfill
(\ref{eq:2.285}). By (\ref{eq:3.259}) $\vec{S}(s_1)$ evolves into
$\vec{S}(s)$ via the spin transfer matrix
$\underline{R}_{II,nrad}$. For $s_1=L/2,s=3L/2$ one has by
(\ref{eq:3.253}),
(\ref{eq:3.254}):
\begin{eqnarray}
&& \rho_1(3L/2;L/2) =
 2\cdot d\cdot\cos(\lambda_0\cdot L/2)\cdot\biggl(
 \cos(\lambda_0\cdot L/2)-1 \biggr) \; , \nonumber\\
&& \rho_2(3L/2;L/2) =
 2\cdot d\cdot\sin(\lambda_0\cdot L/2)\cdot\biggl(
 \cos(\lambda_0\cdot L/2)-1 \biggr) \; . \nonumber\\&&
\label{eq:3.262} 
\end{eqnarray}
The matrix $\underline{g}_{\,s}(3L/2;L/2)$ 
of the linearized formalism (see (\ref{eq:3.263}))
can now be written via
(\ref{eq:3.259}),
(\ref{eq:3.265}),
(\ref{eq:3.262})
as:
\begin{eqnarray}
 \underline{g}_{\,s}(3L/2;L/2) &=& 
       \left( \begin{array}{ccc}
 \rho_1(3L/2;L/2)/a & &
 \rho_2(3L/2;L/2)/\lambda_0 \\
 0 &  & 0 
         \end{array}
         \right)   \; .
\label{eq:3.261} 
\end{eqnarray}
By (\ref{eq:3.261})
the condition for
spin transparency is equivalent to:
\begin{eqnarray}
 && \rho_1(3L/2;L/2) = \rho_2(3L/2;L/2) = 0 \; .
\label{eq:3.266} 
\end{eqnarray}
In this case the matrix 
$\underline{R}_{II,nrad}\biggl(\sigma(s_1),\eta(s_1);3L/2;L/2\biggr)$
reduces, see
(\ref{eq:3.259}) and
(\ref{eq:3.265}),
to the unit matrix so that for Machine II
the spin transparency condition of the linearized formalism
even applies to large deviations of the spin vector from the 
$\vec n_0$-axis. 
By 
(\ref{eq:3.256}),
(\ref{eq:3.257}),
(\ref{eq:3.258}),
(\ref{eq:3.266})
one then gets:
\begin{eqnarray}
&&  \vec{P}^w(\vec z;3L/2) = 
\vec{P}^w\biggl(\exp(-L\cdot\underline{\cal A}_{orb,nrad})\cdot
\vec z;L/2\biggr)\; ,
\end{eqnarray}
so that via
(\ref{eq:2.11}):
\begin{eqnarray}
&& \vec{P}_{tot}^w(3L/2) = 
     \vec{P}_{tot}^w(L/2) \; .
\end{eqnarray}
\section{Summary}
The spin depolarization rate in an electron storage ring is usually
calculated using standard algorithms like SLIM or a Monte-Carlo tracking
program such as SITROS \cite{Kew83,Boe94}. 
SLIM exploits the Derbenev-Kondratenko (DK) formula
\footnote{Other algorithms
exploiting the DK
approach, but at higher order, are SMILE \cite{Man87}
and SODOM \cite{Yok92}.}
and a first order perturbation theory. 
\par Methods based on the DK formalism are only 
applicable once various
transient phenomena have damped away. In this paper I have shown how one
can, instead, apply standard Fokker-Planck methods, but to very simple
model rings. In the process I introduced the polarization density, a
quantity which obeys a universal evolution equation of the Bloch type. This
is a linear equation valid for arbitrary spin distributions and
can therefore be used far from orbital and/or spin 
equilibrium. I also introduced the local polarization vector and its
direction. Both obey Bloch type evolution equations which depend on the
orbital state and are therefore not universal. 
\par In the numerical part of this report I have used the parameters 
(\ref{eq:2.3}) of a typical ring, namely the HERA electron ring, to study the
short and long term behaviour of nine different scenarios, the stochastic
Processes 1,2,3,4 and 5 and the noise free processes 2a, 3a, 4a and 5a. It
is found that Processes 1 and 2 do not lead to complete spin decoherence
and the corresponding equilibrium polarization is calculated. However in
the presence of a Siberian Snake (processes 3,4 and 5) there is complete
spin decoherence, a result which is to some extent counter intuitive
given that snakes tend to stabilize the polarization. 
The asymptotic depolarization rates agree with SLIM estimates and in the
presence of radiation the asymptotic local polarization direction is 
not exactly parallel to the $\vec n$-axis. The radiationless
Green function for the polarization density offers another tool
for studying spin matching.
In processes 2a and 3a the polarization oscillates indefinitely
but in processes 4a and 5a, the spin distribution begins in equilibrium 
and remains there. 
\par  The models considered here are extremely simple but in the next
section I indicate  how the polarization density (and the local polarization
vector and its direction) can be used in 
realistic cases.
\section{Epilogue} 
This paper is the first part of a paper prepared in 1995 but not distributed. 
In the second
part I demonstrate that the polarization density defined on a six-dimensional 
orbital phase space for full three-dimensional spin motion
also obeys a universal evolution equation of Bloch type
with the same linear structure as equations
(\ref{eq:2.64}),
(\ref{eq:3.10}), namely
\begin{eqnarray*}
\frac{\partial\vec{P}^w} {\partial s}   &=&
       L_{FP,orb,gen} \; \vec{P}^w +
\vec{W}_{gen} \wedge \vec{P}^w \; ,
\end{eqnarray*}
where $L_{FP,orb,gen}$ denotes the orbital Fokker-Planck operator and where
the vector $\vec{W}_{gen}$ is determined by the Thomas-BMT equation.
In the more general formulation I use periodic boundary conditions for the
spin phase. 
\footnote{See section 2.3.6.}
The polarization density is appropriate for use with
arbitrarily complicated rings. 
The local polarization vector and its direction
are also useful tools (especially at orbital
equilibrium) and by using these and their Bloch equations
I obtain an expression for the
depolarization rate in terms of the azimuthal and phase space average of
$||\partial \vec P^w_{dir}/\partial\eta||^2\cdot|K|^3$ 
which generalizes the average of
$||\partial \vec n/\partial\eta||^2\cdot|K|^3$ in
equation 6.2 in \cite{DK73}. 
\footnote{Note that $K$ denotes the design orbit curvature, e.g. $K=K_x$
(see section 1).}
By calculating the depolarization rate for Machine II in terms of
$||\partial \vec P^{w_4}_{dir}/\partial\eta||^2\cdot|K|^3$ 
I reproduce the $1/\tau_{spin}$ of (\ref{eq:3.59}).
In fact 
the derivation of the Bloch equation for the general polarization density
constitutes a classical construction for the pure depolarization part of
equation 2 in \cite{DK75} which was obtained by semiclassical methods. 
Using the polarization density 
one can estimate the angle between the $\vec n$-axis and the true
local polarization direction in practical rings.
The Sokolov-Ternov process can be included by
adding in parts of the Baier-Katkov-Strakovenko expression \cite{BKS70}
in an obvious
way. One then has many of the terms in equation 2 in \cite{DK75}. The
remaining terms can only be obtained by a full semiclassical treatment of
the radiation process and this work is in progress and will be published
at a later date together with the classical work on the general
polarization density. 
\par The use of the polarization density obviates the need to begin the 
calculation of the depolarization rate by first diagonalizing the
the combined spin-orbit Hamiltonian as in \cite{DK73}.
\renewcommand{\thesection}{\Alph{section}}
\setcounter{section}{1}
\setcounter{equation}{0}
%
%
\section*{Appendix A}
\addcontentsline{toc}{section}{Appendix A}
\subsection*{A.1}
In this appendix I show that the Thomas-BMT equation 
(\ref{eq:3.4})
for Machine II,
defined in the rotating frame
    $\vec{n}_{0,II},\vec{m}_{0,II},\vec{l}_{0,II}$ of section 3.1, follows
from (\ref{eq:3.1}). First of all I
have to show that the vectors
               $\vec{m}_{0,II}(s),\vec{l}_{0,II}(s),\vec{n}_{0,II}(s)$
solve the closed orbit Thomas-BMT equation:
\begin{eqnarray}
 \vec{\xi}\ '(s) &=&
\vec{\Omega}_{II,0}(s)\wedge \vec{\xi}(s)   \; ,
 \label{eq:A.1}
\end{eqnarray}
where by section 3.1:
\begin{eqnarray*}
&&\vec{\Omega}_{II,0}(s) \equiv
d\cdot\vec{e}_3
+\pi\cdot\delta_{L,per}(s)\cdot\vec{e}_1 \; .
\end{eqnarray*}
I call the region outside the snake `the arcs' and there 
(\ref{eq:A.1})
reduces to a precession around
$\vec{e}_3$, so that on the closed orbit and 
in the machine frame the spin transfer matrix in the arcs reads as:
\begin{eqnarray}
&& \underline{M}_{arc}(s_2,s_1)   \equiv
       \left( \begin{array}{ccc}
\cos\biggl(d\cdot(s_2-s_1)\biggr) &
-\sin\biggl(d\cdot(s_2-s_1)\biggr) & 0 \\
\sin\biggl(d\cdot(s_2-s_1)\biggr) &
\cos\biggl(d\cdot(s_2-s_1)\biggr) & 0 \\
          0           &   0    &  1
                \end{array}
         \right)   \; ,    \;
 \label{eq:A.2}
\end{eqnarray}
i.e. in the arcs:
\begin{eqnarray*}
&& \underline{M}_{arc}(s_2,s_1)\cdot\vec{\xi}(s_1) = \vec{\xi}(s_2) \; .
\end{eqnarray*}
To get the spin transfer matrix for the snake I first
write down 
(\ref{eq:A.1})
for the snake which results in
\begin{eqnarray*}
 \vec{\xi}\ '(s) &=&
 \pi\cdot\delta_{L,per}(s)\cdot\biggl(\vec{e}_1
                    \wedge \vec{\xi}(s)\biggr)
\equiv \delta_{L,per}(s)\cdot
      \underline{M}_0\cdot \vec{\xi}(s) \; ,
\end{eqnarray*}
where in the machine frame $\underline{M}_0$ reads as:
\begin{eqnarray*}
 \underline{M}_0  &\equiv&
       \left( \begin{array}{ccc}
 0 & 0 & 0 \\
 0 & 0 & -\pi \\
 0 & \pi & 0
                \end{array}
         \right)   \; .
\end{eqnarray*}
Because the matrix $\underline{M}_0$ is $s$-independent in the machine frame
one easily finds that in this frame 
the spin transfer matrix for the snake
is given on the closed orbit by:
%
%
%
%
%
%
\begin{eqnarray}
 \underline{M}_{snake}  &=&
\exp(\underline{M}_0) =
 \left( \begin{array}{ccc}
 1 & 0 & 0 \\
 0 &-1 & 0 \\
 0 & 0 &-1
                \end{array}
         \right)   \; .
 \label{eq:A.3}
\end{eqnarray}
The $\vec n_0$-axis of Machine II is given by:
\begin{eqnarray}
 \vec{n}_{0,II}(s)  &\equiv&
\cos\lbrack d\cdot
    (s-L/2-L\cdot {\cal G}(s/L))\rbrack \cdot \vec{e}_1
\nonumber\\ &&
       +\sin\lbrack d\cdot
    (s-L/2-L\cdot {\cal G}(s/L))\rbrack \cdot \vec{e}_2 \; .
 \label{eq:A.5}
\end{eqnarray}
It obviously solves
(\ref{eq:A.1})
in the arcs and is 1-turn periodic in the machine frame. To check if it
solves 
(\ref{eq:A.1})
also at the snake I note that
\begin{eqnarray*}
\lim_{0<s \rightarrow 0}\lbrack\vec{n}_{0,II}(s)\rbrack  &=&
\cos(d\cdot L/2)\cdot\vec{e}_1
-\sin(d\cdot L/2)\cdot\vec{e}_2 \; ,
\nonumber\\
\lim_{0>s \rightarrow 0}\lbrack\vec{n}_{0,II}(s)\rbrack  &=&
\cos(d\cdot L/2)\cdot\vec{e}_1
+\sin(d\cdot L/2)\cdot\vec{e}_2 \; ,
\end{eqnarray*}
from which follows by
(\ref{eq:A.3}):
\begin{eqnarray*}
\lim_{0<s \rightarrow 0}\lbrack\vec{n}_{0,II}(s)\rbrack  &=&
 \underline{M}_{snake}\cdot \lim_{0>s \rightarrow 0}
\lbrack\vec{n}_{0,II}(s)\rbrack\; ,
\end{eqnarray*}
so that in fact $\vec{n}_{0,II}(s)$ solves
(\ref{eq:A.1})
at the snake. Thus $\vec{n}_{0,II}(s)$ is a unit-vector solution of
(\ref{eq:A.1}), which is 1-turn periodic in the machine frame, i.e.
$\vec n_{0,II}$ is the $\vec{n}_0$-axis of Machine II.
\par Now I consider $\vec{l}_{0,II}$ which is defined by:
\begin{eqnarray*}
 \vec{l}_{0,II}(s)  &\equiv& \theta_{2L,per}(s)\cdot\vec{e}_3 \; .
\end{eqnarray*}
One easily finds that $\vec{l}_{0,II}(s)$ solves 
(\ref{eq:A.1})
in the
arcs. To check if it solves
(\ref{eq:A.1})
also at the snake I calculate by
(\ref{eq:A.3}):
\begin{eqnarray*}
&&\lim_{0<s \rightarrow 0}\lbrack\vec{l}_{0,II}(s)\rbrack  =
\vec e_3 = -\underline{M}_{snake}\cdot \vec{e}_3 =
 \underline{M}_{snake}\cdot \lim_{0>s \rightarrow 0}
\lbrack\vec{l}_{0,II}(s)\rbrack\; ,
\end{eqnarray*}
so that in fact $\vec{l}_{0,II}(s)$ solves
(\ref{eq:A.1})
at the snake. Thus I have shown that the dreibein
$\vec{n}_{0,II}(s),\\\vec{m}_{0,II}(s),\vec{l}_{0,II}(s)$ solves 
(\ref{eq:A.1}), where
\begin{eqnarray*}
 \vec{m}_{0,II} &\equiv& \vec{l}_{0,II} \wedge \vec{n}_{0,II} \; .
\end{eqnarray*}
\subsection*{A.2}
Now one can derive 
(\ref{eq:3.4})
from 
(\ref{eq:3.1}). First of all I define
\begin{eqnarray}
 \vec{S}  &\equiv&
       \left( \begin{array}{c}
 \vec{n}_{0,II}^T\cdot\vec{\xi} \\
 \vec{m}_{0,II}^T\cdot\vec{\xi} \\
 \vec{l}_{0,II}^{\;T}\cdot\vec{\xi}
                \end{array}
         \right)   \; ,
 \label{eq:A.6}
\end{eqnarray}
so that (\ref{eq:3.1})
is equivalent to the following equation for
$\vec{S}$:
\begin{eqnarray}
 \vec{S}' &=&
\vec{W}_{II}\wedge \vec{S}   \; ,
 \label{eq:A.7}
\end{eqnarray}
with
\begin{eqnarray*}
\vec{W}_{II}  \equiv
\biggl(
\vec{n}_{0,II}^T\cdot
\lbrack \vec{\Omega}_{II}
          -\vec{U} \rbrack\biggr)\cdot
  \left( \begin{array}{c}
           1       \\
           0       \\
           0
                \end{array}
         \right)
+ \biggl(
\vec{m}_{0,II}^T\cdot
\lbrack \vec{\Omega}_{II}
          -\vec{U} \rbrack\biggr)\cdot
  \left( \begin{array}{c}
           0       \\
           1       \\
           0
                \end{array}
         \right)
+ \biggl(
\vec{l}_{0,II}^{\;T}\cdot
\lbrack \vec{\Omega}_{II}
          -\vec{U} \rbrack\biggr)\cdot
  \left( \begin{array}{c}
           0       \\
           0       \\
           1
                \end{array}
         \right) \; ,
\end{eqnarray*}
where \cite{BHR94a,BHR94b}.
\begin{eqnarray*}
   \vec{U}  &\equiv &\frac{1}{2}\cdot\lbrack
   \vec{n}_{0,II}\wedge \vec{n}_{0,II}'
 + \vec{m}_{0,II}\wedge \vec{m}_{0,II}'
 + \vec{l}_{0,II}\wedge \vec{l}_{0,II}'\rbrack
     =  \vec{\Omega}_{II,0} \; .
\end{eqnarray*}
Therefore
\begin{eqnarray*}
&&\vec{W}_{II} =
d\cdot\eta\cdot\biggl( \vec{e}_3^{\,T}\cdot \vec{l}_{0,II}\biggr)\cdot
  \left( \begin{array}{c}
           0       \\
           0       \\
           1
                \end{array}
         \right)
 =
d\cdot\eta\cdot \theta_{2L,per}\cdot
  \left( \begin{array}{c}
           0       \\
           0       \\
           1
                \end{array}
         \right)  \; .
\end{eqnarray*}
Hence I have shown that 
(\ref{eq:3.4}) follows from
(\ref{eq:3.1}).
\setcounter{section}{2}
\setcounter{equation}{0}
\section*{Appendix B}
\addcontentsline{toc}{section}{Appendix B}
\subsection*{B.1}
In this appendix I calculate the $\vec{n}$-axis 
of Machine II.
\footnote{The $\vec n$-axis for Machine I is given by $\vec e_3$.}
As mentioned in section 3.7.1 the $\vec n$-axis 
is denoted by $\vec n_{II}$ and it obeys
the following radiationless Bloch equation:
\begin{eqnarray}
 \frac{\partial \vec{n}_{II}}{\partial s} &=&
  -a\cdot \eta \cdot
\frac{\partial \vec{n}_{II}}{\partial \sigma}
- b\cdot\sigma \cdot \frac{\partial \vec{n}_{II}}{\partial \eta}
     +   \vec{\Omega}_{II} \wedge   \vec{n}_{II} \; .
\label{eq:B.1}
\end{eqnarray}
In fact, (\ref{eq:B.1}) holds because in the
$(\vec{n}_{0,II},\vec{m}_{0,II},\vec{l}_{0,II})$-frame 
the $\vec n$-axis obeys the radiationless Bloch equation
(\ref{eq:3.11}).
\par I begin by assuming that the $\vec{n}$-axis is horizontal, 
so that I make the ansatz:
\begin{eqnarray}
 \vec{n}_{II}   &\equiv&
\cos(f_{II})\cdot \vec{e}_1  +
\sin(f_{II})\cdot \vec{e}_2  \; ,
\label{eq:B.2}
\end{eqnarray}
where $f_{II}(\sigma,\eta;s)$ needs to be determined.
Then with the ansatz that $f_{II}$ is linear in $\sigma,\eta$ one
finally gets
\begin{eqnarray}
&& f_{II}(\sigma,\eta;s)  =
   g_6(s)  + \sigma\cdot g_{19}(s)  +  \eta\cdot g_{20}(s) \; ,
\label{eq:B.4}
\end{eqnarray}
where $g_6$ is defined in section 3.1 and where:
\begin{eqnarray}
&& g_{19}(s) =
  \frac{d\cdot b}{\lambda_0^2}\cdot
     \frac{1}{1+\cos(\lambda_0\cdot L)}\cdot \biggl\lbrack
\cos\biggl(\lambda_0\cdot\lbrack s-L-L\cdot {\cal G}(s/L)\rbrack\biggr)
\nonumber\\ && \qquad
+ \cos\biggl(\lambda_0\cdot\lbrack s-L\cdot {\cal G}(s/L)\rbrack\biggr)
-\cos(\lambda_0\cdot L) -1 \biggr\rbrack \; , \nonumber\\
&& g_{20}(s) =
\frac{d}{\lambda_0}\cdot
     \frac{1}{1+\cos(\lambda_0\cdot L)}\cdot \biggl\lbrack
\sin\biggl(\lambda_0\cdot\lbrack s-L-L\cdot {\cal G}(s/L)\rbrack\biggr)
+ \sin\biggl(\lambda_0\cdot\lbrack s-L\cdot {\cal G}(s/L)\rbrack\biggr)
                           \biggr\rbrack \; . \nonumber\\&&
\label{eq:B.5}
\end{eqnarray}
\subsection*{B.2}
In this section I show that $\vec n_{II}$, as defined by 
(\ref{eq:B.2}),(\ref{eq:B.4}),(\ref{eq:B.5})
fulfills all properties
of an $\vec n$-axis.
\par Inserting 
(\ref{eq:B.2}) into
(\ref{eq:B.1})
gives the following equation in the arcs:
\begin{eqnarray}
&& \frac{\partial f_{II}}{\partial s} =
-a\cdot \eta \cdot
\frac{\partial f_{II}}{\partial \sigma}
- b\cdot\sigma \cdot \frac{\partial f_{II}}{\partial \eta}
          + d\cdot\eta  +  d \; .
\label{eq:B.3} 
\end{eqnarray}
Also by
(\ref{eq:B.2}),
(\ref{eq:B.4}) one finds that 
$\vec n_{II}$ depends smoothly on the variables $\sigma,\eta$ so that
(\ref{eq:B.1}) reads at the snake as:
\begin{eqnarray}
 \frac{\partial \vec{n}_{II}}{\partial s} &=&
  \pi\cdot\delta_{L,per}\cdot\biggl(\vec{e}_1
                    \wedge \vec{n}_{II}\biggr) \; .
\label{eq:B.8}
\end{eqnarray}
Note that for $0<s<L$ $f_{II}$ reduces by 
(\ref{eq:B.4}),(\ref{eq:B.5})
to:
\begin{eqnarray*}
&& f_{II}(\sigma,\eta;s)  =
   d\cdot
    (s-L/2)
\nonumber\\ && \qquad
   + \frac{d\cdot b\cdot\sigma}{\lambda_0^2}\cdot
     \frac{1}{1+\cos(\lambda_0\cdot L)}\cdot \biggl\lbrack
\cos\biggl(\lambda_0\cdot (s-L) \biggr)
+ \cos(\lambda_0\cdot s)
-\cos(\lambda_0\cdot L) -1 \biggr\rbrack
\nonumber\\ && \qquad
   + \frac{d\cdot \eta}{\lambda_0}\cdot
     \frac{1}{1+\cos(\lambda_0\cdot L)}\cdot \biggl\lbrack
\sin\biggl(\lambda_0\cdot (s-L)\biggr)
+ \sin(\lambda_0\cdot s)
                           \biggr\rbrack \; .
\end{eqnarray*}
It is easily checked that this expression solves 
(\ref{eq:B.3}). One also observes by 
(\ref{eq:B.4}),(\ref{eq:B.5})
that:
\begin{eqnarray*}
\lim_{0<s\rightarrow 0} \lbrack   f_{II}(\sigma,\eta;s)\rbrack
 =
-\lim_{0>s\rightarrow 0} \lbrack
 f_{II}(\sigma,\eta;s)\rbrack  \; .
\end{eqnarray*}
From this follows by
(\ref{eq:A.3}),(\ref{eq:B.2}):
\begin{eqnarray*}
\lim_{0<s \rightarrow 0}\lbrack\vec{n}_{II}(\sigma,\eta;s) \rbrack &=&
 \underline{M}_{snake}\cdot 
\lim_{0>s \rightarrow 0}\lbrack\vec{n}_{II}(\sigma,\eta;s)\rbrack\; ,
\end{eqnarray*}
so that $\vec n_{II}$ obeys
(\ref{eq:B.8}) at the snake.
One also sees that $f_{II}(\sigma,\eta;s)$, given by 
(\ref{eq:B.4}), is 1-turn periodic in
$s$. 
\par Thus I have shown that $\vec{n}_{II}(\sigma,\eta;s)$, given by 
(\ref{eq:B.2}),
(\ref{eq:B.4}), is a
unit-vector solution of the radiationless Bloch 
equation 
(\ref{eq:B.1}), 1-turn periodic in
$s$ in the machine frame. 
It is thus the vector field $\vec n$ of \cite{DK72,HH96}.
\par Note that for $\sigma=\eta=0$ one gets:
\begin{eqnarray*}
  \vec{n}_{II}(\sigma=0,\eta=0;s) &=&
  \vec{n}_{0,II}(s) \; .
\end{eqnarray*}
One also observes that with this ansatz a 
singularity in $\vec{n}_{II}$ occurs if the
fractional part of the orbital tune $Q_s=
(\lambda_0\cdot L)/(2\pi)$ equals $1/2$, i.e. if one is at
a spin-orbit resonance.
\subsection*{B.3}
Denoting the $\vec{n}$-axis in the 
$(\vec{n}_{0,II},\vec{m}_{0,II},\vec{l}_{0,II})$-frame 
by $\hat{\vec{n}}_{II}$ one observes by 
(\ref{eq:3.3}),
(\ref{eq:3.244}),
(\ref{eq:B.2}),
(\ref{eq:B.4}):
\begin{eqnarray}
&& \hat{\vec{n}}_{II}(\sigma,\eta;s) =
    \left( \begin{array}{c}
\vec{n}_{0,II}^T(s)\cdot\vec{n}_{II}(\sigma,\eta;s)
            \\ \\
\vec{m}_{0,II}^T(s)\cdot\vec{n}_{II}(\sigma,\eta;s)
            \\ \\
\vec{l}_{0,II}^{\;T}(s)\cdot\vec{n}_{II}(\sigma,\eta;s)
                \end{array}
         \right)
  = \left( \begin{array}{c}
\cos\biggl(f_{II}(\sigma,\eta;s)-g_6(s)\biggr) \\
\theta_{2L,per}(s)\cdot
\sin\biggl(f_{II}(\sigma,\eta;s)-g_6(s)\biggr) \\
        0
                \end{array}
         \right)
\nonumber\\
&=& \left( \begin{array}{c}
\cos\biggl(g_{19}(s)\cdot\sigma
     + g_{20}(s)\cdot \eta \biggr) \\
\theta_{2L,per}(s)\cdot
\sin\biggl(g_{19}(s)\cdot\sigma
     + g_{20}(s)\cdot \eta \biggr) \\
                      0
                \end{array}
         \right)        \; .
\label{eq:B.6}
\end{eqnarray}
The corresponding formula obtained from the SLIM formalism,
which linearizes spin motion and is therefore only applicable for small
angles between the $\vec n$-axis and the $\vec n_0$-axis, is 
\cite{BHR92,BHR94b}:
\begin{eqnarray}
&& \hat{\vec{n}}_{II,SLIM}(\sigma,\eta;s) =
 \left( \begin{array}{c}
    1 \\
    \theta_{2L,per}(s)\cdot\biggl( g_{19}(s)\cdot\sigma
     + g_{20}(s)\cdot \eta \biggr) \\
                      0
                \end{array}
         \right)        \; .
\label{eq:B.7}
\end{eqnarray}
From the above it is clear that $\hat{\vec{n}}_{II}$ obeys the
radiationless Bloch equation (\ref{eq:3.11}).
\setcounter{section}{3}
\setcounter{equation}{0}
\section*{Appendix C}
\addcontentsline{toc}{section}{Appendix C}
%
%
%
In this Appendix I briefly reconsider Machine I by extending it to 
`Machine III'
which is obtained by including the nonhorizontal component of the spin 
vector, i.e. by considering the 
full three-dimensional spin motion. 
For horizontal spin the dynamics of Machine III is the same as for Machine I.
\par As for Machine I I denote the spin vector in the
$(\vec{m}_{0,I},\vec{l}_{0,I},\vec{n}_{0,I})$-frame by $\vec S$.
To cover the full three-dimensional spin motion
one can employ, as mentioned in section 2.3.6,
spherical coordinates, so that:
\begin{eqnarray}
 \vec{S}  &\equiv& \frac{\hbar}{2}\cdot
  \left( \begin{array}{c}
                      \cos(\psi)\cdot \sin(\theta) \\
                      \sin(\psi)\cdot \sin(\theta) \\
                      \cos(\theta)
                \end{array}
         \right)    \; ,
\label{eq:C.1}
\end{eqnarray}
where $\psi,\theta$ denote the azimuthal and polar angles. The horizontal
spin vector (\ref{eq:1.4}) 
can be obtained from
(\ref{eq:C.1}) 
by setting $\theta=\pi/2$.
Thus for $\theta=\pi/2$ Machine III effectively reduces to Machine I.
\par In dealing with 4 variables, the processes to be studied for Machine III
are denoted by $\vec y(s)$, where:
\begin{eqnarray}
&&\vec{y}(s)\equiv
       \left( \begin{array}{c}
                \sigma(s)         \\
           \eta(s)              \\
                \psi(s)  \\
                \theta(s)               
 \end{array}
         \right)  \; ,
\label{eq:C.3}
\end{eqnarray}
i.e. compared with $\vec x(s)$ they have an additional component.
The Langevin equation for Machine III is then defined by:
\begin{eqnarray}
 d\vec{y}(s) &=& \underline{\check{\cal A}}\cdot\vec{y}(s)\cdot ds
               +\underline{\check{\cal B}}\cdot d\check{\vec{\cal W}}(s) \; ,
 \label{eq:C.4}
\end{eqnarray}
where
\begin{eqnarray*}
\underline{\check{\cal A}} &\equiv&        \left( \begin{array}{cccc}
      0 & a & 0 & 0                            \\
      b & c & 0 & 0                           \\
     0 & d & 0 & 0                           \\
      0 & 0 & 0 & 0 
                \end{array}
         \right) \; , \qquad
    \underline{\check{\cal B}}  \equiv   \sqrt{\omega}\cdot
                              \left( \begin{array}{cccc}
   0 & 0  & 0  & 0                     \\
   0 & 1  & 0  & 0                      \\
   0 & 0  & 0  & 0                     \\
   0 & 0  & 0  & 0
                \end{array}
         \right) \; ,  
\end{eqnarray*}
with
\begin{eqnarray*}
   d\check{\vec{\cal W}}(s) &\equiv &
       \left( \begin{array}{c}
             d\check{\cal W}_1(s)           \\
             d\check{\cal W}_2(s)             \\
             d\check{\cal W}_3(s)           \\
             d\check{\cal W}_4(s)
                \end{array}
         \right)  \; .
\end{eqnarray*}
Here the $\check{\cal W}_k(s)$ are Wiener processes.
One sees by the behaviour of the $\theta$-variable 
in the Langevin equation (\ref{eq:C.4}) that Machine III indeed describes
the same dynamics as Machine I.
Therefore the Fokker-Planck
equation corresponding to the Langevin equation 
(\ref{eq:C.4}) is identical with the Fokker-Planck equation
(\ref{eq:2.8}) for Machine I.
\par For Machine III I adopt standard
boundary conditions in all four variables $\sigma,\eta,\psi,\theta$
so that:
$\check{w}\rightarrow 0$ for $\sigma,\eta,\psi,\theta\rightarrow\pm\infty$, 
where $\check{w}$ denotes the probability density.
Thus the probability density is normalized by:
\begin{eqnarray}
   1 &=& \int_{-\infty}^{+\infty} \; d\sigma
         \int_{-\infty}^{+\infty} \; d\eta
         \int_{-\infty}^{+\infty} \; d\psi
        \int_{-\infty}^{+\infty} \; d\theta\cdot
 \check{w}(\sigma,\eta,\psi,\theta;s) \; .
 \label{eq:C.6}
\end{eqnarray}
The polarization vector is defined by:
\begin{eqnarray}
\vec{P}_{tot}^{\check{w}}(s) &\equiv&
  \int_{-\infty}^{+\infty} \; d\sigma
                             \int_{-\infty}^{+\infty} \; d\eta
                             \int_{-\infty}^{+\infty} \; d\psi\cdot
                            \int_{-\infty}^{+\infty} \; d\theta\cdot
   \check{w}(\sigma,\eta,\psi,\theta;s)
\cdot  
  \left( \begin{array}{c}
                      \cos(\psi)\cdot \sin(\theta) \\
                      \sin(\psi)\cdot \sin(\theta) \\
                      \cos(\theta)
                \end{array}
         \right)    \; .\qquad
 \label{eq:C.7}
\end{eqnarray}
The polarization density is defined by:
\begin{eqnarray}
 \vec{P}^{\check{w}}(\sigma,\eta;s) &\equiv&
                      \int_{-\infty}^{+\infty} \; d\psi \cdot
                     \int_{-\infty}^{+\infty} \; d\theta \cdot
  \check{w}(\sigma,\eta,\psi,\theta;s) \cdot
  \left( \begin{array}{c}
                      \cos(\psi)\cdot \sin(\theta) \\
                      \sin(\psi)\cdot \sin(\theta) \\
                      \cos(\theta)
                \end{array}
         \right)    \; .
 \label{eq:C.8}
\end{eqnarray}
Using the Fokker-Planck equation
(\ref{eq:2.8}) one observes by
(\ref{eq:C.8}) that
the polarization density fulfills the same Bloch equation
(\ref{eq:2.64}) as for Machine I.
\par For a process with only horizontal spin the probability density
has the form:
\begin{eqnarray}
 \check{w}(\sigma,\eta,\psi,\theta;s) &=&
  w(\sigma,\eta,\psi;s)\cdot\delta(\theta-\pi/2+2\pi N) \; ,
 \label{eq:C.9}
\end{eqnarray}
where $N$ denotes an integer and where $w$ 
denotes the probability density 
which arises if the process would be described in 
the framework of Machine I. Note that $\check{w}$, as given by
(\ref{eq:C.9}),
fulfills the Fokker-Planck equation
(\ref{eq:2.8}).
\setcounter{section}{4}
\setcounter{equation}{0}
\section*{Appendix D}
\addcontentsline{toc}{section}{Appendix D}
\subsection*{D.1}
In section 2.9 I observed that the equilibrium behaviour of
Machine I in the presence of radiation
resembles a
Hamiltonian flow.
In this appendix I now briefly reconsider Machine I 
by switching off the radiation effects to obtain a real
Hamiltonian flow.
I call this model `Machine IV'.
The aim is to investigate the existence of an equilibrium spin
distribution and to do that I adopt the usual approach of describing
the system in terms of action-angle variables.
Since I use a canonical formalism I need an even
number of variables. Hence I supplement
$\sigma,\eta,\psi$ by a fourth variable $J$, canonically conjugate
to $\psi$, which is defined by:
\begin{eqnarray}
 J  &\equiv& \frac{\hbar}{2}\cdot\cos(\theta) \; ,
\label{eq:D.22}
\end{eqnarray}
where $\theta$ denotes the polar angle in the spherical coordinate
expression (\ref{eq:2.240}) of the spin vector $\vec S$.
\footnote{Note that for Machine III (see Appendix C)
the variable $\theta$ is used instead of $J$.}
Thus for Machine IV the spin vector in the 
$(\vec{m}_{0,I},\vec{l}_{0,I},\vec{n}_{0,I})$-frame is parametrized as:
\begin{eqnarray}
 \vec{S}  &\equiv&
  \left( \begin{array}{c}
     \sqrt{\hbar^2/4-J^2}\cdot\cos(\psi) \\
     \sqrt{\hbar^2/4-J^2}\cdot\sin(\psi) \\
                      J
                \end{array}
         \right)    \; .
\label{eq:D.1}
\end{eqnarray}
Hence, as for Machine III, I consider the 
full three-dimensional spin motion, i.e. the
nonhorizontal component of the spin vector is included.
One sees that the additional variable describes the projection of the
spin onto the vertical direction so that for processes running with
Machine I one has: $J=0$. But owing to the field geometry in Machine I
one could have carried through the analysis with nonzero $J$ if I had
only been interested in the $\psi$ distribution.
The Poisson brackets are defined by \cite{DK73,Yok86,BHR94a,BHR94b}
\footnote{Note that because of 
(\ref{eq:D.22}),(\ref{eq:D.2})
I use units in Appendix D where $\hbar$ has the dimension of length
\cite{BHR94a,BHR94b}.}:
\begin{eqnarray}
1 &=& \lbrace \sigma, \eta \rbrace =
\lbrace \psi, J \rbrace \; ,
\nonumber\\
0 &=& \lbrace \sigma, \psi \rbrace =
      \lbrace \sigma, J  \rbrace =
      \lbrace \eta,\psi  \rbrace =
      \lbrace \eta, J    \rbrace \; ,
\label{eq:D.2}
\end{eqnarray}
and the Hamiltonian reads as:
\begin{eqnarray}
H_a &\equiv& - \frac{b}{2}\cdot\sigma^2
               + \frac{a}{2}\cdot\eta^2  + d\cdot\eta\cdot J
   \equiv  H_{orb} + H_{spin} \; ,
\label{eq:D.3}
\end{eqnarray}
where $H_{orb}$ is defined in section 2.5 and:
\begin{eqnarray*}
    H_{spin} &\equiv&   d\cdot\eta\cdot J \; .
\end{eqnarray*}
Due to the absence of radiation the Langevin equation for Machine IV
reduces to the following canonical equations of motion:
\begin{eqnarray}
&& \sigma' = a\cdot \eta + d\cdot J \; , 
\qquad \eta' = b\cdot\sigma \; , \qquad \psi' = d\cdot\eta \; , \qquad
J'=0 \; .
\label{eq:D.23}
\end{eqnarray}
One sees that
$H_{spin}$ contributes a  (very small) Stern-Gerlach term
\cite{BHR94a,BHR94b} appearing in the first identity of
(\ref{eq:D.23})
which was neglected for Machine I because
this Stern-Gerlach effect vanishes at $J=0$.
\par Thus Machine III, unlike Machine IV, is lacking the
Stern-Gerlach force and Machine IV, unlike Machine III, is lacking
radiation effects. Both machines involve the 
full three-dimensional spin motion but
they parametrize it in different ways.
\subsection*{D.2}
Coming to the probability density, I now adopt
boundary conditions for the variables $\psi,J$ which are natural for their
role as spherical coordinates. In particular I adopt for $\psi$ the periodic
boundary conditions mentioned in section 2.3.6.
The probability density in the present case also depends on $J$ and
I denote it by $w_{mod,a}$.
The normalization condition 
(\ref{eq:2.206})
for $w_{per}$ translates by using
(\ref{eq:D.22}) into:
\begin{eqnarray}
   1 &=& \int_{-\infty}^{+\infty} \; d\sigma
         \int_{-\infty}^{+\infty} \; d\eta
         \int_{0}^{2\pi} \; d\psi
         \int_{-\hbar/2}^{+\hbar/2} \; d J \cdot
 w_{mod,a}(\sigma,\eta,\psi,J;s) \; .
 \label{eq:D.6}
\end{eqnarray}
For horizontal spin $w_{mod,a}$ has the form:
\begin{eqnarray}
  w_{mod,a}(\sigma,\eta,\psi,J;s) &\equiv&
  w_{per}(\sigma,\eta,\psi;s)\cdot\delta(J) \; .
\label{eq:D.7}
\end{eqnarray}
Due to the absence of radiation
the Fokker-Planck equation for Machine IV is 
the Liouville equation for the phase space evolution associated with the
Hamiltonian 
(\ref{eq:D.3}):
\begin{eqnarray}
&& 0 =
\biggl( \frac{d}{ds}\biggr)_{tot} \,
       w_{mod,a}  =
 \frac{\partial w_{mod,a}}{\partial s} +
      \lbrace w_{mod,a}\; , \; H_a \rbrace  \; .
\label{eq:D.8}
\end{eqnarray}
The total derivative is zero for a Hamiltonian flow \cite{Gol80}.
\subsection*{D.3}
To investigate the matter of equilibrium with the help of the
evolution equation 
(\ref{eq:D.8})
I transform to action-angle variables.
   To come to these I first replace
$\sigma,\eta$ by
the orbital  variables $\phi,J_{orb}$ defined in section
2.5. Thus I treat my system using   the variables
$\phi,J_{orb},\psi,J$ with the Poisson brackets:
\begin{eqnarray}
1 &=& \lbrace \phi, J_{orb} \rbrace =
\lbrace \psi, J \rbrace \; ,
\nonumber\\
0 &=& \lbrace \phi, \psi \rbrace =
      \lbrace \phi, J  \rbrace =
      \lbrace J_{orb},\psi  \rbrace =
      \lbrace J_{orb}, J    \rbrace \; .
\label{eq:D.9}
\end{eqnarray}
The Hamiltonian 
(\ref{eq:D.3})
transforms into
\begin{eqnarray}
H_b &\equiv& -{\sqrt{-ab}}\cdot J_{orb}
 -d\cdot(-\frac{b}{a})^{1/4}\cdot\sqrt{2J_{orb}}\cdot \sin(\phi)
                                                 \cdot J\; .
\label{eq:D.10}
\end{eqnarray}
This Hamiltonian is not yet in action-angle form since it still
contains the phase $\phi$. The probability density $w_{mod,b}$ for the
variables $\phi,J_{orb},\psi,J$ reads as:
\begin{eqnarray}
  w_{mod,b}(\phi,J_{orb},\psi,J;s)  &\equiv&
  w_{mod,a}(\sigma,\eta,\psi,J;s) \; .
\label{eq:D.11}
\end{eqnarray}
Note that $w_{mod,b}$ is periodic in $\phi,\psi$ with period $2\pi$.
For the final step in obtaining
                   the action-angle variables I use the spin-orbit
action-angle formalism  of
\cite{DK73,Yok86,BHR}.
This  allows one to  perform the following  canonical
transformation
\footnote{See equations 4.30-32 in \cite{Yok86}.}
   \\          $\phi,J_{orb},\psi,J\rightarrow
 \phi_{new},J_{orb,new},\psi_{new},J_{new}$:
\setcounter{INDEX}{1}
\begin{eqnarray}
 \phi_{new} &\equiv& \phi + d\cdot
 (-a^3\cdot b)^{-1/4}\cdot(2J_{orb})^{-1/2} \cdot \cos(\phi)\cdot J\;,
\label{eq:D.13a}
         \\
\addtocounter{equation}{-1}
\addtocounter{INDEX}{1}
 J_{orb,new}  &\equiv&   J_{orb} + d\cdot
 (-a^3\cdot b)^{-1/4}\cdot\sqrt{2J_{orb}}\cdot \sin(\phi)\cdot J\;,
\label{eq:D.13b}
         \\
\addtocounter{equation}{-1}
\addtocounter{INDEX}{1}
 \psi_{new} &\equiv& \psi + d\cdot
        (-a^3\cdot b)^{-1/4}\cdot\sqrt{2J_{orb}}\cdot \cos(\phi)\;,
\label{eq:D.13c}
     \\
\addtocounter{equation}{-1}
\addtocounter{INDEX}{1}
 J_{new}  &\equiv&   J  \; ,
\label{eq:D.13d}
\end{eqnarray}
\setcounter{INDEX}{0}%
whereby the terms containing $J$ in 
(\ref{eq:D.13a}),
(\ref{eq:D.13b})
are due to
Stern-Gerlach effects and are very small so that in effect the new
orbital variables are numerically very close to the original
orbital variables.
The new variables have the following Poisson brackets
\footnote{The above action-angle formalism neglects
higher orders in $\hbar$ in a specific way which is made use of in
(\ref{eq:D.14}).}:
\begin{eqnarray}
1 &=& \lbrace \phi_{new}, J_{orb,new} \rbrace =
\lbrace \psi_{new}, J_{new} \rbrace \; ,
\nonumber\\
0 &=& \lbrace \phi_{new}, \psi_{new} \rbrace =
      \lbrace \phi_{new}, J_{new}  \rbrace =
      \lbrace J_{orb,new},\psi_{new}  \rbrace =
      \lbrace J_{orb,new}, J_{new}   \rbrace \; .
\label{eq:D.14}
\end{eqnarray}
The Hamiltonian $H_b$ transforms into:
\begin{eqnarray}
H_c &\equiv& -{\sqrt{-ab}}\cdot J_{orb,new} \; ,
\label{eq:D.15}
\end{eqnarray}
and since it now only contains  an action I finally have the desired
form. Note that this Hamiltonian does not contain the spin action
$J_{new}$. Note also that $\psi_{new}$ is identical to
$\tilde{\psi}$ in section 2.2.   Thus one sees that this canonical
transformation which has removed the spin dependence from the
Hamiltonian is equivalent to the reduction from a three-dimensional
problem to a two-dimensional problem observed by 
(\ref{eq:2.6b}).
This reduction reflects the nonuniqueness
of the equilibrium state already observed in section 2 in the presence
of radiation effects.
Denoting the probability density in these variables by $w_{mod,c}$
one gets:
\begin{eqnarray}
  w_{mod,c}(\phi_{new},J_{orb,new},\psi_{new},J_{new};s)&=&
  w_{mod,b}(\phi,J_{orb},\psi,J;s) \; .
\label{eq:D.16}
\end{eqnarray}
Note that $w_{mod,c}$ is periodic in $\phi_{new},\psi_{new}$ with period
$2\pi$. 
The corresponding Liouville equation reads as:
\begin{eqnarray}
 \frac{\partial w_{mod,c}}{\partial s} &=&
      \lbrace H_c, w_{mod,c} \rbrace  \; .
\label{eq:D.18}
\end{eqnarray}
\subsection*{D.4}
Having obtained action-angle variables one now can discuss
equilibrium.
I define `equilibrium' to mean that
$\partial w_{mod,c}/\partial s$ is zero. Then by 
(\ref{eq:D.18})
the
Poisson bracket $\lbrace H_c, w_{mod,c} \rbrace$ vanishes. Since
$H_c$ is independent of $\phi_{new}$
the probability density $w_{mod,c}$ must be
independent of $\phi_{new}$. However, since 
(\ref{eq:D.15})
does not contain
$J_{new}$ the probability density $w_{mod,c}$ can still depend on
$\psi_{new}$. The interpretation of this is that since the $\psi_{new}$
for each particle is constant (see 
(\ref{eq:2.6b})), the $\psi_{new}$
distribution does not change as $s$ increases and is therefore in
equilibrium.
So although I have a complete transformation to action-angle
variables
   the special form for the Hamiltonian
(\ref{eq:D.15})
means that by insisting on equilibrium one cannot say very
much about the $\psi_{new}$ distribution except that $w_{mod,c}$
has the form
\begin{eqnarray}
 w_{mod,c}(\phi_{new},J_{orb,new},\psi_{new},J_{new};s)&=&
   w_{mod,c}(J_{orb,new},\psi_{new},J_{new}) \; ,
\label{eq:D.19}
\end{eqnarray}
so that $w_{mod,c}$  neither depends on $\phi_{new}$ nor on $s$.
The dependence of $w_{mod,c}$ on $\psi_{new}$
reflects the nonuniqueness of the equilibrium state
already observed in section 2 in the presence
of radiation effects.
If on the contrary the Hamiltonian had contained $J_{new}$ the
$\psi_{new}$ distribution would have had to be uniform.
In the case of horizontal spin one has $J=0$ so that
by 
(\ref{eq:D.13d})
one has: $J_{new}=0$. Then
(\ref{eq:D.19})
simplifies to:
\begin{eqnarray}
 w_{mod,c}(J_{orb,new},\psi_{new},J_{new})&\equiv&
   w_{rest}(J_{orb,new},\psi_{new})
                       \cdot\delta(J_{new}) \; ,
\label{eq:D.20}
\end{eqnarray}
where $w_{rest}$  neither depends  on $\phi_{new},J_{new}$ nor on $s$.
\setcounter{section}{5}
\setcounter{equation}{0}
\section*{Appendix E}
\addcontentsline{toc}{section}{Appendix E}
\subsection*{E.1}
The $s$-dependence of
the first moment of the variables $\sigma,\eta$ allows one to define 
a damping time for every process. 
It turns out that this `orbital damping
time' is independent of the process. It is the same for machines I and II
since they have the same orbital equations of motion.
\par Also it is shown how the orbital damping time is involved in the
orbital correlation matrix (defined below).
\subsection*{E.2}
For a given process $\vec x(s)=(\sigma(s),\eta(s),\psi(s))^T$ 
the stochastic averages of the orbital variables are given via 
(\ref{eq:2.225}),
(\ref{eq:2.231})
by:
\begin{eqnarray}
&& <\vec z(s)> 
   =   \exp\biggl(\underline{\cal A}_{orb}\cdot(s-s_1)\biggr)\cdot
<\vec z(s_1)> \nonumber\\
&& = \frac{i}{2\cdot\lambda}
   \cdot                      \left( \begin{array}{cc}
    g_1(s-s_1)  &\qquad  -a\cdot g_2(s-s_1)  \\
-b\cdot  g_2(s-s_1)  &\qquad  - g_3(s-s_1)
                \end{array}
         \right)\cdot<\vec z(s_1)> \; .
 \label{eq:E.10}
\end{eqnarray}
One observes by 
(\ref{eq:2.232}),
(\ref{eq:E.10})
that $<\vec z(s)>$ contains the exponentially decreasing factor 
$\exp(c\cdot s/2)$ and that the remaining factors are periodic in $s$
with period $2\cdot\pi/\lambda$.
I therefore 
define the orbital damping time, denoted as $\tau_{damp}$, by:
\begin{eqnarray}
\tau_{damp} &\equiv& -\frac{2}{c} = \frac{L}{\alpha_s} \; .
 \label{eq:E.8}
\end{eqnarray}
One sees that $\tau_{damp}$ is independent of the process. 
\subsection*{E.3}
The `orbital correlation matrix'
$\underline{k}_{orb}$ is defined by \cite{Gar85}:
\begin{eqnarray}
    \underline{k}_{orb}(s;s_1)&\equiv &
                              \left( \begin{array}{cc}
  k_{orb,11}(s;s_1)   &  k_{orb,12}(s;s_1)  \\
  k_{orb,21}(s;s_1)   &  k_{orb,22}(s;s_1)
                \end{array}
         \right)        \; ,
\label{eq:E.1}
\end{eqnarray}
where
\begin{eqnarray}
  k_{orb,11}(s;s_1)  &\equiv&
 <\sigma(s)\cdot\sigma(s_1)> -
 <\sigma(s)>\cdot<\sigma(s_1)> \; ,
\nonumber\\
  k_{orb,12}(s;s_1)  &\equiv&
 <\sigma(s)\cdot\eta(s_1)> -
 <\sigma(s)>\cdot<\eta(s_1)> \; ,
\nonumber\\
  k_{orb,21}(s;s_1)  &\equiv&
 <\eta(s)\cdot\sigma(s_1)> -
 <\eta(s)>\cdot<\sigma(s_1)> \; ,
\nonumber\\
  k_{orb,22}(s;s_1)  &\equiv&
 <\eta(s)\cdot\eta(s_1)> -
 <\eta(s)>\cdot<\eta(s_1)> \; . \nonumber\\&&
\label{eq:E.2}
\end{eqnarray}
Using the orbital joint probability density 
$w_{orb,joint}$ given by (\ref{eq:2.236}) one finds for $s_1\leq s$:
\begin{eqnarray}
 && <\sigma(s)\cdot\sigma(s_1)> = 
\int_{-\infty}^{+\infty} d\sigma
\int_{-\infty}^{+\infty} d\sigma_1
\int_{-\infty}^{+\infty} d\eta
\int_{-\infty}^{+\infty} d\eta_1\cdot \sigma\cdot\sigma_1\cdot
w_{orb,joint}(\sigma,\eta;s;\sigma_1,\eta_1;s_1) \; ,
\nonumber\\
&& <\sigma(s)\cdot\eta(s_1)> = 
\int_{-\infty}^{+\infty} d\sigma
\int_{-\infty}^{+\infty} d\sigma_1
\int_{-\infty}^{+\infty} d\eta
\int_{-\infty}^{+\infty} d\eta_1\cdot \sigma\cdot\eta_1\cdot
w_{orb,joint}(\sigma,\eta;s;\sigma_1,\eta_1;s_1) \; ,
\nonumber\\
&& <\eta(s)\cdot\sigma(s_1)>  =
\int_{-\infty}^{+\infty} d\sigma
\int_{-\infty}^{+\infty} d\sigma_1
\int_{-\infty}^{+\infty} d\eta
\int_{-\infty}^{+\infty} d\eta_1\cdot \eta\cdot\sigma_1\cdot
w_{orb,joint}(\sigma,\eta;s;\sigma_1,\eta_1;s_1) \; ,
\nonumber\\
&& <\eta(s)\cdot\eta(s_1)> =
\int_{-\infty}^{+\infty} d\sigma
\int_{-\infty}^{+\infty} d\sigma_1
\int_{-\infty}^{+\infty} d\eta
\int_{-\infty}^{+\infty} d\eta_1\cdot \eta\cdot\eta_1\cdot
w_{orb,joint}(\sigma,\eta;s;\eta_1,\eta_1;s_1) \; . \nonumber\\&&
\label{eq:E.9}
\end{eqnarray}
Because of 
(\ref{eq:2.225}),
(\ref{eq:2.220}),
(\ref{eq:2.236}),
(\ref{eq:E.2}),
(\ref{eq:E.9})
one has for $s_1\leq s$:
\begin{eqnarray}
\frac{\partial\underline{k}_{orb}(s;s_1)}{\partial s} &=&
\underline{\cal A}_{orb}\cdot\underline{k}_{orb}(s;s_1) \; .
 \label{eq:E.4}
\end{eqnarray}
Also by
(\ref{eq:2.17}),
(\ref{eq:E.2}) one has:
\begin{eqnarray}
    \underline{k}_{orb}(s_1;s_1)&=&
  \underline{\sigma}_{orb}(s_1) \; ,
\label{eq:E.3}
\end{eqnarray}
where $\underline{\sigma}_{orb}$ denotes the orbital covariance matrix.
From 
(\ref{eq:E.4}),
(\ref{eq:E.3}) it follows for $s_1\leq s$ that:
\begin{eqnarray}
&& \underline{k}_{orb}(s;s_1) =
\exp\biggl(\underline{\cal A}_{orb}\cdot (s-s_1)\biggr)\cdot
  \underline{\sigma}_{orb}(s_1) \; ,
 \label{eq:E.5}
\end{eqnarray}
and from (\ref{eq:2.231}),(\ref{eq:E.5}) I have for $s_1\leq s$:
\begin{eqnarray}
&& \underline{k}_{orb}(s;s_1) =
 \frac{i}{2\cdot\lambda}
   \cdot                      \left( \begin{array}{cc}
    g_1(s-s_1)  &\qquad  -a\cdot g_2(s-s_1)  \\
-b\cdot  g_2(s-s_1)  &\qquad  - g_3(s-s_1)
                \end{array}
         \right)\cdot\underline{\sigma}_{orb}(s_1) \; .
 \label{eq:E.6}
\end{eqnarray}
Then by
(\ref{eq:2.232}),(\ref{eq:E.6}) one finds for $s_1\leq s$ that
the matrix elements of the orbital correlation
matrix contain the exponentially decreasing factor $\exp(c\cdot s/2)$.
The remaining factors are periodic in $s$
with period $2\cdot\pi/\lambda$.
Therefore $\tau_{damp}$ is not only the orbital damping time, but also 
plays the role of an `orbital correlation time'.
\section*{Guide for the reader}
\addcontentsline{toc}{section}{Guide for the reader}
Please note the following conventions used in this paper: 
\begin{itemize}
\item The modulus of a real or complex number $v$ is denoted by $|v|$.
The real part of a complex number $v$ is denoted by $\Re e\lbrace v\rbrace$.
\item The transpose of a matrix is denoted by ${}^T$.
\item The symbol $\cdot$ denotes either
matrix multiplication or scalar multiplication of matrices
(this includes the multiplication of scalars).
\item Objects $\vec v$ which are denoted with an arrow (e.g. $\vec x,\vec z$)
are {\it column} vectors, i.e.
$n\times 1$ matrices. Thus $\vec v=(v_1,...,v_n)^T$, where $v_1,...,v_n$
are the components of $\vec v$. The norm $||\vec v||$ of a vector $\vec v$ is 
defined by $||\vec v||\equiv\sqrt{v_1^2+...+v_n^2}$.
\item The vector product is denoted by $\wedge$.
\item A necessary ingredient of a Gaussian probability density, is that
the resulting covariance matrix is nonsingular.
\item The starting azimuth of a process is denoted by $s_0$. 
Thus the domain of the azimuthal variable is given by 
$\lbrack s_0\, ,+\infty)$. For processes 1,2,3,4 and 5 I have chosen $s_0=0$,
i.e. the domain is given by the nonnegative real numbers.
\end{itemize}
The following table helps to find some
of the main results on processes 1,2,3,4 and 5:
\begin{center}
\begin{tabular}{||l|c|c|c|c|c||}\hline\hline
name of the process & Process 1 & Process 2 & Process 3 & Process 4
                                                        & Process 5
                                                      \\\hline\hline
Langevin equation        & (\ref{eq:2.1})
  & (\ref{eq:2.1}) & 
 (\ref{eq:3.5}) & 
 (\ref{eq:3.5}) & 
  (\ref{eq:3.5})  
\\\hline
Fokker-Planck equation   
& (\ref{eq:2.8}) & (\ref{eq:2.8}) 
& (\ref{eq:3.7}) & (\ref{eq:3.7}) 
& (\ref{eq:3.7})
                                                          \\\hline
probability density 
&  (\ref{eq:2.23})
& (\ref{eq:2.38}), 
&  (\ref{eq:3.17}),
& (\ref{eq:3.73}), 
& (\ref{eq:3.101}), \\
& 
& (\ref{eq:2.44}) 
&  (\ref{eq:3.18}) 
& (\ref{eq:3.245}) 
& (\ref{eq:3.246}) 
                                                          \\\hline
characteristic function
& (\ref{eq:2.269}) & (\ref{eq:2.228}) 
& (\ref{eq:3.236}) & (\ref{eq:3.240}) 
& (\ref{eq:3.242})              
                                                             \\\hline
Bloch equation for       &        &       &        &   &   \\
the polarization density & 
(\ref{eq:2.64})  & (\ref{eq:2.64}) & 
(\ref{eq:3.10})  & (\ref{eq:3.10}) & 
(\ref{eq:3.10})  
\\\hline
polarization density      & 
(\ref{eq:2.68})  &
(\ref{eq:2.78})  & 
(\ref{eq:3.68})  &  
(\ref{eq:3.247})  &  
(\ref{eq:3.248})  
\\\hline
polarization vector      & 
(\ref{eq:2.30})  &
(\ref{eq:2.57})  & 
(\ref{eq:3.20})  &  
(\ref{eq:3.79})  &  
(\ref{eq:3.107})  
\\\hline
complete decoherence     &        &       &        &  &  \\
of spin                  & no & no    & yes    & yes  & yes
                                                           \\\hline
                                                             \hline
\end{tabular}
\end{center}
The following table helps to find some of the main abbreviations:
\begin{center}
\begin{tabular}{||l|c||}\hline\hline
$g_1(s),g_2(s),g_3(s),g_4(s),g_5(s)$          & section 2.4            
       \\\hline
$\lambda$                    & section 2.4                   \\\hline
$\lambda_0,\sigma_{\sigma},\sigma_{\eta},\sigma_{\psi}$
                             & section 2.5                   \\\hline
${\cal G}(s),\delta_{L,per}(s),\theta_{2L,per}(s),\hat{d}(s)$
                             & section 3.1                \\\hline
$\vec{n}_{0,II}(s),\vec{m}_{0,II}(s),\vec{l}_{0,II}(s)$
& section 3.1                \\\hline
$g_6(s)$                     & section 3.1                   \\\hline
$g_{7}(s),g_{8}(s),
  g_{9}(s),g_{10}(s),g_{11},g_{12}(s)$  & section 3.5  \\
 $g_{13}(s),g_{14}(s),
   g_{15},g_{16}(s),g_{17},g_{18}$  &  \\\hline
$g_{21}(s),g_{22}(s),g_{23}(s)$                  & section 3.7       \\\hline
$g_{19}(s),g_{20}(s)$        & Appendix B                   \\\hline
\hline
\end{tabular}
\end{center}
\section*{Acknowledgements}
\addcontentsline{toc}{section}{Acknowledgements}
I wish to thank Desmond P.\ Barber for stimulating discussions and valuable
remarks on the manuscript and G. Ripken
for useful discussions. 
I wish to thank Prof. A.W. Chao for pointing out a simple way to arrive
at the result for $\sigma_{\psi}$
of Process 1 given in section 2.5.

\begin{thebibliography}{42}
\addcontentsline{toc}{section}{References}
%
\bibitem[Abr61]{Abr61}
A. Abragam, ``The principles of nuclear magnetism'', Oxford (1961).
%
\bibitem[ACDO91]{ACDO91}
M. Antoine, A. Comtet, J. Desbois, S. Ouvry, J. Phys. A: Math. Gen. {\bf 24}, 
p.2581 (1991).
%
\bibitem[Arn73]{Arn73}
L. Arnold, ``Stochastische Differentialgleichungen'', M\"unchen (1973).
%
\bibitem[BKS70]{BKS70}
V.N.Baier, V.M.Katkov, V.M.Strakhovenko, Phys. Lett. {\bf 31A}, p.198 (1970).
%
\bibitem[BKRRS85]{BKRRS85}
D.P. Barber, J. Kewisch, G. Ripken, R. Rossmanith, R. Schmidt, Part. Accel. 
{\bf 17}, p.243 (1985).
%
%
\bibitem[BHMR91]{BHMR91}
D.P. Barber, K. Heinemann, H. Mais, G. Ripken, DESY {\bf 91-146} (1991).
%
\bibitem[BHR92]{BHR92}
D.P. Barber, K. Heinemann, G. Ripken, DESY {\bf M-92-04} (1992).
%
\bibitem[BBHMR94a]{BBHMR94a}
D.P. Barber, M. B\"oge, K. Heinemann, H. Mais, G. Ripken,
Proc. 11th Int. Symp. High Energy Spin Physics, Bloomington,
Indiana (1994).
%
\bibitem[BBHMR94b]{BBHMR94b}
D.P. Barber, M. B\"oge, K. Heinemann, H. Mais, G. Ripken, DESY {\bf M-94-13}
(1994). \\
I  use this opportunity to mention some typing errors on page 7 of
[BBHMR94b]. \\
Line 10: replace $p_{\sigma}$ by $\sigma_{p_{\sigma}}$.
Line 11: replace $\sigma_{p_{\sigma}}$ by $\sigma_{\psi}$.
Line 31: replace $\sigma_{p_{\sigma}}$ by $\sigma_{\psi}$. \\
Corresponding corrections should be made in \cite{BBHMR94a}.
%
\bibitem[BHR94a]{BHR94a}
D.P. Barber, K. Heinemann, G. Ripken, Z. Phys. {\bf C64}, p.117 (1994).
%
\bibitem[BHR94b]{BHR94b}
D.P. Barber, K. Heinemann, G. Ripken, Z. Phys. {\bf C64}, p.143 (1994).
%
\bibitem[Bar96]{Bar96}
D.P. Barber, Proc. 12th Int. Symp. High Energy Spin Physics, Amsterdam,
The Netherlands (1996).
%
\bibitem[Bar97]{Bar97}
D.P. Barber, private communication.
%
\bibitem[BHR]{BHR}
D.P. Barber, K. Heinemann, G. Ripken: A paper on the construction of
combined spin-orbit action-angle variables in preparation.
%
\bibitem[BMT59]{BMT59}
V. Bargmann, L. Michel, V.L. Telegdi,
Phys. Rev. Lett. {\bf 2}, p.435 (1959).
%
\bibitem[Boe94]{Boe94}
M. Boege, Ph.D. thesis, DESY {\bf 94-87} (1994).
%
\bibitem[Cha81]{Cha81}
A.W. Chao, Nucl. Instr. Meth. {\bf 180}, p.29 (1981).
%
\bibitem[DK72]{DK72}
Ya.S. Derbenev, A.M. Kondratenko,
Sov. Phys. JETP {\bf 35}, p.230 (1972).
%
\bibitem[DK73]{DK73}
Ya.S. Derbenev, A.M. Kondratenko,
Sov. Phys. JETP {\bf 37}, p.968 (1973).
%
\bibitem[DK75]{DK75}
Ya.S. Derbenev, A.M. Kondratenko, Sov. Phys. Dokl. {\bf 19}, p.438 (1975).
%
\bibitem[DK78]{DK78}
Y.S. Derbenev, A.M. Kondratenko, in:
American Institute of Physics Conference Proceedings 51, Argonne (1978).
%
\bibitem[Gar85]{Gar85}
C.W. Gardiner, ``Handbook of stochastic methods for
physics, chemistry and the natural sciences'', Berlin (1985).
%
\bibitem[GS71]{GS71}
I.I. Gichman, A.W. Skorochod, ``Stochastische
Differentialgleichungen'', Berlin (1971).
%
\bibitem[Gol80]{Gol80}
H. Goldstein, ``Classical Mechanics'', Reading (1980).
%
\bibitem[GV88]{GV88}
J.M. Gracia-Bondia, J.C. Varilly, J. Phys. {\bf A21}, p.L879 (1988).
%
\bibitem[GV89]{GV89}
J.M. Gracia-Bondia, J.C. Varilly, Ann. Phys. {\bf 190}, p.107 (1989).
%
\bibitem[HH96]{HH96}
K. Heinemann, G.H. Hoffst\"atter, Phys. Rev. {\bf E54}, p.4240 (1996).
%
\bibitem[Hei96]{Hei96}
K. Heinemann, DESY {\bf 96-229} (1996).
%
\bibitem[Jow85]{Jow85}
J.M. Jowett, ``Introductory Statistical Mechanics for
 Electron Storage Rings'', in:
American Institute of Physics Conference Proceedings 153, Stanford (1985).
%
\bibitem[Kew83]{Kew83}
J. Kewisch, DESY {\bf 83-032} (1983).
%
\bibitem[Kou91]{Kou91}
J.P. Koutchouk, CERN Note {\bf SL}/{\bf AP}-{\bf 16} (1991).
%
\bibitem[Lig59]{Lig59}
M.J. Lighthill,``Introduction to Fourier analysis and generalised
           functions'', Cambridge (1959).
%
\bibitem[Man87]{Man87}
S.R. Mane: Phys. Rev. {\bf A36}, p.105, p.120 (1987).
%
\bibitem[Mon84]{Mon84}
B.W. Montague, Phys. Reports {\bf 113}, p.1 (1984).
%
\bibitem[Ris89]{Ris89}
H. Risken, ``The Fokker-Planck equation. Methods of solution and
                 applications'', Berlin (1989).
%
\bibitem[ST64]{ST64}
A.A. Sokolov, I.M. Ternov, Sov. Phys. Doklady {\bf 8}, p.1203 (1964).
%
%
\bibitem[Str57]{Str57}
R.L. Stratonovich, Sov. Phys. JETP {\bf 4}, p.891 (1957).
%
\bibitem[Tho27]{Tho27}
L.H. Thomas, Phil. Mag. {\bf 3}, p.1 (1927).
%
\bibitem[Van81]{Van81}
N.G. Van Kampen, ``Stochastic processes in Physics and Chemistry'',
Amsterdam (1981).
%
%
\bibitem[Yok86]{Yok86}
K. Yokoya, DESY {\bf 86-57} (1986).
%
\bibitem[Yok92]{Yok92}
K. Yokoya, KEK {\bf 92-6} (1992).
%
\end{thebibliography}
\end{document}